\documentclass[mnsc,nonblindrev]{informscopy} 
\newcommand{\majorrevcolor}[1]{{\color{black}#1}}

\usepackage{mathtools}
\usepackage[T1]{fontenc}
\usepackage[utf8]{inputenc}
\usepackage{lmodern}

\usepackage{comment}
\usepackage[font=small]{caption}

\DeclareSymbolFont{extraup}{U}{zavm}{m}{n}

\DeclareMathSymbol{\vardiamond}{\mathalpha}{extraup}{87}

\TheoremsNumberedThrough 
\EquationsNumberedThrough

\usepackage[breakable, theorems, skins]{tcolorbox}
\definecolor{cornellred}{rgb}{0.7, 0.11, 0.11}
\definecolor{dgreen}{rgb}{0.0, 0.7, 0.0}
\definecolor{ballblue}{rgb}{0.13, 0.67, 0.8}
\definecolor{royalblue(web)}{rgb}{0.25, 0.41, 0.88}
\definecolor{bleudefrance}{rgb}{0.19, 0.55, 0.91}
\definecolor{royalazure}{rgb}{0.0, 0.22, 0.66}
\definecolor{forestgreen}{HTML}{009B55}

\usepackage{hyperref}
\hypersetup{
	colorlinks = true,
	linkcolor=royalazure,
	citecolor=royalazure,
	urlcolor=black,
	linkbordercolor = {white}
}

\usepackage{algorithm, algorithmic}%
 \usepackage{caption}
\usepackage{enumitem}
\usepackage{subfiles}
\usepackage{bm}
\usepackage{xfrac}
\usepackage{bbm}
\usepackage{csquotes}
\usepackage{pst-all}
\usepackage{accents}
\usepackage{subcaption}
\usepackage{perpage}
\usepackage{footnote}
\usepackage{xcolor}

\makeatletter
\renewenvironment*{displayquote}
  {\begingroup\setlength{\leftmargini}{0cm}\csq@getcargs{\csq@bdquote{}{}}}
  {\csq@edquote\endgroup}
\makeatother
\DeclareRobustCommand{\mybox}[2][gray!15]{
\begin{tcolorbox}[ 
        colback=white,      
        colframe=gray,  
        boxrule=0.2pt,      
        arc=2pt,outer arc=2pt,
        left=12pt,
        right=12pt,
        top=5pt,
        bottom=5pt,
        width=1.07\linewidth,
        enlarge left by=-0.55cm,
        before upper=\renewcommand{\baselinestretch}{1.3}\selectfont,
        after upper=\normalfont
        ]
 #2
 \end{tcolorbox}
}

\usepackage{natbib}
 \bibpunct[, ]{(}{)}{,}{a}{}{,}%
 \def\bibfont{\small}%

\usepackage{thm-restate}

\setcitestyle{authoryear}

\newcommand{\Timeidx}{t}
\newcommand{\Tauidx}{\tau}
\newcommand{\TotalTime}{T}
\newcommand{\Type}{A}
\newcommand{\Arrival}{\mathbf{\Type}}
\newcommand{\ArrivalSupp}{\mathcal{\Type}}
\newcommand{\Weight}{w}
\newcommand{\Weightvec}{\mathbf{\Weight}}
\newcommand{\WeightEach}{\Weight_{\Timeidx, \Locidx}}
\newcommand{\Locidx}{i}
\newcommand{\Target}{i^\dagger}
\newcommand{\Locnum}{m}
\newcommand{\Simplex}{\Delta_\Locnum}
\newcommand{\OnlineALG}{\pi}
\newcommand{\Decision}{z}
\newcommand{\FeasibleSet}[1]{\mathcal{Z}({#1})}
\newcommand{\BasisVec}[1]{\mathbf{e}_{#1}}
\newcommand{\DecisionVec}{\mathbf{\Decision}}
\newcommand{\DecisionEach}{\Decision_{\Timeidx,  \Locidx}}
\newcommand{\Capacity}{c}
\newcommand{\CapacityVec}{\mathbf{\Capacity}}
\newcommand{\CapRatio}{\rho}
\newcommand{\CapRatioVec}{\boldsymbol{\CapRatio}}
\newcommand{\Indicator}{\mathbbm{1}}
\newcommand{\Service}{s}
\newcommand{\ServiceVec}{\mathbf{\Service}}
\newcommand{\ServiceSlack}{\epsilon}
\newcommand{\ServiceSlackVec}{\boldsymbol{\ServiceSlack}}
\newcommand{\ServiceEach}{\Service_{\Timeidx, \Locidx}}
\newcommand{\BuildUp}{b}
\newcommand{\BuildUpVec}{\mathbf{\BuildUp}}
\newcommand{\BuildUpti}[2]{\BuildUp_{{#1}, {#2}}}
\newcommand{\ServiceRate}{r}

\newcommand{\PositivePart}[1]{\left(#1\right)_{+}}
\newcommand{\History}{\mathcal{H}}
\newcommand{\OverCost}{\alpha}
\newcommand{\BuildUpCost}{\gamma}
\newcommand{\ALG}{\textup{\textsf{ALG}}}
\newcommand{\ALGObj}[3]{\textup{\textsf{ALG}}^{#1}({#2}, {#3})}
\newcommand{\ArrivalDist}{\mathcal{F}}
\newcommand{\OPT}{\textup{\textsf{OPT}}}

\newcommand{\Dual}{\textsf{Dual}}
\newcommand{\PerDual}{g}
\newcommand{\Regret}{\textup{\textsf{Regret}}_{\TotalTime}}
\newcommand{\DualOver}{\theta}
\newcommand{\DualOverVec}{\boldsymbol{\DualOver}}
\newcommand{\DualHard}{\lambda}
\newcommand{\DualHardVec}{\boldsymbol{\DualHard}}
\newcommand{\DualBuild}{\beta}
\newcommand{\DualBuildVec}{\boldsymbol{\DualBuild}}
\newcommand{\SurrogateDual}{\textup{\textsf{Dual}}^{\textup{\textsf{S}}}}
\newcommand{\SurrogatePrimal}{
{\textup{\textsf{Primal}}}^{\textup{\textsf{S}}}
}
\newcommand{\Norm}[2]{
\lVert{#1} \rVert_{#2}}
\newcommand{\Stopping}{T_A}

\newcommand{\SinglePrimal}{\textsf{P}^\star}
\newcommand{\SingleDual}{\textsf{D}^\star}
\newcommand{\MRate}{\mu}
\newcommand{\MRateVec}{\boldsymbol{\MRate}}
\newcommand{\NetReward}{\textup{\textsf{NMR}}}
\newcommand{\size}{n}
\newcommand{\allocation}{matching}
\newcommand{\Allocation}{Matching}

\newcommand{\backlog}{backlog}
\newcommand{\Backlog}{Backlog}

\newcommand{\CA}{\texttt{CA-DL}}
\newcommand{\CAM}{$\texttt{CA-DL}^{\texttt{M}}$}
\newcommand{\CO}{\texttt{CO-DL}}
\newcommand{\COM}{$\texttt{CO-DL}^{\texttt{M}}$}

\newcommand{\DP}{\textup{\textsf{DP}}}

\newcommand{\R}{\mathbb{R}}
\newcommand{\E}{\mathbb{E}}
\renewcommand{\P}{\mathbb{P}}

\DeclareMathOperator{\Var}{Var}

\newcommand{\BigO}{\mathcal{O}}

\definecolor{maroon}{rgb}{0.5, 0.0, 0.0}

\newcommand{\revcolor}[1]{{\color{black}#1}}

\usepackage[suppress]{color-edits}

\addauthor{VM}{black}
\addauthor{SL}{red} 
\addauthor{RN}{purple}
\addauthor{EP}{black}
\addauthor{KB}{black}
\addauthor{EC}{black} 
\addauthor{EAAMO}{black} 

\usepackage{booktabs} 
\usepackage{multirow} 
\usepackage{makecell} 
\usepackage{subcaption} 

\newcommand{\Cut}[1]{{#1}}
\newcommand{\EC}[1]{{#1}}
\newcommand{\ECKEEP}[1]{{#1}}

\newcommand{\ECCUT}[1]{\ignorespacesafterend}

\usepackage{color-edits}

\addauthor{VM}{purple}
\addauthor{SL}{red} 
\addauthor{RN}{green}
\addauthor{EP}{orange}
\addauthor{KB}{brown}
\addauthor{EC}{black} 

\usepackage{cleveref}
\OneAndAHalfSpacedXI

\begin{document}

\TITLE{Dynamic Matching with Post-allocation Service and its Application to Refugee Resettlement}

\RUNAUTHOR{Bansak et al.}

\RUNTITLE{Dynamic Matching with Post-allocation Service}

\ARTICLEAUTHORS{%
\AUTHOR{Kirk Bansak}
\AFF{University of California, Berkeley, CA, \EMAIL{kbansak@berkeley.edu}} 
\AUTHOR{Soonbong Lee}
\AFF{Yale School of Management, New Haven, CT, \EMAIL{soonbong.lee@yale.edu}}
\AUTHOR{Vahideh Manshadi}
\AFF{Yale School of Management, New Haven, CT, \EMAIL{vahideh.manshadi@yale.edu}}
\AUTHOR{Rad Niazadeh}
\AFF{University of Chicago Booth School of Business, Chicago, IL, \EMAIL{rad.niazadeh@chicagobooth.edu}}
\AUTHOR{Elisabeth Paulson}
\AFF{
 Harvard Business School, Boston, MA,
\EMAIL{epaulson@hbs.edu}
}
}

\ABSTRACT{
Motivated by our collaboration with a major refugee resettlement agency in the U.S., we study a dynamic matching problem where each new arrival (a refugee case) must be matched immediately and irrevocably to one of the static resources (a location with a fixed annual quota).
In addition to consuming the static resource, each case requires {\em post-allocation} service from a server, such as a translator. {Given the time-consuming nature of service,} a server may not be available at a given time, thus we refer to it as a {\em dynamic resource}. Upon matching, the case will wait to avail service in a first-come-first-serve manner. 
Bursty matching to a location may result in undesirable congestion at its corresponding server. 
Consequently, the central planner (the agency) faces a dynamic matching problem with an objective that combines the matching reward (captured by pair-specific employment outcomes) with the cost for congestion for dynamic resources and over-allocation for the static ones. 
Motivated by the observed fluctuations in the composition of refugee pools across the years, we design algorithms that do not rely on distributional knowledge {constructed based on past years' data}. To that end, we develop learning-based algorithms that are asymptotically optimal in certain regimes, easy to interpret, and computationally fast. Our design is based on learning the dual variables of the underlying optimization problem; however, the main challenge lies in the time-varying nature of the dual variables associated with dynamic resources. To overcome this challenge, our theoretical development brings together techniques from Lyapunov analysis, adversarial online learning, and stochastic optimization. 
On the application side, when tested on real data from our partner agency and \majorrevcolor{incorporating practical considerations}, our method outperforms existing ones making it a viable candidate for replacing the current practice upon experimentation.


}



\maketitle

\vspace{-0.5cm}
\section{Introduction}\label{sec:intro}

According to the United Nations High Commissioner for Refugees (UNHCR), the forcibly displaced population worldwide exceeded 123 million by the end of 2024. This number includes approximately 8 million asylum seekers and more than 37 million refugees \citep{UNHCR-stat}. The process of providing assistance to these individuals often involves making high-stakes decisions in the face of complex operational intricacies. Data-driven and algorithmic approaches that capture the unique features of these processes, yet yield easy-to-implement solutions, can significantly enhance these operations and improve the well-being of these vulnerable communities. In this paper, we demonstrate this potential within the context of refugee resettlement in the United States.



This work is motivated by our collaboration with a major refugee resettlement agency in the U.S. Refugee resettlement, recognized as a durable solution to the global refugee crisis, is a process largely overseen by the UNHCR. In this process, refugees are relocated to participating host countries and granted long-term or permanent residence. The U.S. resettles tens of thousands of refugees each year---the largest number among all host countries~\citep{UNHCR-USresettlement}---through ten national non-profit resettlement agencies, one of which is our partner. In what follows, we first provide background on the resettlement process, highlighting key features that motivate our modeling framework and research questions. We then provide an overview of our contributions, followed by a discussion of related work. 


\subsection{Background on Refugee Resettlement}\label{subsec:intro+background}
In the following, we briefly describe the refugee resettlement process in the U.S., which is relevant to our partner agency. We note that while exact details vary by country, refugee resettlement and/or asylum procedures of many host countries (e.g., Switzerland, Netherlands, Sweden, etc.) share important commonalities with the process described below.

After the UNHCR allocates a \emph{refugee case}\footnote{\majorrevcolor{A case typically includes multiple individuals, i.e., members of a family. For simplicity of mathematical exposition, we mostly consider cases with ``size one.'' We incorporate this and other practical considerations later in the paper.}} to the U.S. for resettlement, the case is vetted with the help of the federal government\footnote{In particular, three federal agencies: (i) the Department of Homeland Security's office of U.S. Citizenship and Immigration Services (USCIS), (ii) the U.S. Department of State's Bureau of Population, Refugees and Migration, and (iii) the U.S. Department of Health and Human Services.} to ensure that the refugee case is eligible for resettlement in the U.S.  and meets all security and legal criteria before admission and proceeding with the resettlement process. Upon admission to the U.S., the case is handed to one of the national resettlement agencies. The corresponding agency then assigns the case to a local service provider (often referred to as an {\em affiliate}) within its network. These affiliates typically provide job search and vocational services, as well as assistance in obtaining financial literacy and access to community resources.

Hereafter, for consistency, we refer to this assignment as a \emph{matching}. Importantly, there is strong evidence that the initial matching significantly impacts finding employment within 90 days after resettlement~\citep{bansak2018improving}, which is the key integration metric tracked and reported to Congress in the U.S. Additionally, recent work has shown that employment outcomes for any case-affiliate pair can be reasonably predicted using machine learning (ML) models~\citep{bansak2018improving, bansak2024learning}. Equipped with these ML models, a resettlement agency can use the predictions to inform matching decisions.


In using ML predictions to find better matches between refugees and affiliates, the resettlement agency faces several operational considerations:
\begin{enumerate}[label=(\roman*)]
    \item If a refugee has {\em U.S. ties} (such as family or close friends) in a particular locality, the agency is required to match the case to the affiliate associated with that locality~\citep{bruno2017reception}.

    \item Each affiliate has a target annual {\em quota} for the number of refugees it receives in the matching, which is approved by the U.S.  Department of State with input from the resettlement agencies.  
\majorrevcolor{
These annual quotas serve as soft upper-bounds on the number of refugees to be resettled in a given year.
} \ECedit{In light of this, the relevant decision-making horizon for the agency is one year}.

    \item \majorrevcolor{
When new cases (physically) arrive at matched affiliates, they must receive short-term onboarding services. For example, Global Refuge, a leading U.S. resettlement agency, notes that \emph{“case managers help new arrivals navigate everything from enrolling in English classes and schools to securing jobs and learning how to use public transportation.”}~\citep{globalrefuge2024} These services require a nontrivial time investment, and each affiliate has a limited number of case managers. Furthermore,
resettlement agencies are expected to provide these services in specific time frames after a case's arrival, and they must report on the whether these standards were met. As a result, bursty placement of refugees at a single affiliate can overburden case workers and make it challenging to complete the required onboarding services as required.
\footnote{\label{footnote:post-allocation-service}\majorrevcolor{Our first-hand communication with the partner agency indicates that avoiding such congestion is a first-order concern: placement officers, though unsystematically, often have an eye toward balancing workloads across affiliates. Beyond our partner, congestion has been a significant operational challenge in other resettlement programs as well. For instance, in Switzerland, localities have at times suspended new placements due to temporary overloads in reception infrastructure~\citep{swiss2024asylverteilung}.}}}

\item Finally, cases are admitted to the U.S. and handed to agencies  over time. As such, the agency must make matching decisions {\em upon arrival} and without foreknowledge of future arrivals.
\end{enumerate}

The lack of information about future arrivals presents a major challenge in optimizing matching decisions. 
\ECedit{One approach to addressing this challenge is to assume that the current year's arrival pool resembles that of previous years, and use \EPedit{that} as ``distributional knowledge'' to simulate future arrivals.} 
As we discuss later, several recent papers adopt such an approach; see, for example, the work of \citet{bansak2024outcome} and \citet{ahani2023dynamic}. Although natural, this approach falters if the pool's composition changes across years---a phenomenon observed in reality.

\begin{figure}[t]
        \centering
\includegraphics[scale=0.45]{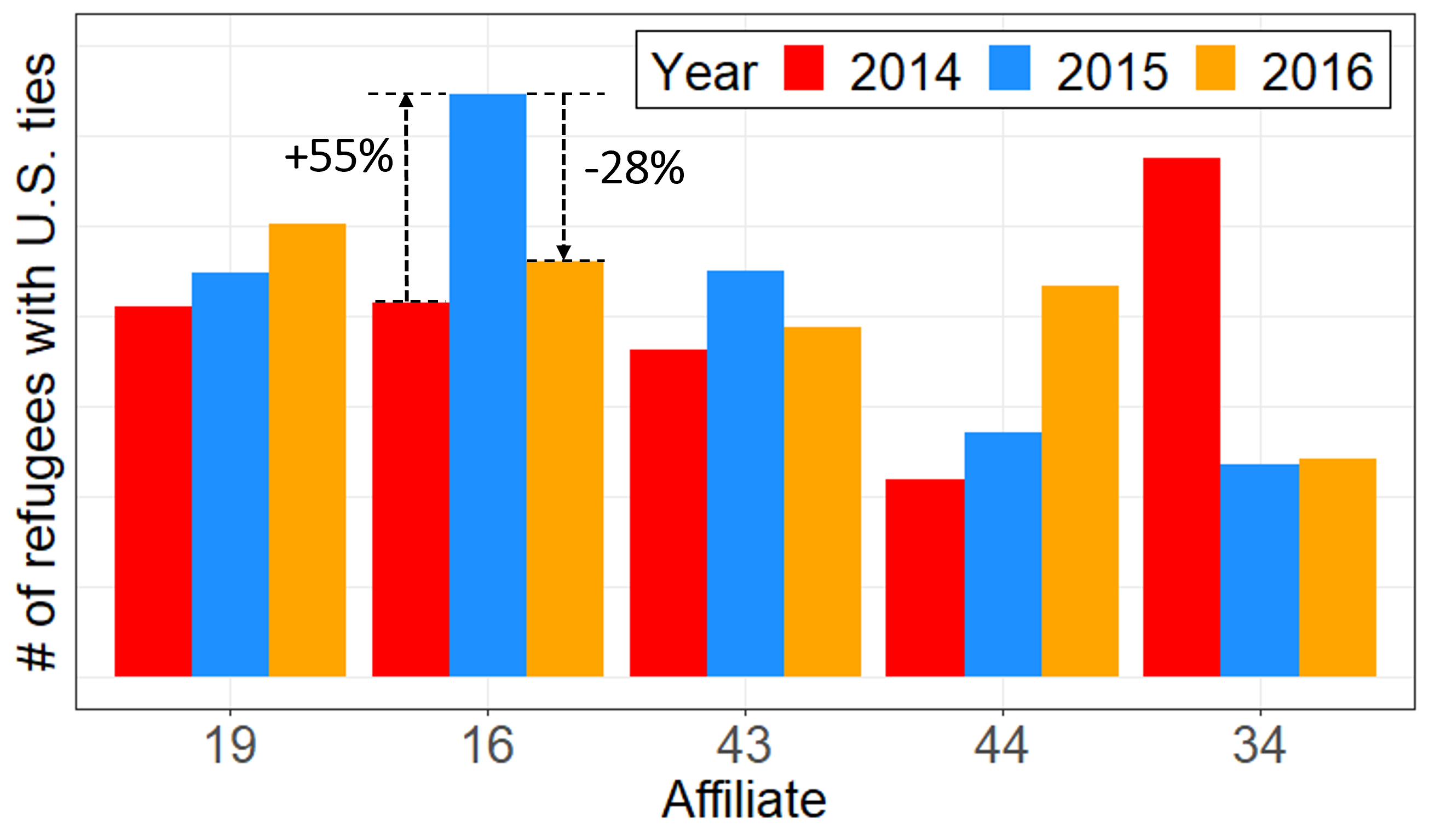}
        \caption{The number of refugees with U.S. ties at five affiliates normalized by the total number of arrivals. }
    \label{fig:tied+proportion}
    \end{figure}

To illustrate this, in Figure~\ref{fig:tied+proportion} we focus on five selected affiliates of our partner agency. For each affiliate, we plot the (normalized) number of cases with U.S. ties to this affiliate over the years 2014-2016. We observe a substantial fluctuation in the number of such cases over these years; for example, the portion of cases tied to affiliate $16$ increased by $55\%$ in 2015 compared to 2014. This fluctuation can significantly impact matching decisions. As a toy example, consider this affiliate $16$ and suppose that we are making decisions in 2015. If we use data from 2014 to simulate the number of cases with U.S. ties to this affiliate, we substantially underestimate this number. Given that these cases can only be matched to affiliate $16$, we may reach the affiliate's quota well before the end of the year.\footnote{
\label{footnote:within+year+variation}\ECedit{In \Cref{apx+within+year+variations}, we further elaborate on the distinction between within-year and across-year variation and highlight how this distinction motivates our distribution-free approach to algorithm design for this application.}} 
    
The above background and observations motivate our main research question: 
\begin{displayquote}
\mybox{\em{
{How can we design a dynamic matching algorithm that optimizes employment outcomes without relying on past years' data, while  respecting affiliates' quotas and {managing congestion for services}?}}}
\end{displayquote}


\subsection{Our Contributions}\label{subsec:contribution}
Motivated by the above question, we introduce a framework for dynamic matching with post-allocation service, and develop two learning-based algorithms that do not rely on prior distributional knowledge, such as historical data. Under the assumption that the arrival sequence throughout the year is independent and identically distributed (i.i.d.) from an \emph{unknown} distribution, 
we show that both algorithms achieve asymptotic optimality in certain regimes. In addition, using data from our partner agency, we conduct a case study demonstrating that our method outperforms existing approaches used by our partner agency or proposed in the literature for similar dynamic refugee matching problems. 

\smallskip
{\bf A Model for Dynamic Matching with Post-allocation Service (Section \ref{sec:model}):}
Following our motivating application, we use the terminology of refugee resettlement to introduce our model. We consider a discrete-time, finite-horizon model (e.g., one year) where exactly one refugee case arrives in each period.\footnote{\label{footnote:batching} \ECedit{In some contexts, an agency may encounter batches of arrivals. In \Cref{apx+batching}, we discuss modifications to our framework and algorithm to handle batched arrivals, and numerically explore their impact in our case study.}}
Each case is either ``tied'' to a specific target affiliate\footnote{\EAAMOedit{Since the resettlement agency is mandated to settle refugees with U.S. ties at \emph{the most proximate} affiliate to theirties~\citep{bansak2018improving}, we assume that each tied case has exactly one target affiliate.}} (reflecting the U.S. ties described in \Cref{subsec:intro+background}) or ``free,'' meaning it can be matched to any affiliate. Upon arrival, the agency also observes the \emph{rewards} of matching the case to each affiliate, representing the employment probabilities predicted by the ML models. Given these observed rewards and knowing whether the case is free or tied, a dynamic matching algorithm must irrevocably match the arriving case to an affiliate (or leave it unmatched),\footnote{{In practice, all cases are matched. In our theoretical model, however, free cases may go  unmatched, while we match all tied cases to their target affiliates. In our case study, aligned with practice, we require \emph{all cases} to be matched.
}} while considering the resource  limitations of the affiliates, which we describe next.

We model each affiliate as a \emph{static} resource with fixed capacity, representing its annual quota. Each case uses one unit of the static resource.\footnote{
\majorrevcolor{Our model can incorporate multiple-unit consumption (e.g., when a case is a family of multiple individuals, as alluded to earlier) and multiple types of static resources (e.g., school enrollment slots or temporary housing); see \Cref{subsec:model+assumption}.}}  
The most novel aspect of our model is the introduction of \emph{post-allocation} service, capturing the need for service \emph{after matching}, as observed in applications such as refugee resettlement. To model this, we endow each affiliate with a server, also referred to as a \emph{dynamic resource}, whose availability follows an i.i.d. Bernoulli process with a given service rate.\footnote{As we discuss in \Cref{sec:model}, this model is equivalent to having geometric service times with the same service rate.} 
Upon becoming available, the server serves a matched case  
(if any) on a first-come-first-served (FCFS) basis. We measure server congestion via its {\em backlogs} throughout the horizon, representing the number of matched cases awaiting service at the affiliate in each period. The described service process resembles a classical queueing system, with one crucial distinction: the arrival rate is {\em endogenously} determined by matching decisions. Despite this distinction, a congested system is still undesirable.

Given the above resource limitations, the agency pursues multiple objectives: maximizing the total matching reward while avoiding excessive congestion for dynamic resources and over-allocation for static resources (i.e., exceeding an affiliate's annual quota).\EC{\footnote{Due to the presence of tied cases, some over-allocation may be unavoidable (see \Cref{sec:model}).}}
To capture these objectives simultaneously, we define an objective function consisting of the total reward penalized by time-averaged backlog (dynamic resource) and total over-allocation (static resource); see \eqref{ALG+obj}. Motivated by significant year-to-year fluctuations in the refugee arrival pool (\Cref{fig:tied+proportion}), we study this dynamic matching problem with \emph{no knowledge} of the underlying arrival distribution, though we assume that the arrival sequence is i.i.d.\ from an unknown distribution---and thus potentially learnable over the horizon.

To evaluate the performance of any dynamic matching algorithm, we compare it to a properly defined benchmark (see \Cref{def:offline}) and introduce the notion of regret relative to such a benchmark. \majorrevcolor{Within this framework, in Sections \ref{sec:alg+design} and \ref{subsec:alg+surrogate+primal}, we design two algorithms with sub-linear regrets in the horizon length---which are asymptotically optimal or near-optimal in almost all regimes (as established in \Cref{apx:regret+lower})}. Importantly, our regret bounds hold when the congestion penalty parameter  is ``small enough,'' meaning that it grows sub-linearly with the horizon. We complement these positive results with an impossibility result (\Cref{prop:impossibility}), showing that focusing on this regime is indeed a fundamental requirement: no algorithm can achieve sub-linear regret outside of this regime.


Before providing an overview of our algorithms, we note that, in the absence of dynamic resources, the above problem reduces to online resource allocation with static resources---a topic extensively explored in the literature (see \Cref{subsec:lit+review} 
for further discussion). However, introducing congestion for dynamic resources changes the nature of the problem (see \Cref{prop:offline} and related discussions). Consequently, designing and analyzing algorithms requires new ideas and techniques. In the following, we highlight some of these main challenges and new ideas needed to address them.

\smallskip
{\bf Design and Analysis of Algorithms (Sections \ref{sec:alg+design} \& \ref{subsec:alg+surrogate+primal}):}\label{subsec:intro+contribution}
The first step in designing our algorithm is to consider the omniscient offline benchmark for the problem---that is, the matching that maximizes our objective function, knowing the entire sample path of arrivals and server availabilities in advance. In \Cref{def:offline}, we formulate this optimization problem as a simple convex program. In particular, the program includes time-dependent constraints, one for each affiliate at each time period, to track the  backlog dynamics (see equation \eqref{const:build-up}), as well as a simple capacity constraint for each static resource. Consequently, the corresponding dual program has the time-varying dual variables for dynamic resources, in contrast to the time-invariant dual variables for static ones (see \Cref{prop:offline}). Had we known the ``optimal'' values of these dual variables, we could use them and directly design a \emph{score-based} optimal algorithm for the primal problem that simply matches each arriving case to the affiliate with the maximum dual-adjusted reward. In the absence of this knowledge, we aim to \emph{learn} these dual variables while making matching decisions online.

The dynamic nature of backlogs, which leads to time-varying dual variables, presents fundamental challenges in designing and analyzing learning-based algorithms, and marks a major departure from the literature on online resource allocation (see \Cref{subsec:lit+review} 
for further discussions). In our first algorithm (\Cref{ALG+Surrogate+D}), we address these challenges by introducing a \emph{surrogate dual} problem (\Cref{def:dual+surrogate}), which imposes a time-invariance constraint on dual variables associated with dynamic resources. This algorithm maintains estimates of the solution to the surrogate dual problem and makes matching decisions based on the dual-adjusted scores. The key ingredient is then to properly update these dual variables to learn the optimal solution. For dual variables associated with static resources, we rely on classical techniques from online adversarial learning; however, for those associated with dynamic resources, we exploit a crucial structural property linking a specific learning rule to backlog dynamics. The resulting adjustment has an interpretable form: we penalize each affiliate's reward by \emph{a scaled version of its current backlog} (see equation~\eqref{eq:beta+dynamics} and related discussions).

As our first technical contribution, we prove that this algorithm achieves a sub-linear regret in all regimes of interest (see \Cref{thm:ALG+Surrogate+Dual} and \Cref{near+part+a}). To establish this result, we combine techniques from adversarial online learning and the drift-plus-penalty method \citep{tassiulas1990stability,neely2006energy, neely2008fairness, neely2022stochastic}, defining an appropriate potential function and pseudo-rewards to analyze regret (see \Cref{subsec:ALG+D+pfsketch}). This novel integration of the two frameworks\ECKEEP{, which to the best of our knowledge is the first of its kind,} enables us to obtain a simultaneous guarantee on the algorithm's total matching reward (\Cref{lemma:ALG+D+reward}) and the average backlog in expectation (\Cref{Lemma+ALG+D+Build}). \majorrevcolor{We complement our results by establishing matching regret lower bounds for any online algorithm, demonstrating the asymptotic optimality of the regret upper bounds achieved by our algorithm in nearly all parameter regimes (see \Cref{apx:regret+lower} for technical details).}


We call our first algorithm ``\emph{congestion-aware} dual learning'' (\CA{}) because it requires exact information on the current backlog. Although this requirement is not a practical concern for our partner agency (see \Cref{sec:numerics}), such information may not be readily available in other contexts. For example, some agencies may only have access to unreliable or delayed backlog information, rendering it unsuitable for matching decisions. Motivated by this limitation, we develop a second algorithm called ``\emph{congestion-oblivious} dual-learning'' (\CO{}), which learns the dual variables without relying on knowledge of the current backlogs. This algorithm is based on a \emph{surrogate primal} problem (\Cref{def:primal+surrogate}), which disregards the congestion penalty in its objective. We design a dual-adjusted score-based algorithm (\Cref{ALG+Surrogate+P}) for this problem, similar to \CA{},  but with subtle differences in the learning process, including the use of a \emph{time-varying learning rate}, to control the drift of the backlog.


Somewhat surprisingly, we show that even though \CO{} does use backlog information, it still achieves a sub-linear regret under most regimes, subject to mild regularity assumptions (\Cref{thm:ALG+Surrogate+Primal}). The main technical challenge is analyzing \CO{}'s backlog (\Cref{Lemma+ALG+P+Build}), which in turn requires studying the ``endogenous'' arrival rates induced by  this algorithm. We overcome this challenge by utilizing a high-probability last-iterate convergence property of the dual variables constructed by $\CO{}$ (\Cref{prop:harvey}). Specifically, the dual learner employed by \CO{} is a particular variant of online stochastic mirror descent (OSMD) with an appropriately chosen \emph{time-varying} learning rate. Although our proof involves several intricate steps, it crucially exploits the time-varying learning rate, and adapts a recent result on high-probability last-iterate convergence of the stochastic gradient descent from \citet{harvey2019tight} to our OSMD variant (see \Cref{subsec:ALG+P+static+dual} for technical details).
While both \texttt{CA-DL} and \texttt{CO-DL} achieve sub-linear regret in many regimes, we also show that under certain conditions (specifically, slow service rates and certain ranges of penalty parameters), \texttt{CO-DL} cannot achieve a sub-linear regret (\Cref{thm:near+critical}) unlike \texttt{CA-DL}, implying the inherent advantage of explicitly accounting for the backlog.

Finally, we note that our algorithms are computationally fast, and their score-based structure makes them easy to communicate to practitioners (including the managers of our partner agency). Additionally, as we discuss next,  our learning-based method outperforms existing approaches (relying on past years data) in the context of our collaboration, when tested on actual data from our partner agency.


\smallskip
{\bf Case Study on Refugee Resettlement Data (\Cref{sec:numerics}):} To demonstrate the practical effectiveness of our learning-based approach, we conduct a case study using data from our partner agency. In \Cref{subsec:data}, we detail the construction of primitives and explain how to adapt our learning-based approach to this specific context to comply with practical considerations of our partner agency. In \Cref{subsec:results}, we show that, had the agency used our proposed algorithm on the \emph{actual} sequence of arrivals, it would have led to substantial improvements in terms of predicted employment outcomes and average backlog. We further compare our proposal with another recent proposal based on \citet{bansak2024outcome}, which relies on data from past year to predict future arrivals. Compared to this alternative proposal, we show that ours improves the objective function (combining total employment, backlog, and over-allocation) across a broad range of penalty parameters without negatively impacting any of the three outcomes (see \Cref{fig:performance} and \Cref{fig:more+case}). These promising numerical results, together with the practical advantages of our method discussed earlier, position our method as a strong contender for experimentation to replace the current system.

Beyond refugee resettlement, post-allocation service may arise in assignment problems in other contexts, such as foster care and healthcare, where individuals not only consume long-term resources with fixed capacities (e.g., beds), but also require time-consuming on-boarding services (e.g., initial screenings). In these contexts, undesirable backlog may emerge due to shortages of dynamic resources\majorrevcolor{---see \Cref{sec:discussions} for a broader discussion on potential applications of our model.} 
Our framework offers managers flexibility to balance allocation rewards and resource costs (both static and dynamic)  to varying degrees. Moreover, the computational efficiency of our algorithms allows policymakers to understand the trade-offs between these objectives by experimenting with different penalty parameters.

{
\subsection{Related Literature}\label{subsec:lit+review}
Our work relates and contributes to the literature on refugee matching, online resource allocation,
and queueing system control. Below, we highlight
the most closely related work, referring the readers to \Cref{subsec:lit+review+further} for a detailed literature review.

In refugee matching, the most relevant comparison is with \citet{bansak2024outcome}, who also consider an objective that penalizes congestion but do not consider tied cases---and thus over-allocation is not included in their objective. We adapt their proposed sampling-based algorithm (relying on data from the past year) to handle tied cases in our case study (\Cref{sec:numerics}), and show that our prior-free algorithms substantially outperform their approach, thanks to its robustness. Unlike previous work including \citet{bansak2024outcome}, we also show theoretical performance guarantees of our method.

Methodologically, our study is closely related to \citet{agrawal2014fast} and \citet{balseiro2023best}, who examine online resource allocation with objectives depending only on total allocation. In this special case, the dual problem only has time-invariant variables, which can be learned fast over time to ensure vanishing regret. However, this is not the case in our setting with the average backlog in the objective function, as we have time-varying dual variables associated with backlog dynamics that cannot be learned fast using similar methods. This critical difference necessitates new analytical techniques beyond these papers (and similar papers in the queuing literature). Specifically, our analysis of \CA{} integrates drift-plus-penalty methods and adversarial online learning (\Cref{subsec:ALG+D+pfsketch}, \Cref{apx:lyapunov}), which is the first of its kinds to the best of our knowledge. For analyzing \CO{}, we exploit high-probability last-iterate convergence of stochastic mirror descent (\Cref{prop:harvey}), an aspect less explored in prior literature on online resource allocation. 

\section{Model and Preliminaries}\label{sec:model}

\revcolor{We start by formally introducing our model for dynamic refugee matching with unknown i.i.d.\ arrivals and post-allocation service. Specifically, we describe two types of resource limitations in this setting, which we refer to as static and dynamic resource limitations. We then present our objective function combining employment outcomes, over-allocation, and congestion. Next, we define an offline benchmark based on this objective, along with the notion of regret used for performance evaluation.  \majorrevcolor{A summary of the main notation and key modeling assumptions is provided in \Cref{apx+notation}.}



\smallskip
{\bf Notation:} We use $[m]$ to denote $\{1, 2,..., m\}$ for any $m\in\mathbb{N}$. We use bold cases to denote vectors. The non-negative orthant in the $m$-dimensional Euclidean space is denoted by $\R_{+}^m$ , with $\BasisVec{\Locidx} \in \R^m_{+}$ denoting the standard basis vector at coordinate $\Locidx$. We use $\Simplex := \{\mathbf{z} \in \R^m_{+}: \sum_{\Locidx=1}^m z_i \leq 1\}$ to denote the $m$-dimensional standard simplex. The positive part of $x\in\mathbb{R}$ is denoted by  $\PositivePart{x}$, and an indicator function is denoted by $\Indicator[\cdot]$. Finally, we adopt the standard asymptotic notation. For functions $f,g:\R \to \R$, we write $f(x) = \BigO\left(g(x)\right)$ (resp. $f = \Omega\left(g(x)\right)$) if $\lvert f(x)\rvert$ is upper-bounded (resp. lower-bounded) by a positive constant multiple of $\lvert g(x)\rvert $ for all sufficiently large values of $x$. If $f(x)=\BigO\left(g(x)\right)$ and $f(x)=\Omega\left(g(x)\right)$, we write it as $f(x)=\Theta\left(g(x)\right)$. Similarly, we write $f(x) = o\left(g(x)\right)$ if $\tfrac{f(x)}{g(x)} \to 0$ as $x \to \infty$. 
Finally, throughout, a ``constant'' refers to any scalar independent of $\TotalTime$.}



 {\bf Problem Setup:} The problem consists of $\Locnum$ affiliates and an arrival sequence of $\TotalTime$ refugee cases, where case $t$ arrives at the beginning of period $t$. Upon arrival of case $\Timeidx$, the agency observes its type denoted by $\Arrival_\Timeidx := (\Weightvec_\Timeidx, \Target_\Timeidx)$, which we assume is an i.i.d. random variable drawn from an \emph{unknown} distribution $\mathcal{F}$. The first component of the type is a vector  $\Weightvec_\Timeidx = (\WeightEach)_{\Locidx \in [\Locnum]}$, where  $\WeightEach$ is a case-affiliate pairwise \emph{reward} from matching case $t$ to affiliate $i$. \revcolor{This reward $\WeightEach$ can be equivalently thought of as the predicted employment outcome if case $t$ is matched to affiliate $i$.} Without loss of generality, we assume $\WeightEach\in[0,1]$. The second component of the type is a \emph{target affiliate} 
denoted by $\Target_\Timeidx$, which is a predetermined affiliate to which the case must be matched. If $\Target_\Timeidx \in [m]$, arrival $\Timeidx$ is a \emph{tied} case targeted to affiliate $\Target_\Timeidx$. In contrast, we use $\Target_\Timeidx=0$ to denote a \emph{free} case, which does not have any target affiliate.

Upon observing $\Arrival_\Timeidx$, the agency makes an irrevocable matching decision denoted by $\DecisionVec_\Timeidx = (\DecisionEach)_{\Locidx\in[\Locnum]}$.
To succinctly represent the feasibility set for both tied and free cases, for a given target affiliate $\Target$, we define a \emph{type-feasibility} set $\FeasibleSet{\Target}$ as $\FeasibleSet{\Target} = \Simplex$ if $\Target = 0$ and $\FeasibleSet{\Target} =\{\BasisVec{\Target}\}$ otherwise. 
Some comments are in order. First, while the type-feasibility set allows for fractional allocations, our algorithms always make \revcolor{integral decisions}. Second, although inaction is not permitted for tied cases, we allow it for free cases for technical reasons. {(In our case study in \Cref{sec:numerics}, consistent with practice, we do not allow inaction and ensure that every refugee is matched to an actual affiliate.) 
For ease of exposition, we interpret inaction as matching the case to a dummy affiliate with zero reward and unlimited resources. 

\smallskip
\textbf{Resource \revcolor{Limitations}}: \revcolor{Each affiliate is endowed with two types of resources: static and dynamic.\ECKEEP{ In the following, we elaborate on each type of resources separately,  and formally explain their associated constraints and their corresponding penalty terms in the objective function of the agency.}}

\EC{\smallskip}
\noindent{\emph{(i) Static Resource (Capacity)}}: 
Each affiliate $\Locidx$ is endowed with capacity $\Capacity_\Locidx$, referred to as a \emph{static} resource, which represents its annual quota. Matching a case to an affiliate consumes one unit of its static resource. 
We further define $\CapRatio_\Locidx= {\Capacity_\Locidx}/{\TotalTime}$, referred to as the \emph{capacity ratio}. We use $\underline{\CapRatio}= \min_{\Locidx\in[\Locnum]}\CapRatio_\Locidx$ to denote the minimum capacity ratio and impose two mild assumptions on these ratios: (i)
$\sum_{\Locidx=1}^\Locnum \CapRatio_\Locidx \leq 1$\footnote{\revcolor{This assumption is merely for ease of exposition and without loss of generality, and all of our main results extend to problem instances with arbitrary capacities.}} and ({ii}) there exists a constant $d>0$ such that $\CapRatio_\Locidx - \P[\Target_\Timeidx = \Locidx]\geq d $ for all $\Locidx \in [\Locnum]$, meaning that, in expectation, each affiliate has capacity $\Theta(T)$ for free cases.\footnote{
\label{footnote:constant+rho}
Note that this implies that $\boldsymbol{\CapRatio}$ is also a constant. In practice, \revcolor{capacities are determined by first creating a proportionality index across affiliates} and hence scale with $\TotalTime$ by construction.}

The agency aims to respect the endowed capacities as much as possible. Without tied cases, we can use a standard packing constraint to ensure $\sum_{\Timeidx=1}^\TotalTime \DecisionEach \leq \Capacity_\Locidx$ for all $\Locidx \in [\Locnum]$. However, due to uncertainty in the number of tied cases, such a constraint is overly stringent, as tied cases can cause over-allocation. 
To address this, we introduce two constraints in our model. First, we introduce a (weaker) hard constraint to capture that, at any time period, over-allocation can occur only due to the matching of tied cases. Formally, for any $\Timeidx \in [\TotalTime]$ and $\Locidx \in [\Locnum]$, any feasible matching decision must satisfy:
\begin{equation}
    \sum_{\Tauidx=1}^\Timeidx \Indicator[\Target_\Tauidx = 0]\Decision_{
    \Tauidx, \Locidx
    } \leq \Big(\Capacity_\Locidx - \sum_{\Tauidx=1}^\Timeidx \Indicator[\Target_\Tauidx = \Locidx]\Decision_{
    \Tauidx, \Locidx}\Big)_{+}, \quad \forall \Locidx \in [\Locnum]. \tag{\textsf{Capacity Feasibility-$\Timeidx$}}
    \label{hard+constraint:t}
\end{equation}
Because $\mathbf{0} \in \mathcal{Z}({0})$ by our earlier assumption, this constraint can always be satisfied for any arrival sequence. Moreover, it reduces to the aforementioned standard packing constraint for affiliates that do not face tied cases. Second, we introduce a soft constraint through a penalty $\OverCost\geq 0$ per unit of over-allocation. Formally, for each affiliate $\Locidx$, the agency incurs an over-allocation cost given by
$\OverCost \Big(\sum_{\Timeidx=1}^\TotalTime \DecisionEach - \Capacity_\Locidx \Big)_{+}$. 
Here, $\OverCost$ is a penalty parameter whose magnitude represents the agency's tolerance or physical cost of exceeding capacity.\footnote{For instance, the State Department provides per-capita funding to resettlement agencies in proportion to the approved capacity \citep{PRM+capacity}. Hence, the over-allocation penalty parameter $\OverCost$ directly captures the burden of securing additional funding per refugee case beyond the agency's initial budget. \label{footnote:over-allocation}} 

  Throughout the paper, we often use the (total) \emph{net \allocation{} reward} to refer to the total reward minus the over-allocation cost, that is, $\sum_{\Timeidx=1}^\TotalTime \sum_{\Locidx=1}^\Locnum \WeightEach\DecisionEach - \OverCost\sum_{\Locidx=1}^{\Locnum} \Big(\sum_{\Timeidx=1}^\TotalTime \DecisionEach - \Capacity_\Locidx \Big)_{+}$. It is important to highlight that, with knowledge of the entire arrival sequence in advance, we would never over-allocate unless the total number of tied cases for an affiliate exceeds its capacity.


\smallskip
\noindent{\emph{ (ii) Dynamic Resource (Server):}} Each case further requires \emph{post-allocation} service. To model this service process, we endow each affiliate with a \emph{server}. Upon matching, the case will wait to receive the service. Each server is a \emph{dynamic} resource, as its availability changes stochastically over time. In particular, we assume that the service availability of affiliate $\Locidx$ follows an i.i.d.\ Bernoulli process with success probability $\ServiceRate_\Locidx\in (0,1)$, denoted by $s_{t,i}\sim \texttt{Ber}(\ServiceRate_\Locidx)$. We also refer to $\ServiceRate_\Locidx$ as the \emph{service rate}.
At the end of period $\Timeidx$, if the server is available ($s_{t,i} = 1$), it processes one waiting case (if any) in a FCFS manner. This outlined process is equivalent to a (more conventional) service model with random service times,  where each service duration at affiliate $\Locidx$ is independently drawn from a geometric distribution with mean $1/\ServiceRate_\Locidx$. 
During each period, if the server is available, it initiates service for a new case and remains unavailable until that service is completed. If the server becomes available but no case is waiting, we assume it begins serving an ``outside" task (e.g., external cases or other operational task within the agency) with the same service time distribution. Consequently, the generated service token does not carry over to the next time period when a new case arrives.\footnote{\label{footnote:iid+service}\majorrevcolor{For instance, case workers at our partner agency may engage in administrative tasks or assist other teams when not actively managing refugee cases. 
In other contexts, however, service staff may remain idle once they become available but find no waiting cases, allowing them to serve the next arrival immediately. This alternative service model is indeed more relaxed compared to our base model, allowing more cases to be served in every sample path. In \Cref{apx+server+idleness}, we describe how our model and analysis can be modified to obtain similar results in such a setting. In particular, see \Cref{sec:alternative-model} for this new model. We then show that our algorithms obtain \emph{exactly} the same performance guarantees (\Cref{subsec:upper+idle}), and this setting, in some sense, does not make the problem any easier (\Cref{sec:alternative-lower-bound}).} 
}

As motivated in the introduction, server congestion is undesirable. To formalize this notion, we define the \emph{\backlog{}}, denoted by $\BuildUpVec_\Timeidx = (\BuildUpti{\Timeidx}{\Locidx})_{\Locidx \in [\Locnum]}$, where $\BuildUpti{\Timeidx}{\Locidx}$ represents the number of cases waiting for service at affiliate $\Locidx$ at the end of period $\Timeidx$. Formally, the \backlog{} process evolves recursively as:
\begin{equation}
\forall t\in[T]:~~~~\BuildUpti{\Timeidx}{\Locidx}= 
(\BuildUpti{\Timeidx - 1}{\Locidx}  + \DecisionEach -  \ServiceEach)_{+} 
\label{const:build-up}
\end{equation}
with initial \backlog{} set to zero ($\BuildUpVec_{0} = \mathbf{0}$). The \backlog{} grows large when an affiliate becomes ``congested'' due to bursty matching patterns. Therefore, we use the time-average \backlog{}, that is $\frac{1}{\TotalTime}\sum_{\Timeidx=1}^\TotalTime \sum_{\Locidx=1}^\Locnum \BuildUpti{\Timeidx}{\Locidx}$, as our overall measure of congestion. We further assume that the agency incurs a cost $\BuildUpCost\geq 0$ for each unit of the (time-)average \backlog{}. Hence, at the end of the horizon, the agency pays a total congestion cost given by $\frac{\BuildUpCost}{\TotalTime}\sum_{\Timeidx=1}^\TotalTime \sum_{\Locidx=1}^\Locnum \BuildUpti{\Timeidx}{\Locidx}$. Similar to the over-allocation penalty parameter $\OverCost$, the congestion penalty parameter $\BuildUpCost$ represents the agency's tolerance for congestion.


We assume that the service rate satisfies $\ServiceRate_\Locidx \geq  \CapRatio_\Locidx + \ServiceSlack$ for all $\Locidx \in [\Locnum]$, where $\ServiceSlack \geq 0$ is referred to as the \emph{service slack}.\footnote{Since $r_i < 1$, we implicitly focus on service slack values \( \epsilon < 1 - \rho_i \) for all \( i \in [m] \).}  Assuming $\ServiceRate_{\Locidx} \geq \CapRatio_{\Locidx}$ is well-motivated both technically and practically. From a technical perspective, this assumption comes from the following ``stability sanity check'': suppose that we match all cases to affiliates without over-allocation. Then, the total number of ``arrivals'' into affiliate $\Locidx$ is $\CapRatio_\Locidx \TotalTime$. To have no backlog at the end of the horizon in expectation, we must have at least $\CapRatio_\Locidx \TotalTime$ periods when the server is available, requiring $\ServiceRate_\Locidx \geq \CapRatio_\Locidx$. From a practical perspective, each affiliate's quota is partly determined based on its service resource availability, making the quota roughly proportional to the available service resources~\citep{bansak2024outcome}. In evaluating our proposed algorithms, we distinguish two parameter regimes for the service slack $\ServiceSlack$, formally defined as follows.


\begin{definition}[{Stable vs. Near-Critical Regimes}]\label{def:eps+regimes}
Given the service slack $\epsilon\geq0$, we categorize the problem into the ``stable'' regime if $\ServiceSlack=\Omega(1)$, and the ``near-critical'' regime if $\ServiceSlack = \BigO(1/\sqrt{\TotalTime})$.
\end{definition}

\Cref{def:eps+regimes} is inspired by queueing theory, which distinguishes between regimes where the service ``slack'' (i.e., the difference between service and arrival rates) remains constant (stable regime) and those where it vanishes (heavy-traffic regime). These two regimes often exhibit distinct behaviors. We adopt a parallel distinction in \Cref{def:eps+regimes}. As we show in Sections~\ref{sec:alg+design} and~\ref{subsec:alg+surrogate+primal}, our separate analyses under these regimes reveal similarly distinct behaviors.

\smallskip
\textbf{Information Setting:} Before formalizing the agency's objective, we describe the information available for decision-making. When matching case~$\Timeidx$ to an affiliate, the exact realization of server availability~$s_{t,i}$ is \emph{not known}, reflecting uncertainty in service times. Regarding backlog information, we distinguish between two settings. In the \emph{congestion-aware} setting (\Cref{sec:alg+design}), the agency observes the current \backlog{}~$\BuildUpVec_{\Timeidx-1}$ at the time of decision making. By contrast, in the \emph{congestion-oblivious} setting (\Cref{subsec:alg+surrogate+primal}), the agency does not have any backlog information.


{\smallskip}
{\bf Objective:} We now formally define the agency's matching process and its objective. The agency employs a dynamic matching algorithm $\pi$, which upon the arrival of case $\Timeidx$, makes an immediate and irrevocable \allocation{} decision $\DecisionVec_\Timeidx^{\OnlineALG}$. To maintain consistency with the literature, we hereafter refer to such a dynamic matching algorithm $\pi$ as an \emph{online} algorithm. Formally, an online algorithm $\pi$ is a mapping from an observable history $\History_{\Timeidx-1}$ and current arrival type $\mathbf{A}_t$ to a matching decision $\mathbf{z}^\pi_t$. In light of the above two information settings, the observable history is given by $\History_{\Timeidx-1}:=\{\Arrival_\Tauidx, \ServiceVec_\Tauidx, \DecisionVec_\Tauidx^\OnlineALG\}_{\Tauidx=1}^{\Timeidx-1}$ (resp. 
$\{\Arrival_\Tauidx, \DecisionVec_\Tauidx^\OnlineALG\}_{\Tauidx=1}^{\Timeidx-1}$) in the congestion-aware (resp. congestion-oblivious) setting. 
For a given penalty parameters $\OverCost$ and $\BuildUpCost$, the agency's objective is given by:\footnote{
\label{footnote+post+allocation+service}{
We highlight that managing congestion and respecting the annual quota are two completely distinct and incomparable goals. The former is captured by the final term in \eqref{ALG+obj}, while the latter is captured by the hard constraints \ref{hard+constraint:t} for all $\Timeidx \in [\TotalTime]$ and the second term in \eqref{ALG+obj}, which functions as a soft constraint.}
}

\begin{equation}
    \ALGObj{\OnlineALG}{\OverCost} {\BuildUpCost} := 
\sum_{\Locidx=1}^\Locnum \sum_{\Timeidx=1}^\TotalTime\WeightEach \DecisionEach^{\OnlineALG}
    - \OverCost\sum_{\Locidx=1}^{\Locnum} \PositivePart{\sum_{\Timeidx=1}^\TotalTime \DecisionEach^{\OnlineALG} - \Capacity_\Locidx }
    - \frac{\BuildUpCost}{\TotalTime} \sum_{\Locidx=1}^{\Locnum}\sum_{\Timeidx=1}^\TotalTime \BuildUpti{\Timeidx}{\Locidx}^{\OnlineALG} \tag{\textsf{Objective}}\label{ALG+obj}
\end{equation}
where $\{\mathbf{b}_t^{\pi}\}_{t=1}^T $ are the backlog vectors induced by the matching decisions $\{\DecisionVec_\Timeidx^{\OnlineALG}\}_{\Timeidx=1}^\TotalTime$ through the backlog dynamics in   \eqref{const:build-up}.
We refer to $\{\DecisionVec_\Timeidx^{\OnlineALG}\}_{\Timeidx=1}^\TotalTime$ as a \emph{\allocation{} profile}. 
As we highlighted earlier, a feasible matching profile in our model must satisfy {(i) $\DecisionVec_\Timeidx^{\OnlineALG} \in \FeasibleSet{\Target_\Timeidx}$ and (ii) constraints  \ref{hard+constraint:t}, for all $\Timeidx \in [\TotalTime]$.} 

\smallskip
{\bf Performance Metric:} We compare the performance of any online algorithm against an \emph{optimal offline} benchmark. This benchmark, formally defined below, solves the same optimization problem as online algorithms, but with full foreknowledge of the sequence of arrivals and service availabilities.

\begin{definition}[Optimal Offline Benchmark]\label{def:offline}
Given knowledge of the sample-path $\{\Arrival_\Timeidx, \ServiceVec_\Timeidx\}_{\Timeidx=1}^{\TotalTime}$ of arrival and service availability sequences, the optimal offline benchmark solves this  convex program:\footnote{
{By introducing auxiliary variables, this convex program can equivalently be formulated as a linear program.}
\label{footnote:linear+program}
}
\begin{alignat}{4}
&\OPT(\OverCost, \BuildUpCost) :=   
& 
\! \max_{\substack{
\DecisionVec_\Timeidx \in 
\FeasibleSet{\Target_\Timeidx} 
\\
\BuildUpVec_\Timeidx \geq \mathbf{0} 
}}
& 
\sum_{\Locidx=1}^\Locnum\sum_{\Timeidx=1}^\TotalTime \WeightEach \DecisionEach
    - \OverCost\sum_{\Locidx=1}^{\Locnum} 
    \left(\sum_{\Timeidx=1}^\TotalTime \DecisionEach - \Capacity_\Locidx \right)_{+}
 & 
\!\!\!\!\!\!\!\!
    -\frac{\BuildUpCost}{\TotalTime} \sum_{\Locidx=1}^{\Locnum}\sum_{\Timeidx=1}^\TotalTime \BuildUpti{\Timeidx}{\Locidx}
\nonumber  \\
&
&\textrm{s.t.} 
\quad
& {\sum_{\Timeidx=1}^\TotalTime \Indicator [\Target_\Timeidx  = 0]\DecisionEach \leq \Big(\Capacity_\Locidx - \sum_{\Timeidx=1}^\TotalTime \Indicator[\Target_\Timeidx=\Locidx] \DecisionEach\Big)_{+}}
 &
\!\!\!\!\!\!\!\!
{\forall \Locidx \in [\Locnum]}\tag{\textsf{Capacity Feasibility-$\TotalTime$}}\label{line:hard+constraint} \\
& 
& 
& 
\BuildUpti{\Timeidx}{\Locidx} \geq \BuildUpti{\Timeidx-1}{\Locidx} + \DecisionEach - \ServiceEach
 &
\!\!\!\!\!\!\!\!
{\forall t\in[T], i\in [m]} \tag{\textsf{Backlog Inequality}}\label{line:const+build+temp}
\end{alignat}
where we define $b_{0,\Locidx}=0$ for all $\Locidx \in [\Locnum]$ by convention. 
\end{definition}

Some comments are in order. First, note that we impose the capacity constraint \emph{only at $\Timeidx = \TotalTime$} when defining the offline benchmark. Hence, the above convex program is a relaxation of our original problem, that is, for any given sample path, $\ALGObj{\OnlineALG}{\OverCost} {\BuildUpCost} \leq \textsf{OPT}(\OverCost,\BuildUpCost)$. Second, it is straightforward to see that in any optimal solution, at least one of these constraints is binding: $\BuildUpti{\Timeidx}{\Locidx}\geq 0$ or $\BuildUpti{\Timeidx}{\Locidx} \geq \BuildUpti{\Timeidx-1}{\Locidx} + \DecisionEach - \ServiceEach$. Therefore, the optimal solution for $\BuildUpti{\Timeidx}{\Locidx}$ must satisfy the backlog dynamics in \eqref{const:build-up}.

We evaluate an algorithm based on its worst-case performance across all instances, where each instance $\mathcal{I}$ comprises (i) the set of affiliates $[\Locnum]$, (ii) the capacity ratios $\boldsymbol{\CapRatio}$, and (iii) the (unknown) arrival type distribution $\ArrivalDist$. With the above benchmark, we define \emph{regret} as the worst-case difference between the expected objective value achieved by the algorithm and that of the optimal offline benchmark.


\begin{definition}[Worst-case Regret]\label{def:regret} The regret of an online algorithm $\OnlineALG$ is given by
\begin{equation}
   \Regret^\OnlineALG := \sup_{\mathcal{I}}~ \E\left[\OPT(\OverCost,\BuildUpCost) - \ALGObj{\OnlineALG}{\OverCost} {\BuildUpCost}\right].
\end{equation}
where the expectation is over the arrival distribution $\ArrivalDist$, the Bernoulli service process with a service rate vector $\mathbf{\ServiceRate}$, and (potential) randomness of the algorithm itself.\footnote{
For brevity, we omit the dependence of expectation on the arrival and service distributions.
}
\end{definition} 
Given our benchmark and notion of regret, our goal is to design algorithms that are asymptotically optimal, meaning that their regret grows sub-linearly with $\TotalTime$, ideally at a (near) optimal rate.\footnote{
\label{footnote:competitive-ratio}
\majorrevcolor{Another common performance measure in the online algorithms literature is the \emph{competitive ratio}—the worst-case ratio of an algorithm's objective to that of an offline benchmark. Under our framework, sublinear regret implies a competitive ratio of $1 - o(1)$ as long as the expected value of the offline benchmark (\Cref{def:offline}) linearly grows in $T$.}}

\majorrevcolor{
\subsection{Modeling Assumptions and Generalizations with Practical Considerations}\label{subsec:model+assumption}
Our baseline model imposes the following assumption:
(i) each affiliate is endowed with a single type of static resource, and
(ii) each case consumes exactly one unit of that resource. These assumptions align with current practice. For example, the annual quota is the only static resource explicitly and currently tracked by our partner agency. Moreover, certain assignment procedures---such as an ongoing pilot for algorithm-assisted resettlement in Switzerland~\citep{bansak2024outcome}---track capacities at the case level, for which our baseline model remains directly applicable.

Although these assumptions were adopted for simplicity of exposition, our framework can easily be extended to relax them. In \Cref{apx+non+unit+size}, we show how our framework can accommodate \emph{multiple knapsack constraints}, allowing each affiliate to manage several types of static resources (e.g., school enrollment slots in addition to annual quota). Moreover, we show extensions where each case consumes a \emph{varying amount of each resource}. This generalization is particularly relevant to the U.S. refugee resettlement context, where cases often consist of multiple family members, and may consume multiple units of an affiliate’s annual quota. We also numerically explore this aspect of varying family sizes---alongside other practical considerations relevant to our collaboration---in our case study in \Cref{subsec:more+numerics}.

}


\section{Algorithm Design for Congestion-Aware Setting}\label{sec:alg+design}

In this section, we propose our first learning-based algorithm for the congestion-aware setting. In \Cref{subsec:motivation}, we first study the dual program of the optimal offline (\Cref{def:offline}) and introduce our learning-based approach. Next, in \Cref{subsec:alg+surrogate+dual}, we formally present the congestion-aware dual-learning algorithm (\CA{})  and show its (asymptotically optimal) sub-linear regret guarantees in both stable and near-critical regimes,  provided the congestion penalty parameter satisfies $\BuildUpCost = o(\TotalTime)$. We also show an impossibility result (\Cref{prop:impossibility}), establishing that the condition $\gamma = o(\TotalTime)$ is indeed unavoidable to achieve sub-linear regret. Finally, in \Cref{subsec:ALG+D+pfsketch}, we provide an overview of our proof technique.


\subsection{Motivation: Surrogate Dual Problem}\label{subsec:motivation}
{
Our first step is to study the optimal offline benchmark through duality. Observe that, due to our assumption $\CapRatio_\Locidx - \P[\Target_\Timeidx = \Locidx] = \Omega(1)$, the total number of tied cases during $\TotalTime$ periods at each affiliate does not exceed the capacity with high probability for sufficiently large $\TotalTime$.\footnote{{Specifically, by the Azuma-Hoeffding inequality and union bound over all affiliates, this event occurs with probability $1-\BigO(\Locnum\exp(-\TotalTime))$. Note that our assumption $\underline{\CapRatio}=\Theta(1)$ implies $m 
= \Theta(1)$ and hence we suppress dependences on $\Locnum$.} } Furthermore, under this event, the capacity constraint of the offline benchmark, i.e., \eqref{line:hard+constraint} in \Cref{def:offline}, reduces to the standard packing constraint $\sum_{\Timeidx=1}^\TotalTime \Decision_{\Timeidx,\Locidx} \leq \Capacity_\Locidx,  \forall \Locidx \in [\Locnum]$.  The following proposition, which we prove in \Cref{apx+prop+offline}, characterizes the dual program of the optimal offline under this high probability event.} 


\begin{proposition}[{Dual of Optimal Offline}] \label{prop:offline}
For any given sample path $\{\Arrival_\Timeidx, \ServiceVec_\Timeidx\}_{\Timeidx=1}^\TotalTime$, consider the following dual program:
\begin{eqnarray}
    \displaystyle\emph{\Dual}(\OverCost, \BuildUpCost) := &\displaystyle\underset{\DualOverVec, \DualHardVec,\DualBuildVec_\Timeidx\geq \mathbf{0}}{\min} &\displaystyle\sum_{\Timeidx=1}^\TotalTime  \Big\{
    \max_{\DecisionVec_\Timeidx \in \FeasibleSet{\Target_\Timeidx}}{\left(\mathbf{w}_t - \DualOverVec - \DualHardVec - \DualBuildVec_\Timeidx\right)\cdot\DecisionVec_\Timeidx} +
    \CapRatioVec\cdot \DualOverVec +
    \CapRatioVec\cdot \DualHardVec + 
    \ServiceVec_\Timeidx \cdot \DualBuildVec_\Timeidx \Big\}\nonumber\\
     &\textrm{s.t.}& \displaystyle\DualOver_\Locidx \leq \alpha \quad \forall i \in [m]
        \label{const:theta}\\
     && \displaystyle \DualBuild_{\Timeidx,\Locidx} - \DualBuild_{\Timeidx+1,\Locidx} \leq \frac{\BuildUpCost}{\TotalTime} \quad \forall \Timeidx \in [\TotalTime-1], \Locidx\in[\Locnum]\label{const:Delta.1},
     \quad \quad \DualBuild_{\TotalTime,\Locidx} \leq \frac{\BuildUpCost}{\TotalTime} \quad \forall\Locidx\in[\Locnum]
\end{eqnarray}
{Then we have the following strong duality: $ \emph{\OPT}(\OverCost, \BuildUpCost)\Indicator[G_\TotalTime] = \emph{\Dual}(\OverCost, \BuildUpCost)\Indicator[G_\TotalTime]$ where $G_T$ is the event that sample path $(\Arrival_t, \ServiceVec_t)_{t=1}^T$ satisfies $\sum_{t=1}^\TotalTime \Indicator[\Target_t = i] \leq c_{i}, \forall i\in[m]$.}
\end{proposition}


In \Cref{prop:offline}, the dual variables ($\DualOverVec$,  $\DualHardVec$) are associated  with static resources, corresponding respectively to the over-allocation penalty in the objective (via Fenchel duality) and the hard constraint \eqref{line:hard+constraint} (after Lagrangifying this constraint). The vector sequence $\{\DualBuildVec_\Timeidx\}_{\Timeidx=1}^{\TotalTime}$ consists of dual variables associated with dynamic resources, each corresponding to constraint \eqref{line:const+build+temp} in \Cref{def:offline} (after Lagrangifying these constraints). Notably, the sequence $\DualBuildVec_\Timeidx$ is time-dependent, in contrast to $(\DualOverVec, \DualHardVec)$, which arises from constraints \eqref{const:Delta.1} that are coupling consecutive $\DualBuildVec_\Timeidx$ and $\DualBuildVec_{\Timeidx+1}$.\footnote{Indeed, by complementary slackness, we can show that the optimal dual variables satisfy $\DualBuild_{\Timeidx,\Locidx}^* -\DualBuild_{\Timeidx+1,\Locidx}^*>0$ if the backlog of the optimal offline at period $\Timeidx$ and affiliate $\Locidx$ is nonzero.}

\Cref{prop:offline} implies a simple method for optimal matching decisions given the optimal dual variables $(\DualOverVec^*, \DualHardVec^*, \{\DualBuildVec^*_\Timeidx\}_{\Timeidx=1}^\TotalTime)$: at time $\Timeidx$, choose the decision $\DecisionVec_\Timeidx \in \FeasibleSet{\Target_\Timeidx}$ that maximizes the \emph{dual-adjusted score} {${(\mathbf{w}_t - \DualOverVec^* - \DualHardVec^* - \DualBuildVec^*_\Timeidx)\cdot\DecisionVec_\Timeidx}$}. \Cut{Motivated by this observation, we adopt a \emph{learning-based} approach,  iteratively updating the dual variables and choosing the matching decisions that maximize the dual-adjusted score based on the learned values. Due to the time invariance of $(\boldsymbol{\theta}^*, \boldsymbol{\lambda}^*)$ and the stationarity of arrivals and service availabilities, we can efficiently learn the static dual variables, similar to \citet{agrawal2014fast}, using a first-order method that updates via the (stochastic) gradient of the dual function upon each arrival. However, directly learning the time-varying sequence $\{\DualBuildVec^*_\Timeidx\}_{\Timeidx=1}^\TotalTime$ subject to the inter-temporal constraints \eqref{const:Delta.1} appears challenging, if not infeasible. To overcome this challenge, we introduce the following surrogate dual problem.}

\begin{definition}[\textup{Surrogate Dual Problem}] \label{def:dual+surrogate}
For any given sample path $\{\Arrival_\Timeidx, \ServiceVec_\Timeidx\}_{\Timeidx=1}^\TotalTime$, define the per-period surrogate dual function for period $\Timeidx$ as
\begin{equation}
{
    \PerDual_\Timeidx(\DualOverVec, \DualHardVec,\DualBuildVec) := 
    \max_{\DecisionVec_\Timeidx \in \FeasibleSet{\Target_\Timeidx}}
    \Big\{(\mathbf{w}_t - \DualOverVec - \DualHardVec - \DualBuildVec)\cdot\DecisionVec_\Timeidx\Big\} +
    \CapRatioVec\cdot \DualOverVec +
    \CapRatioVec\cdot \DualHardVec + 
    \ServiceVec_\Timeidx \cdot \DualBuildVec.} \label{line:per+dual}
\end{equation}
The surrogate dual problem is then defined by 
\begin{eqnarray}
    \displaystyle{\SurrogateDual}(\OverCost) := &\displaystyle\underset{\DualOverVec, \DualHardVec,\DualBuildVec\geq \mathbf{0}}{\min} &\displaystyle\sum_{\Timeidx=1}^\TotalTime  \PerDual_\Timeidx(\DualOverVec, \DualHardVec,\DualBuildVec) \quad \displaystyle
     \textrm{s.t.} 
     \quad \displaystyle\DualOver_\Locidx \leq \alpha \quad \forall i \in [m] 
     \label{const:theta+2} 
\end{eqnarray}
\end{definition}
Compared to the original dual program, the surrogate dual problem addresses the time-varying nature of $\{\DualBuildVec^*_\Timeidx\}_{\Timeidx=1}^{\TotalTime}$ by considering a restricted program in which these dual variables remain \emph{identical} across periods. As we show later, this restriction leads to an important advantage in controlling the congestion. Specifically, it enables us to uncover a crucial structural connection between the \backlog{} dynamics of each affiliate and the trajectory of an online projected subgradient descent used to learn the time-invariant dual solution $\DualBuildVec^*$ of this surrogate problem.

\subsection{Algorithm and Analysis}\label{subsec:alg+surrogate+dual}

\begin{algorithm}[t]
\caption{Congestion-Aware Dual-Learning (\texttt{CA-DL}) Algorithm}
\label{ALG+Surrogate+D}
\begin{algorithmic}[1]
\STATE \textbf{Input:} $\TotalTime$, $\CapRatioVec$, $\eta$, and $\zeta$
\STATE Initialize $\DualOver_{1,\Locidx} \leftarrow \exp(-1)$, $\DualHard_{1,\Locidx} \leftarrow \exp(-1)$, $\Capacity_{0,\Locidx} \leftarrow \CapRatio_\Locidx \TotalTime$, and $\BuildUpti{0}{\Locidx} \leftarrow 0$ for all $\Locidx \in [\Locnum]$
\FOR{each arrival $\{\Arrival_\Timeidx\}_{\Timeidx=1}^\TotalTime$}
    \STATE \label{line:primal+begin}%
    \textbf{if} $\Target_\Timeidx \in [\Locnum]$ \textbf{then} set $\DecisionVec_\Timeidx = \BasisVec{\Target_\Timeidx}$ 
    \STATE \textbf{else if} $\min_{\Locidx\in[\Locnum]} \Capacity_{\Timeidx-1,\Locidx} > 0$ \textbf{then} set $\DecisionVec_\Timeidx \in \argmax_{\mathbf{z} \in \Simplex} (\Weightvec_\Timeidx - \DualOverVec_\Timeidx - \DualHardVec_\Timeidx -\zeta \BuildUpVec_{\Timeidx-1})\cdot \DecisionVec$ \label{line:primal+decision}
    \STATE \label{line:primal+end} \textbf{else} set $\DecisionVec_\Timeidx = \mathbf{0}$ 

    \STATE Update for all $\Locidx \in [\Locnum]$: $\Capacity_{\Timeidx,\Locidx} \leftarrow \Capacity_{\Timeidx-1,\Locidx} - \DecisionEach$, \quad $\BuildUp_{\Timeidx,\Locidx} \leftarrow (\BuildUp_{\Timeidx-1,\Locidx} + \DecisionEach - \ServiceEach)_+$
    \STATE \label{dual:begin}%
    Update dual variables for all $\Locidx \in [\Locnum]$:
    \begin{equation}
    \DualOver_{\Timeidx+1,\Locidx} \leftarrow \min\{\DualOver_{\Timeidx,\Locidx} \exp(\eta(\DecisionEach - \CapRatio_\Locidx)), \OverCost\}, \quad
    \DualHard_{\Timeidx+1,\Locidx} \leftarrow \min\left\{\DualHard_{\Timeidx,\Locidx} \exp(\eta(\DecisionEach - \CapRatio_\Locidx)), \frac{1+2\OverCost}{\underline{\CapRatio}}\right\}
    \label{eq:update+theta}
    \end{equation}
\noindent\texttt{\textcolor{black}{//
The dual variable for the dynamic resource is implicitly updated as 
$\smash{\beta_{t+1,i} = (\beta_{t,i} + \zeta(z_{t,i} - s_{t,i}))_+}$ 
with $\beta_{0,i} = 0$, so $\beta_{t+1,i} = \zeta b_{t,i}$.}}
\ENDFOR
\end{algorithmic}
\end{algorithm}

We formally present our first algorithm, called \CA{}, in \Cref{ALG+Surrogate+D}. At a high level, \CA{} maintains dual variables for the surrogate dual problem, upon which it bases its primal decision.  
\Cut{Specifically, the algorithm uses primal decisions to obtain gradient information for updating dual variables via online learning. It explicitly updates dual variables $(\DualOverVec, \DualHardVec)$ associated with static resources using online mirror descent, and implicitly updates dual variables $\DualBuildVec$ for dynamic resources (in the surrogate dual problem) through \backlog{} dynamics. 
Consequently, the scaled backlog  effectively serves as the ``correct'' dual variable for the dynamic resource.} We also note that the algorithm is initialized with certain dual values, and also and learning rates $\eta \geq 0$ for static resources and $\zeta \geq 0$ for dynamic resources. 
For each arrival $\Timeidx$, the algorithm proceeds in two phases.

\EC{\smallskip}
{\bf Primal Phase (line \ref{line:primal+begin}-\ref{line:primal+end}):} 
The algorithm selects the decision that maximizes the dual-adjusted score based on the current dual variables $(\DualOverVec_\Timeidx, \DualHardVec_\Timeidx, \zeta\BuildUpVec_{\Timeidx-1})$, subject to the capacity constraint and $\DecisionVec_\Timeidx \in \FeasibleSet{\Target_\Timeidx}$. Specifically, the algorithm matches the arriving free case to the affiliate $\Locidx$ with the highest dual-adjusted score $\Weight_{\Timeidx,\Locidx} - \DualOver_{\Timeidx,\Locidx} - \DualHard_{\Timeidx,\Locidx} - \zeta\BuildUp_{\Timeidx-1,\Locidx}$ (\SLdelete{breaking ties arbitrarily}\majorrevcolor{with arbitrary selection among all affiliates that achieve the maximum}), provided that (i) the chosen affiliate's adjusted score is non-negative and (ii) every affiliate has remaining capacity.\footnote{This second condition is mostly needed for a technical reason in our theoretical analysis. See \Cref{remark+no+inaction}.} Otherwise, the case is matched to the dummy affiliate. By construction, our algorithm satisfies \eqref{hard+constraint:t} for all $\Timeidx \in [\TotalTime]$.

\EC{\smallskip}

{\bf Dual Phase (line \ref{dual:begin}):} 
Based on the primal decision $\DecisionVec_\Timeidx$, we update the dual variables using online learning. Focusing first on the update rule \eqref{eq:update+theta}, note that $\CapRatioVec - \DecisionVec_\Timeidx$ is a subgradient of the per-period surrogate dual function $g_t$ (see equation \eqref{line:per+dual} in \Cref{def:dual+surrogate}) with respect to  $\DualOverVec$, evaluated at the current assignment of dual variables $(\DualOverVec_\Timeidx, \DualHardVec_\Timeidx, \zeta\BuildUpVec_{\Timeidx-1})$. We use this gradient to perform the multiplicative weight update \citep{freund1997using}, which is a special case of online mirror descent. The same principle applies to the update rule for $\DualHardVec$.\footnote{Specifically, the multiplicative dual update in equation \eqref{eq:update+theta} is a special case of mirror descent with the mirror map $h(\mathbf{x})$ being the negative entropy function (i.e., $h(\mathbf{x}) = \sum_{i} x_i \log(x_i)$). See \cite{bubeck2011introduction} for details.}
The upper bound on $\DualOverVec$ comes from the domain of $\DualOverVec$ in the surrogate dual problem \eqref{const:theta+2}. The upper bound on $\DualHardVec$ is imposed for technical reasons.\footnote{Since negative entropy is strongly convex only within a bounded domain, we impose an upper bound on $\boldsymbol{\lambda}$.}



In addition to explicitly tracking the dual variables $(\DualOverVec,\DualHardVec)$ for the static resource, our algorithm also \emph{implicitly} tracks the dual variable $\DualBuildVec$ (in the surrogate dual problem) for dynamic resources via the \backlog{} dynamics. To see this, suppose we update the sequence $\{\DualBuildVec_\Timeidx\}_{\Timeidx=1}^\TotalTime$, initialized with $\DualBuildVec_1 = \mathbf{0}$, using a slightly different (and simpler) update rule---specifically, vanilla projected gradient descent~\citep{zinkevich2003online} with the non-negative orthant $\DualBuildVec_\Timeidx \geq \mathbf{0}$ as the constraint set.\footnote{This update rule is a special case of mirror descent when the mirror map $h(\mathbf{x})$ is the squared  Euclidean norm.}
Note that $\ServiceVec_\Timeidx - \DecisionVec_\Timeidx$ is a subgradient of $g_\Timeidx$ with respect to $\DualBuildVec$. Given a learning rate $\zeta$, we obtain:
\begin{equation}
\DualBuild_{\Timeidx+1, \Locidx} = (\DualBuild_{\Timeidx,\Locidx} + \zeta(\DecisionEach - \ServiceEach))_{+}. \label{eq:beta+dynamics}
\end{equation}

We observe that, up to a scalar factor $\zeta$, the update rule  \eqref{eq:beta+dynamics} is equivalent to the backlog dynamics \eqref{const:build-up}. By induction over period $\Timeidx$, we can show that $\DualBuild_{\Timeidx, \Locidx} = \zeta \BuildUpti{\Timeidx-1}{\Locidx}$ for all $\Timeidx \in [\TotalTime]$ and $\Locidx \in [\Locnum]$. In other words, the scaled \backlog{} acts as the dual variable for dynamic resources. Intuitively, \CO{} manages congestion by ``penalizing'' a scaled version of the current \backlog{}.\footnote{
\SLedit{Our algorithm design also has an intimate connection to Lyapunov optimization, particularly the drift-plus-penalty method \citep{neely2022stochastic}. We elaborate on this connection in \Cref{apx:lyapunov}.}
}


We now formally analyze the regret of \CA{}, first focusing on the stable regime (i.e.,  $\ServiceSlack=\Omega(1)$).


{\begin{theorem}[Regret of \CA{} under Stable Regime]\label{thm:ALG+Surrogate+Dual}
Let $\eta = \Theta(1/\sqrt{\TotalTime})$ and $\zeta = \Theta(1/\sqrt{\TotalTime})$. The regret (\Cref{def:regret}) of \CA{} is $\BigO(\sqrt{\TotalTime} + \frac{\BuildUpCost}{\ServiceSlack})$ for any service slack parameter $\ServiceSlack > 0$. In particular, under the stable regime (\Cref{def:eps+regimes}), the regret of \CA{} is $\BigO(\sqrt{\TotalTime} + \BuildUpCost)$ .
\end{theorem}}

We sketch the proof of \Cref{thm:ALG+Surrogate+Dual} in \Cref{subsec:ALG+D+pfsketch}. {For the near-critical regime (i.e., $\ServiceSlack = \BigO(1/\sqrt{\TotalTime})$), we complement \Cref{thm:ALG+Surrogate+Dual} with the following corollary, proved in \Cref{apx+near+a}. This corollary establishes that \CA{} also achieves sub-linear regret in this regime (with a different choice of $\zeta$).}


\begin{corollary}[Regret of \CA{} under Near-Critical Regime]\label{near+part+a}
Let $\eta = \Theta(1/\sqrt{\TotalTime})$ and $\zeta = \Theta(\sqrt{\BuildUpCost/\TotalTime})$. Under the near-critical regime (\Cref{def:eps+regimes}), the regret (\Cref{def:regret}) of \CA{} is $\BigO(\sqrt{\BuildUpCost \TotalTime})$.
\end{corollary}

{\Cref{thm:ALG+Surrogate+Dual} and \Cref{near+part+a} imply that \CA{}'s regret is sub-linear in $\TotalTime$ (in both stable and near-critical regimes) unless $\BuildUpCost = \Omega(\TotalTime)$. A natural question, therefore, is whether sub-linear regret is achievable when $\BuildUpCost = \Omega(\TotalTime)$. We rule out this possibility with the following lower bound result, concluding that \CA{} attains sub-linear regret for \emph{all} feasible values of $\BuildUpCost$ for which such a small regret is achievable.}



\begin{proposition}[Lower Bound on Achievable Regret]\label{prop:impossibility}
For $\BuildUpCost = \Omega(\TotalTime)$ and any given $\ServiceSlack \geq 0$ such that the resulting service rates satisfy \( r_i \in [\rho_i + \epsilon,\ 1) \) for all \( i \in [m] \), there exists an instance for which the regret (\Cref{def:regret}) of any online algorithm {(even in the congestion-aware setting)} is $\Omega(\TotalTime)$.  
\end{proposition}\par 

\Cut{We provide a formal proof of \Cref{prop:impossibility} in \Cref{apx+hardness}. To sketch the proof, we construct an instance with a single affiliate with capacity ratio $\CapRatio = 0.5$ and service rate $\ServiceRate = \CapRatio + \ServiceSlack$, where $\ServiceSlack \in [0, 1-\CapRatio)$.  Each arriving case $\Timeidx$ has type $\Arrival_\Timeidx = (1, 0)$---that is, each arrival is a free case with a reward of one. Since there is only one actual affiliate, the decision reduces to whether or not to match each case to this affiliate.  In this instance, the offline optimal benchmark can leverage its foreknowledge of service availability to avoid \backlog{}s, matching cases to the affiliate only when the server is available. In the full proof, we show that this offline strategy yields an expected total reward of at least $0.5\TotalTime - \Theta(\sqrt{\TotalTime})$ with zero average \backlog{} (note that the maximum {expected} total reward is $0.5\TotalTime$). In contrast, because online algorithms do not have information about current service availability, they inevitably incur \backlog{} without achieving greater reward. Based on this intuition, we show that any online algorithm attaining $\Theta(T)$ reward must incur congestion cost of $\Theta(\BuildUpCost)$, leading to the desired result.}

We dedicate the next subsection to outlining the proof of \Cref{thm:ALG+Surrogate+Dual}. The proof of \Cref{near+part+a} follows similar ideas. The main challenge in establishing the regret guarantee arises from the objective function defined in \eqref{ALG+obj}, which consists of two fundamentally different components: the net matching reward and the time-average backlog. The key innovation of our analysis is to (i) define pseudo-rewards that implicitly incorporate both components and (ii) combine techniques from the celebrated drift-plus-penalty method~\citep{neely2022stochastic} and adversarial online learning to bound these pseudo-rewards in terms of the optimal offline solution (\Cref{def:offline}), and thus establish
low regret. Roughly speaking, the two terms appearing in our regret bound correspond respectively to separate bounds we establish for the net matching reward and congestion cost (see Lemmas \ref{lemma:ALG+D+reward} and \ref{Lemma+ALG+D+Build} in \Cref{subsec:ALG+D+pfsketch}).

\majorrevcolor{
\begin{remark}[\textbf{Matching Lower Bounds}]\label{remark:lower+bdd}
While our regret definition uses $\OPT(\alpha, \gamma)$ as the benchmark, the proof of \Cref{thm:ALG+Surrogate+Dual} establishes a stronger result: the regret bound holds even when compared to $\OPT(\alpha, 0)$, which does not incur the   congestion penalty and solely maximizes the net matching reward. 
{In \Cref{apx:regret+lower}, we provide matching lower bounds for this stronger benchmark---corresponding to the upper bounds in \Cref{thm:ALG+Surrogate+Dual} (the stable regime) and \Cref{near+part+a} (the near-critical regime when $\epsilon=0$)---that (almost) establish the asymptotic optimality of our upper-bounds. We also provide a near-matching lower bound for the weaker benchmark $\OPT(\alpha, \gamma)$} in the stable regime.
\end{remark}}
\begin{remark}[\textbf{Stopping Time}]\label{remark+no+inaction}
For technical reasons, we assume that \Cref{ALG+Surrogate+D} matches free cases to a dummy affiliate once any affiliate's capacity is depleted. In \Cref{apx+stopping+exp}, we prove that this ``stopping time'' occurs near the end of the process. Furthermore, we show that \Cref{thm:ALG+Surrogate+Dual} remains unchanged even if the dummy affiliate is removed before this stopping time (see \Cref{apx+no+inaction}).
\end{remark}




\subsection{Proof Sketch of \texorpdfstring{\Cref{thm:ALG+Surrogate+Dual}}{}}\label{subsec:ALG+D+pfsketch}

We first introduce some notation. Hereafter, a sample path refers to a sequence of arrivals and  service availabilities over the horizon, i.e., $\{\Arrival_\Timeidx, \ServiceVec_\Timeidx\}_{\Timeidx=1}^\TotalTime$. We use $\{\DecisionVec_\Timeidx\}_{\Timeidx=1}^\TotalTime$ and $\{\DecisionVec^*_\Timeidx\}_{\Timeidx=1}^\TotalTime$ to denote the matching profile of \CA{} and the optimal offline benchmark (\Cref{def:offline}), respectively. Let $\NetReward(\cdot;\OverCost)$  denote the net \allocation{} reward of a feasible matching profile given the over-allocation penalty parameter $\OverCost$:

\begin{equation}
    \NetReward(\{\hat{\DecisionVec}_\Timeidx\}_{\Timeidx=1}^\TotalTime;\OverCost) = \sum_{\Timeidx=1}^\TotalTime \sum_{\Locidx=1}^\Locnum \Weight_{\Timeidx,\Locidx}\hat{\Decision}_{\Timeidx,\Locidx} - \OverCost\sum_{\Locidx=1}^\Locnum\PositivePart{\sum_{\Timeidx=1}^\TotalTime \hat{\Decision}_{\Timeidx,\Locidx} - \Capacity_\Locidx}.\label{line:net+reward}
\end{equation}
Because the congestion cost is non-negative, $ \NetReward(\{\DecisionVec^*_\Timeidx\}_{\Timeidx=1}^\TotalTime;\OverCost) \geq \OPT(\OverCost, \BuildUpCost)$ for every sample path and every penalty parameter pair $(\OverCost, \BuildUpCost)$. Hence, for any arrival distribution $\ArrivalDist$, we have this decomposition:
\begin{equation}
\begin{split}
&\E[\OPT(\OverCost, \BuildUpCost) 
- \ALGObj{\CA{}}{\OverCost} {\BuildUpCost}]
\leq
\underbrace{
\vphantom{\sum_{\Timeidx=1}^\TotalTime}
\E[\NetReward(\{\DecisionVec^*_\Timeidx\}_{\Timeidx=1}^\TotalTime; \OverCost)]
-
\E[\NetReward(\{\DecisionVec_\Timeidx\}_{\Timeidx=1}^\TotalTime; \OverCost)]
}_{\mathclap{\textsf{(A)=Loss of net \allocation{} reward}}} + 
\BuildUpCost\cdot \underbrace{\E\Big[\frac{1}{\TotalTime} 
\sum_{\Timeidx=1}^{\TotalTime}\Norm{\BuildUpVec_\Timeidx}{1}\Big]}_{\mathclap{\textsf{(B)=Average \backlog{}}}}.    \label{line:regret+decomp}
\end{split}
\end{equation}
In light of this decomposition, it suffices to separately upper bound terms $\textsf{(A)}$ and $\textsf{(B)}$. The following two key lemmas establish the desired upper bounds for each term.

\begin{lemma}[Bounding Loss of Net \Allocation{} Reward]\label{lemma:ALG+D+reward}
For any arrival distribution $\ArrivalDist$ and $\ServiceSlack \geq 0$, we have
\begin{equation}
\E[{\NetReward}(\{\DecisionVec^*_\Timeidx\}_{\Timeidx=1}^\TotalTime; \OverCost)]
-
\E[{\NetReward}(\{\DecisionVec_\Timeidx\}_{\Timeidx=1}^\TotalTime; \OverCost)]
\leq \BigO(\sqrt{\TotalTime}).
\end{equation}  
\end{lemma}

\begin{lemma}[Bounding Average \Backlog{}]\label{Lemma+ALG+D+Build} For any arrival distribution $\ArrivalDist$ and $\ServiceSlack>0$, we have
    \begin{equation}
    \E\Big[\frac{1}{T}\sum_{t=1}^T \lVert \mathbf{b}_t \rVert_1 \Big]\leq \BigO\Big(\frac{1}{\ServiceSlack}\Big).
    \end{equation}
\end{lemma}
Combined with the decomposition \eqref{line:regret+decomp}, the two lemmas imply \Cref{thm:ALG+Surrogate+Dual}. The remainder of this section is devoted to sketching the proofs of Lemmas \ref{lemma:ALG+D+reward} and \ref{Lemma+ALG+D+Build}. The analysis consists of three steps: In Step~1, we define a pseudo-reward---a stochastic process designed to facilitate the comparison between the objective values of \CA{} and the optimal offline benchmark. \revcolor{This pseudo-reward accounts not only for the immediate reward of a match, but also for the opportunity costs of using a unit of capacity at each affiliate, the immediate marginal cost of over-allocation, and the marginal effect of a new match on congestion at the affiliates}. 
In Step~2, we establish lower and upper bounds on the pseudo-rewards that crucially use the design of \CA{}. Finally, Step~3,  we combine these bounds to complete the proof.

\textbf{Step 1: Defining a Pseudo-reward.} 
Consider the following  potential function, which is a commonly used Lyapunov function in the literature to show stability in dynamical systems (e.g., switched queuing networks; see~\cite{stolyar2004maxweight, eryilmaz2007fair, neely2008fairness}), and its drift.
\begin{equation}
    \psi(\BuildUpVec_\Timeidx) := \frac{1}{2}\Norm{\BuildUpVec_\Timeidx}{2}^2, \quad \textrm{}\quad 
    D_\Timeidx := \psi(\BuildUpVec_\Timeidx) - \psi(\BuildUpVec_{\Timeidx-1}).\label{psi}
\end{equation}
One simple yet powerful property of this potential function $\psi$ is the following lemma.
\begin{lemma}[{Drift Lemma}]\label{lemma:drift}
   For all sample paths, $${\BuildUpVec_{\Timeidx-1} \cdot (\DecisionVec_\Timeidx - \ServiceVec_\Timeidx)\leq} D_{\Timeidx} \leq \BuildUpVec_{\Timeidx-1} \cdot (\DecisionVec_\Timeidx - \ServiceVec_\Timeidx) + \BigO(1).$$
\end{lemma}
We prove \Cref{lemma:drift} in \Cref{apx+drift}. \Cref{lemma:drift} states that the drift of the potential function $\psi$ can be upper and lower bounded by a linear function of the current \backlog{}. With this background, we are ready to define the pseudo-reward at time $\Timeidx$, denoted by $K_\Timeidx$, as follows:
\begin{equation}
    K_\Timeidx := \Weightvec_\Timeidx\cdot\DecisionVec_\Timeidx + \DualOverVec_\Timeidx\cdot(\CapRatioVec-\DecisionVec_\Timeidx) + \DualHardVec_\Timeidx\cdot(\CapRatioVec-{\DecisionVec_\Timeidx}) - \zeta D_\Timeidx \label{eq:K_t}
\end{equation}
To better understand our design of pseudo-rewards, we observe that \Cref{lemma:drift} implies
{\begin{equation}
K_\Timeidx = (\Weightvec_\Timeidx - \DualOverVec_\Timeidx - {\DualHardVec_\Timeidx} - \zeta \BuildUpVec_{\Timeidx-1})\cdot \DecisionVec_\Timeidx + \DualOverVec_\Timeidx\cdot \CapRatioVec + \DualHardVec_\Timeidx\cdot \CapRatioVec + \zeta \BuildUpVec_{\Timeidx-1}\cdot \ServiceVec_\Timeidx - \BigO(\zeta) \label{eq:pseudo+drift}
\end{equation}}
for every sample path. The first term $(\Weightvec_\Timeidx - \DualOverVec_\Timeidx - {\DualHardVec_\Timeidx} - \zeta \BuildUpVec_{\Timeidx-1})\cdot \DecisionVec_\Timeidx$ is the adjusted score that \CA{} maximizes (in the primal phase), allowing us to use the optimality criterion of \CA{} to lower bound the sum of pseudo-rewards. Moreover, setting $\DualBuildVec_\Timeidx=\zeta \BuildUpVec_{\Timeidx-1}$ and ignoring the additive error $\BigO(\zeta)$, the pseudo-reward is \emph{exactly} $g_t(\DualOverVec_\Timeidx, \DualHardVec_\Timeidx,\DualBuildVec_\Timeidx)$, the per-period surrogate dual function evaluated at the current estimate of dual variables (see equation \eqref{line:per+dual} in \Cref{def:dual+surrogate}). This connection enables us to use online learning (in the dual phase) to upper bound the sum of pseudo-rewards. By carefully comparing these lower and upper bounds, we establish the desired upper bounds on both the loss in net matching reward (\Cref{lemma:ALG+D+reward}) and the average backlog (\Cref{Lemma+ALG+D+Build}). We elaborate on these steps next.

\textbf{Step 2: Lower and Upper Bounding the Pseudo-rewards.}
We provide a lower bound on the expected cumulative pseudo-reward up to the last time that no hard resource constraint, i.e.,  \eqref{hard+constraint:t} for $\Timeidx \in [\TotalTime]$, is binding for \CA{}. Formally, we define the \emph{stopping time} as follows:
\begin{equation}
{\Stopping := \min\Big\{\Timeidx \leq \TotalTime: \sum_{\Tauidx=1}^{\Timeidx} \Decision_{\Tauidx, \Locidx} \geq \Capacity_\Locidx \ \text{for some $\Locidx \in [\Locnum]$}\Big\}.}
\label{def:stopping}
\end{equation}

For all $1\leq \Timeidx\leq \Stopping$, the primal phase of \CA{} (lines \ref{line:primal+begin}-\ref{line:primal+end}) can be succinctly written as
\begin{equation}
    \DecisionVec_\Timeidx \SLedit{\in}  \argmax_{\DecisionVec\in \FeasibleSet{\Target_\Timeidx} } (\Weightvec_\Timeidx - \DualOverVec_\Timeidx - {\DualHardVec_\Timeidx} - \zeta \BuildUpVec_{\Timeidx-1})\cdot \DecisionVec \label{line:optimality}.
\end{equation}
The following lemma provides a lower bound on the total expected pseudo-reward up to the stopping time, expressed (in part) in terms of (i) the expected net \allocation{} reward of {\emph{any} matching profile subject to the same constraint as the offline benchmark (\Cref{def:offline})} and (ii) the backlog of \CA{}. 
\begin{lemma}[Lower Bound on Pseudo-Rewards]\label{lemma:pseudo+lower}
{For any feasible \ECKEEP{\allocation{} profile} $\{ \hat{\DecisionVec}_{\Timeidx}\}_{\Timeidx=1}^\TotalTime$ that satisfies $\hat{\DecisionVec}_\Timeidx \in \FeasibleSet{\Target_\Timeidx}$ for all $\Timeidx \in [\TotalTime]$ and \eqref{line:hard+constraint}}, we have
\begin{equation}
\E\Big[\sum_{\Timeidx=1}^{\Stopping} K_t\Big] 
\geq 
\E[\NetReward(\{ \hat{\DecisionVec}_{\Timeidx}\}_{\Timeidx=1}^\TotalTime; \OverCost)]
+ \zeta\ServiceSlack
    \E\Big[\sum_{\Timeidx=1}^{\Stopping - 1} \Norm{\BuildUpVec_\Timeidx}{1} \Big] - (\TotalTime-\Stopping) - \BigO(\zeta T)
\end{equation}
\end{lemma}
 We prove \Cref{lemma:pseudo+lower} in \Cref{apx+pseudo+lower}. The proof crucially  relies on the optimality criterion \eqref{line:optimality} of \CA{}. Specifically, we compare the decision of \CA{} at time $\Timeidx$ to that of a \emph{static} control (see Claim~\ref{fluid} in \Cref{apx+pseudo+lower}), whose expected value serves as an upper bound on the per-period net \allocation{} reward of any matching profile subject to the same constraint as the offline benchmark. 

The following lemma upper bounds the pseudo-rewards via the net \allocation{} reward of \CA{}.

\begin{lemma}[Upper Bound on Pseudo-Rewards]\label{lemma:pseudo+upper}
For every sample path, we have
    \begin{equation}
\sum_{\Timeidx=1}^{\Stopping} K_t \leq 
\NetReward(\{{\DecisionVec}_{\Timeidx}\}_{\Timeidx=1}^\TotalTime; \OverCost)
- (T- \Stopping) - \zeta\psi(\BuildUpVec_{\Stopping}) + \BigO(\sqrt{\TotalTime}) 
    \end{equation}
\end{lemma}
The proof of \Cref{lemma:pseudo+upper} is presented in \Cref{apx:pseudo+upper}. Similar to \citet{agrawal2014fast} and \citet{balseiro2023best}, our proof leverages the adversarial online learning regret guarantee of the online mirror descent algorithm used to update the dual variables for static resources (see update rules \eqref{eq:update+theta} and related discussion). Importantly, this regret guarantee holds for \emph{every} sample path and is oblivious to the primal decisions made by \CA{}, allowing us to directly apply this guarantee in our analysis.

\textbf{Step 3: Putting Everything Together.}
Combining \Cref{lemma:pseudo+lower} and \Cref{lemma:pseudo+upper} and taking expectations, we arrive at the following crucial inequality: For any \ECKEEP{\allocation{} profile} $\{ \hat{\DecisionVec}_{\Timeidx}\}_{\Timeidx=1}^\TotalTime$ that satisfies $\hat{\DecisionVec}_\Timeidx \in \FeasibleSet{\Target_\Timeidx}$ for all $\Timeidx \in [\TotalTime]$ and constraint \eqref{line:hard+constraint}, 
\begin{equation}
\underbrace{
\vphantom{\sum_{\Timeidx=1}^\TotalTime}
\E[\NetReward(\{ \hat{\DecisionVec}_{\Timeidx}\}_{\Timeidx=1}^\TotalTime; \OverCost)]
-
\E[\NetReward(\{{\DecisionVec}_{\Timeidx}\}_{\Timeidx=1}^\TotalTime; \OverCost)]
}_{\mathclap{\textsf{(A): \Cref{lemma:ALG+D+reward}}}}+ 
\underbrace{
\zeta\ServiceSlack\E\Big[\sum_{\Timeidx=1}^{\Stopping - 1} \Norm{\BuildUpVec_\Timeidx}{1} \Big] 
+\zeta\E[\psi(\BuildUpVec_{\Stopping})] }_{
\mathclap{\textsf{(B'): \Cref{Lemma+ALG+D+Build}}}
}
\leq \BigO(\sqrt{\TotalTime} + \zeta\TotalTime) \label{ineq:important}
\end{equation}

The remaining steps of the proof are (i) establishing upper bounds on terms $\textsf{(A)}$ and $\textsf{(B')}$ by choosing an appropriate feasible matching profile $\{\hat{\DecisionVec}_\Timeidx\}_{\Timeidx=1}^\TotalTime$, leading to Lemmas~\ref{lemma:ALG+D+reward} and \ref{Lemma+ALG+D+Build}, respectively, and (ii) controlling the \backlog{} accrued after the stopping time. We elaborate these steps in \Cref{apx+finish+ALG+1+build}.

\SLcomment{Shortened version}

\section{Algorithm Design for Congestion-Oblivious Setting}\label{subsec:alg+surrogate+primal}
As evident from \Cref{ALG+Surrogate+D}, \CA{} requires access to up-to-date backlog information. While this is not a practical concern for our partner agency (see \Cref{sec:numerics}), such information might not be available in other contexts. Motivated by this, we consider the congestion-oblivious setting and propose a dual-learning algorithm (\CO{}) that does not rely on backlog information. In \Cref{subsec:ALG+P+motivation}, we motivate our technical approach. In \Cref{subsec:ALG+P+design}, we formally introduce \CO{} and establish its sub-linear regret in the stable regime under mild conditions. We further show that in certain regimes where \CA{} still achieves sub-linear regret, \CO{} does not---highlighting the inherent benefit of accounting for backlog.

\subsection{Motivation: Surrogate Primal Problem}\label{subsec:ALG+P+motivation}
To motivate our approach, consider a \emph{fluid} approximation of the offline optimum, where we replace random arrivals and service availabilities with their expectations---or equivalently, relax the capacity and backlog constraints to hold in expectation, replace rewards with their expected values, and assume that a $\ServiceRate_\Locidx$ fraction of matched cases is deterministically processed at each affiliate $i$ in every period. Under this approximation, the stationary (fractional) solution $\Decision^*_{\Timeidx,\Locidx}=\CapRatio_\Locidx$ is feasible and yields zero backlog. This observation motivates our investigation of a sample-path-based \emph{surrogate primal} program that \emph{disregards} the backlog penalty. Our goal is to design an algorithm that competes with this benchmark in net reward while maintaining bounded average backlog.


{\begin{definition}[{Surrogate Primal Problem}]\label{def:primal+surrogate} The surrgoate primal problem is defined by
\begin{eqnarray}
    \displaystyle \SurrogatePrimal(\OverCost) 
\displaystyle  := &\underset{\mathbf{\DecisionVec_\Timeidx} \in 
\FeasibleSet{\Target_\Timeidx}}{\max} 
\displaystyle &\sum_{\Timeidx=1}^\TotalTime\sum_{\Locidx=1}^\Locnum \Weight_{\Timeidx,\Locidx} \Decision_{\Timeidx,\Locidx}
    - \OverCost\sum_{\Locidx=1}^{\Locnum} \PositivePart{\sum_{\Timeidx=1}^\TotalTime \DecisionEach - \Capacity_\Locidx } \displaystyle \nonumber \quad \\
    \displaystyle &\text{s.t.} \displaystyle &\sum_{\Timeidx=1}^\TotalTime \Indicator [\Target_\Timeidx  = 0]\DecisionEach \leq \Big(\Capacity_\Locidx-\sum_{\Timeidx=1}^\TotalTime \Indicator[\Target_\Timeidx = \Locidx]\Decision_{\Timeidx,\Locidx}\Big)_{+} \quad \displaystyle \forall \Locidx \in [\Locnum] \nonumber
\end{eqnarray}
\end{definition}}
Similar to the surrogate dual problem (\Cref{def:dual+surrogate}), the surrogate primal also addresses the time-varying nature of optimal dual variables for dynamic resources in the dual of the optimal offline  (see \Cref{prop:offline}). \revcolor{This time,} it does so by disregarding the \backlog{} penalty in the objective function. \revcolor{Consequently,} a learning-based algorithm for the surrogate primal ignores dual variables associated with dynamic resources and focuses solely on maintaining the dual variables for static resources.

This simple idea---ignoring backlogs and relying on surrogate primal problems---faces important technical obstacles. Prior literature (e.g., \cite{agrawal2014fast}) has explored online-learning-based approaches for solving problems similar to our surrogate primal.  However, despite the fluid approximation having no backlog, it is unclear whether a learning-based algorithm can effectively control backlog under random arrivals and service availability. Controlling backlog requires the algorithm-induced arrival rates (a function of matching decisions of the algorithm) to have sufficiently strong convergence to the stationary solution $\boldsymbol{\CapRatio}$. In particular, the convergence must be (i) fast enough, (ii) almost uniform over time, and (iii) hold with high probability (we will formalize these requirements later). Achieving such strong convergence is not immediately obvious, but we show that it is possible using a \emph{proper time-dependent learning rate}, as discussed next.


\subsection{Algorithm and Analysis }\label{subsec:ALG+P+design}

We now introduce \CO{}, our congestion-oblivious dual-learning algorithm, formally described in \Cref{apx+P+pseudo+code} (\Cref{ALG+Surrogate+P}).  The algorithm's structure closely follows that of \CA{}. For brevity, we only highlight the main differences compared to \CA{} in the following.


\EC{\smallskip}
{\bf Primal Phase}: 
{\CO{} matches the arriving case $t$ to the affiliate $i_t$ that maximizes the new adjusted score, that is, $\DecisionVec_t\in \argmax_{\mathbf{z} \in \Simplex} ({\Weightvec_\Timeidx- \DualOverVec_\Timeidx-\DualHardVec_\Timeidx})\cdot\DecisionVec_t$, which no longer depends on backlog.}


\EC{\smallskip}
{\bf Dual Phase:} Similar to \CA, updates the dual variables for static resources using the multiplicative weight method~\citep{freund1997using}. The key difference is that \CO{} uses a \emph{time-varying} step size $\{\eta_{\Timeidx}\}_{\Timeidx=1}^{\TotalTime}$ to achieve the strong convergence requirements discussed earlier
(and further detailed in \Cref{subsec:ALG+P+static+dual}). Specifically, the update rules \eqref{eq:update+theta} from \Cref{ALG+Surrogate+D} are now replaced by:
\begin{equation}
    \DualOver_{\Timeidx+1, \Locidx} = \min\{\DualOver_{\Timeidx,\Locidx}\exp(\eta_\Timeidx(\DecisionEach-\CapRatio_i)), \OverCost\}, 
    \quad
    \DualHard_{\Timeidx+1,\Locidx} = \min\Big\{\DualHard_{\Timeidx,\Locidx}\exp(\eta_\Timeidx({\DecisionEach-\CapRatio_\Locidx})), \frac{1+2\OverCost}{\underline{\CapRatio}}\Big\}
    \label{line:update+CO}
\end{equation} 
where $\eta_\Timeidx= k/\sqrt{\Timeidx}$ is the time-varying learning rate with an input parameter $k>0$. 

\EC{\smallskip}
We now present our main result for \CO{} in Theorems~\ref{thm:ALG+Surrogate+Primal}. This result relies on the following regularity assumption on the reward distribution, whose implications are discussed later in our analysis.

\begin{assumption}[Lipschitz Continuous Reward Distribution]\label{assump:PDF}
    The PDF of the reward vector is $L$-Lipschitz continuous with respect to $\ell_1$-norm for some constant $L>0$. 
\end{assumption}
\begin{theorem}[Regret of \CO{} Under Stable Regime]\label{thm:ALG+Surrogate+Primal} Let $\eta_\Timeidx= k/\sqrt{\Timeidx}$ with a constant $k>0$. Under Assumption \ref{assump:PDF} and stable regime (\Cref{def:eps+regimes}),  the regret  of \CO{} is $\BigO(\sqrt{\TotalTime}+\frac{\BuildUpCost}{\ServiceSlack})$.
\end{theorem}


\Cref{thm:ALG+Surrogate+Primal} shows that, under the stable regime, \CO{} matches the performance of \CA{}, achieving the same order of regret despite being oblivious to backlog information. However, the following theorem establishes that \CO{} cannot achieve the same regret performance as \CA{} in the near-critical regime.

\begin{theorem}[Lower bound on Regret of \CO{} under Near-critical Regime]\label{thm:near+critical} There exists a constant $q>0$ and an instance such that for any given $\ServiceSlack \leq \frac{q}{\sqrt{\TotalTime}}$, the regret of $\CO{}$ is $\Omega(\gamma\sqrt{\TotalTime})$.
\end{theorem}
We prove \Cref{thm:near+critical} by constructing a simple single-affiliate instance similar to \Cref{example:stable} (will be discussed shortly). The detailed proof, along with numerical illustrations, is presented in \Cref{apx+near+thm+proof}.

\begin{remark}[\textbf{Benefit of Backlog Information}]
\label{remark:CO-in-near-critical} 
Together with \Cref{near+part+a} (see \Cref{subsec:ALG+D+pfsketch}), which states that \CA{} achieves a regret of $\BigO(\sqrt{\BuildUpCost\TotalTime})$ in the near-critical regime, \Cref{thm:near+critical} implies that \CA{} outperforms \CO{} in terms of regret by at least a factor of $\sqrt{\BuildUpCost}$. In particular, \CO{} fails to provide a sub-linear regret if $\BuildUpCost = \Theta(\TotalTime^{\delta})$ with $\delta \in [1/2, 1)$. Intuitively, this difference arises because \CA{} can adapt its matching decisions by explicitly penalizing the adjusted scores via (scaled version of) backlogs, underscoring the inherent advantage of explicitly accounting for congestion.
\end{remark}


We now turn our focus to the proof of \Cref{thm:ALG+Surrogate+Primal}. The key step is to establish that \CO{} achieves a constant average \backlog{} in expectation under the stable regime, even though \emph{completely} ignores backlog information. As the main building block, we show that a sufficient condition for this result is for the dual variables to converge to their optimal values \emph{in the last iteration (i.e., in every iteration) with high probability}, which we formalize and verify for \CO{} in \Cref{subsec:ALG+P+static+dual}. The following example helps in building an intuition about why this strong convergence guarantee is indeed useful.

\smallskip
\begin{example}\label{example:stable}
Consider an instance with $T =5000$, a single affiliate, and no tied cases. For notational convenience, we omit the subscript $i=1$. The rewards are uniformly distributed on $(0,1)$, the capacity ratio is $\CapRatio=0.5$, and the service slack is $\ServiceSlack=0.1$. For this instance, \CO{}’s matching decision reduces to $\Decision_\Timeidx = \Indicator[\Weight_\Timeidx \geq \phi_\Timeidx]$, where $\phi_\Timeidx \triangleq \DualOver_\Timeidx + \DualHard_\Timeidx$. The optimal dual variable $\hat{\phi}$ of the surrogate primal is the median of $(w_1,..,w_\TotalTime)$, \footnote{For this arrival sequence, the dual of the surrogate primal problem is $\min_{\phi \geq 0} D(\phi) := \sum_{\Timeidx=1}^\TotalTime \{(w_\Timeidx - \phi)_{+} + 0.5\phi\}$. Its derivative (when differentiable) is $-\sum_{\Timeidx=1}^\TotalTime \Indicator[w_\Timeidx \geq \phi] + 0.5\TotalTime$, which is negative (positive) if $\phi$ is below (above) the sample median. Hence, the sample median of $(w_1,..,w_\TotalTime)$ minimizes this dual function. \label{footnote:sample+median}} 
which is ``highly concentrated'' around $0.5$ (the median of reward distribution). With $10,000$ sample paths, the $5\%$ quantile of $\hat{\phi}$ is $0.5 \pm 0.01$. Thus, we let $\hat{\phi} = 0.5$ for ease of exposition.



\begin{figure}[t]
    \centering
    \includegraphics[scale=0.5]{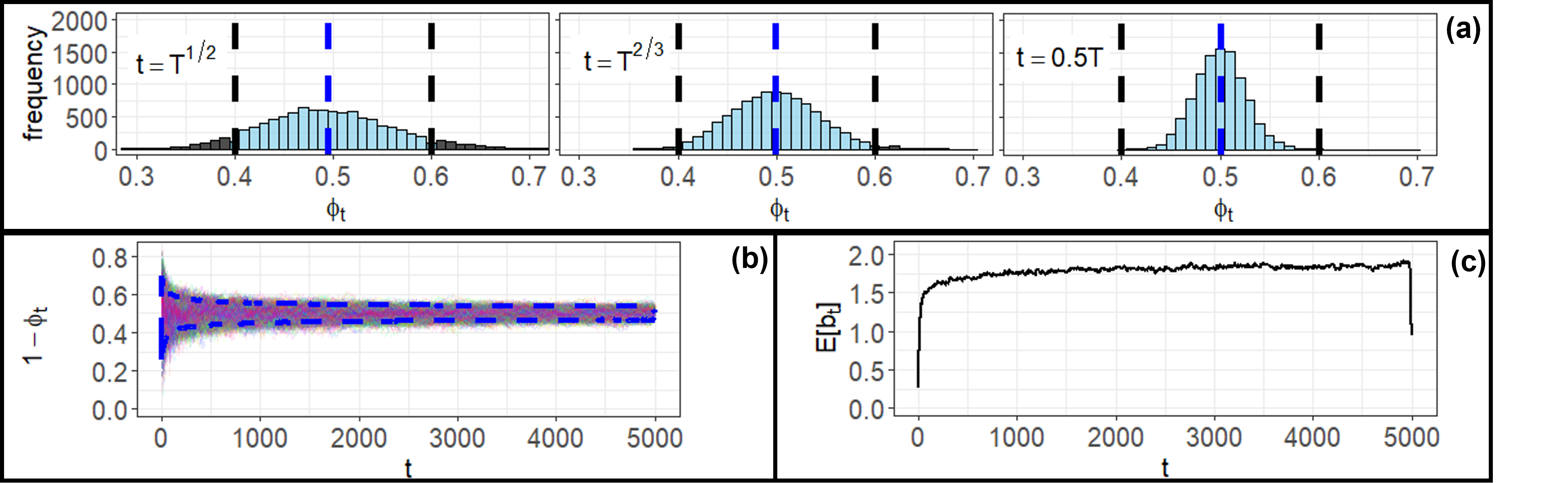}
    \caption{Numerical illustrations for \Cref{example:stable} with $\boldsymbol{\eta_\Timeidx = 1/\sqrt{\Timeidx}}$, $\boldsymbol{\TotalTime=5000}$, and 10,000 sample paths. Panel (a) shows histograms of $\boldsymbol{\phi_\Timeidx}$ for three values of $\boldsymbol{\Timeidx}$. Panel (b) displays sample paths of the endogenous arrival rates $\boldsymbol{1-\phi_\Timeidx}$, with 5\% and 95\% quantiles (blue dashed lines). Panel (c) depicts the expected backlog for each $\boldsymbol{\Timeidx}$.
  }
    \label{fig:stable}
\end{figure}

\smallskip
With post-allocation service and stochastic service time, we can interpret the above example as a simple queueing system where the arrival rate at each period is time-varying and determined \emph{endogenously} by \CO{}'s matching decisions. Specifically, the conditional arrival rate is given by $\E[\Decision_\Timeidx|\History_{\Timeidx-1}] = \E[\Indicator[w_\Timeidx \geq \phi_\Timeidx]|\History_{\Timeidx-1}]=1-\phi_\Timeidx$. Suppose for a moment we choose decisions according to $\hat{\Decision}_\Timeidx = \Indicator[w_\Timeidx \geq 0.5]$. It is straightforward to show that the resulting average backlog remains constant (in expectation), as its backlog process $\hat{b}_\Timeidx$ behaves like a non-negative random walk with negative drift at each period $\Timeidx$.\footnote{{More precisely, using the drift lemma (\Cref{lemma:drift}), we have, for any $\Timeidx$, that $\E[\psi(\hat{b}_\Timeidx) - \psi(\hat{b}_{\Timeidx-1})|\History_{\Timeidx-1}] \leq b_{\Timeidx-1}\E[\hat{z}_\Timeidx - \Service_\Timeidx|\History_{\Timeidx-1}] + \BigO(1) \leq -0.1 b_{\Timeidx-1} + \BigO(1)$. Summing this inequality over all periods $\Timeidx \in [\TotalTime]$ and taking the expectations, we deduce that the average backlog is $\BigO(1)$ in expectation.}}


Returning to \CO{}, although $\phi_\Timeidx$ is not exactly equal to 0.5,  we have designed its learning process (particularly the time-varying learning rates) so that $\phi_\Timeidx$ highly concentrates around $0.5$ for sufficiently large $t$, and gets even \emph{more} concentrated as time progresses. To illustrate this convergence, \Cref{fig:stable}-(a) shows the distribution of $\phi_\Timeidx$ at three different periods $\Timeidx \in \{\TotalTime^{1/2}, \TotalTime^{2/3}, 0.5\TotalTime\}$. 
The strong concentration around 0.5 implies that the induced arrival rate $\E[\Decision_\Timeidx|\History_{\Timeidx-1}] = 1-\phi_\Timeidx$ consistently stays close to $0.5$, which is well below the service rate $\ServiceRate=0.6$, with high probability (as illustrated in \Cref{fig:stable}-(b)). This ensures, with high probability, a negative drift in backlog process at almost every period, which is sufficient to show a constant expected average backlog (as confirmed in \Cref{fig:stable}-(c)).

\end{example}

Building on these intuitions, in \Cref{subsec:ALG+P+static+dual}, we first formalize a high probability last-iterate convergence property for the dual variables constructed by \CO{}. Then, in \Cref{subsubsec:ALG+P+pf+sketch}, we show that this property ensures a negative drift in backlog for \CO{} with high probability at nearly every period. This enables us to finish the proof \Cref{thm:ALG+Surrogate+Primal} by showing the constant average \backlog{} of \CO{}.


\subsubsection{High Probability Last-iterate Convergence of Dual Variables }\label{subsec:ALG+P+static+dual}\hfill\\
To establish our desired convergence of the dual variables in \CO{}, we first define a \emph{static dual problem}. 

\begin{definition}[Static Dual Problem \& Dual-based Decision]\label{def:staic+dual}
The static dual problem and its optimal solution $\boldsymbol{\nu^*}$ are defined as follows.
\begin{equation*}
\begin{split}
D(\boldsymbol{\nu}):= \E\Big[
\max_{\DecisionVec \in \FeasibleSet{\Target}}({\Weightvec - \boldsymbol{\DualOver} - \boldsymbol{\DualHard}})\cdot \DecisionVec + \CapRatioVec\cdot \boldsymbol{\DualOver} + \CapRatioVec\cdot \boldsymbol{\DualHard}\Big],
\quad
    D(\boldsymbol{\nu}^*):= \min_{\boldsymbol{\nu} \in \mathcal{V}} 
D(\boldsymbol{\nu})
\end{split}
\end{equation*}
where $\boldsymbol{\nu} := (\DualOverVec,\DualHardVec)$, $\mathcal{V} := [0, \alpha]^m \times [0, \frac{1+2\OverCost}{\underline{\CapRatio}}]^m$, and the expectation is taken over the stochastic arrival $\Arrival =(\Weightvec, \Target)\sim \ArrivalDist$.  
Further, for \ECKEEP{a given }dual variable $\boldsymbol{\nu}\in\mathcal{V}$ and arrival type $\Arrival\in\ArrivalSupp$, we define its corresponding \ECKEEP{(primal) }decision as 
\begin{equation*}
 \tilde{\mathbf{z}}(\boldsymbol{\nu}, \Arrival) := \argmax_{\mathbf{z} \in \FeasibleSet{\Target}} \{({\Weightvec - \DualOverVec - \DualHardVec})\cdot \DecisionVec \}.
\end{equation*}
\end{definition}
\EC{\smallskip}
The following proposition (formally proven in detail in \Cref{apx+harvey}) establishes the desired high probability last-iterate convergence of dual variables $\boldsymbol{\nu}_\Timeidx$ constructed by \CO{} to  $\boldsymbol{\nu^*}\in \argmin_{\boldsymbol{\nu} \in \mathcal{V}} 
D(\boldsymbol{\nu})$.


\begin{proposition}[High-probability Last Iterate Convergence of Dual Variables] \label{prop:harvey} 
Let $\Stopping$ be the stopping time of \CO{}.\footnote{
The stopping $\Stopping$ for \CO{} is identically defined as the one for \CA{} given in equation \eqref{def:stopping}.
} For any fixed $\Timeidx \leq \Stopping$ and $\delta \in (0,1)$, we have
\begin{equation*}
\P\left[D(\boldsymbol{\nu}_\Timeidx) - D(\boldsymbol{\nu}^*) \leq 
    \BigO\Big(
\frac{\log(\Timeidx)\log(1/\delta)}{\sqrt{\Timeidx}}
    \Big) \right] \geq 1 -\delta.
\end{equation*}
\end{proposition}

We sketch the proof of \Cref{prop:harvey}, which relies on the connection between the dual update rule of \CO{} in \eqref{line:update+CO} and the online stochastic mirror descent (OSMD) \citep{nemirovski2009robust} for solving the static dual problem (\Cref{def:staic+dual}). Specifically, the gradient of the static dual function $D(\cdot)$ is $
{\nabla D(\boldsymbol{\nu}) = (\E_{\mathbf{A}}[\CapRatioVec - \tilde{\DecisionVec}(\boldsymbol{\nu}, \Arrival)], \E_{\mathbf{A}}[\CapRatioVec - \tilde{\DecisionVec}(\boldsymbol{\nu}, \Arrival)])}$. Therefore, up to $\Timeidx \leq \Stopping$, the dual update in \eqref{line:update+CO} is a mirror descent update, with negative entropy as the mirror map $h(\boldsymbol{\nu}) = \sum_{\Locidx=1}^{\Locnum} \DualOver_\Locidx\log(\DualOver_\Locidx) + \sum_{\Locidx=1}^{\Locnum} \DualHard_\Locidx\log(\DualHard_\Locidx)$:
\begin{equation}
    \boldsymbol{\nu}_{t+1} = \argmin_{\boldsymbol{\nu} \in \mathcal{V}}~\mathbf{\hat{g}}_t \cdot \boldsymbol{\nu} + \frac{1}{\eta_t}V_h(\boldsymbol{\nu}, \boldsymbol{\nu}_t),
\end{equation}
where $V_h(\cdot,\cdot)$ is the Bregman distance with respect to $h$~\citep{bubeck2015convex}. Here, the stochastic gradient is ${\mathbf{\hat{g}}_t = (\boldsymbol{\rho} - \tilde{\mathbf{z}}(\boldsymbol{\nu}_t, \mathbf{A}_t), \boldsymbol{\rho} - \tilde{\mathbf{z}}(\boldsymbol{\nu}_t, \mathbf{A}_t) )}$, which is an unbiased estimator for the true gradient $\nabla D(\boldsymbol{\nu}_\Timeidx)$ for all $\Timeidx \leq \Stopping$. This is because $\mathbb{E}[\mathbf{\hat{g}}_{\Timeidx}|\History_{\Timeidx-1}] = \nabla D(\boldsymbol{\nu}_{\Timeidx})$ for all $\Timeidx \leq \Stopping$, as (i) each arrival $\mathbf{A}_\Timeidx$ is an i.i.d. sample from the distribution $\ArrivalDist$ and (ii) $\boldsymbol{\nu}_\Timeidx$ is $\History_{\Timeidx-1}$-measurable. Given this connection, the key step in our proof is establishing the last-iterate high probability convergence for our variant of OSMD. For this, we closely follow the approach of \cite{harvey2019tight}, who proved an analogous result for stochastic gradient descent---a special case of OSMD---and adapt their proof to our variant of OSMD.



\majorrevcolor{
\begin{remark}[\textbf{Fixed versus Time-varying Learning Rates}]\label{remark:fixed+step}
So far, we have established the convergence of $\boldsymbol{\nu}_t$ at \emph{any} $t$ (up to the stopping time), using the time-varying learning rate $\eta_t = \Theta(1/\sqrt{t})$. In \Cref{subsubsec:ALG+P+pf+sketch}, we leverage this uniform-in-time convergence to prove a constant average backlog for \CO{} via drift analysis. At a high level, this drift analysis involves comparing the endogenous arrival rate---determined by $\boldsymbol{\nu}_t$---to the service rate $\rho_i + \epsilon$ at each period $t$, thus requiring consistent concentration of $\boldsymbol{\nu}_t$ throughout (nearly) the entire horizon. Such uniform per-period drift control is standard in queueing theory, where stability typically follows from uniform-in-time drift bounds. By contrast, using a fixed learning rate (e.g., $\eta = \Theta(1/\sqrt{T})$) only guarantees convergence at the \emph{final} time $T$, which is insufficient for our analysis. Hence, the time-varying step size is crucial for our results.
\end{remark}}

\subsubsection{Proof Sketch of \texorpdfstring{\Cref{thm:ALG+Surrogate+Primal}}{}}\label{subsubsec:ALG+P+pf+sketch}\hfill\\
With all the ingredients discussed in \Cref{subsec:ALG+P+static+dual}, we are now ready to sketch the proof of \Cref{thm:ALG+Surrogate+Primal}. As highlighted earlier, the main part is the following lemma, which establishes an upper bound on the expected average backlog in the stable regime via drift analysis (formally proven in \Cref{apx+thm+2+pf}).

\begin{restatable}[Bounding Average \Backlog{} of \CO{} Under Stable Regime]{lemma}{COBacklog}
\label{Lemma+ALG+P+Build} Under Assumption \ref{assump:PDF} and stable regime (\Cref{def:eps+regimes}), for any arrival distribution $\ArrivalDist$ and \revcolor{service slack} $\ServiceSlack > 0 $, we have $    \E\Big[\dfrac{1}{T}\sum_{t=1}^T \lVert \mathbf{b}_t \rVert_1 \Big]\leq \BigO\Big(\dfrac{1}{\ServiceSlack}\Big)$.
\end{restatable}

With the above lemma, the remaining step is to bound the loss of the net matching reward (\Cref{lemma:ALG+P+reward}, \Cref{apx+P+reward}), similar to \Cref{lemma:ALG+D+reward} for \CA{}. We defer proof details of \Cref{thm:ALG+Surrogate+Primal} to \Cref{apx+thm+2+pf}, and dedicate the remainder of this section to outlining the proof of \Cref{Lemma+ALG+P+Build}.

The crux of the proof is establishing an $\BigO(1/\ServiceSlack)$ bound on the average backlog up to the stopping time. 
As alluded to earlier, it suffices to show that the backlog has a negative drift with high probability in (almost) every period before the stopping time. To that end, define the following ``good event'' $ \mathcal{B}_{\TotalTime}$:
\begin{equation}
    \mathcal{B}_{\TotalTime} := \Big\{ \E_\Arrival[\tilde{z}_i(\boldsymbol{\nu}_\Timeidx, \Arrival)] - (\rho_i+\ServiceSlack)\leq -\frac{\epsilon}{2} \ \text{for all $\sqrt{\TotalTime} \leq \Timeidx \leq \Stopping $ and $\Locidx \in [\Locnum]$}\Big\} ~.\label{line:good+event+def} 
\end{equation}
First, let us unpack why the occurrence of the good event $ \mathcal{B}_{\TotalTime}$ is sufficient to establish a constant average backlog. Recall that the endogenous arrival rate at period $\Timeidx\leq \Stopping$ is given by $\E[\DecisionVec_\Timeidx|\History_{\Timeidx-1}] = \E_\Arrival[\tilde{\mathbf{z}}(\boldsymbol{\nu}_\Timeidx, \Arrival)]$, while the service rate for each affiliate $\Locidx$ is given by ${\CapRatio}_\Locidx + \ServiceSlack $. Therefore, $\E_\Arrival[\tilde{z}_i(\boldsymbol{\nu}_\Timeidx, \Arrival)] - (\CapRatio_\Locidx +\ServiceSlack)$ roughly serves as the drift of the backlog process for each affiliate $\Locidx$, which is always negative under the good event. In the proof of \Cref{lemma:B-1} (a required technical lemma, \Cref{apx+B-1}), we use a Lyapunov argument with the quadratic potential function $\psi$ (equation \eqref{psi}) to formally show that such negative drift leads to an $\BigO(1/\ServiceSlack)$ bound on the average backlog up to the stopping time.

The remainder of the proof of \Cref{Lemma+ALG+P+Build} involves establishing that the good event occurs with high probability, as formalized in \Cref{lemma:B}. The proof is fairly intricate and here we only provide a high-level overview. Our proof crucially relies on Assumption \ref{assump:PDF} and the convergence result of \Cref{prop:harvey}. Under Assumption \ref{assump:PDF}, we can translate the high probability convergence of $D(\boldsymbol{\nu}_\Timeidx)$ to $D(\boldsymbol{\nu}^*)$ (\Cref{prop:harvey}) into convergence of $\E_\Arrival[\tilde{\mathbf{z}}(\boldsymbol{\nu}_\Timeidx, \Arrival)]$ to $\E_\Arrival[\tilde{\mathbf{z}}(\boldsymbol{\nu}^*, \Arrival)]$. Furthermore, as shown in \Cref{apx+fluid+dual}, the optimal solution $\boldsymbol{\nu}^*$ of the static dual problem (\Cref{def:staic+dual}) satisfies $\E_{\Arrival}[\tilde{\DecisionVec}(\boldsymbol{\nu}^*, \Arrival)] \leq \CapRatioVec$ (with equality whenever $\DualOver^*_\Locidx >0$). Hence, the endogenous arrival rates $\E_\Arrival[\tilde{\mathbf{z}}(\boldsymbol{\nu}_\Timeidx, \Arrival)]$ concentrate around a value at most $\CapRatioVec$ for (almost) every period $\Timeidx \leq \Stopping$, ensuring the good event occurs with high probability.
 

%

\section{Case Study on Refugee Resettlement Data}
\label{sec:numerics}

In this section, we numerically evaluate our learning-based approach using data from our partner agency. We first describe the construction of primitives and minor algorithmic adaptations made to meet practical requirements. We then present numerical results showing that our algorithm outperforms both the current practice and other algorithms proposed in the refugee matching literature.



\subsection{Data and Benchmarks}\label{subsec:data}


{\bf Data and Primitives.} We use data on working-age refugees (18 to 64 years old) resettled by our partner agency during 2014--2016. 
In 2015, our partner resettled $\TotalTime = 3819$ cases across $m=45$ affiliates. We set the affiliate capacities to the actual number of cases resettled in each affiliate (institutional restrictions prevent us from accessing exact knowledge of initial capacities used by our partner).
We use the \emph{actual sequence} of refugee arrivals to construct a sequence of type arrivals $\Arrival_\Timeidx = (\Weightvec_\Timeidx, \Target_\Timeidx)$. For the reward vector $\Weightvec_\Timeidx$, we follow the methodology of \citet{bansak2018improving}, using a supervised machine learning model to predict employment probabilities for each case--affiliate pair based on demographic characteristics. \majorrevcolor{Specifically, our partner has trained a gradient boosting model on historical data from prior years, to estimate these probabilities for 2015 arrivals.}\footnote{
\label{footnote:ML+1}
\majorrevcolor{We emphasize that historical data from prior years is used solely to estimate employment probabilities, not to predict future arrival patterns. As previously discussed (e.g., \Cref{fig:tied+proportion}) and further illustrated later (e.g., \Cref{fig:across+years}), predicting future arrivals is considerably more challenging since the composition of arrivals varies  from year to year.}
} Our data also includes indicators for cases with U.S. ties, which must be matched to predetermined affiliates. For these tied cases, we set the target affiliate to be the one where the case was actually placed. The remaining cases are considered free and can be matched to any affiliate.
To measure \backlog{}, following the approach of \citet{bansak2024outcome}, we adopt a deterministic service flow with service rates equal to $\boldsymbol{\CapRatio}$. This particular backlog measure is a metric currently used by our partner agency to assess congestion. We apply the same procedure to construct primitives for 2016 ($T = 4950$). \majorrevcolor{For both 2015 and 2016, roughly 70\% of arrivals were tied cases.} Data from 2014 is reserved for parameter tuning, as will be discussed shortly.



\smallskip
{\bf Algorithms under Evaluation.} We compare the performance of the following policies:\par

{\emph{(i) Robust Online Learning-based Algorithm (\texttt{RO-Learning})}.} 
This policy is an adaptation of our \CA{} (\Cref{ALG+Surrogate+D}), modified to remove the dummy affiliate entirely. Specifically, each arriving free case is matched to an actual affiliate $\Locidx$ with the highest adjusted score (possibly negative) among those with remaining capacity, and dual variables are updated according to line \eqref{eq:update+theta} of \Cref{ALG+Surrogate+D}. Under our primitives, since $\sum_{\Locidx=1}^\Locnum \CapRatio_\Locidx =1$, there is always at least one affiliate with remaining capacity at each period $\Timeidx \in [\TotalTime]$. This adaptation mirrors real-world heuristics, as each case  must be matched to an actual affiliate (see also \Cref{remark+no+inaction} on incorporating this adaptation into our theoretical analysis). \CA{} relies on two learning rates, $\eta$ and $\zeta$. We tune them in a data-driven manner to achieve reasonable ranges of over-allocation and average backlog, using 2014 data. Further details are provided in \Cref{apx+case+ALG+D}.\par

\majorrevcolor{ 
{\emph{(ii) Congestion-Oblivious Learning-based Algorithm (\CO{})}.}  
This policy is a minor modification of \CO{} (Sections \ref{subsec:alg+surrogate+primal} and \ref{apx+case+ALG+D}, \Cref{ALG+Surrogate+P}), similarly ensuring that cases are matched only to actual affiliates. The step size $\eta_t$ in line~\eqref{line:update+CO} is tuned using the same data-driven approach for \texttt{RO-Learning}.\par
}

{\emph{(iii) Sampling-based Algorithm (\texttt{Sampling})}}. As motivated in \Cref{sec:intro}, the existing proposals in the refugee matching literature are based on a \emph{sampling} approach, which simulates future arrivals using past years' data. We use (an adaptation of) the algorithm from \citet{bansak2024outcome} as our main benchmark based on this approach. Their original algorithm is developed for the case with no tied cases. See Sections~\ref{subsec:lit+review} and \ref{apx+case+Sampling}, detailing how we modify this algorithm to handle tied cases.

\texttt{Sampling} was also designed to maximize the same objective \eqref{ALG+obj} as ours, with the current congestion level as a penalty term scaled with $\BuildUpCost$. However, a crucial difference is that \EPedit{\texttt{Sampling}} relies on sample trajectories of future arrivals: to match case $\Timeidx$, \texttt{Sampling} solves an offline problem for periods $\Timeidx$ and onward by sampling future arrivals from a given sampling pool. Following \citet{bansak2024outcome}, we use arrivals from \emph{the previous year} as this sampling pool. For example, when tested on the 2015 data, \texttt{Sampling} uses 2014 arrivals to simulate future cases. Thus, the performance of \texttt{Sampling} intuitively depends on how representative the sampling pool is for the current arrival sequence.


{\emph{(iv) Actual Placements} (\texttt{Actual}).} Under current practice, case officers at our partner agency manually determine initial case placements. \ECedit{As highlighted in \Cref{sec:intro}, the agency provides essential services (such as case management) to refugees after placement. Beyond respecting annual quotas, preventing congestion in these services is a critical consideration for the agency. Hence, current placements have been primarily driven by workload balancing and capacity constraints, without systematically accounting for employment outcomes \citep{bansak2024outcome}.} We use actual case officer decisions as a benchmark to assess how our learning-based approach can improve the agency's current practice.

\SLdelete{We have also run \CO{} (\Cref{subsec:alg+surrogate+primal}) with the same primitives calibrated by our dataset as above. In \Cref{apx+CO}, we show that \CO{} performs worse than \CA{} based on the results of our simulations using our partner agency's data. As a result, our main algorithm proposed to our partner agency, which is also the main algorithm evaluated numerically in this section, is \texttt{RO-Learing} (adaptation of \CA{}).}

\subsection{\revcolor{Simulation Results}}\label{subsec:results}

\begin{table}[t]
    \centering
    \caption{Results for year 2015 (2016, resp.) for penalty parameters $\boldsymbol{\OverCost=3}$ and $\boldsymbol{\BuildUpCost=5}$.  The total number of cases is $\boldsymbol{T=3819}$ (4950, resp.) for year 2015 (2016, resp.). For \texttt{Sampling}, we take an average over five simulations. 
    }
    \label{fig:performance}
    \footnotesize
\begin{tabular}{@{}lccc@{}}
    \toprule
    & \textbf{Employment Rate (\%)} 
    & \textbf{Total Over-allocation} 
    & \textbf{Average Backlog}  \\
    \midrule
    \texttt{Actual}        & 37.3 (37.7)   & 0 (0)         & 226.6 (323.5)    \\
    \texttt{Sampling}      & 45.0 (46.3)   & 107.6 (99.8)  & 180.1 (225.2)  \\
    \texttt{RO-Learning}   & 44.6 (46.0)   & 71 (81)       & 151 (199)       \\
    \texttt{\CO{}}         & 44.4 (46.3)   & 123 (84)      & 202.5 (251.0)   \\
    \bottomrule
\end{tabular}
\end{table}
\textbf{Main Result:} \Cref{fig:performance} reports results for penalty parameters $\OverCost = 3$ and $\BuildUpCost = 5$. The first column shows the employment rate, measured as the percentage of total employment outcomes (including tied cases) relative to total arrivals $\TotalTime$. The second and third columns show the total over-allocation (summed across all affiliates) and the average \backlog{}, respectively. Since affiliate capacities were set based on actual placements, \texttt{Actual} incurs no over-allocation by construction. Nevertheless,  \texttt{RO-Learning} significantly improves upon \texttt{Actual} in both employment rates and the average \backlog{}---for example, by roughly {20}\% and 33\%, respectively, in 2015. We also highlight that the employment rate in \Cref{fig:performance} includes both tied and free cases. If we exclude tied cases (for which all algorithms make the same decision), \texttt{RO-Learning} improves employment rate of free cases by roughly 50\% compared to \texttt{Actual}.\footnote{\label{footnote:caution}That said, we remark that \texttt{Actual} may have been influenced by certain unobserved constraints not accounted for in this case study (e.g., unobserved incompatibility between certain free cases and affiliates). Therefore, the improvement over \texttt{Actual} should not be interpreted at face value but rather as an indication of the potential benefits that \texttt{RO-Learning} could offer when implemented in real-world scenarios.}



Moving beyond the comparison to actual placements, we now evaluate \texttt{RO-Learning} against the more sophisticated  \texttt{Sampling} benchmark. For $\OverCost=3$ and $\BuildUpCost=5$, \texttt{RO-Learning} improves the combined objective (employment, backlog, and over-allocation; see \eqref{ALG+obj}) by roughly {48}\% for 2015 ({20}\% for 2016). To understand this improvement further, we compare each of the three outcomes separately in \Cref{fig:performance}. \texttt{RO-Learning} significantly reduces both over-allocation and average \backlog{} in both years, with minimal impact (less than 1\%) on employment outcomes. For instance, in 2015, \texttt{RO-Learning} achieves roughly  {34}\% lower total over-allocation and {16}\% lower average \backlog{} compared to \texttt{Sampling}.
\majorrevcolor{
Lastly, although \CO{} does not use backlog information, it still substantially improves employment outcomes (by 19.0\%–22.8\%) and average backlog (by 10.6\%–22.4\%) relative to \texttt{Actual}. However, it underperforms \texttt{RO-Learning}, incurring higher over-allocation and average backlog.\footnote{
\majorrevcolor{In \Cref{apx+CO}, we further show that \texttt{RO-Learning} consistently outperforms \texttt{CO-DL} over a wide range of penalty parameters, although this performance gap narrows as the congestion penalty parameter $\gamma$ decreases.}}   Thus, our primary recommendation to the partner agency is \texttt{RO-Learning} (an adaptation of \CA{}).
}
\begin{figure}[t]
\majorrevcolor{    
    \centering
\includegraphics[width=0.85\linewidth]{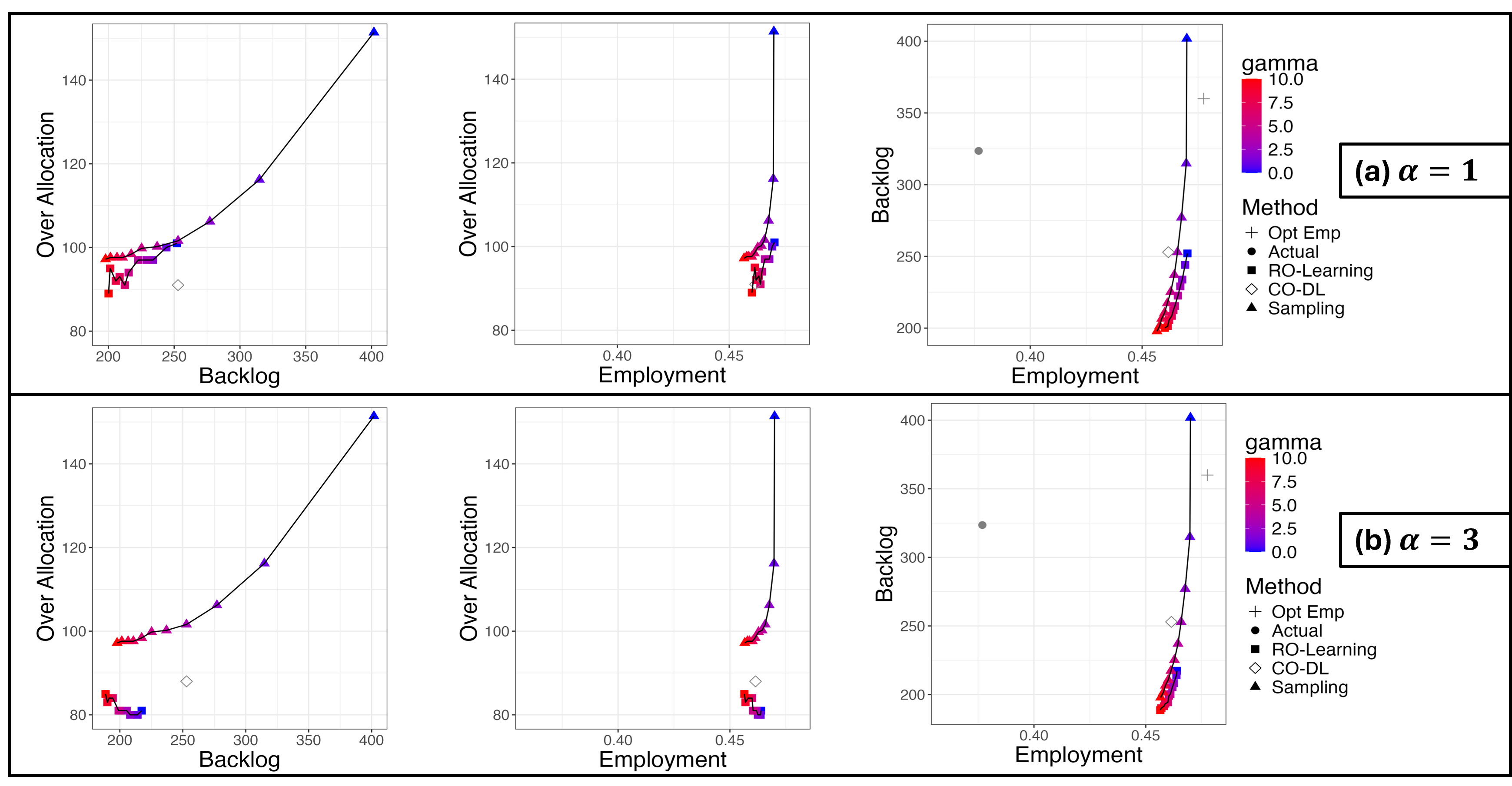}
  \caption{
    \majorrevcolor{Numerical results for year 2016 with $\boldsymbol{\OverCost \in \{1,3\}}$  and $\boldsymbol{\BuildUpCost \in \{0,\ldots,10\}}$. 
  }}
  \label{fig:frontier-2016-summary}
}
\end{figure}


To further understand how penalty parameters impact the three outcomes under different algorithms, we simulate over a range of values $\OverCost \in \{1,2,..,5\}$ and $\BuildUpCost \in \{0,1,...,10\}$. In \Cref{fig:frontier-2016-summary}, we present trajectories of the three outcomes for 2016 with $\OverCost = 1$ and $3$ while varying $\BuildUpCost$, deferring a more comprehensive analysis to Section~\ref{appx:full-param-sweep}. Each subfigure includes three panels, each comparing two of the three key outcomes. Results are shown for all of the benchmarks introduced in \Cref{subsec:data}. We also include  \texttt{Opt Emp}, which corresponds to the surrogate primal problem in \Cref{def:primal+surrogate} and represents the sample-path-based maximum employment outcome attainable by any (offline or online) algorithm. Note that, by construction of our arrival sequence, \texttt{Opt Emp} incurs no over-allocation.

Focusing on the right panels, we first observe that the employment outcomes for both \texttt{RO-Learning} and \texttt{Sampling} are close to \texttt{Opt Emp}, indicating that the accounting for backlogs does not necessarily compromise employment outcomes significantly. However, increasing the backlog penalty parameter $\BuildUpCost$ slightly reduces the employment rate. Additionally, compared to \texttt{Actual}, both \texttt{RO-Learning} and \texttt{Sampling} achieve substantially higher employment rates while maintaining lower average backlogs.

Comparing \texttt{RO-Learning} and  \texttt{Sampling}, we observe that the trajectory of the outcome metrics for \texttt{RO-Learning} generally dominates that of \texttt{Sampling}. For example, given a specific employment outcome, \texttt{RO-Learning} consistently achieves a lower over-allocation and average \backlog{}. Similar trends are evident in the tradeoff between over-allocation and backlog. Furthermore, over-allocation for \texttt{RO-Learning} tends to decrease as $\OverCost$ increases. Taken together, these improvements lead to a consistently better overall objective value~\eqref{ALG+obj} for \texttt{RO-Learning} over \texttt{Sampling}, with a median improvement of 37\% in 2015 and 17\% in 2016, across parameters $\alpha \in \{1,2,\ldots,5\}$ and $\gamma \in \{0,1,\ldots,10\}$.

\smallskip
{\bf Performance Improvement through Robustness:} 
To understand the primary source of performance gains, we highlight the \emph{robustness} inherent in our learning-based approach. \texttt{Sampling} heavily relies on a sampling pool of arrivals from the previous year, making it vulnerable to discrepancies between this pool and actual current-year arrivals (as evidenced in \Cref{fig:tied+proportion} in \Cref{subsec:intro+background}). In contrast, \texttt{RO-Learning} avoids this pitfall by adaptively learning arrival patterns directly from current-year data. 

\smallskip
{\bf Practical Benefits of \texttt{RO-Learning}:}
We highlight additional practical advantages of adversarial online learning employed by \texttt{RO-Learning}. First, the algorithm is inherently balanced due to its \emph{self-correcting} nature. Specifically, beyond explicitly penalizing \backlog{}, the dual-variable update rule (line~\eqref{eq:update+theta} in \Cref{ALG+Surrogate+D}) naturally prevents bursty matchings to each affiliate, as dual variables corresponding to an affiliate automatically  increase when a case is matched to that affiliate, and decrease otherwise. This self-correction tends to align the algorithm's endogenous arrival rate closely with $\boldsymbol{\CapRatio}$. In contrast, \texttt{Sampling} would maintain this property \emph{only} if the sampling pool accurately represents the true underlying distribution, which is not the case in our data. Indeed, as shown in the right panels of \Cref{fig:more+case}, \texttt{RO-Learning} maintains a relatively low backlog even when $\BuildUpCost=0$, comparable to \texttt{Sampling} at higher values of $\BuildUpCost$. Second, \texttt{RO-Learning} is significantly advantageous over \texttt{Sampling} in terms of computational efficiency; Thanks to online learning, unlike \texttt{Sampling}, it does not need to solve any (heavy) auxiliary optimization problem in each period---which can be valuable since, in practice, policymakers may need to run multiple simulations for parameter tuning or back-testing on various datasets.
Lastly, the simplicity and interpretability of \texttt{RO-Learning}'s score-based rule makes it easy to communicate with the practitioners at our partner agency.


\majorrevcolor{
\subsection{Additional Numerical Results with  Practical Considerations}\label{subsec:more+numerics}
In this subsection, to enhance the practical relevance of our proposed algorithm, we extend our case study to incorporate additional practical considerations relevant to our partner agency.

{\bf (i) Beyond Aggregate Over-allocation.} 
While resettlement agencies aim to adhere to the annual quotas, they are not enforced as hard constraints: affiliates are typically allowed to exceed their stated capacity by up to 10\% without additional approval~\citep{ahani2021placement}. However, any resettlement exceeding 110\% of an affiliate's capacity requires formal approval by the State Department~\citep{dos2011coopagreement}, incurring extra administrative costs for the  agency.\footnote{
\label{footnote:over-allocation-2}
\majorrevcolor{That said, over-allocation within the 10\% threshold is still costly, as the annual quota is directly linked to the per-capita funding provided by the government (See \Cref{footnote:over-allocation}). Thus, exceeding quotas---even within the permitted range---still places a burden on affiliate resources.}} For this reason, it is important to examine not only total over-allocation (aggregated across all affiliates, as discussed in \Cref{sec:model}), but also how it is distributed---particularly whether, and to what extent, specific affiliates are ``flagged'' for exceeding the 10\% threshold. Table~\ref{table:relative-overallocation} reports the over-allocation for affiliates exceeding 110\% of their capacity, using the same penalty parameters as in \Cref{fig:performance}. We focus on comparing \texttt{RO-Learning} (our main proposal) with \texttt{Sampling}. 
The first column lists the number of flagged affiliates. While \texttt{RO-Learning} flags more affiliates than \texttt{Sampling} in 2016 (5 vs. 3), the largest over-allocation above threshold among these flagged affiliates is smaller. Furthermore, summing all placements above threshold---each requiring formal reporting---\texttt{RO-Learning} yields a lower total. Thus, \texttt{RO-Learning} not only incurs less total over-allocation (as previously shown in \Cref{fig:performance}), but also reduces the administrative burden associated with exceeding the 10\% threshold.

\begin{table}[t]
    \centering
\caption{\majorrevcolor{Over-allocation outcomes for affiliates exceeding 110\% of their capacity for year 2015 (2016, resp.). We use the same penalty parameters in \Cref{fig:performance} and take an average over five simulations for \texttt{Sampling}.}}\label{table:relative-overallocation}
\begingroup
\majorrevcolor{
\footnotesize
    \begin{tabular}{@{}lccc@{}}
        \toprule
        \textbf{Method} & 
        \makecell{\# Flagged Affiliates \\ with >{}10\% Over-allocation} & 
        \makecell{Max Over-allocation \\ above 110\% Capacity } & 
        \makecell{Total Over-allocation \\ above 110\% Capacity} \\
        \midrule
        \texttt{Sampling}      & 2 (3) & 15.2 (5.4) & 19.8 (11.7) \\
        \texttt{RO-Learning}   & 2 (5) & 10.8 (4.10)  & 12.6 (9.3)     \\
        \bottomrule
    \end{tabular}\
    }
\endgroup

\end{table}

{\bf (ii) Varying Case Size.} In the U.S., annual quotas are tracked at the individual level, meaning that each refugee case may consume multiple units of capacity depending on family size. Here, we extend our simulations to incorporate varying case sizes. As noted in \Cref{subsec:model+assumption}, our algorithmic framework can accommodate this variation via multiple knapsack constraints (\Cref{apx+non+unit+size}). We build on this generalized model and algorithms in Section~\ref{apx+non+unit+size}. Specifically, our data includes the number of family members per case. Based on this, we augment each case~$t$ with a family size~$n_t$. We retain the same case-level employment probability vector~$\mathbf{w}_t$ as described in \Cref{subsec:data}.\footnote{
\label{footnote:ML+2}\majorrevcolor{Our partner agency's in-house machine learning model predicts employment probabilities for each individual family  member of each refugee case~$t$ at affiliate~$i$. Following the approach of \cite{bansak2018improving}, we calculate the case-level employment probability~$w_{t,i}$ as the probability that at least one family member secures employment at affiliate~$i$, assuming independence across family members’ employment outcomes.}}
Affiliate capacity~$c_i$ is set to the actual number of individuals placed at affiliate~$i$, with each case consuming~$n_t$ units. Table~\ref{table:varying-size-simulation} reports the results under the same penalty parameters as in  \Cref{fig:performance}. \texttt{RO-Learning} continues to substantially improve employment outcomes and reduce average backlog compared to the current practice (\texttt{Actual}). Relative to \texttt{Sampling}, it reduces over-allocation by 22--47\% and backlog by over 20\%, with only a small loss in employment (within 5\%). Overall, these findings show that our proposed algorithms remain effective even when accounting for varying case sizes.

\begin{table}[t]
    \centering
\begingroup
\majorrevcolor{
    \caption{
    \majorrevcolor{Results with varying case size for 2015 (2016, resp.) under the same penalty parameter in \Cref{fig:performance}.} 
    }
    \label{table:varying-size-simulation}
    \footnotesize
    \begin{tabular}{@{}lccc@{}}
        \toprule
        & \makecell{\textbf{Employment}  \textbf{Rate (\%)}} 
        & \makecell{\textbf{Total} \textbf{Over-} \textbf{allocation}} 
        & \makecell{\textbf{Average}  \textbf{Backlog}} \\
        \midrule
        \texttt{Actual}        & 37.3 (37.9) & 0 (0)             & 641.2 (996.8) \\
        \texttt{Sampling}      & 45.7 (47.5) & 308.4 (366.2)     & 526.9 (792.6) \\
        \texttt{RO-Learning}   & 44.3 (45.2) & 239 (194)         & 414.3 (615.5) \\
        \CO{} & 45.5 (45.2) & 284 (314) & 542.9 (852.7) \\
        \bottomrule
    \end{tabular}
    }
\endgroup
\end{table}

{\bf (iii) Mid-Year Disruptions and Capacity Revisions.} 
While our main numerical study uses data from 2014--2016, recent years have seen significant disruptions in U.S. refugee resettlement policy. Typically, the federal government sets an admissions ceiling at the start of each fiscal year, based on which the agencies plan their total number of arrivals $T$ and annual quotas. However, this ceiling---and accordingly $T$ and capacities---can change mid-year due to policy shifts. Recently, such disruptions have also been accompanied by major shifts in the \emph{composition} of arrivals. For example, in 2017, an executive order reduced the admissions ceiling  and banned free-case admissions from several countries~\citep{howe2017travelban,advancingjustice2019muslimban}. To assess robustness under such disruptions, in \Cref{apx+disrpution}, we provide simulation results based on a synthetic dataset that mimics the 2017 mid-year disruption. Our results indicate that our proposal continues to outperform other benchmarks.}

\section{Conclusion and Future Directions}\label{sec:discussions}

Motivated by our collaboration with a major U.S. refugee resettlement agency, we introduced a new dynamic matching problem capturing key practical considerations: resettling refugees consumes static resources (e.g., annual quotas) and requires dynamic, post-matching services (e.g., translation). The agency must therefore balance matching rewards against congestion costs of dynamic resources, while respecting constraints on static resources. Drawing insights from our partner's data, we developed prior-free, learning-based algorithms that do not rely on historical arrival data and provide provable performance guarantees. As demonstrated in Section~\ref{sec:numerics}, our robust approach significantly improves outcomes and offers several additional practical advantages.

From a theoretical perspective, while our framework shares features with models studied in online (static) resource allocation and matching queues, it differs fundamentally from both. As a result, designing and analyzing low-regret online algorithms required developing novel technical ideas, potentially of independent interest in other problems. Leveraging these new techniques, we established sub-linear regret guarantees against a strong benchmark whenever possible (i.e., when $\BuildUpCost = o(\TotalTime)$). Exploring weaker benchmarks is an interesting theoretical direction for future research. 



\majorrevcolor{{\bf Beyond Refugee Resettlement.}
Our framework can be applied to general dynamic resource allocation problems sharing two structural features: (i) allocation of long-term resources with fixed capacities; and (ii) the need for short-term, local onboarding services as an essential component of such allocation. For example, in hospital admission control, incoming patients must be assigned to one of the  specialized units (e.g., cardiac medicine or general surgery)---each with a fixed number of beds that serve as long-term resources---or be rejected. Each admitted patient then requires local onboarding services, such as initial testing upon arrival to the unit. 
In these contexts, undesirable backlog may arise due to a shortage of dynamic resources. (In the context of emergency departments, see \citet{shi2016models}, which  discusses post-bed-allocation delay caused by factors such as nurse shortages.) 

A similar structure arises in adult foster care: when individuals in need of long-term care are discharged from hospitals, discharge coordinators must assign them to adult foster care facilities, each with a limited intake capacity~\citep{bartle2025faster}. Each assignment initiates local onboarding processes, such as an initial assessment at the assigned facility. Our framework can help managers balance the reward from consuming static resources against the congestion of dynamic resources. Furthermore, in the aforementioned applications, historical data from prior planning horizons may be unreliable. For example, patient flows in hospitals can fluctuate due to seasonal illnesses or emergent events. Our prior-free approach offers robust and computationally efficient decision support by not relying on solving auxiliary optimization problems.}

\section*{Acknowledgment}      
Vahideh Manshadi and Rad Niazadeh gratefully acknowledge the Simons Institute for the Theory of Computing, as this work was
done in part while attending the program on Data Driven Decision Processes. The authors thank Global Refuge for access to data and guidance. The data used in this study were provided under a collaboration research agreement with Global Refuge, which requires that these data be not transferred or disclosed. Kirk Bansak and Elisabeth Paulson are Faculty Affiliates of the Immigration Policy Lab (IPL) at Stanford University and ETH Zurich. This work is associated with IPL's GeoMatch project, for which the authors acknowledge funding from the Charles Koch Foundation, Stanford Institute for Human-Centered Artificial Intelligence, Google.org, and Stanford Impact Labs. The authors extend their gratitude to the entire GeoMatch team for helpful feedback. Soonbong Lee and Vahideh Manshadi also extend their gratitude to the Management Science and Engineering Department at Stanford University for hosting them during part of this research. Rad Niazadeh’s research is partially supported by the Asness Junior Faculty Fellowship at Chicago Booth School of Business.  

\bibliographystyle{plainnat}
\OneAndAHalfSpacedXI
{\footnotesize
\bibliography{refs}}
\newpage
\renewcommand{\theHchapter}{A\arabic{chapter}}
\renewcommand{\theHsection}{A\arabic{section}}
\ECSwitch
\majorrevcolor{
\section{Summary of Notation and Assumptions}\label{apx+notation}

\begin{table}[htbp]
\centering
\begingroup
\majorrevcolor{
\caption{\majorrevcolor{Summary of Main Notation}}
\label{tab:notation}
\begin{tabular}{p{3.5cm} p{11cm}}
\toprule
\textbf{Symbol} & \textbf{Description} \\
\midrule
$T$ & Total number of arrivals. \\
$m$ & Number of affiliates. \\
$\Simplex$ & $m$-dimensional standard simplex ($\{\mathbf{z} \in \mathbb{R}_+^m : \sum_{i=1}^m z_i \leq 1\}$). \\
$x_+ := \max\{x, 0\}$ & Positive part of scalar $x \in \mathbb{R}$. \\
$t \in [T]$ & Index for arrival period (one arrival per period). \\
$i \in [m]$ & Index for affiliates. \\
$\mathbf{w}_t \in [0,1]^m$ & Reward vector for arrival $t$, where $w_{t,i}$ indicates employment outcome (estimated probability of employment) at affiliate $i$. \\
$i_t^\dagger \in [m] \cup \{0\}$ & Target affiliate for arrival $t$: $i_t^\dagger = 0$ indicates a free case; $i_t^\dagger = i \in [m]$ indicates a tied case for affiliate $i$. \\
$\mathbf{A}_t = (\mathbf{w}_t, i_t^\dagger)$ &  Type of arrival $t$. \\
$\mathbf{z}_t \in \mathbb{R}^m$ & Matching decision for arrival $t$; $z_{t,i}$ indicates matching to affiliate $i$. \\
$\mathcal{Z}(i_t^\dagger)$ & Type feasibility set for arrival $t$: equals the simplex $\Simplex$ if $i_t^\dagger = 0$ (free case), and the singleton set $\{\mathbf{e}_{i_t^\dagger}\}$ if $i_t^\dagger \in [m]$ (tied case). \\
$\{\mathbf{z}_1, \dots, \mathbf{z}_T\}$ & Matching profile over the horizon. A profile is \emph{feasible} if each $\mathbf{z}_t \in \mathcal{Z}(i_t^\dagger)$ and \eqref{hard+constraint:t} holds for all $t \in [T]$.  \\
$c_i$ & Total capacity (annual quota) of affiliate $i$. \\
$\rho_i = c_i / T$ & Normalized capacity ratio of affiliate $i$. \\
$\underline{\rho} = \min_i \rho_i$ & Minimum capacity ratio across affiliates. \\
$r_i$ & Service rate of affiliate $i$. \\
$s_{t,i} \sim \text{Ber}(r_i)$ & Service availability indicator for affiliate $i$ at time $t$, modeled as an i.i.d.\ Bernoulli random variable with success probability $r_i$. \\
$b_{t,i}$ & Backlog at affiliate $i$ at time $t$. \\
$\epsilon$ & Service slack parameter: $r_i \geq \rho_i + \epsilon$ {(Definition~\ref{def:eps+regimes})}. \\
$\alpha$ & Penalty parameter for over-allocation. \\
$\gamma$ & Penalty parameter for average backlog (congestion). \\
$\OPT(\alpha,\gamma)$ & Offline benchmark {(Definition~\ref{def:offline})}. \\
$\ALG^\pi(\alpha,\gamma)$ & Objective achieved by algorithm $\pi$ (equation~\eqref{ALG+obj}). \\
$\textsf{Regret}_T^\pi$ & Regret of algorithm $\pi$  {(Definition~\ref{def:regret})}. \\
$\boldsymbol{\theta} \in \mathbb{R}_{+}^m$ & Dual variables corresponding to over-allocation cost of static resources {(\Cref{prop:offline})}. \\
$\boldsymbol{\lambda} \in \mathbb{R}_{+}^m$ & Dual variables corresponding to the capacity constraint for static resources in the offline benchmark {(\Cref{prop:offline})}. \\
$\boldsymbol{\beta}_t \in \mathbb{R}_{+}^m$ & Time-varying dual variables associated with backlog dynamics at time $t$ {(\Cref{prop:offline})}. \\
$\eta$ (resp. $\zeta$) & Learning rates for dual variables for static (resp. dynamic resources)~ (\Cref{ALG+Surrogate+D}). \\
$\psi(\mathbf{b}_t)$ & Lyapunov function for backlog: $\psi(\mathbf{b}_t) = \frac{1}{2} \|\mathbf{b}_t\|_2^2$ (see \Cref{subsec:ALG+D+pfsketch}). \\
$D(\boldsymbol{\nu})$ & Objective value of the static dual problem given dual variable for static resources $\boldsymbol{\nu} = (\boldsymbol{\theta}, \boldsymbol{\lambda})$~(\Cref{def:staic+dual}). \\
$\boldsymbol{\nu}^* = (\boldsymbol{\theta}^*, \boldsymbol{\lambda}^*)$ & Optimal dual variables for static resources solving the static dual problem~(\Cref{def:staic+dual}). \\ 
$\tilde{\mathbf{z}}(\boldsymbol{\nu}, \mathbf{A})$ & $\argmax_{\mathbf{z} \in \mathcal{Z}(\Target)} (\mathbf{w} - \boldsymbol{\theta} - \boldsymbol{\lambda}) \cdot \mathbf{z}$---primal decision in the static dual problem~(\Cref{def:staic+dual}) given dual variable $\boldsymbol{\nu}$ for static resources and arrival type $\mathbf{A} = (\mathbf{w}, \Target)$. \\
\bottomrule
\end{tabular}
}
\endgroup
\end{table}

\begin{table}[htbp]
\centering
\begingroup
\majorrevcolor{
\caption{\majorrevcolor{Summary of Key  Assumptions}}
\label{tab:assumptions}
\begin{tabular}{p{1.2cm} p{5.0cm} p{9.0cm}}
\toprule
\textbf{Label} & \textbf{Description} & \textbf{Justification / Interpretation} \\
\midrule
\textbf{A1} & \textbf{Unknown i.i.d. Arrivals:} Each arrival $A_t = (\mathbf{w}_t, i_t^\dagger)$ is drawn i.i.d.\ from an unknown distribution. & Stationarity enables learning of arrival patterns (see \Cref{apx+within+year+variations} for empirical support). The unknown distribution is motivated by year-to-year variation in arrival composition (see Figures~\ref{fig:tied+proportion}). \\[1ex]
\addlinespace

\textbf{A2} & \textbf{Service Slack:} There exists $\epsilon \geq 0$ such that $r_i \geq \rho_i + \epsilon$ for all $i \in [m]$. & Non-negativity of $\epsilon$ is required for backlog stability. Parameter $\epsilon$ distinguishes  stable ($\epsilon = \Omega(1)$) vs.\ near-critical ($\epsilon = O(1/\sqrt{T})$) regimes~(\Cref{def:eps+regimes}). \\[1ex]
\addlinespace

\textbf{A3} & \textbf{Bernoulli Service Availability:} Server availability $s_{t,i}$ follows an i.i.d.\ Bernoulli process with success probability $r_i$. & This assumption  is analogous to assuming that service times follow a geometric distribution with mean $1/r_i$. A practical motivation in the context of our application is provided in \Cref{footnote:iid+service}. An extension allowing for server idleness is discussed in \Cref{apx+server+idleness}. \\[1ex]
\addlinespace

\textbf{A4} & \textbf{Buffer for Free Cases:} For each $i \in [m]$, we assume $\rho_i - \mathbb{P}[i_t^\dagger = i] \geq d$ for some constant $d > 0$. & Ensures each affiliate has capacity for free cases with high probability (used in \Cref{prop:offline}). \\[1ex]
\addlinespace

\textbf{A5} & \textbf{Reward Regularity:} The distribution of reward vectors $\mathbf{w}_t$ has a Lipschitz-continuous density (Assumption~\ref{assump:PDF}). & Technical assumption required to connect the convergence of dual variables to that of endogenous arrival rates for \CO{} (see \Cref{thm:ALG+Surrogate+Primal} and \Cref{subsubsec:ALG+P+pf+sketch}). \\
\bottomrule
\end{tabular}
}
\endgroup 
\end{table}

}
\section{Further Related Work}\label{subsec:lit+review+further}
In this section, we provide a more extensive review of related literature to our work. 

{\bf Matching Policies for Refugee Resettlement:} 
{There has been growing academic interest in designing matching policies for refugee resettlement, leading to the two main approaches. The first approach is to consider a centralized \emph{outcome-based} system, focusing on integration outcomes such as employment within 90 days, and design matching algorithms that target such objectives. Central to the outcome-based approach is the development of ML models that predict employment outcomes for a pair of case and affiliate~ \citep{bansak2018improving, bansak2024learning}. Equipped with such models, several recent papers, including ours, aim to optimize matching decisions to improve employment, taking into account some of the operational considerations  discussed in or fairness criteria \Cref{subsec:intro+background} \citep{golz2019migration, ahani2021placement,
ahani2023dynamic,
bansak2024outcome, freund2024group}. The second approach focuses on designing a centralized \emph{preference-based} matching system that  respects the preferences of refugees and/or affiliates (see the work of \citet{nguyen2021stability} and \citet{delacretaz2023matching} and references therein). {While this approach has promise, most of the current programs, including the U.S. system, lack systematic data on preferences. Instead, they operate as a centralized outcome-based system.} 
}

\SLedit{Among the existing works, the most relevant work to ours is \citet{bansak2024outcome} and we highlight key differences from this work in \Cref{subsec:lit+review}.} Beyond this comparison, our paper departs from prior work in several important ways: (i) {W}hile these work{s} focused on designing effective heuristics, we take a theoretical approach and propose algorithms with provable guarantees; 
{(ii) {O}ur work focuses on operational and algorithmic intricacies due to post-allocation service and dynamics of backlogs, which have not been studied before;}
(iii) {T}he existing work designed algorithms that crucially rely on a pool of past years' data to either estimate the arrival distribution or simulate future arrivals. However, as elaborated in \Cref{subsec:intro+background}, such an approach can be \emph{non-robust} to discrepancies between the data from past years and the arrivals from the current year. In contrast, our algorithms are robust to such discrepancies, as they do not rely on data from previous years; and (iv) {A}s a result of such robustness, and in addition to enjoying theoretical guarantees, our method outperforms its counterpart in practical situations, as we discuss next.

{\bf Online Resource Allocation:}
{Online resource allocation problems have been extensively studied in operations and computer science. Given the broad scope of the literature, we only highlight the stream most closely related to our setting. For an overview of other streams of work in the literature, we refer the readers to \cite{mehta2007adwords} and \cite{huang2024online} for an informative survey.

As mentioned in \Cref{subsec:intro+contribution}, without dynamic resources, our setting reduces to the online resource allocation problem with an unknown i.i.d. arrival model. Several papers designed and analyzed algorithms for such settings. 
The primary intuition guiding the design is that, under the stochastic input model, we can \emph{learn} the arrival pattern by observing a small portion of the arrival data. Building on this intuition, earlier studies have explored primal methods~\citep{kesselheim2018primal} that periodically solve a linear program using the observed data, or dual-based methods that seek to efficiently learn the dual variables of the associated offline problem and avoid the computational costs associated with resolving. These methods work by one-time dual learning~\citep{devanur2009adwords, feldman2010online,molinaro2014geometry} (which is similar in nature to the ``predict-then-optimize'' approach, e.g., \cite{elmachtoub2022smart}), multiple-time dual learning~\citep{agrawal2014dynamic, li2022online}, or using adversarial online learning throughout the horizon~\citep{devanur2011near, agrawal2014fast, wang2014close, gupta2016experts, badanidiyuru2018bandits, dughmi2021bernoulli,balseiro2023best}. {All of these methods have also found further applications in various operational scenarios~\citep{chen2017revenue,zhalechian2022online}.} \SLedit{As highlighted in \Cref{subsec:lit+review}, our paper departs from this literature due to the time-varying nature of the dual problems introduced by the dynamic resources.}

Another marginally related line of work is the literature on minimizing regret for Bayesian online resource allocation (see, for example, the work of  \cite{arlotto2019uniformly} and \cite{vera2019bayesian}). Our work diverges from this line of work as the arrival distribution is assumed to be known in these models, and the main focus of the algorithms is to control the error propagation of decisions due to stochastic arrivals instead of learning the arrival. Our dynamic resources are also similar to the reusable resources studied in recent literature (see, for example, \cite{gong2022online,feng2024near,delong2024online,goyal2025asymptotically,feng2021robustness} and references therein). However, the models in this stream of work do not allow one to wait for such resources. {As such, there is no notion of congestion or desire for a smooth matching.}

\par 
{\bf Dynamic Matching and Control of Queueing Systems:} 
In queuing literature, a stream of work studies matching queues; in these systems, agents arrive over time and wait to be matched according to a deterministic compatibility structure \citep{aouad2020dynamic, aveklouris2021matching, afeche2022optimal, gupta2022greedy, hu2022dynamic, kerimov2023dynamic, wei2023constant} or a random one \citep{ashlagi2013kidney, anderson2017efficient, ashlagi2019matching}. While most of these papers focus on maximizing total (match-dependent) rewards, the work of \citet{aveklouris2021matching} considers an objective similar to ours that penalizes the matching reward by the cost of waiting. However, the crucial distinction of our paper from the aforementioned works is the {\em reverse order of matching and waiting}. As discussed in \Cref{subsec:intro+contribution}, our model can be viewed as a queuing system with an endogenous arrival rate controlled by the matching process. Once again, this subtle difference presents technical challenges in, for example, controlling the size of the backlog \revcolor{while performing learning and optimization.}

From a methodological standpoint, the design and analysis of our first algorithm (\Cref{ALG+Surrogate+D}) partly rely on the celebrated drift-plus-penalty method \citep{tassiulas1990stability,lyapunov1992general,neely2006energy,neely2008fairness,neely2022stochastic} for stochastic network optimization and stability.  Our work complements this approach by combining the drift-plus-penalty method with adversarial online learning. Furthermore, our first algorithm implicitly utilizes the scaled backlog as a dual variable, establishing a crucial structural connection to the projected subgradient method \citep{zinkevich2003online}. A similar connection was observed in \cite{ashlagi2022price}  {and \cite{kanoria2023blind}}, albeit in a context different from the setup of our paper. We also note that, in a different context, \cite{wei2023constant} also integrates the drift-plus-penalty method with the resolving techniques.
\section{Proof of \texorpdfstring{\Cref{prop:offline}}{}}
\label{apx+prop+offline}
Fix a sample path $\{\Arrival_\Timeidx, \ServiceVec_\Timeidx\}_{\Timeidx=1}^\TotalTime$ {under the event $G_\TotalTime$}. Given the matching profile $\{\DecisionVec_\Timeidx\}_{\Timeidx=1}^\TotalTime$, let us denote its objective value as 
\begin{equation*}
    f(\{ \DecisionVec_\Timeidx\}_{\Timeidx=1}^\TotalTime) =
\sum_{\Timeidx=1}^\TotalTime\sum_{\Locidx=1}^\Locnum \WeightEach \DecisionEach
    - \OverCost\sum_{\Locidx=1}^{\Locnum} \PositivePart{\sum_{\Timeidx=1}^\TotalTime \DecisionEach - \Capacity_\Locidx }
    - \frac{\BuildUpCost}{\TotalTime} \sum_{\Timeidx=1}^\TotalTime\sum_{\Locidx=1}^{\Locnum} \BuildUpti{\Timeidx}{\Locidx}
\end{equation*}
where $\BuildUpti{\Timeidx}{\Locidx} = (\BuildUpti{\Timeidx}{\Locidx} + \DecisionEach - \ServiceEach)_{+}$ with $b_{0,\Locidx}=0$ for all $ t\in[T]$ and $i\in [m]$.\footnote{Note that this is equivalent to our original objective in \Cref{def:offline} by taking the maximum of \eqref{ALG+obj} over the decision variables $\{\BuildUpVec_\Timeidx\}_{\Timeidx=1}^\TotalTime$ for any given $\{\DecisionVec_\Timeidx\}_{\Timeidx=1}^\TotalTime$ subject to $\BuildUpti{\Timeidx}{\Locidx}\geq 0$ and $\BuildUpti{\Timeidx}{\Locidx} \geq \BuildUpti{\Timeidx-1}{\Locidx} + \DecisionEach - \ServiceEach$ for all $\Timeidx \in [\TotalTime]$ and $\Locidx \in [\Locnum]$.} For brevity, we also define a primal and dual feasible set as:
\begin{equation*}
\begin{aligned} 
    \mathcal{P} &:= \Bigl\{\{\DecisionVec_\Timeidx\}_{\Timeidx=1}^\TotalTime \ | \ \DecisionVec_\Timeidx \in \FeasibleSet{\Target_\Timeidx}, \forall \Timeidx \in [\TotalTime]\Bigr\} \\ 
     \mathcal{D} &:= \Bigl\{ 
(\DualOverVec, \DualHardVec, \{\DualBuildVec_\Timeidx\}_{\Timeidx=1}^\TotalTime) \geq \mathbf{0} \ | \
\ \DualOver_\Locidx \leq \alpha \ \forall \Locidx \in [\Locnum], \ \DualBuild_{\Timeidx,\Locidx} - \DualBuild_{\Timeidx+1,\Locidx} \leq \frac{\BuildUpCost}{\TotalTime} \ \text{$\forall \Timeidx \in [\TotalTime-1]$ $\Locidx\in[\Locnum]$}, 
   \\ 
  & 
  \ \DualBuild_{\TotalTime, \Locidx} \leq \frac{\BuildUpCost}{\TotalTime} \ \forall \Locidx \in [\Locnum]
  \Bigr\}
    \end{aligned}
    \end{equation*}
Consider the following Lagrangian function.
\begin{equation*}
L\Big(\{\DecisionVec_\Timeidx\}_{\Timeidx=1}^{\TotalTime},
\DualOverVec, \DualHardVec, \{\DualBuildVec_\Timeidx\}_{\Timeidx=1}^\TotalTime\Big)
= 
\sum_{\Timeidx=1}^\TotalTime 
\Weightvec_\Timeidx\cdot \DecisionVec_\Timeidx + 
\underbrace{
\sum_{\Timeidx=1}^\TotalTime\DualOverVec\cdot(\CapRatioVec-\DecisionVec_\Timeidx)}_{\mathclap{\textsf{CS}(\DualOverVec)}} + 
\underbrace{
\sum_{\Timeidx=1}^\TotalTime\DualHardVec\cdot(\CapRatioVec-\DecisionVec_\Timeidx)}_{\mathclap{\textsf{CS}(\DualHardVec)}} + 
\underbrace{
\sum_{\Timeidx=1}^\TotalTime
\DualBuildVec_\Timeidx\cdot(\ServiceVec_\Timeidx-\DecisionVec_\Timeidx)}_
{\mathclap{\textsf{CS}(\{\DualBuildVec_\Timeidx\}_{\Timeidx=1}^\TotalTime)}}.
\end{equation*}
Here, each $\textsf{CS}(\cdot)$ term can be seen as a complementary slackness term corresponding to each dual variable. We will show that, {under event $G_\TotalTime$,}
\begin{align}
{\Dual}(\OverCost, \BuildUpCost) 
&=\min_{
(\DualOverVec, \DualHardVec, \{\DualBuildVec_\Timeidx\}_{\Timeidx=1}^\TotalTime)\in \mathcal{D} 
} \ 
\max_{(\{\DecisionVec_\Timeidx\}_{\Timeidx=1}^\TotalTime)\in \mathcal{P}} L\Big(\{\DecisionVec_\Timeidx\}_{\Timeidx=1}^{\TotalTime},
\DualOverVec, \DualHardVec, \{\DualBuildVec_\Timeidx\}_{\Timeidx=1}^\TotalTime\Big) \nonumber \\
&=\max_{(\{\DecisionVec_\Timeidx\}_{\Timeidx=1}^\TotalTime)\in \mathcal{P}}\min_{
(\DualOverVec, \DualHardVec, \{\DualBuildVec_\Timeidx\}_{\Timeidx=1}^\TotalTime)\in \mathcal{D} 
}L\Big(\{\DecisionVec_\Timeidx\}_{\Timeidx=1}^{\TotalTime},
\DualOverVec, \DualHardVec, \{\DualBuildVec_\Timeidx\}_{\Timeidx=1}^\TotalTime\Big)
\label{line:max+min+sub}\\
&= \textsf{OPT}(\OverCost, \BuildUpCost).\label{line:max+min+main}
\end{align}
The first line follows from the definition of the dual function $\Dual(\OverCost,\BuildUpCost)$ in the main statement. The second line is the consequence of Sion's min-max theorem \citep{sion1958general}.\footnote{More specifically, the Lagrangian is a linear function of the primal and dual variables. The primal feasible set $\mathcal{P}$ is convex and compact and the dual feasible set is also convex. Hence, we can switch the minimum and maximum operator by applying Sion's min-max theorem.} To show the last equality, we show that 
\begin{equation}
\begin{split}
\min_{
(\DualOverVec, \DualHardVec, \DualBuildVec_\Timeidx) \in \mathcal{D}}L\Big(\{\DecisionVec_\Timeidx\}_{\Timeidx=1}^{\TotalTime},
\DualOverVec, \DualHardVec, \{\DualBuildVec_\Timeidx\}_{\Timeidx=1}^\TotalTime\Big) \quad =\begin{cases}
 - \infty &\text{if $\sum_{\Timeidx=1}^\TotalTime \DecisionVec_\Timeidx > \mathbf{\Capacity}$} \\
 f(\{ \DecisionVec_\Timeidx\}_{\Timeidx=1}^\TotalTime) \quad &\text{otherwise}.
\end{cases} \label{line:max-min}
\end{split}
\end{equation}

Note that line \eqref{line:max+min+main} directly follows from line \eqref{line:max-min} {because the capacity constraint of the offline benchmark (\Cref{def:offline}) is equivalent to $\sum_{\Timeidx=1}^\TotalTime \Decision_{\Timeidx,\Locidx} \leq \Capacity_\Locidx $ for all $\Locidx \in [\Locnum]$ under the event $G_\TotalTime$}. Hence, it suffices to show line \eqref{line:max-min}. Recall that $\mathbf{c} = \TotalTime\CapRatioVec$ by definition of $\CapRatioVec$. If $\sum_{\Timeidx=1}^\TotalTime \DecisionVec_\Timeidx > \mathbf{\Capacity}$, then the optimal solution $\DualHardVec^* \geq \mathbf{0}$ of the inner minimization problem in line \eqref{line:max+min+sub} can grow arbitrarily large, hence making $\textsf{CS}(\DualHardVec)$ unbounded below. Otherwise, it is straightforward to see $\textsf{CS}(\DualHardVec^*)= 0$ at the optimal $\DualHardVec^*$ of the inner minimization problem in line \eqref{line:max+min+sub}. We now investigate the minimum value of $\textsf{CS}(\DualOverVec)$ and $\textsf{CS}(\{\DualBuildVec_\Timeidx\}_{\Timeidx=1}^\TotalTime)$ over the dual feasible set $\mathcal{D}$. We first focus on the former. Note that the inner minimization problem in line \eqref{line:max+min+sub} is separable over $\DualOverVec$ and $\{\DualBuildVec_\Timeidx\}_{\Timeidx=1}^\TotalTime$. Hence, we have: 
\begin{equation*}
\min_{\DualOverVec \in [0,\OverCost]^\Locnum} 
\textsf{CS}(\DualOverVec) = 
-\max_{\DualOverVec \in [0,\OverCost]^\Locnum} \sum_{\Locidx=1}^\Locnum \DualOver_\Locidx \left(
\sum_{\Timeidx=1}^\TotalTime \DecisionEach - \Capacity_\Locidx
\right) =  
- \OverCost\sum_{\Locidx=1}^{\Locnum} \PositivePart{\sum_{\Timeidx=1}^\TotalTime \DecisionEach - \Capacity_\Locidx }.
\end{equation*}
The the last inequality follows from the fact that $\alpha (x)_{+}=\max_{\theta \in [0,\alpha]} \theta x$ for any $x \in \R$. In the similar way, 
\begin{align*}
&\min_{\{\DualBuildVec\}_{\Timeidx=1}^\TotalTime \geq \mathbf{0}} 
\textsf{CS}(\{\DualBuildVec_\Timeidx\}_{\Timeidx=1}^\TotalTime) \quad \text{s.t.} \quad \DualBuild_{\Timeidx,\Locidx} - \DualBuild_{\Timeidx+1,\Locidx} \leq \frac{\BuildUpCost}{\TotalTime} \ \text{$\forall \Timeidx \in [\TotalTime-1]$ and $\Locidx\in[\Locnum]$}, \ \DualBuild_{\TotalTime, \Locidx} \leq \frac{\BuildUpCost}{\TotalTime} \ \forall \Locidx \in [\Locnum] \nonumber \\
&= -\left[\max_{\{\DualBuildVec_\Timeidx\}_{\Timeidx=1}^\TotalTime\geq \mathbf{0}}
\sum_{\Timeidx=1}^\TotalTime\sum_{\Locidx=1}^\Locnum \DualBuild_{\Timeidx,\Locidx}(\DecisionEach - \ServiceEach) \quad \text{s.t} \quad \DualBuild_{\Timeidx,\Locidx} - \DualBuild_{\Timeidx+1,\Locidx} \leq \frac{\BuildUpCost}{\TotalTime} \ \text{$\forall \Timeidx \in [\TotalTime-1]$ and $\Locidx\in[\Locnum]$}, \ \DualBuild_{\TotalTime, \Locidx} \leq \frac{\BuildUpCost}{\TotalTime} \ \forall \Locidx \in [\Locnum] \right]\nonumber \\ 
&= - \left[\frac{\BuildUpCost}{\TotalTime}\min_{\{{\BuildUpVec}_\Timeidx\}_{\Timeidx=1}^\TotalTime \geq \mathbf{0}}
\sum_{\Timeidx=1}^\TotalTime \sum_{\Locidx=1}^\Locnum{\BuildUp}_{\Timeidx,\Locidx} \quad \text{s.t.} \quad {{\BuildUp}_{\Timeidx,\Locidx} \geq {\BuildUp}_{\Timeidx-1,\Locidx}} + \DecisionEach - \ServiceEach \quad
\forall \Timeidx \in [\TotalTime], \Locidx\in[\Locnum]\right]  \\ 
&= -\frac{\BuildUpCost}{\TotalTime}\sum_{\Timeidx=1}^\TotalTime\sum_{\Locidx=1}^\Locnum (\BuildUp_{\Timeidx-1,\Locidx} + \DecisionEach - \ServiceEach)_{+}.
\end{align*}
    In the third line, we used the strong duality of the linear program. Note that, at the optimal solution of the third line, either of ${\BuildUp}_{\Timeidx,\Locidx}\geq 0$  or ${{\BuildUp}_{\Timeidx,\Locidx} \geq {\BuildUp}_{\Timeidx-1,\Locidx} + \DecisionEach - \ServiceEach}$ must bind. Hence, the last line follows. Combining, we have proved line \eqref{line:max-min}. This completes the proof. 


\section{Proof of \Cref{prop:impossibility}}\label{apx+hardness}

We consider an instance with $m=1$. For brevity, we omit the subscript for $\Locidx$. There are $\TotalTime$ arrivals with deterministic reward $\Weight_\Timeidx=1$ for all $\Timeidx \in [\TotalTime]$. There are no tied cases and the capacity is $\Capacity = 0.5\TotalTime$. The service rate is $\ServiceRate = 0.5 + \ServiceSlack$ where $\ServiceSlack \in [0, 0.5)$. 
Note that $\alpha$ does not play any role here. Hence, we denote the expected objective value of offline benchmark (\Cref{def:offline}) by $\OPT(\BuildUpCost)$. In the following, we first obtain a lower bound on $\OPT(\BuildUpCost)$ and then an upper bound on the objective value of any online algorithm. \par 
\begin{description}[leftmargin=0cm]
\item{\noindent\textbf{Step 1: Lower-bounding Objective Value of Optimal Offline.}}\par 
We first analyze the objective value of the offline.
\begin{claim}\label{claim:hardness+opt}
For any $\epsilon \geq 0$, 
\begin{equation*}
\E[\emph{\OPT}(\BuildUpCost)] \geq 0.5T - \Theta(\sqrt{\TotalTime}) \label{line:opt}
\end{equation*}
\end{claim}
{\bf Proof of Claim \ref{claim:hardness+opt}}.
Consider the following feasible solution of the offline program: set $z_{t} = \Indicator[s_{t}=1]$ while satisfying the capacity constraint (i,e, $\sum_{\Timeidx=1}^\TotalTime z_{t} \leq 0.5\TotalTime$). Let $S_{T}^{\epsilon}$ be a binomial random variable with $T$ trials and success probability $0.5+\epsilon$. The proposed feasible solution never incurs the \backlog{}, while obtaining $\min(S^{\epsilon}_{T}, 0.5T)$ reward. Hence, we have $\textsf{OPT}(\gamma) \geq \min(S^{\epsilon}_{T}, 0.5T)$ for every sample path of arrivals and services. Taking the expectation of the preceding inequality, we have 
\begin{equation*}
    \E[\textsf{OPT}(\gamma)] \geq \E[\min(S^{\epsilon}_{T}, 0.5T)] \geq \E[\min(S^{0}_{T}, 0.5T)]
\end{equation*}
where the last inequality is because $S_T^{\epsilon}$ stochastically dominates $S_T^{0}$ for all $\epsilon \geq 0$ in the first order.%
\footnote{
Random variable $X$ stochastically dominates (in the first-order) if any only if $\E[u(X)] \geq \E[u(Y)]$ for all non-decreasing function $u$. Furthermore, let $X$ and $Y$ be binomial random variables with the respective success probability $p_X$ and $p_Y$ and the common number of trials $T$. If $p_X \geq p_Y$, $X$ stochastically dominates $Y$ in the first order \citep{shaked2007stochastic}.} We now lower-bound the right-hand side of the above inequality by $0.5\TotalTime - \Theta(\sqrt{\TotalTime})$. Let us define $Z_\TotalTime^0:= \dfrac{S_\TotalTime^0 - 0.5\TotalTime}{\sqrt{\Var[S_\TotalTime^0]}}$. Then we have
\begin{equation*}
    \E[\min(S_T^0, 0.5T)] = 0.5T - \E[(0.5T - S_T^0)_{+}] = 0.5T - \sqrt{\Var[S_T^0]}\E[(-Z_\TotalTime^0)_{+}] = 0.5T - \Theta(\sqrt{T}).
\end{equation*}
where the last equality is because $\E[(-Z_\TotalTime^0)_{+}]$ has a constant mean for large $T$ by the central limit theorem. 
 This completes the proof of Claim \ref{claim:hardness+opt}. 
\hfill \halmos
\endproof
\item{\noindent\textbf{Step 2: Upper-bounding Objective Value of Any Online Algorithm.}}\par 
We now turn our attention to upper-bounding the objective value of any online algorithm, denoted by $\textsf{ALG}$. Fix an online algorithm and let $\Decision_{\Timeidx}$ denote the decision at time $\Timeidx$. Without loss, we assume that $\E[\sum_{\Timeidx=1}^\TotalTime \Decision_\Timeidx] \geq \frac{\TotalTime}{4}$ because otherwise, we have $\E[\textsf{OPT}(\BuildUpCost)] - \E[\textsf{ALG}] \geq 0.25\TotalTime - \Theta(\sqrt{\TotalTime}) = \Omega(\TotalTime)$ from Claim \ref{claim:hardness+opt}. 
\begin{claim}\label{claim:hardness+online}
The expected value of any online algorithm for which $\E[\sum_{\Timeidx=1}^\TotalTime \Decision_\Timeidx] \geq \frac{T}{4}$ satisfies: 
\begin{equation*}
\E[\emph{\textsf{ALG}}]\leq 0.5\TotalTime - 0.25\BuildUpCost(0.5-\ServiceSlack)
\end{equation*}
\end{claim}
{\bf Proof of Claim \ref{claim:hardness+online}}.
We first lower-bound the total cumulative \backlog{} by
\begin{equation}
    \E\left[\sum_{\Timeidx=1}^T \BuildUp_{t} \right]\geq  \E\left[\sum_{\Timeidx=1}^\TotalTime \Decision_\Timeidx\right]\left(0.5 - \ServiceSlack\right).\label{ineq:total+build}
\end{equation}
To see this, we first observe that
\begin{equation*}
    \BuildUp_{\Timeidx} \geq \Indicator[\Service_{\Timeidx}=0]\Decision_{\Timeidx}
\end{equation*}
for every sample path. More specifically, if $\Service_{\Timeidx}=0$, then we have at least $\Decision_{\Timeidx}$ amount of \backlog{} at the end of period $\Timeidx$. Otherwise, the bound is trivial. Taking the expectation of the preceding inequality conditional on history $\History_{\Timeidx-1}$, we have 
\begin{equation*}
\begin{split}
    \E[ \BuildUp_{\Timeidx}|\History_{\Timeidx-1}] &\geq \E[\Indicator[\Service_{\Timeidx}=0]\Decision_{\Timeidx}|\History_{\Timeidx-1}]\\
    &= 
    \E\left[\Indicator[\Service_{\Timeidx}=0] | \mathcal{H}_{t-1}\right] \E[\Decision_{\Timeidx}|\mathcal{H}_{t-1}
    ]\\
    &= \left(0.5-\epsilon\right)  \E[z_{t}|\mathcal{H}_{t-1}]
\end{split}    
\end{equation*}
For the second line, we used the following fact: since any online algorithm does not know the realization of $\Service_{\Timeidx}$ at the time of making the decision $\Decision_{\Timeidx}$, the random variables $\Decision_{\Timeidx}$ and $\Service_{\Timeidx}$ are conditionally independent given $\History_{\Timeidx-1}$. The last line follows because $\Service_{\Timeidx}$ is an i.i.d. Bernoulli random variable with mean $0.5 + \epsilon$. Taking outer expectation of the preceding inequality, we obtain that 
\begin{equation*}
    \E[\BuildUp_{\Timeidx}] \geq \left(0.5-\ServiceSlack\right) \E[\Decision_{\Timeidx}]
\end{equation*}
Summing up for all $\Timeidx \in [\TotalTime]$ gives the inequality \eqref{ineq:total+build}. Hence, we conclude that
\begin{equation*}
    \E[\textsf{ALG}] = \E\left[
    \sum_{\Timeidx=1}^\TotalTime \Decision_{\Timeidx}
    \right]  - \frac{\BuildUpCost}{\TotalTime}\E\left[
    \sum_{\Timeidx=1}^\TotalTime \BuildUp_{\Timeidx}
    \right] 
    \leq \E\left[
    \sum_{\Timeidx=1}^\TotalTime \Decision_{\Timeidx}
    \right] - \frac{\BuildUpCost}{\TotalTime}(0.5-\ServiceSlack)\E\left[\sum_{\Timeidx=1}^\TotalTime \Decision_\Timeidx\right]
    \leq 0.5\TotalTime - 0.25\BuildUpCost(0.5-\ServiceSlack).
\end{equation*}
In the last inequality, we used that $0.25\TotalTime \leq \E[\sum_{\Timeidx=1}^\TotalTime \Decision_\Timeidx]$ and the fact that capacity is $0.5\TotalTime$.
\hfill \Halmos
\endproof

\item{\textbf{Step 3: Putting Everything Together. }}

From Claim \ref{claim:hardness+opt} and Claim \ref{claim:hardness+online}, we conclude that 
\begin{equation*}
    \E[\textsf{OPT}] - \E[\textsf{ALG}] \geq
     0.25\BuildUpCost(0.5-\ServiceSlack) - \Theta(\sqrt{T}).
\end{equation*}
for any online algorithm such that $\E[\sum_{\Timeidx=1}^\TotalTime \Decision_\Timeidx] \geq 0.25\TotalTime$. Hence, whenever $\gamma = \Omega(T)$, the regret must be at least $\Omega(T)$. This completes the proof of \Cref{prop:impossibility}.
\end{description}


{\color{black}
\section{Alternative Interpretation of \CA{} (\texorpdfstring{\Cref{ALG+Surrogate+D}}{})}\label{apx:lyapunov}

In this appendix, we describe how our design idea of \Cref{ALG+Surrogate+D} is connected to the theory of Lyapunov optimization, in particular drift-plus-penalty method~\citep{neely2022stochastic}. To understand the connection of our algorithm to this method, it is helpful to think of two different policies designed for each of the following extremes. First, suppose we just want to control the average backlog without considering the net matching reward. The idea of Lyapunov optimization is that one can design a policy that can upper-bound an objective function by (i) defining a Lyapunov potential function (that is closely related to the original objective function) (ii) choosing a control that minimizes an \emph{upper bound} of the expected drift (conditional on the history) of the potential function --- often called Lyapunov drift. In our context, the potential function is the sum of squared backlogs, i.e., $\psi(\mathbf{b}{_t})= \frac{1}{2}\Norm{\mathbf{b}_t}{2}^2$ (as commonly used). Due to our drift lemma (\Cref{lemma:drift}), we know that $\psi(\mathbf{b}_{\Timeidx})- \psi(\mathbf{b}_{\Timeidx-1}) \leq  \mathbf{b}_{\Timeidx-1}\cdot (\mathbf{z}_t - \mathbf{s}_t ) + \BigO(1)$. Because the service rate of each affiliate $i$ is $\rho_i + \epsilon$, the Lyapunov drift of $\psi(\cdot)$ is then upper bounded by
\begin{equation}
\E[\psi(\mathbf{b}_{\Timeidx})- \psi(\mathbf{b}_{\Timeidx-1})|\History_{\Timeidx-1}] \leq \mathbf{b}_{\Timeidx-1}\cdot (\E[\DecisionVec_{\Timeidx}|\History_{\Timeidx-1}] - \CapRatioVec - \ServiceSlack\mathbf{1} ) + \BigO(1).
\end{equation}
where $\mathbf{1}$ is a vector of ones. Hence, a policy that minimizes the upper bound of the Lyapunov drift is simply assigning a case $\Timeidx$ to an affiliate $\Locidx$ with the \emph{minimum} current backlog $\BuildUp_{\Timeidx-1,\Locidx}$. Note that this policy is equivalent to using an adjusted score $-\BuildUp_{\Timeidx-1,\Locidx}$ and matching a case to the affiliate with the maximum adjusted score. In fact, if we replace line \ref{line:primal+decision} of \Cref{ALG+Surrogate+D} with such a policy, one can show that the expected average backlog of such policy is $\BigO(1/\ServiceSlack)$.

Now consider the other extreme where the goal is to maximize the net matching reward (without considering the average backlog). The literature of online resource allocation (see \Cref{subsec:lit+review} for a review), and in particular the scoring techniques therein,  suggests that there exists a time-invariant dual variable $\boldsymbol{\theta}^*$ (for the over-allocation cost) and $\boldsymbol{\lambda}^*$ (for the capacity constraint) where we can obtain the optimal net matching reward by assigning case $\Timeidx$ to affiliate $\Locidx$ with the \emph{maximum} adjusted score $(\WeightEach-{\theta}^*_{\Locidx} - {\lambda}^*_{\Locidx})$. Furthermore, under the i.i.d. arrival model, one can learn the dual variables by employing adversarial online learning. Importantly, the expected loss (regret) stemming from such learning is $\BigO(\sqrt{\TotalTime})$ \citep{agrawal2014fast, balseiro2023best}.

A natural way of interpolating both extremes is to employ a scoring policy with the adjusted score $\Weightvec_\Timeidx - \boldsymbol{\theta}^* - \boldsymbol{\lambda}^* - \zeta\BuildUpVec_{\Timeidx-1}$ where $\zeta$ is a parameter that encodes the ``weight'' on the objective of minimizing the average backlog compared to the other. A keen reader would note that this is reminiscent of the \emph{drift-plus-penalty} method, a general technical framework designed to minimize the cumulative penalty function (in our context, a negative of the total net matching reward) of a queueing network while stabilizing the queues (see \cite{neely2022stochastic} and references therein for a more detailed review). The method, though developed in different contexts, similarly chooses a control action that greedily minimizes a linear combination of the penalty function and the Lyapunov drift of the queues. 

The main novelty of our algorithm is that, because we do not know $(\boldsymbol{\theta}^*, \boldsymbol{\lambda}^*)$ a priori, we employ adversarial online learning techniques to adaptively learn (update) them. In this sense, our algorithm can be viewed as a novel combination of the drift-plus-penalty method with the adversarial online learning method. We remark that, while the proof involves many intricacies, the regret bound in \Cref{thm:ALG+Surrogate+Dual} takes a natural combination of $\BigO(1/\ServiceSlack)$ on the average backlog and $\BigO(\sqrt{\TotalTime})$ on the loss of the net matching reward --- which is consistent with each of the two aforementioned extreme objective functions.}





\section{Missing Proofs of 
\texorpdfstring{\Cref{sec:alg+design}}{}}\label{apx+thm+surrogate+D}
Before going forward, we first present some technical preliminaries that we will frequently use. The following claim is the standard result of online mirror descent. 
\begin{restatable}[{Online Mirror Descent} \citep{bubeck2015convex}]{claim}{osmd}\label{fact:average+conv}
Let $\{f_t: \mathcal{D}\to \R\}$ be a sequence of convex functions and $\boldsymbol{\nu}_t$ be a sequence of iterates such that
\begin{equation}
\begin{split}
   \nabla h(\tilde{\boldsymbol{\nu}}_{t+1}) &=  \nabla h(\boldsymbol{\nu}_t) - \eta_t \mathbf{\hat{g}}_t \\
   {\boldsymbol{\nu}}_{t+1} &= \argmin_{\mathbf{x}\in \mathcal{D}} V_h(\mathbf{x},  \tilde{\boldsymbol{\nu}}_{t+1})
\end{split}\label{line:update+temp}
\end{equation}
where (1) $\mathcal{D}$ is a bounded convex set, (2) $\E[\mathbf{\hat{g}}_t|\History_{t-1}] = \nabla f_t(\boldsymbol{\nu}_t)$, and (3) $h$ is a $\sigma$-strongly convex function with respect to $\lVert \cdot \rVert$ norm on set $\mathcal{D}$,. For \underline{any} $\boldsymbol{\nu} \in \mathcal{D}$ and positive integers $(k, s)$, we have
\begin{equation}
    \sum_{t=k}^{s}( f_t(\boldsymbol{\nu}_t)- f_t(\boldsymbol{\nu})) \leq \sum_{t=k}^s \frac{\eta_t \lVert \mathbf{\hat{g}}_t\rVert_{*}^2}{2\sigma} + \frac{1}{\eta_k}V_h(\boldsymbol{\nu},\boldsymbol{\nu}_k) + 
    \sum_{t=k+1}^s \left(\frac{1}{\eta}_t - \frac{1}{\eta_{t-1}}\right)V_h(\boldsymbol{\nu}, \boldsymbol{\nu}_t) + 
    \sum_{t=k}^s \mathbf{\hat{u}}_t\cdot(\boldsymbol{\nu}_t - \boldsymbol{\nu}) 
\end{equation}
where $\mathbf{\hat{u}}_t := \E[\mathbf{\hat{g}}_t|H_{t-1}] - \mathbf{\hat{g}}_t $ (i.e., noise of the gradient) and $V_h(\mathbf{x}, \mathbf{y}) = h(\mathbf{x}) - h(\mathbf{y}) - \nabla h (\mathbf{y})\cdot(\mathbf{x}-\mathbf{y})$ is the Bregman divergence with respect to $h$.
\end{restatable}

The stated theorem is slightly different from the one from \cite{bubeck2015convex} because we allow for the step size to be time-varying in Section \ref{subsec:alg+surrogate+primal}. For the sake of completeness, we reproduce the proof of  Claim \ref{fact:average+conv} in Appendix \ref{apx+OMD+pf}.\par 
The following claim shows that we can always ``linearize'' the total over-allocation cost, which will be crucial throughout the proof.
\begin{claim}[{Dual Representation of Over-allocation Cost}]\label{fact:over}
There exists $\boldsymbol{\theta}^\star \in [0,\alpha]^m$ for which
\begin{equation}
\OverCost \sum_{\Locidx=1}^\Locnum \left(\sum_{\Timeidx=1}^\TotalTime \DecisionEach - \Capacity_\Locidx\right)_{+} = \sum_{\Timeidx=1}^\TotalTime \DualOverVec^\star \cdot ( \DecisionVec_\Timeidx - \CapRatioVec)
\end{equation}
\end{claim}
{\bf Proof of Claim \ref{fact:over}.}
The proof simply follows from $\alpha \cdot (x)_{+} = \max_{\theta \in [0,\alpha]} \theta x$.
\hfill \halmos


\subsection{Proof of \Cref{lemma:drift}}\label{apx+drift}
We prove both the lower and upper bound on the drift separately. 
\paragraph{Lower-bound on drift:} To show the lower-bound on the drift $\phi_i(b_{t,i})-\phi_i(b_{t-1,i})$, note that the potential function $\psi(x)=\frac{1}{2}\lVert \mathbf{x}\rVert_2^2=\sum_{i}\phi_i(x)$, where each $\phi_i(x)\triangleq\frac{1}{2}x_i^2$ is convex. Now, either $b_{t-1,i}=0$ and in that case $\phi_i(b_{t,i})-\phi_i(b_{t-1,i})\geq 0=b_{t-1,i}(z_{t,i}-s_{t,i})$, or $b_{t-1,i}\neq 0$ and therefore $b_{t,i} = b_{t-1,i}+z_{t,i}-s_{t,i}$.\footnote{The equation is always valid as long as $z_{t,i} \in \{0,1\}$, which is true for \Cref{ALG+Surrogate+D}.} 
In this case, as $\phi_i(x)$ is convex, we have:
$$
\phi_i(b_{t,i})-\phi_i(b_{t-1,i})\geq \phi'_i(b_{t-1,i})(b_{t,i}-b_{t-1,i})=b_{t-1,i}(z_{t,i}-s_{t,i})~.
$$
Putting two cases together and summing over all $i$, we conclude that
$$
 \psi(\BuildUpVec_\Timeidx) - \psi(\BuildUpVec_{\Timeidx-1})
        \geq \BuildUpVec_{\Timeidx-1}\cdot \left(\DecisionVec_\Timeidx - \ServiceVec_{\Timeidx}\right)
$$

{\paragraph{Upper-bound on drift:} To show the upper-bound on the drift $\phi_i(b_{t,i})-\phi_i(b_{t-1,i})$, }we first note that $\BuildUpti{\Timeidx}{\Locidx} = (\BuildUpti{\Timeidx-1}{\Locidx} +\DecisionEach - \ServiceEach)_{+} \leq |\BuildUpti{\Timeidx-1}{\Locidx} +\DecisionEach - \ServiceEach| $. Hence, we always have $\BuildUpti{\Timeidx}{\Locidx}^2 \leq (\BuildUpti{\Timeidx-1}{\Locidx} +\DecisionEach - \ServiceEach)^2$. Rearranging this inequality, we have
\begin{align*}
    \frac{\BuildUpti{\Timeidx}{\Locidx}^2 - \BuildUpti{\Timeidx-1}{\Locidx}^2}{2} \leq  \BuildUpti{\Timeidx-1}{\Locidx}(\DecisionEach - \ServiceEach) + \frac{(\DecisionEach - \ServiceEach)^2}{2}.
\end{align*}
Summing the inequality over $\Locidx \in [\Locnum]$, we have
\begin{align*}
    \psi(\BuildUpVec_\Timeidx) - \psi(\BuildUpVec_{\Timeidx-1})
       & \leq \BuildUpVec_{\Timeidx-1}\cdot \left(\DecisionVec_\Timeidx - \ServiceVec_{\Timeidx}\right) + \frac{\lVert \DecisionVec_\Timeidx - \ServiceVec_{\Timeidx}\rVert_2^2 }{2}.
\end{align*}
{We highlight that while we proved the above inequality from first principles, we could derive it directly using the fact that $\frac{1}{2}\lVert x\rVert_2^2$ is a 1-Lipschitz smooth function.} 
Finally, the squared norm $\lVert \DecisionVec_\Timeidx - \ServiceVec_{\Timeidx}\rVert_2^2$ is trivially upper-bounded by $1 + m = \BigO(1)$. This completes the proof.

\subsection{Proof of \Cref{lemma:pseudo+lower}}\label{apx+pseudo+lower}

Let $\DualBuildVec_\Timeidx = \zeta\BuildUpVec_{\Timeidx-1}$ (see equation \eqref{eq:beta+dynamics} in \Cref{subsec:alg+surrogate+dual} and related discussions), which is the implicit dual variables for dynamic resources. We recall from line \eqref{eq:pseudo+drift} that
\begin{equation}
\begin{split}
    K_\Timeidx &:= \Weightvec_\Timeidx\cdot\DecisionVec_\Timeidx + \DualOverVec_\Timeidx\cdot(\CapRatioVec-\DecisionVec_\Timeidx) + \DualHardVec_\Timeidx\cdot(\CapRatioVec- \DecisionVec_\Timeidx) - \zeta D_\Timeidx  \\
    &\geq (\Weightvec_\Timeidx - \DualOverVec_\Timeidx - \DualHardVec_\Timeidx - \DualBuildVec_{\Timeidx})\cdot \DecisionVec_{\Timeidx} + \DualOverVec_\Timeidx\cdot \CapRatioVec + \DualHardVec_\Timeidx\cdot \CapRatioVec + \DualBuildVec_{\Timeidx}\cdot \ServiceVec_\Timeidx - \BigO(\zeta)
\end{split} \label{eq:lower}
\end{equation}
where the last inequality follows from \Cref{lemma:drift}. Furthermore, we note that for all $t \leq \Stopping$, we recall from equation \eqref{line:optimality} that 
\begin{equation}
    \DecisionVec_\Timeidx \in  \argmax_{\DecisionVec\in \FeasibleSet{\Target_\Timeidx} } (\Weightvec_\Timeidx - \DualOverVec_\Timeidx - \DualHardVec_\Timeidx - \DualBuildVec_\Timeidx)\cdot \DecisionVec \label{eq:optimality}
\end{equation}

In the similar spirit of \citet{talluri1998analysis}, 
{the following claim shows that there always exists a static control whose value  (roughly) upper-bounds the net \allocation{} reward of the offline benchmark (\Cref{def:offline}) (part (c)) with a marginal matching rate at most $\CapRatioVec$ in expectation (part (b)).}

{
\color{black}
\begin{claim}[Static Control]\label{fluid}
Consider the following static control problem:
\begin{equation}
\begin{split}
{\SinglePrimal_{\mathcal{Z}}}&:= \max_{\substack{\DecisionVec(\Arrival) \in \FeasibleSet{\Target} \\
\forall \Arrival \in \ArrivalSupp}} \E[\Weightvec\cdot \DecisionVec(\Arrival)] \quad \text{s.t.} \ 
\E\left[\DecisionVec(\Arrival)\right] \leq \CapRatioVec 
\end{split}
\end{equation}
where the expectation is with respect to $\mathbf{A} = (\Weightvec, \Target)\sim \ArrivalDist$. Let $\bar{\DecisionVec}: \ArrivalSupp \to \Simplex$ be the optimal static control that solves the above program. Then we have
\begin{enumerate}[label=(\alph*)]
    \item $\bar{\DecisionVec}(\Arrival) \in \FeasibleSet{\Target}$ for any $\Arrival = (\Weightvec, \Target)$ \label{fluid:feasible}
    \item $\E_{\Arrival}[\bar{\DecisionVec}(\Arrival)] \leq \CapRatioVec$ \label{fluid:rate}
    \item $\SinglePrimal_{\mathcal{Z}} = \E_{\Arrival}[\Weightvec \cdot \bar{\DecisionVec}(\Arrival)] \geq \dfrac{\E[
    \NetReward(\{ \hat{\DecisionVec}_\Timeidx\}_{\Timeidx=1}^\TotalTime;\OverCost) 
    ]}{T} - \BigO(1/\TotalTime^2)
    $ for and any \allocation{} profile $\{ \hat{\DecisionVec}_\Timeidx\}_{\Timeidx=1}^\TotalTime$ that satisfies 
    \begin{equation}
\hat{\DecisionVec}_\Timeidx \in \FeasibleSet{\Target_\Timeidx} \quad \forall \Timeidx \in [\TotalTime], \quad \sum_{\Timeidx=1}^\TotalTime \Indicator[\Target_\Timeidx = 0] \Decision_{\Timeidx,\Locidx}
    \leq \left(\Capacity_\Locidx - \sum_{\Timeidx=1}^\TotalTime \Indicator[\Target_t = \Locidx]\Decision_{\Timeidx,\Locidx}\right)_{+} \quad \forall \Locidx \in [\Locnum].\label{line:feasible+profile}
    \end{equation} 
    \label{fluid:opt}
\end{enumerate} 
\end{claim}
}
{Note that, due to our assumption that $\CapRatio_\Locidx \geq \P[\Target=\Locidx]$, the program is always feasible and therefore $\SinglePrimal_{\mathcal{Z}}$ is well-defined.} We defer the proof of Claim \ref{fluid} to the end of this section, but first we complete the proof of \Cref{lemma:pseudo+lower} building on this claim.  Based on the static control $\bar{\DecisionVec}(\cdot)$ defined in Claim \ref{fluid}, we define an alternative \allocation{} decision as
\begin{equation}
\bar{\DecisionVec}_\Timeidx := \bar{\DecisionVec}(\Arrival_\Timeidx).\label{line:ref+decision}
\end{equation} 


By line \eqref{eq:lower} and the optimality criterion \eqref{eq:optimality}, for $t \leq \Stopping$, we have
\begin{equation*}
    K_t \geq (\Weightvec_\Timeidx - \DualOverVec_\Timeidx - \DualHardVec_\Timeidx - \DualBuildVec_{\Timeidx})\cdot \bar{\DecisionVec}_{\Timeidx} + \DualOverVec_\Timeidx\cdot \CapRatioVec + \DualHardVec_\Timeidx\cdot \CapRatioVec + \DualBuildVec_{\Timeidx}\cdot \ServiceVec_\Timeidx - \BigO(\zeta)
\end{equation*}
for every sample path. Taking the expectation conditional on history $\History_{\Timeidx - 1}= \{\Arrival_\Tauidx, \ServiceVec_{\Tauidx}\}_{\Tauidx=1}^{\Timeidx - 1}$, we obtain the following: for any \allocation{} profile $\{ \hat{\DecisionVec}_\Timeidx\}_{\Timeidx=1}^\TotalTime$ that satisfies \eqref{line:feasible+profile}, 
\begin{align}
\E[K_t|\History_{\Timeidx-1}]
&\geq \E[(\Weightvec_\Timeidx - \DualOverVec_\Timeidx - \DualHardVec_\Timeidx - \DualBuildVec_{\Timeidx})\cdot \bar{\DecisionVec}_{\Timeidx} + \DualOverVec_\Timeidx\cdot \CapRatioVec + \DualHardVec_\Timeidx\cdot \CapRatioVec + \DualBuildVec_{\Timeidx}\cdot \ServiceVec_\Timeidx - \BigO(\zeta)|\History_{\Timeidx-1}] \nonumber \\
&= \E[\Weightvec_\Timeidx\cdot\bar{\DecisionVec}_\Timeidx + \DualOverVec_\Timeidx\cdot(\CapRatioVec-\bar{\DecisionVec}_\Timeidx) + \DualHardVec_\Timeidx\cdot(\CapRatioVec-\bar{\DecisionVec}_\Timeidx) + \DualBuildVec_{\Timeidx}\cdot(\ServiceVec_\Timeidx - \bar{\DecisionVec}_{\Timeidx}) - \BigO(\zeta)| \History_{\Timeidx - 1}]\nonumber \\
\begin{split}
&=\E[\Weightvec_\Timeidx\cdot\bar{\DecisionVec}_\Timeidx|\History_{\Timeidx - 1}] + 
\DualOverVec_\Timeidx\cdot\E[(\CapRatioVec-\bar{{\DecisionVec}}_\Timeidx)|\History_{\Timeidx - 1}] + 
\DualHardVec_\Timeidx\cdot\E[(\CapRatioVec-\bar{{\DecisionVec}}_\Timeidx)|\History_{\Timeidx - 1}] +\\
&\quad \DualBuildVec_\Timeidx\cdot\E[(\ServiceVec_\Timeidx-\bar{\DecisionVec}_\Timeidx)|\History_{\Timeidx - 1}]- \BigO(\zeta)
\end{split}
\label{line:third} \\
&\geq \frac{\E[    \NetReward(\{ \hat{\DecisionVec}_\Timeidx\}_{\Timeidx=1}^\TotalTime;\OverCost)]}{T}  {-\BigO\Big(\frac{1}{\TotalTime^2}\Big)} 
+ 
\zeta\ServiceSlack \Norm{\BuildUpVec_{\Timeidx-1}}{1} - \BigO(\zeta) \label{line:fluid}
\end{align}
Line \eqref{line:third} is because $(\DualOverVec_\Timeidx,\DualHardVec_\Timeidx, \DualBuildVec_{\Timeidx})$ are $\History_{\Timeidx-1}$-measurable (that is, knowing the exact realizations of $\History_{\Timeidx-1}$ determines the value of the dual variables). In line \eqref{line:fluid}, we used the property of the static control $\bar{\DecisionVec}(\cdot)$ stated in Claim \ref{fluid}. 
To be precise, we first explain how we relate $\E[\Weightvec_\Timeidx\cdot\bar{\DecisionVec}_\Timeidx|\History_{\Timeidx - 1}]$ of line \eqref{line:third} to 
$ \tfrac{\E[    \NetReward(\{ \hat{\DecisionVec}_\Timeidx\}_{\Timeidx=1}^\TotalTime;\OverCost)]}{T} $ in line \eqref{line:fluid}. 
Note that the static control $\bar{\DecisionVec}(\cdot)$ only depends on the arrival distribution and the current arrival type. Hence, combined with the i.i.d. nature of the arrival process, we have $\E[\Weightvec_\Timeidx \cdot \bar{\DecisionVec}_\Timeidx|\History_{\Timeidx-1}] = \E[\Weightvec_\Timeidx \cdot \bar{\DecisionVec}_\Timeidx] = \SinglePrimal_{\mathcal{Z}}$. Furthermore, from Claim \ref{fluid}-\ref{fluid:opt}, we have $\E[\Weightvec_\Timeidx \cdot \bar{\DecisionVec}_\Timeidx] \geq \frac{\E[ \NetReward(\{ \hat{\DecisionVec}_\Timeidx\}_{\Timeidx=1}^\TotalTime;\OverCost)]}{T}{ -\BigO({1}/{\TotalTime^2})}
$. From the same line of reasoning, we can apply Claim \ref{fluid}-\ref{fluid:rate} to 
 show that $\E[(\CapRatioVec-\bar{{\DecisionVec}}_\Timeidx)|\History_{\Timeidx - 1}] \geq \mathbf{0}$. Lastly, Claim \ref{fluid}-\ref{fluid:rate} again implies $\E[(\ServiceVec_\Timeidx-\bar{\DecisionVec}_\Timeidx)|\History_{\Timeidx - 1}] \geq \epsilon \mathbf{1}$ where $ \mathbf{1}$ is the vector of ones. Combining these lower-bounds with $\DualOverVec_\Timeidx\geq \mathbf{0}$, $\DualHardVec_\Timeidx \geq \mathbf{0}$ and $\DualBuildVec_\Timeidx = \zeta \BuildUpVec_{\Timeidx-1}\geq \mathbf{0}$, we obtain the final line \eqref{line:fluid}.\par  
We now sum line \eqref{line:fluid} for $t \in [\Stopping]$ to deduce that, for any feasible \allocation{} $\{\hat{\DecisionVec}_\Timeidx\}_{\Timeidx=1}^\TotalTime$ and filtration $\{\History_\Timeidx\}_{\Timeidx=1}^\TotalTime$:
\begin{align*}
\sum_{\Timeidx=1}^{\Stopping}
\E[K_t|\History_{\Timeidx-1}] &\geq \frac{\Stopping \E[    \NetReward(\{ \hat{\DecisionVec}_\Timeidx\}_{\Timeidx=1}^\TotalTime;\OverCost)]}{T} {-\BigO\Big(\frac{\Stopping}{\TotalTime^2}\Big)}
+ \zeta \ServiceSlack\sum_{\Timeidx=1}^{\Stopping-1}
\Norm{\BuildUpVec_{\Timeidx}}{1} - \BigO(\Stopping \zeta) \\
&= 
\E[    \NetReward(\{ \hat{\DecisionVec}_\Timeidx\}_{\Timeidx=1}^\TotalTime;\OverCost)] - \frac{T-\Stopping}{T}\E[    \NetReward(\{ \hat{\DecisionVec}_\Timeidx\}_{\Timeidx=1}^\TotalTime;\OverCost)] {-\BigO\Big(\frac{\Stopping}{\TotalTime^2}\Big)}
+
\zeta \ServiceSlack\sum_{\Timeidx=1}^{\Stopping-1}
\Norm{\BuildUpVec_{\Timeidx}}{1} 
- \BigO(\Stopping\zeta)\\
&\geq \E[    \NetReward(\{ \hat{\DecisionVec}_\Timeidx\}_{\Timeidx=1}^\TotalTime;\OverCost)] - (T - \Stopping) + 
\zeta \ServiceSlack\sum_{\Timeidx=1}^{\Stopping-1}
\Norm{\BuildUpVec_{\Timeidx}}{1} - \BigO(T\zeta).
\end{align*}
In the last line, we used the fact that (i) $ \NetReward(\{ \hat{\DecisionVec}_\Timeidx\}_{\Timeidx=1}^\TotalTime;\OverCost)\leq \TotalTime$ (because the per-period reward is at most one) and (ii) $\Stopping \leq \TotalTime$ for every sample path. {(note that $\BigO(\Stopping/\TotalTime^2) = \BigO(1/\TotalTime)$, which is absorbed by $\BigO(\TotalTime\zeta)$)}.
We finally take the outer expectation of the preceding inequality over the entire history to obtain
\begin{align}
\E\left[
\sum_{\Timeidx=1}^{\Stopping}
\E[K_t|\History_{\Timeidx-1}] 
\right] \geq \E[    \NetReward(\{ \hat{\DecisionVec}_\Timeidx\}_{\Timeidx=1}^\TotalTime;\OverCost)] - 
\E\left[(T - \Stopping)\right] + \zeta \ServiceSlack
\E\left[\sum_{\Timeidx=1}^{\Stopping-1}
\Norm{\BuildUpVec_{\Timeidx}}{1}\right] - \BigO(T\zeta).
\end{align}
It only remains to prove that the left-hand side is equivalent to 
\begin{equation}
\E\left[
\sum_{\Timeidx=1}^{\Stopping}
\E[K_t|\History_{\Timeidx-1}] 
\right]  = \E\left[\sum_{\Timeidx=1}^{\Stopping} K_t\right].\label{line:optional}
\end{equation}
To prove this, define a stochastic process $Y_t = K_t - \E[K_t|\History_{\Timeidx -1 }]$ and $X_t = \sum_{\tau=1}^{t} Y_t$ with $X_0 := 0$. We observe that $Y_t$ is $\History_\Timeidx$-measurable and $\E[Y_t|\History_{\Timeidx - 1}] = 0$. Hence, $\{Y_\Timeidx\}$ is a Martingale difference sequence with respect to filtration $\{\History_{\Timeidx}\}$. Furthermore, we note that $\Stopping$ is a bounded stopping time with respect to the filtration $\{\History_{\Timeidx}\}$. Hence, the optional stopping theorem implies that $\E[X_{\Stopping}] = X_0 = 0$, which implies line \eqref{line:optional}. This completes the proof of \Cref{lemma:pseudo+lower}.

\color{black}
{\bf Proof of Claim \ref{fluid}}. 
For ease of reference, we rewrite the definition of the static control defined in the claim.
\begin{equation}
\begin{split}
{\SinglePrimal_{\mathcal{Z}}}&:= \max_{\substack{\DecisionVec(\Arrival) \in \FeasibleSet{\Target} \\
\forall \Arrival \in \ArrivalSupp}} \E[\Weightvec\cdot \DecisionVec(\Arrival)] \quad \text{s.t.} \ 
\E\left[\DecisionVec(\Arrival)\right] \leq \CapRatioVec  \label{eq:P}
\end{split}
\end{equation}
where the expectation is with respect to $\mathbf{A} = (\Weightvec, \Target)\sim \ArrivalDist$. This optimization problem can be viewed as the stochastic program where the decision is to choose a function (control) $\DecisionVec: \ArrivalSupp \to \Simplex$ subject to $\DecisionVec(\Arrival) \in \FeasibleSet{\Target}
$. That is, for each arrival type $\Arrival = (\Weightvec, \Target)$, we choose the matching rate $\DecisionVec(\Arrival) \in \Simplex$ if $\Target = 0$ (free case) or we set it as $\DecisionVec(\Arrival) = \BasisVec{\Target}$ otherwise. The objective is to maximize the functional $\E[\Weightvec\cdot\DecisionVec(\Arrival)]$, which is the expected per-period \allocation{} reward given the control $\DecisionVec(\cdot)$, subject to {the (ex-ante) capacity constraint on the matching rate}. 
Define $\bar{\DecisionVec}(\cdot)$ as the optimal mapping that solves the problem in \eqref{eq:P}. In the following, we show that $\bar{\DecisionVec}(\cdot)$ satisfies the condition \ref{fluid:feasible}-\ref{fluid:opt} stated in Claim \ref{fluid}.\par 
\noindent{\bf Proof of Claim \ref{fluid} \ref{fluid:feasible} and \ref{fluid:rate}.} This trivially follows from the feasibility condition of $\bar{\DecisionVec}(\cdot)$ in line \eqref{eq:P}.

\noindent{\bf \bf Proof of Claim \ref{fluid} \ref{fluid:opt}.} Let $\{\hat{\DecisionVec}_\Timeidx\}_{\Timeidx=1}^\TotalTime$ be an arbitrary (potentially random) sequence that satisfies
    \begin{equation}
\hat{\DecisionVec}_\Timeidx \in \FeasibleSet{\Target_\Timeidx} \quad \forall \Timeidx \in [\TotalTime], \quad \sum_{\Timeidx=1}^\TotalTime \Indicator[\Target_\Timeidx = 0] \Decision_{\Timeidx,\Locidx}
    \leq \left(\Capacity_\Locidx - \sum_{\Timeidx=1}^\TotalTime \Indicator[\Target_\Timeidx = \Locidx]\Decision_{\Timeidx,\Locidx}\right)_{+} \quad \forall \Locidx \in [\Locnum].\label{line:off+feasible}
    \end{equation} 

It suffices to prove
\[
\SinglePrimal_{\mathcal{Z}}
\geq 
\E\Big[
\frac{\NetReward({\{\hat{\DecisionVec}_\Timeidx\}}_{\Timeidx=1}^\TotalTime; \OverCost)}{\TotalTime}
\Big] - \BigO(1/T^2).
\]

To prove this, we now consider the dual problem of $\SinglePrimal_\mathcal{Z}$, defined as:
\begin{equation}
  \textsf{D}_{\mathcal{Z}} (\boldsymbol{\phi}):= \E\left[
\max_{\DecisionVec(\Arrival) \in \FeasibleSet{\Target}}(\Weightvec - \boldsymbol{\phi})\cdot \DecisionVec(\Arrival) + \CapRatioVec\cdot \boldsymbol{\phi}
    \right], \quad 
\SingleDual_{\mathcal{Z}} := \min_{\boldsymbol{\phi}\geq \mathbf{0}} \textsf{D}(\boldsymbol{\phi}).  \label{line:def+D_z}
\end{equation}

By the compactness and convexity of $\FeasibleSet{\Target}$ for all $\Target \in \{0\}\cup[m]$, a straightforward application of Sion's minmax theorem \citep{sion1958general} implies that $\SingleDual_{\mathcal{Z}} = \SinglePrimal_{\mathcal{Z}}$. Furthermore, as we show in Claim \ref{claim:arrival+rate} (see \Cref{apx+fluid+dual}), we can restrict the domain to $\Norm{\boldsymbol{\phi}}{\infty}\leq 1$ without loss of optimality. In summary, we have
\begin{equation}
\SinglePrimal_{\mathcal{Z}} = \SingleDual_{\mathcal{Z}} = \min_{\boldsymbol{\phi}\geq \mathbf{0}} \textsf{D}_{\mathcal{Z}}(\boldsymbol{\phi}) = \min_{\boldsymbol{\phi}\geq \mathbf{0}, \Norm{\boldsymbol{\phi}}{\infty}\leq 1} \textsf{D}_{\mathcal{Z}}(\boldsymbol{\phi}) \label{line:property}
\end{equation}

Let $N_{\TotalTime,\Locidx} := \sum_{\Timeidx=1}^\TotalTime \Indicator[\Target=\Locidx]$ denote the total number of tied cases at affiliate $\Locidx$ over the horizon. Define event $G_{\TotalTime}$ as 
\begin{equation}
    G_{\TotalTime} := \{N_{\TotalTime,\Locidx} \leq \Capacity_\Locidx, \ \forall \Locidx\in[\Locnum]\}.
\end{equation}

Note that, due to our assumption that $\CapRatio - \P[\Target = \Locidx] = \Omega(1)$, a straightforward application of Hoeffding's inequality and union bound implies that $\P[G_{\TotalTime}] \geq 1 - \BigO(\exp(-\TotalTime)) \geq 1-\BigO(1/\TotalTime^2)$. Given these ingredients, we show that $\SinglePrimal_{\mathcal{Z}}\geq \E[\tfrac{\NetReward(\{\hat{\DecisionVec}_\Timeidx\}_{\Timeidx=1}^\TotalTime; \OverCost)}{\TotalTime}] - \BigO(1/\TotalTime^2)$ in the following. 
\begin{align}
&\E[\NetReward(\{\hat{\DecisionVec}_\Timeidx\}_{\Timeidx=1}^\TotalTime; \OverCost)] \nonumber  \\
&\leq \E\left[
\max_{
\substack{
\DecisionVec_\Timeidx \in \mathcal{Z}(\Target_\Timeidx),\\
\Timeidx\in [\TotalTime]}}
\sum_{\Timeidx=1}^\TotalTime
\Weightvec_\Timeidx \cdot \DecisionVec_\Timeidx
    - \OverCost\sum_{\Locidx=1}^{\Locnum} \PositivePart{\sum_{\Timeidx=1}^\TotalTime \DecisionEach - \Capacity_\Locidx } 
    \text{s.t.} \ \sum_{\Timeidx=1}^\TotalTime \Indicator [\Target_\Timeidx  = 0]\DecisionEach \leq \left(\Capacity_\Locidx - \sum_{\Timeidx=1}^\TotalTime \Indicator[\Target_\Timeidx = \Locidx]\Decision_{\Timeidx,\Locidx} \right)_{+}, \forall \Locidx \in [\Locnum]
\right] 
    \label{line:relax+cap}\\
&\leq 
\begin{aligned}
&\E\Biggl[
\Biggl\{\max_{
\substack{
\DecisionVec_\Timeidx \in \mathcal{Z}(\Target_\Timeidx),\\
\Timeidx\in [\TotalTime]}}
\sum_{\Timeidx=1}^\TotalTime
\Weightvec_\Timeidx \cdot \DecisionVec_\Timeidx
    - \OverCost\sum_{\Locidx=1}^{\Locnum} \PositivePart{\sum_{\Timeidx=1}^\TotalTime \DecisionEach - \Capacity_\Locidx } 
    \\ &
    \quad \text{s.t.} \ \sum_{\Timeidx=1}^\TotalTime \Indicator [\Target_\Timeidx  = 0]\DecisionEach \leq \left(\Capacity_\Locidx - \sum_{\Timeidx=1}^\TotalTime \Indicator[\Target_\Timeidx = \Locidx]\Decision_{\Timeidx,\Locidx} \right)_{+}, \forall \Locidx \in [\Locnum]
\Biggr\}\cdot \Indicator[G_\TotalTime]
\Biggr]  +  \BigO(1/\TotalTime)
\end{aligned}
     \label{line:G_T}\\
&= \E\left[
\left(
\max_{\DecisionVec_\Timeidx \in \mathcal{Z}(\Target_\Timeidx), \Timeidx\in [\TotalTime]}
\sum_{\Timeidx=1}^\TotalTime
\Weightvec_\Timeidx \cdot \DecisionVec_\Timeidx
    \quad 
    \text{s.t.}\ \sum_{\Timeidx=1}^\TotalTime \DecisionEach \leq \Capacity_\Locidx, \forall \Locidx \in [\Locnum]\right)\cdot\Indicator[G_\TotalTime]\right] + \BigO(1/\TotalTime)
    \label{line:packing+cap}\\
&\leq \E\left[
\left(
\max_{\DecisionVec_\Timeidx\in \mathcal{Z}(\Target_\Timeidx), \Timeidx\in [\TotalTime]}
\sum_{\Timeidx=1}^{\TotalTime}
    \Weightvec_\Timeidx \cdot \DecisionVec_\Timeidx
- \boldsymbol{\phi}\cdot\left(\sum_{\Timeidx=1}^\TotalTime  \DecisionVec_\Timeidx -\CapacityVec 
\right)
 \right)\cdot\Indicator[G_\TotalTime]   \right]
+ \BigO(1/\TotalTime)
\quad \forall \boldsymbol{\phi}\geq\mathbf{0}, \ \Norm{\boldsymbol{\phi}}{\infty}\leq 1 
\label{line:lagragify}
\\
&= \sum_{\Timeidx=1}^\TotalTime \E\left[ 
\max_{\DecisionVec_\Timeidx \in \mathcal{Z}(\Target_\Timeidx)} 
(\Weightvec_\Timeidx - \boldsymbol{\phi})\cdot \DecisionVec_\Timeidx
+ \boldsymbol{\phi}\cdot \CapRatioVec
\Big|G_\TotalTime\right]\P[G_\TotalTime] + \BigO(1/\TotalTime)
\quad \forall \boldsymbol{\phi}\geq\mathbf{0}, \ \Norm{\boldsymbol{\phi}}{\infty}\leq 1 
\label{line:separable}
\\
&\leq 
\sum_{\Timeidx=1}^\TotalTime \left\{\textsf{D}_{\mathcal{Z}}(\boldsymbol{\phi}) + \BigO(1/\TotalTime^2) \right\}+ \BigO(1/\TotalTime) 
\quad \forall \boldsymbol{\phi}\geq\mathbf{0}, \ \Norm{\boldsymbol{\phi}}{\infty}\leq 1 
\label{line:conditional+exp}\\ 
&= \TotalTime \textsf{D}_{\mathcal{Z}}(\boldsymbol{\phi}) + \BigO(1/\TotalTime)
\quad \forall \boldsymbol{\phi}\geq\mathbf{0}, \ \Norm{\boldsymbol{\phi}}{\infty}\leq 1 
\label{line:iid}
\\
&\leq \TotalTime \SingleDual_{\mathcal{Z}} + \BigO(1/\TotalTime) \label{line:take+minimum}
\end{align}
We elaborate on each line in the following. 
Line \eqref{line:relax+cap} follows from maximizing the net matching reward subject to line \eqref{line:off+feasible}. 
In line \eqref{line:G_T}, we bounded the expected total net matching reward under the event $G_\TotalTime^c$ by using that  that $\P[G_\TotalTime] \geq 1-\BigO(1/\TotalTime^2)$ and the total net matching reward is $\BigO(\TotalTime)$. 
Line \eqref{line:packing+cap} is because the capacity constraint in line \eqref{line:G_T} reduces to $\sum_{\Timeidx=1}^\TotalTime \DecisionEach \leq \Capacity_\Locidx$ under event $G_\TotalTime$. 
Lines \eqref{line:lagragify} follows from weak-duality.

To understand line \eqref{line:conditional+exp}, let $\hat{D}_{\mathcal{Z}}(\boldsymbol{\phi};\Arrival_\Timeidx) := \max_{\DecisionVec_\Timeidx \in \mathcal{Z}(\Target_\Timeidx)} 
(\Weightvec_\Timeidx - \boldsymbol{\phi})\cdot \DecisionVec_\Timeidx
+ \boldsymbol{\phi}\cdot \CapRatioVec$. For any $\boldsymbol{\phi}$ such that $\Norm{\boldsymbol{\phi}}{\infty}\leq 1$, arrival $\Arrival_\Timeidx = (\Weightvec_\Timeidx, \Target_\Timeidx)$, and $\DecisionVec_\Timeidx \in \mathcal{Z}(\Target_\Timeidx)$, 
\begin{equation}
| (\Weightvec_\Timeidx - \boldsymbol{\phi})\cdot \DecisionVec_\Timeidx
+ \boldsymbol{\phi}\cdot \CapRatioVec | \leq 
\Norm{\Weightvec_\Timeidx}{\infty}\Norm{\DecisionVec_\Timeidx}{1} +  \Norm{\boldsymbol{\phi}}{\infty}\Norm{\DecisionVec_\Timeidx}{1} + 
\Norm{\boldsymbol{\phi}}{\infty}\Norm{\CapRatioVec}{1} \leq 3.
\end{equation}
The first inequality is due to Cauchy-Schwartz inequality, and the last inequality is because of our assumption that the reward is at most one and $\sum_{\Locidx=1}^\Locnum \CapRatio_\Locidx \leq 1$ (which was without loss of generality). Hence, we must have $|\hat{D}_{\mathcal{Z}}(\boldsymbol{\phi};\Arrival_\Timeidx)| \leq 3$ for any arrival $\Arrival_\Timeidx$
and $\Norm{\boldsymbol{\phi}}{\infty}\leq 1$. We now use this fact to derive \eqref{line:conditional+exp} as follows:
\begin{align}
    \textsf{D}_{\mathcal{Z}}(\boldsymbol{\phi}
    )&= \E[\hat{D}_{\mathcal{Z}}(\boldsymbol{\phi};\Arrival_\Timeidx)] \\ 
    &= \E[\hat{D}_{\mathcal{Z}}(\boldsymbol{\phi};\Arrival_\Timeidx)|G_{\TotalTime}]\P[G_\TotalTime] + \E[\hat{D}_{\mathcal{Z}}(\boldsymbol{\phi};\Arrival_\Timeidx)|G_{\TotalTime}^c]\P[G^c_\TotalTime] \\ 
    &\geq \E[\hat{D}_{\mathcal{Z}}(\boldsymbol{\phi};\Arrival_\Timeidx)|G_{\TotalTime}]\P[G_\TotalTime] - \BigO(1/\TotalTime^2). \label{line:imply}
\end{align}
The first line is by definition of $\textsf{D}_{\mathcal{Z}}(\cdot)$ (see line \eqref{line:def+D_z}) and the arrival is i.i.d. The second line is just re-writing the expectation. The last line is because $|\hat{D}_{\mathcal{Z}}(\boldsymbol{\phi};\Arrival_\Timeidx)| \leq 3 = \BigO(1)$ for all arrival $\Arrival_\Timeidx$ and $\Norm{\boldsymbol{\phi}}{\infty}\leq 1$, along with $\P[G_\TotalTime^c] \leq \BigO(1/\TotalTime^2)$. Hence, from line \eqref{line:imply}, we directly deduce line \eqref{line:conditional+exp}.

Finally, line \eqref{line:take+minimum} follows from taking minimum over $\boldsymbol{\phi}$ of the inequality \eqref{line:iid}, along with line \eqref{line:property}. The proof is complete because $\SingleDual_{\mathcal{Z}} = \SinglePrimal_{\mathcal{Z}}$ by line \eqref{line:property}. \hfill \halmos

\color{black}

\subsection{Proof of \Cref{lemma:pseudo+upper}}\label{apx:pseudo+upper}
By the definition of the pseudo-rewards, we have the following:
\begin{align}
\sum_{\Timeidx=1}^{\Stopping} K_t &= \sum_{\Timeidx=1}^{\Stopping} \{ \Weightvec_\Timeidx\cdot\DecisionVec_\Timeidx + \DualOverVec_\Timeidx\cdot(\CapRatioVec-\DecisionVec_\Timeidx) + \DualHardVec_\Timeidx\cdot(\CapRatioVec-\DecisionVec_\Timeidx) - \zeta D_\Timeidx  \} \nonumber \\
&= \sum_{\Timeidx=1}^{\Stopping}
\Weightvec_\Timeidx\cdot\DecisionVec_\Timeidx + \sum_{\Timeidx=1}^{\Stopping}\DualOverVec_\Timeidx\cdot(\CapRatioVec-\DecisionVec_\Timeidx) + 
\sum_{\Timeidx=1}^{\Stopping}\DualHardVec_\Timeidx\cdot(\CapRatioVec-\DecisionVec_\Timeidx) - \zeta\psi(\BuildUpVec_{\Stopping})\label{line:b0=0} \\
\begin{split} 
&=\sum_{\Timeidx=1}^{\Stopping}
\Weightvec_\Timeidx\cdot\DecisionVec_\Timeidx  - \OverCost \sum_{\Locidx=1}^\Locnum \PositivePart{\sum_{\Timeidx=1}^\TotalTime \DecisionEach - \Capacity_\Locidx} + 
\sum_{\Timeidx=1}^{\Stopping}\DualOverVec_\Timeidx\cdot(\CapRatioVec-\DecisionVec_\Timeidx) - \sum_{\Timeidx=1}^{\TotalTime}\DualOverVec^\star\cdot(\CapRatioVec-\DecisionVec_\Timeidx) + \\
&\quad \sum_{\Timeidx=1}^{\Stopping}\DualHardVec_\Timeidx\cdot(\CapRatioVec-\DecisionVec_\Timeidx) - \zeta\psi(\BuildUpVec_{\Stopping})\label{line:add+subtract} 
\end{split} \\ 
\begin{split}
&\leq
\underbrace{
\sum_{\Timeidx=1}^{\TotalTime}
\Weightvec_\Timeidx\cdot\DecisionVec_\Timeidx  - \OverCost \sum_{\Locidx=1}^\Locnum \PositivePart{\sum_{\Timeidx=1}^\TotalTime \DecisionEach - \Capacity_\Locidx}}_{
\NetReward(\{\DecisionVec_\Timeidx\}_{\Timeidx=1}^\TotalTime)
}
+ 
\underbrace{
\sum_{\Timeidx=1}^{\TotalTime}\DualOverVec_\Timeidx\cdot(\CapRatioVec-\DecisionVec_\Timeidx) - \sum_{\Timeidx=1}^{\TotalTime}\DualOverVec^\star\cdot(\CapRatioVec-\DecisionVec_\Timeidx)
}_{R_{\DualOver}} + \\
&\quad  
\underbrace{
\sum_{\Timeidx=1}^{\Stopping}\DualHardVec_\Timeidx\cdot(\CapRatioVec-\DecisionVec_\Timeidx)}_{R_{\DualHard}} + 2\OverCost(\TotalTime - \Stopping) 
- \zeta\psi(\BuildUpVec_{\Stopping}) 
\label{line:cauchy}
\end{split}
\end{align}
Line \eqref{line:b0=0} is because of the telescoping sum of $D_\Timeidx$ and $\BuildUpVec_0 = \mathbf{0}$. In line \eqref{line:add+subtract}, we added and subtracted the over-allocation cost and used Claim \ref{fact:over}. In line \eqref{line:cauchy}, we used the Cauchy-Schwartz inequality to bound $\sum_{\Timeidx=1}^{\Stopping}\DualOverVec_\Timeidx\cdot(\CapRatioVec-\DecisionVec_\Timeidx) \leq \sum_{\Timeidx=1}^{\TotalTime}\DualOverVec_\Timeidx\cdot(\CapRatioVec-\DecisionVec_\Timeidx) + 2\OverCost(\TotalTime - \Stopping)$ by using that (i) $\DualOverVec \in [0,\OverCost]^\Locnum$, (ii) $\Norm{\CapRatioVec - \DecisionVec_{\Timeidx}}{1} \leq \Norm{\CapRatioVec}{1} + \Norm{\DecisionVec_\Timeidx}{1} \leq 2$ (by definition of $\DecisionVec_\Timeidx$ and our assumption that $\Norm{\CapRatioVec}{1} \leq  1$). \par 

In the following claims, we show that $R_\DualOver \leq \BigO(\sqrt{\TotalTime})$ and $R_{\DualHard} \leq \BigO(\sqrt{\TotalTime}) - (1+2\OverCost)(\TotalTime - \Stopping)$. Plugging the bound of $R_{\DualOver}$ and $R_{\DualHard}$ into inequality \eqref{line:cauchy} completes the proof of \Cref{lemma:pseudo+upper}.
\begin{claim}\label{claim:theta}
$R_\DualOver \leq \BigO(\sqrt{\TotalTime})$ for every sample path.
\end{claim}
{\bf Proof of Claim \ref{claim:theta}}.
We first describe how we can view  $\{\DualOverVec_\Timeidx\}_{\Timeidx=1}^{\TotalTime}$ as a sequence of the online mirror descent iterates for properly defined primitives stated in Claim \ref{fact:average+conv}.  Let
$f_t(\DualOverVec) := \boldsymbol{\DualOver} \cdot (\CapRatioVec - \DecisionVec_{\Timeidx})$, $\mathcal{D}:= [0,\OverCost]^\Locnum$, $\hat{\mathbf{g}}_t :=  \CapRatioVec - \DecisionVec_{\Timeidx}$, and $h(\boldsymbol{\DualOver}):= \sum_{\Locidx=1}^{\Locnum} \DualOver_i\log(\DualOver_i)$. Note that the gradient has no noise (i.e., $\mathbf{\hat{u}}_t = \mathbf{0}$). With these primitives, it is straightforward to see that the update rule \eqref{line:update+temp} is reduced to that of line \eqref{eq:update+theta} (for dual variables $\DualOverVec$) in \Cref{ALG+Surrogate+D}. Furthermore, $\Norm{\mathbf{\hat{g}}}{1} \leq \Norm{\DecisionVec_{\Timeidx}}{1} + \Norm{ \CapRatioVec}{1} \leq 2$ and $h(\cdot)$ is $(1/\OverCost)$-strongly convex with respect to $\Norm{\cdot}{\infty}$ in $\mathcal{D}$. Hence, we invoke Claim \ref{fact:average+conv} with the fixed step size $\eta_t= \eta$:
\begin{equation}
R_{\DualOver} = \sum_{\Timeidx=1}^{\TotalTime} (f_\Timeidx(\DualOverVec_\Timeidx) - f_\Timeidx(\DualOverVec^\star))\leq 
2\OverCost \eta \TotalTime + \frac{1}{\eta}V_h(\DualOverVec^\star,\DualOverVec_1) \label{line:R+theta}
\end{equation}
The bregman distance $D(\boldsymbol{\DualOver}^\star, \boldsymbol{\DualOver}_1)$ is bounded because $\DualOverVec^\star$ and $\DualOverVec_1$ are all in $\mathcal{D}$ by definition. Hence, with $\eta = \Theta(1/\sqrt{\TotalTime})$, the right-hand side of \eqref{line:R+theta} is $\BigO(\sqrt{\TotalTime})$.
\hfill \halmos

{
\begin{claim}\label{claim:lambda}
$R_{\DualHard} \leq \BigO(\sqrt{\TotalTime}) -(1+2\OverCost)(\TotalTime - \Stopping)$ for every sample path.
\end{claim}
{\bf Proof of Claim \ref{claim:lambda}}. Let us first denote the upper-bound of $\DualHard$ as $\bar{\DualHard} := \frac{1+2\OverCost}{\underline{\CapRatio}}$.
Similar to the proof of Claim \ref{claim:theta}, we define $f_t(\DualHardVec) := \DualHardVec \cdot (\CapRatioVec -\DecisionVec_{\Timeidx})$, $\mathcal{D}:= [0,\bar{\DualHard}]^\Locnum$, $\hat{\mathbf{g}}_t := \CapRatioVec - \DecisionVec_{\Timeidx}$, and $h(\DualHardVec):= \sum_{\Locidx=1}^{\Locnum} \DualHard_i\log(\DualHard_i)$. It is straightforward to see that the update rule \eqref{line:update+temp} is reduced to that of line \eqref{eq:update+theta} (for dual variable $\DualHardVec$) in \Cref{ALG+Surrogate+D}. Furthermore, $\Norm{\mathbf{\hat{g}}}{1} \leq 2$ and $h(\cdot)$ is $1/\bar{\DualHard}$-strongly convex with respect to $\Norm{\cdot}{\infty}$ in $\mathcal{D}$. Hence, we invoke Claim \ref{fact:average+conv} with the fixed step size $\eta_t = \eta$ to obtain that, for any $\DualHardVec^\star \in [0, \bar{\DualHard}]^\Locnum$,
\begin{align*}
R_{\DualHard}  &\leq \underbrace{
\vphantom{\sum_{\Timeidx=1}^\TotalTime}
2\bar{\DualHard} \eta \TotalTime + \frac{1}{\eta}V_h(\DualHardVec^\star,\DualHardVec_1)}_{\vardiamond} +
\underbrace{\sum_{\Timeidx=1}^{\Stopping}\DualHardVec^\star \cdot (\CapRatioVec - \DecisionVec_{\Timeidx})}_{\diamondsuit} 
\end{align*}
Since both $\DualHardVec^\star$ and $\DualHardVec_1$ are bounded, the term $\vardiamond$ is $\BigO(\sqrt{T})$ with $\eta = \Theta(1/\sqrt{\TotalTime})$. We now show that, by properly choosing $\DualHardVec^\star$, the term $\diamondsuit$ is upper bounded by $(1+2\OverCost)(\Stopping - \TotalTime)$. We consider two cases. First, if $\Stopping = \TotalTime$, then we take $\DualHardVec^\star = \boldsymbol{0}$ and the bound is trivial. Otherwise, $\Stopping < \TotalTime$ and there exists $i$ such that $\sum_{\Timeidx=1}^{\Stopping} \DecisionEach \geq \CapRatio_i \TotalTime$ (see line \eqref{def:stopping} in \Cref{subsec:ALG+D+pfsketch}). Take that coordinate $i$ and let $\DualHardVec^\star = \frac{1+2\OverCost}{\CapRatio_i}\BasisVec{i}$. Note that $\DualHardVec^\star \in [0, \bar{\DualHard}]^\Locnum$ by definition of $\bar{\DualHard}$. The term $\diamondsuit$ is then given by 
\begin{equation*}
\begin{split}
\diamondsuit &= \frac{1+2\OverCost}{\CapRatio_i}\sum_{\Timeidx=1}^{\Stopping}(\CapRatio_i - \DecisionEach)\\
&\leq \frac{1+2\OverCost}{\CapRatio_i}((\Stopping - \TotalTime)\CapRatio_i)\\
&= (1+2\OverCost)(\Stopping - \TotalTime)
\end{split}
\end{equation*}
The second line is by the definition of the coordinate $i$ we have chosen. The last line simply follows from $\underline{\CapRatio} \leq \CapRatio_\Locidx$.
Hence, taking the worst case over $\DualHardVec^* \in \{\mathbf{0}, \frac{1+2\OverCost}{\CapRatio_1}\BasisVec{1}, \frac{1+2\OverCost}{\CapRatio_2}\BasisVec{2},...,\frac{1+2\OverCost}{\CapRatio_1}\BasisVec{\Locnum}\}$, the term $\diamondsuit$ is always at most $(1+2\alpha)(\Stopping-\TotalTime)$. This completes the proof. 
}
\hfill\halmos

\subsection{Missing Details for Step 3: Finishing the Proof of Lemmas \texorpdfstring{\ref{lemma:ALG+D+reward}}{} and \texorpdfstring{\ref{Lemma+ALG+D+Build}}{}}\label{apx+finish+ALG+1+build}

With inequality \eqref{ineq:important} in place to prove Lemma \ref{lemma:ALG+D+reward}, we set $\hat{\DecisionVec}_\Timeidx = \DecisionVec^*_\Timeidx$ for all $\Timeidx \in [\TotalTime]$ (i.e., the optimal offline allocation {in \Cref{def:offline}}) and 
observe that $\textsf(\textsf{B}') \geq 0$ because the 
\backlog{} is non-negative. The proof is complete by plugging $\zeta = \Theta(1/\sqrt{\TotalTime})$ as stated in \Cref{thm:ALG+Surrogate+Dual}. 

To prove  \Cref{Lemma+ALG+D+Build}, we set $\{ \hat{\DecisionVec}_\Timeidx\}_{\Timeidx=1}^\TotalTime$ to be the matching profile that maximizes the net \allocation{} reward subject to $\hat{\DecisionVec}_\Timeidx \in \FeasibleSet{\Target_\Timeidx}$ for all $\Timeidx \in [\TotalTime]$ and \eqref{line:hard+constraint}. By definition of such $\{ \hat{\DecisionVec}_\Timeidx\}_{\Timeidx=1}^\TotalTime$, term $\textsf{(A)}$ is nonnegative and hence, term $\textsf{(B')}$ is upper bounded by $\BigO(\sqrt{\TotalTime} + \zeta \TotalTime)$. 
{Hence, the average backlog \emph{up to} the stopping time is $\BigO(1/\ServiceSlack)$. The remaining step is to show that the average backlog \emph{after} the stopping time is not dominant. Observe that (i) by construction of \CA{}, only tied cases can arrive after $\Stopping$ and (ii) the arrival rate of the tied case is strictly smaller than $\CapRatioVec$. Therefore, the backlog process is a random walk with a negative drift. In the following, we combine these observations with a drift analysis of the potential function $\psi(\cdot)$ to show that the average backlog accrued after the stopping time is $\BigO(1)$. 

We begin by decomposing the average \backlog{} for the entire horizon as 
\begin{equation}
\frac{1}{\TotalTime}\E\left[\sum_{\Timeidx=1}^{\TotalTime} \Norm{\BuildUpVec_\Timeidx}{1}\right] = \frac{1}{\TotalTime}\E\left[\sum_{\Timeidx=1}^{{\Stopping}-1
} \Norm{\BuildUpVec_\Timeidx}{1}\right]  + \frac{1}{\TotalTime}\E\left[\sum_{\Timeidx={\Stopping}}^\TotalTime \Norm{\BuildUpVec_\Timeidx}{1}\right]
\end{equation}
The first (second, resp.) term is the contribution of the \backlog{} until (after, resp.) the stopping time. We first bound the first term. Recall from the inequality \eqref{ineq:important} that
\begin{equation}
\zeta\ServiceSlack\E\left[\sum_{\Timeidx=1}^{\Stopping - 1} \Norm{\BuildUpVec_\Timeidx}{1} \right] 
+\zeta\E[\psi(\BuildUpVec_{\Stopping})] 
\leq \BigO(\sqrt{\TotalTime} + \zeta\TotalTime). \label{line:important+2}
\end{equation}
Noting that $\psi(\cdot)$ is always non-negative, we straightforwardly obtain
\begin{equation}
\frac{1}{\TotalTime}\E\left[\sum_{\Timeidx=1}^{{\Stopping}-1
} \Norm{\BuildUpVec_\Timeidx}{1}\right] \leq \BigO\Big(\frac{1}{\ServiceSlack}(1+\frac{1}{\zeta\sqrt{\TotalTime}})\Big) \leq \BigO\Big(\frac{1}{\epsilon}\Big). \label{ineq:1/eps}
\end{equation}
where the last inequality is because we set $\zeta = \Theta(1/\sqrt{\TotalTime})$ in the stable regime.

We now turn our attention to bounding the contribution of the \backlog{} after $\Timeidx \geq {\Stopping}$. Toward this goal, we first obtain a bound of the \backlog{} at the stopping time. 
\begin{claim}\label{claim:build+up+stopping}
$\E[\Norm{\BuildUpVec_{\Stopping}}{1}^2] \leq \BigO(\TotalTime)$
\end{claim}
\textbf{Proof of Claim \ref{claim:build+up+stopping}}.
The proof is the direct consequence of line \eqref{line:important+2}. To be precise, because the \backlog{} is non-negative, the line \eqref{line:important+2} implies that
\begin{equation*}
\E[\psi(\BuildUpVec_{\Stopping})] \leq \BigO\left(\TotalTime + \frac{\sqrt{\TotalTime}}{\zeta}\right) = \BigO(T) 
\end{equation*}
where the last inequality follows since $\zeta = \Theta(1/\sqrt{\TotalTime})$. We finally recall the definition of $\psi(\BuildUpVec_{\Stopping}) = \frac{1}{2}\Norm{\BuildUpVec_{\Stopping}}{2}^2$. By Cauchy-Schwarz inequality, we have $\Norm{\BuildUpVec_{\Stopping}}{1}^2 \leq m \Norm{\BuildUpVec_{\Stopping}}{2}^2 $ (for every sample path). Hence, $\BigO(T)$ bound on $\E[\psi(\BuildUpVec_{\Stopping})]$ is equivalent of showing the same order of bound on $\E[\Norm{\BuildUpVec_{\Stopping}}{1}^2]$.\footnote{
Because of our assumption that the capacity grows linearly with $\TotalTime$, the number of affiliates must be $O(1)$.
} This completes the proof.
\hfill \halmos

Building on this claim and the fact that only the tied cases can arrive at each affiliate after $\Timeidx\geq \Stopping$, we now prove that the average backlog after the stopping time is $\BigO(1)$, which completes the proof of \Cref{Lemma+ALG+D+Build}.
{\begin{claim}\label{claim:Backlog+after+stopping}
$\frac{1}{\TotalTime}\E\left[\sum_{\Timeidx={\Stopping}}^\TotalTime \Norm{\BuildUpVec_\Timeidx}{1}\right] \leq \BigO(1)$
\end{claim}
\textbf{Proof of Claim \ref{claim:Backlog+after+stopping}}.
For each affiliate $\Locidx$ and $\Timeidx \geq \Stopping + 1$, we have
\begin{align}
\E[\BuildUp_{\Timeidx, \Locidx}^2 - \BuildUp_{\Timeidx-1, \Locidx}^2 | \History_{\Timeidx-1}] &\leq \BuildUp_{\Timeidx-1,\Locidx}(\E[\Decision_{\Timeidx,\Locidx}-\Service_{\Timeidx,\Locidx}|\History_{\Timeidx-1,\Locidx}]) + \BigO(1) \\
&= \BuildUp_{\Timeidx-1,\Locidx}(\P[\Target_\Timeidx = \Locidx] -\CapRatio_\Locidx - \ServiceSlack) + \BigO(1)\\
&\leq -(d+\ServiceSlack)\BuildUp_{\Timeidx-1,\Locidx} + \BigO(1)
\end{align}
The first line follows from (one-dimensional version of) \Cref{lemma:drift}. For the second line, we used the fact that only tied cases can arrive at each affiliate after the stopping time, along with the i.i.d. nature of the arrival and service process. For the last line, we used our assumption that there exists constant $d >0$ such that $\CapRatio_\Locidx - \P[\Target=\Locidx] \geq d$. Summing up the previous inequalities over $\Stopping+1 \leq \Timeidx \leq \TotalTime$ (and using the similar Martingale argument used to justify line \eqref{line:optional}), we deduce that
\begin{equation}
    \E[\BuildUp_{\TotalTime,\Locidx}
    ^2-\BuildUp_{\Stopping,\Locidx}^2] \leq - (d+\ServiceSlack)\E\Big[\sum_{\Timeidx=\Stopping}^{\TotalTime-1}\BuildUp_{\Timeidx,\Locidx}\Big] + O(\E[\TotalTime-\Stopping]).
\end{equation}
Combined with Claim \ref{claim:build+up+stopping}, the above inequality directly implies that $\E\Big[\sum_{\Timeidx=\Stopping}^{\TotalTime-1}\BuildUp_{\Timeidx,\Locidx}\Big] \leq \BigO(\TotalTime)$ and $\E[\BuildUp_{\TotalTime,\Locidx}] \leq \BigO(\sqrt{\TotalTime})$. Summing these two bounds across all $\Locidx \in [\Locnum]$, we conclude that
\begin{equation}
\frac{1}{\TotalTime}\E\left[\sum_{\Timeidx={\Stopping}}^\TotalTime \Norm{\BuildUpVec_\Timeidx}{1} \right] \leq \BigO(1).
\end{equation}
This completes the proof. 
\hfill\halmos
}
{\subsection{Some Remarks on \texorpdfstring{\Cref{ALG+Surrogate+D}}{}}\label{apx+remark+ALG+D}
\subsubsection{Bound on the Expected Stopping Time}\label{apx+stopping+exp}\hfill \\
For technical reasons, our analysis relied on taking inaction for free cases after the first time an affiliate reaches its capacity (i.e. the stopping time $\Stopping$ defined in line \eqref{def:stopping}). In the following, under a mild assumption, we show that the stopping time is near the end of the horizon in expectation. To introduce this assumption, we recall the static control introduced in Claim \ref{fluid}. 
\begin{equation}
{\SinglePrimal_{\mathcal{Z}}}:= \max_{\substack{\DecisionVec(\Arrival) \in \FeasibleSet{\Target} \\
\forall \Arrival \in \ArrivalSupp}} \E[\Weightvec\cdot \DecisionVec(\Arrival)] \quad \text{s.t.} \ 
\E\left[\DecisionVec(\Arrival)\right] \leq \CapRatioVec
\end{equation}
We make a mild assumption that the value of the static control is a strictly positive constant as well as the penalty cost parameter $\OverCost$ (for over-allocation) 

\begin{assumption}\label{ass:single+primal} \hfill 
\begin{enumerate}[label=(\alph*)]
    \item The penalty cost parameter for over-allocation is $\OverCost >1$ (that is, greater than the maximum reward of each case).
    \item The arrival distribution $\ArrivalDist$ is such that $\SinglePrimal_{\mathcal{Z}} = \Omega(1)$.\footnote{A sufficient condition for this assumption is, for example, that the reward distribution satisfies $\E[\Weightvec\cdot \CapRatioVec] = \Omega(1)$.}
\end{enumerate}
\end{assumption}

The following proposition shows that, under the above assumption, the stopping time is very near the end of the horizon in expectation. 
\begin{proposition}\label{prop:exp+stop}
Under the stable regime (\Cref{def:eps+regimes}) and Assumption \ref{ass:single+primal}, we have $\E[\Stopping] \geq \TotalTime - \BigO(\sqrt{\TotalTime})$.
\end{proposition}
{\bf Proof of \Cref{prop:exp+stop}}. The proof follows similar steps of that for \Cref{lemma:pseudo+lower}. First, we establish a lower bound on the expected pseudo-reward as follows.

\begin{lemma}\label{claim:stopping+pseudo+lower}
$\E\Big[\sum_{\Timeidx=1}^{\Stopping} K_\Timeidx \Big] \geq \SinglePrimal_{\mathcal{Z}} \TotalTime - \E[\TotalTime-\Stopping] - \BigO(\zeta\TotalTime)$. 
\end{lemma}
{\bf Proof of \Cref{claim:stopping+pseudo+lower}}. The proof only requires a minor modification of the steps taken in \Cref{apx+pseudo+lower}. Specifically, we have
\begin{align*}
\E[K_\Timeidx | \History_{\Timeidx -1}] 
&\geq\E[\Weightvec_\Timeidx\cdot\bar{\DecisionVec}_\Timeidx|\History_{\Timeidx - 1}] + 
\DualOverVec_\Timeidx\cdot\E[(\CapRatioVec-\bar{{\DecisionVec}}_\Timeidx)|\History_{\Timeidx - 1}] + 
\DualHardVec_\Timeidx\cdot\E[(\CapRatioVec-\bar{{\DecisionVec}}_\Timeidx)|\History_{\Timeidx - 1}] +
\DualBuildVec_\Timeidx\cdot\E[(\ServiceVec_\Timeidx-\bar{\DecisionVec}_\Timeidx)|\History_{\Timeidx - 1}]- \BigO(\zeta) \\
&\geq \SinglePrimal_{\mathcal{Z}} - \BigO(\zeta)
\end{align*}
where the first line follows from line \eqref{line:third} and the second line is because of Claim \ref{fluid}. Summing the inequality over $\Timeidx \in [\Stopping]$ and taking expectation, we obtain:
\begin{equation}
    \E\Big[\sum_{\Timeidx=1}^{\Stopping} K_\Timeidx \Big] \geq \SinglePrimal_\mathcal{Z}\E[\Stopping] - \BigO(\zeta\TotalTime) = \SinglePrimal_\mathcal{Z}\TotalTime - \SinglePrimal_\mathcal{Z}\E[\TotalTime-\Stopping] - \BigO(\zeta\TotalTime) \geq \SinglePrimal_\mathcal{Z}\TotalTime -\E[\TotalTime-\Stopping] - \BigO(\zeta\TotalTime)
\end{equation}
 where the last equality is because the per-period reward is at most one. This completes the proof.\hfill\halmos

We now establish an upper bound on the expected pseudo-rewards in terms of $\E[\Stopping]$. 
\begin{lemma}\label{claim:stopping+pseudo+upper}
$\E\Big[\sum_{\Timeidx=1}^{\Stopping} K_\Timeidx \Big] \leq \E[\Stopping]\SinglePrimal_{\mathcal{Z}} - \E[\TotalTime-\Stopping] + \BigO(\sqrt{\TotalTime})$    
\end{lemma}
{\bf Proof of \Cref{claim:stopping+pseudo+upper}.} 
The proof is a modification of the steps taken in \Cref{apx:pseudo+upper}.
Recall the dual program of the static control problem defined in Claim \ref{fluid}:
\begin{equation*}
  \textsf{D}_{\mathcal{Z}}(\boldsymbol{\phi}):= \E\left[
\max_{\DecisionVec(\Arrival) \in \FeasibleSet{\Target}}(\Weightvec - \boldsymbol{\phi})\cdot \DecisionVec(\Arrival) + \CapRatioVec\cdot \boldsymbol{\phi}
    \right], \quad 
\SingleDual_{\mathcal{Z}} := \min_{\boldsymbol{\phi}\geq \mathbf{0}} \textsf{D}_{\mathcal{Z}}(\boldsymbol{\phi}). \label{dual+Z} 
\end{equation*}

By the compactness and convexity of $\FeasibleSet{\Target}$ for all $\Target \in \{0\}\cup[m]$, a straightforward application of Sion's minmax theorem \citep{sion1958general} implies that $\SingleDual_{\mathcal{Z}} = \SinglePrimal_{\mathcal{Z}}$. 

Equipped with the definition of $\SingleDual_{\mathcal{Z}}$, we now obtain the following:
\begin{align}
\sum_{\Timeidx=1}^{\Stopping} K_t &= \sum_{\Timeidx=1}^{\Stopping} \{ \Weightvec_\Timeidx\cdot\DecisionVec_\Timeidx + \DualOverVec_\Timeidx\cdot(\CapRatioVec-\DecisionVec_\Timeidx) + \DualHardVec_\Timeidx\cdot(\CapRatioVec-\DecisionVec_\Timeidx) - \zeta D_\Timeidx  \} \nonumber \\
&= \sum_{\Timeidx=1}^{\Stopping}
\Weightvec_\Timeidx\cdot\DecisionVec_\Timeidx + \sum_{\Timeidx=1}^{\Stopping}\DualOverVec_\Timeidx\cdot(\CapRatioVec-\DecisionVec_\Timeidx) + 
\sum_{\Timeidx=1}^{\Stopping}\DualHardVec_\Timeidx\cdot(\CapRatioVec-\DecisionVec_\Timeidx) - \zeta\psi(\BuildUpVec_{\Stopping}) \nonumber \\  
&\leq \sum_{\Timeidx=1}^{\Stopping}
\Weightvec_\Timeidx\cdot\DecisionVec_\Timeidx + 
\underbrace{\sum_{\Timeidx=1}^{\Stopping}\DualOverVec_\Timeidx\cdot(\CapRatioVec-\DecisionVec_\Timeidx)}_{R_{\DualOver}} +
\underbrace{\sum_{\Timeidx=1}^{\Stopping}\DualHardVec_\Timeidx\cdot(\CapRatioVec-\DecisionVec_\Timeidx)}_{R_{\DualHard}}\label{eq:temp}
\end{align}

The second line is due to the telescoping sum of $D_\Timeidx$, along with $\BuildUpVec_0 = \mathbf{0}$. The last line follows from the non-negativity of the \backlog{}. We now bound $R_{\DualOver}$ and $R_{\DualHard}$ as follows. For the former, define $\boldsymbol{\phi}^* \geq \mathbf{0}$ such that $\textsf{D}_{\mathcal{Z}}(\boldsymbol{\phi}^*) = \SingleDual_{\mathcal{Z}}$. In Claim \ref{claim:arrival+rate}-\ref{fluid+dual+bound} (see \Cref{apx+fluid+dual}), we establish that  $\Norm{\boldsymbol{\phi}^*}{\infty} \leq 1$ and therefore $\Norm{\boldsymbol{\phi}^*}{\infty} \leq \OverCost$ (by Assumption \ref{ass:single+primal}). Hence, by mirroring the arguments in Claim \ref{claim:theta}, we invoke Claim \ref{fact:average+conv} to obtain
\begin{equation}
    R_{\theta} \leq \sum_{\Timeidx=1}^{\Stopping} \boldsymbol{\phi}^*\cdot(\CapRatioVec-\DecisionVec_\Timeidx) + \BigO(\sqrt{\TotalTime}). \label{bound:theta}
\end{equation}

We now bound $R_{\DualHard}$. Let $\bar{\DualHard} = \frac{1+2\OverCost}{\underline{\CapRatio}}$. By mirroring the arguments for the proof of Claim \ref{claim:lambda}, we invoke Claim \ref{fact:average+conv} to obtain that, for any $\DualHardVec^\star \in [0, \bar{\DualHard}]^\Locnum$, $
R_{\DualHard}  \leq \BigO(\sqrt{\TotalTime}) +
\sum_{\Timeidx=1}^{\Stopping}\DualHardVec^\star \cdot (\CapRatioVec - \DecisionVec_{\Timeidx})
$. We now show that, by properly choosing $\DualHardVec^\star$, the second term $\sum_{\Timeidx=1}^{\Stopping}\DualHardVec^\star \cdot (\CapRatioVec - \DecisionVec_{\Timeidx})$ is upper bounded by $(\Stopping - \TotalTime)$. We consider two cases. First, if $\Stopping = \TotalTime$, then we take $\DualHardVec^\star = \boldsymbol{0}$ and the bound is trivial. Otherwise, $\Stopping < \TotalTime$ and there exists $i$ such that $\sum_{\Timeidx=1}^{\Stopping} \DecisionEach \geq \CapRatio_i \TotalTime$ (see line \eqref{def:stopping} in \Cref{subsec:ALG+D+pfsketch}). Take that coordinate $i$ and let $\DualHardVec^\star = \frac{1}{\CapRatio_i}\BasisVec{i}$. Note that $\DualHardVec^\star \in [0, \bar{\DualHard}]^\Locnum$ by definition of $\bar{\DualHard}$. Hence, the term $\diamondsuit$ is then given by
\begin{equation}
\frac{1}{\CapRatio_i}\sum_{\Timeidx=1}^{\Stopping}(\CapRatio_i - \DecisionEach) = \frac{1}{\CapRatio_i}\sum_{\Timeidx=1}^{\Stopping}(\CapRatio_i - \DecisionEach)\\
\leq \frac{1}{\CapRatio_i}((\Stopping - \TotalTime)\CapRatio_i)\\
= \Stopping - \TotalTime. \nonumber
\end{equation}
The first inequality is by the definition of the coordinate $i$ we have chosen. Taking the worst case over $\DualHardVec^* \in \{\mathbf{0}, \frac{1}{\CapRatio_1}\BasisVec{1}, \frac{1}{\CapRatio_2}\BasisVec{2},...,\frac{1}{\CapRatio_1}\BasisVec{\Locnum}\}$ of the above inequality, the term $\diamondsuit$ is always at most $\Stopping-\TotalTime$. Hence, we have deduced that
\begin{equation}
    R_{\DualHard} \leq \BigO(\sqrt{\TotalTime}) + \Stopping-\TotalTime \label{bound:lambda}
\end{equation}

Plugging the bounds \eqref{bound:theta} and \eqref{bound:lambda} into line \eqref{eq:temp} and taking expectation, we have

\begin{align*}
\E\Big[\sum_{\Timeidx=}^{\Stopping} K_\Timeidx \Big] &\leq \Big[\sum_{\Timeidx=1}^{\Stopping}
(\Weightvec_\Timeidx - \boldsymbol{\phi}^*)\cdot\DecisionVec_\Timeidx + \CapRatioVec\cdot\boldsymbol{\phi}^* 
\Big] - \E[\TotalTime - \Stopping] + \BigO(\sqrt{\TotalTime})\nonumber \\
&\leq \Big[\sum_{\Timeidx=1}^{\Stopping}\max_{\DecisionVec_\Timeidx \in \FeasibleSet{\Target_\Timeidx}}
(\Weightvec_\Timeidx - \boldsymbol{\phi}^*)\cdot\DecisionVec_\Timeidx + \CapRatioVec\cdot\boldsymbol{\phi}^* 
\Big] - \E[\TotalTime-\Stopping] + \BigO(\sqrt{\TotalTime}) \\
&= \E[\Stopping]\SingleDual_{\mathcal{Z}} - \E[\TotalTime-\Stopping] + \BigO(\sqrt{\TotalTime})
\end{align*}

In the last line, we used Wald's equation and the i.i.d. nature of the arrival process. The proof is complete by noting that $\SingleDual_{\mathcal{Z}} = \SinglePrimal_{\mathcal{Z}}$.\hfill\halmos

To complete the proof of \Cref{prop:exp+stop}, we combine Lemmas \ref{claim:stopping+pseudo+lower} and \ref{claim:stopping+pseudo+upper} to obtain
\begin{equation}
\SinglePrimal_{\mathcal{Z}}\E[\TotalTime - \Stopping] \leq \BigO(\sqrt{\TotalTime} + \zeta\TotalTime) \leq \BigO(\sqrt{\TotalTime})
\end{equation}
where the last inequality is because $\zeta = \BigO(1/\sqrt{\TotalTime})$ under the stable regime. The proof is complete by dividing both sides of the inequality by $\SinglePrimal_{\mathcal{Z}}$, which is justified by Assumption \ref{ass:single+primal}.  \hfill\halmos

\subsubsection{Removing Dummy Affiliate}\label{apx+no+inaction}\hfill\\
In the main body of our paper, we have allowed inaction for free cases throughout the entire horizon. In this appendix, we partially relax this assumption. Specifically, we consider a variant of \Cref{ALG+Surrogate+D} which cannot take inaction for free cases before the first time an affiliate reaches its capacity. Formally, for a given target affiliate $\Target$, define a type-feasibility set (without any inaction) as 
\begin{align}
    \mathcal{X}(\Target_{\Timeidx}) = \begin{cases}
    \{\DecisionVec \in \R_{+}:\sum_{\Locidx=1}^\Locnum z_\Locidx = 1 \} \quad &\text{if $\Target_\Timeidx = 0$}\\
    \{\BasisVec{\Target}\} \quad &\text{otherwise}.
    \end{cases} 
\end{align}

For $\Timeidx \leq \Stopping$ (see line \eqref{def:stopping} for its definition), we now modify \ref{line:primal+decision} of \Cref{ALG+Surrogate+D} as 
\begin{equation}
    \DecisionVec_{\Timeidx} \in \argmax_{\DecisionVec \in \mathcal{X}(\Target_{\Timeidx})} (\Weightvec_\Timeidx - \DualOverVec_\Timeidx - \DualHardVec_\Timeidx - \zeta\BuildUpVec_{\Timeidx-1})\cdot \DecisionVec.
\end{equation}

The rest of the algorithm remains unchanged (including line \ref{line:primal+end} that leaves free cases unmatched once any of the affiliates exhaust its capacity). We call this algorithm $\overline{\CA{}}$. In the following, we show that $\overline{\CA{}}$ has the same theoretical guarantee as \CA{} under the stable regime {and an additional assumption that $\sum_{\Locidx=1}^\Locnum \CapRatio_\Locidx = 1$.}

\begin{corollary}\label{cor:CA-DL-bar}
Let $\eta = \Theta(1/\sqrt{\TotalTime})$ and $\zeta = \Theta(1/\sqrt{\TotalTime})$. Under the stable regime (\Cref{def:eps+regimes}) and {$\sum_{\Locidx=1}^\Locnum \CapRatio_\Locidx = 1$}, the regret (\Cref{def:regret}) of $\overline{\CA{}}$ is $\BigO(\sqrt{\TotalTime} + \frac{\BuildUpCost}{\ServiceSlack})$. 
\end{corollary}
{\bf Proof of \Cref{cor:CA-DL-bar}.} The proof mirrors that of \Cref{thm:ALG+Surrogate+Dual} except for a minor change of Claim \ref{fluid} (See \Cref{apx+pseudo+lower}). Specifically, we need to modify the claim by defining a static control problem \emph{without any inaction}, and prove that the value of the optimal static control is an upper bound on the net matching value of the offline benchmark (\Cref{def:offline}).
\begin{claim}[Static Control: Modification of Claim \ref{fluid}]\label{claim:static+no+rejection}
Consider the following static control problem:
\begin{equation}
\begin{split}
{\SinglePrimal_{\mathcal{X}}}&:= \max_{\substack{\DecisionVec(\Arrival) \in \mathcal{X}(\Target) \\
\forall \Arrival \in \ArrivalSupp}} \E[\Weightvec\cdot \DecisionVec(\Arrival)] \quad \text{s.t.} \ 
\E\left[\DecisionVec(\Arrival)\right] \leq \CapRatioVec 
\end{split}
\end{equation}
where the expectation is with respect to $\mathbf{A} = (\Weightvec, \Target)\sim \ArrivalDist$. Let $\tilde{\DecisionVec}: \ArrivalSupp \to \Simplex$ be the optimal static control that solves the above program. Then we have
\begin{enumerate}[label=(\alph*)]
    \item $\tilde{\DecisionVec}(\Arrival) \in \mathcal{X}(\Target)$ for any $\Arrival = (\Weightvec, \Target)$ 
    \item $\E_{\Arrival}[\tilde{\DecisionVec}(\Arrival)] \leq \CapRatioVec$ 
    \item $\SinglePrimal_{\mathcal{X}} = \E_{\Arrival}[\Weightvec \cdot \tilde{\DecisionVec}(\Arrival)] \geq \dfrac{\E[
    \NetReward(\{ \hat{\DecisionVec}_\Timeidx\}_{\Timeidx=1}^\TotalTime;\OverCost) 
    ]}{T} 
    $ for and any \allocation{} profile $\{ \hat{\DecisionVec}_\Timeidx\}_{\Timeidx=1}^\TotalTime$ that satisfies 
    \begin{equation*}
\hat{\DecisionVec}_\Timeidx \in \FeasibleSet{\Target_\Timeidx} \quad \forall \Timeidx \in [\TotalTime], \quad \sum_{\Timeidx=1}^\TotalTime \Indicator[\Target_\Timeidx = 0] \Decision_{\Timeidx,\Locidx}
    \leq \left(\Capacity_\Locidx - \sum_{\Timeidx=1}^\TotalTime \Indicator[\Target_\Timeidx = \Locidx]\Decision_{\Timeidx,\Locidx}\right)_{+} \quad \forall \Locidx \in [\Locnum].
    \end{equation*} 
\end{enumerate} 
\end{claim}
{\bf Proof of Claim \ref{claim:static+no+rejection}}. Parts (a) and (b) are trivial. The only minor modification of the proof is for part (c). We first recall the definition of $\SinglePrimal_{\mathcal{Z}}$ from line \eqref{eq:P} (which allowed for inaction only for free cases). Because of our assumption that $\sum_{\Locidx=1}^\Locnum \CapRatio_\Locidx =  1$, the optimal static control $\SinglePrimal_{\mathcal{Z}}$ matches every case to the actual affiliates, and hence we have $\SinglePrimal_{\mathcal{Z}} = \SinglePrimal_{\mathcal{X}}$. The rest of the proof follows the identical steps taken in the proof of Claim \ref{fluid}-(c).
\hfill\halmos

Using this claim, one can follow the steps from line \eqref{line:ref+decision} onward to prove the desired result. Specifically, we replace $\bar{\mathbf{z}}_t$ in line \eqref{line:ref+decision} with $\tilde{\DecisionVec}_\Timeidx := \tilde{\DecisionVec}(\Arrival_\Timeidx).$ and follow the same steps from from line \eqref{line:ref+decision} onward. For brevity, this part of the proof is omitted. \hfill\halmos

Finally, we also show that, the stopping time of $\overline{\CA{}}$ is near the end of the horizon in expectation. 
\begin{corollary}\label{cor:stopping+no+rejection}
Under Assumption \ref{ass:single+primal} and $\sum_{\Locidx=1}^\Locnum \CapRatio_\Locidx = 1$, the stopping time $\Stopping$ (see line \eqref{def:stopping} for definition) of $\overline{\CA{}}$ satisfies $\E[\Stopping] \geq \TotalTime-\BigO(\sqrt{\TotalTime})$ under the stable regime (\Cref{def:eps+regimes}). 
\end{corollary}
{\bf Proof of \Cref{cor:stopping+no+rejection}}. The proof again follows the identical steps taken for the proof of \Cref{prop:exp+stop}. We only point out a minor change in the proof of \Cref{claim:stopping+pseudo+upper}. We define a static dual problem with type-feasibility set $\mathcal{X}$ as 
\begin{equation}
  \textsf{D}_{\mathcal{X}}(\boldsymbol{\phi}):= \E\left[
\max_{\DecisionVec(\Arrival) \in \mathcal{X}(\Target)}(\Weightvec - \boldsymbol{\phi})\cdot \DecisionVec(\Arrival) + \CapRatioVec\cdot \boldsymbol{\phi}
    \right], \quad 
\SingleDual_{\mathcal{X}} := \min_{\boldsymbol{\phi}\geq \mathbf{0}} \textsf{D}_{\mathcal{X}}(\boldsymbol{\phi}).  
\end{equation}

By strong duality, we have $\SinglePrimal_{\mathcal{X}} = \SingleDual_{\mathcal{X}}$. To completely extend the proof of \Cref{claim:stopping+pseudo+upper} to $\overline{\CA{}}$, we need to show that there exists $\boldsymbol{\phi}^* \in \argmin_{\boldsymbol{\phi}\geq\mathbf{0}} \textsf{D}_{\mathcal{X}}$ such that $\Norm{\boldsymbol{\phi}^*}{\infty} \leq \OverCost$ (and the rest of the proof follows the identical steps from line \eqref{eq:temp} and onwards). We claim that such $\boldsymbol{\phi}^*$ indeed exists. To see this, we first recall the static dual problem under the type-feasibility set $\mathcal{Z}(\cdot)$ from line \eqref{line:def+D_z}: 
\begin{equation*}
  \textsf{D}_{\mathcal{Z}}(\boldsymbol{\phi}):= \E\left[
\max_{\DecisionVec(\Arrival) \in \FeasibleSet{\Target}}(\Weightvec - \boldsymbol{\phi})\cdot \DecisionVec(\Arrival) + \CapRatioVec\cdot \boldsymbol{\phi}
    \right], \quad 
\SingleDual_{\mathcal{Z}} := \min_{\boldsymbol{\phi}\geq \mathbf{0}} \textsf{D}_{\mathcal{Z}}(\boldsymbol{\phi}).
\end{equation*}
Let $\boldsymbol{\phi}^*\in \argmin_{\boldsymbol{\phi}\geq\mathbf{0}} \textsf{D}_{\mathcal{Z}}$. As we prove in Claim \ref{claim:arrival+rate}-\ref{fluid+dual+bound} (see \Cref{apx+fluid+dual}), we have $\Norm{\boldsymbol{\phi}^*}{\infty} \leq \OverCost$ under Assumption \ref{ass:single+primal}. We now observe that
\begin{equation}
    \textsf{D}_{\mathcal{X}}(\boldsymbol{\phi}^*) 
    \overset{\text{(a)}}{\leq} 
    \textsf{D}_{\mathcal{Z}}(\boldsymbol{\phi}^*)
    = \SinglePrimal_{\mathcal{Z}}
    \overset{\text{(b)}}{=}  \SinglePrimal_{\mathcal{X}} =
    \SingleDual_{\mathcal{X}}
\end{equation}

Inequality (a) is because $\mathcal{X}(\cdot) \subseteq \mathcal{Z}(\cdot)$. In equation (b), we used our assumption that $\sum_{\Locidx=1}^\Locnum \CapRatio_\Locidx = 1$ (see the proof of claim  \ref{claim:static+no+rejection}). Because $\textsf{D}_{\mathcal{X}}(\boldsymbol{\phi}^*) \geq \SingleDual_{\mathcal{X}}$ by definition, the above inequality implies that $\textsf{D}_{\mathcal{X}}(\boldsymbol{\phi}^*) = \SingleDual_{\mathcal{X}}$, or   $\boldsymbol{\phi}^* \in \argmin_{\boldsymbol{\phi}\geq\mathbf{0}} \textsf{D}_{\mathcal{X}}$ equivalently. This proves the desired claim. \hfill\halmos
}
\subsection{Proof of \texorpdfstring{\Cref{near+part+a}}{}}\label{apx+near+a}
Our first step is to show that whenever the \backlog{} is above $1/\zeta$, we have a constant negative drift on the expected \backlog{}.
\begin{claim}\label{claim:b+neg+drift} Let $t \leq {\Stopping}$.
Whenever $b_{t-1,i} \geq 1/\zeta$, we have
$\E[b_{t,i} - b_{t-1,i} | \History_{t-1}] \leq - (\rho_i - \P[\Target_\Timeidx = i])$
\end{claim}
{\bf Proof of Claim \ref{claim:b+neg+drift}}.
From \Cref{prop:impossibility}, we assume that $\BuildUpCost = o(\TotalTime)$ without loss. Because $1/\zeta = \sqrt{T/\gamma}$ is super-constant under this assumption, we also have $1/\zeta >1$ without loss. Then, $b_{t,i}- b_{t-1,i} = z_{t,i} - s_{t,i}$ whenever $b_{t-1,i} \geq 1/\zeta$. Furthermore, we can write $z_{t,i} = \mathbbm{1}[\Target_\Timeidx=i]$ whenever $b_{t-1,i} \geq  1/\zeta$. To see this, whenever $\Target_\Timeidx = 0$ (a free case) and $b_{t-1,i}\geq 1/\zeta$, we observe that $w_{t,i}- \theta_{t,i} - \lambda_{t,i} - \zeta b_{t-1,i}<0$. Combining this with the fact that the adjusted score of the dummy affiliate is always zero (by definition), we have  $z_{t,i}=0$ whenever $b_{t-1,i}>1/\zeta$. On the other hand, if the arrival $\Timeidx$ is a tied case to location $j \neq i$, we again have $z_{t,i}=0$. Combining, we have
\begin{align*}
    \E[b_{t,i}- b_{t-1,i}|H_{t-1}] = \E[z_{t,i}- s_{t,i}|H_{t-1}] = \P[\Target_\Timeidx=i] - \rho_i - \epsilon.
\end{align*}
The proof is complete because $\ServiceSlack \geq 0$.
\hfill \halmos

We now prove the following lemma, which establishes $\BigO\Big(\frac{1}{\zeta} + 1\Big)$ bound on the average backlog (in expectation) for any $\ServiceSlack \geq 0$.
\begin{lemma}[Upper Bound on \Backlog{} for \Cref{ALG+Surrogate+D}]\label{lemma+near+critical+a}
For any given $\zeta \geq 0$, the expected average \backlog{} of \Cref{ALG+Surrogate+D} is
\begin{equation*}
\frac{1}{\TotalTime}\E\Big[\sum_{\Timeidx=1}^\TotalTime\Norm{\BuildUpVec_\Timeidx}{1}\Big] \leq \BigO\Big(\frac{1}{\zeta} + 1\Big).
\end{equation*}
\end{lemma}

{\bf Proof of \Cref{lemma+near+critical+a}}. We apply the following result extracted from \cite{wei2023constant} and \cite{gupta2022greedy}. 
\begin{claim}[Lemma 5 of \cite{wei2023constant}]\label{claim:gupta} Let $\Psi(t)$ be an $\{\mathcal{H}_t\}$-adapted stochastic process satisfying:
\begin{itemize}
    \item Bounded variation: $\lvert \Psi(t+1) - \Psi(t) \rvert \leq K$ 
    \item Negative drift: $\E[\Psi(t+1) - \Psi(t)|\mathcal{H}_t] \leq - d$ whenever $\Psi(t) \geq D$
    \item $\Psi(0) \leq K+D$
\end{itemize}
Then, we have 
\begin{align*}
    \E[\Psi(t)] \leq K\left(1+ 
    \lceil \frac{D}{K}\rceil \right) + K\left(\frac{K-d}{2d}\right).
\end{align*}
\end{claim}

To apply Claim \ref{claim:gupta} to our setting, let $\Psi(t) := b_{t,i}$ let $d_i = \rho_i - \P[\Target_\Timeidx = i]$. We recall that $d_i > 0$ is a positive constant by our assumption (see \Cref{sec:model}). Furthermore, we have $b_{0,i} =0$ and $|b_{t,i} - b_{t-1,i}| \leq 1$. From Claim \ref{claim:b+neg+drift}, we have $\E[b_{t,i}- b_{t-1,i}|H_{t}] \leq -d_i$ whenever $b_{t-1,i} \geq 1/\zeta$. Applying Claim \ref{claim:gupta} with, $K=1$, and $D = 1/\zeta $, we have 
\begin{align}
    \E[b_{t,i}] \leq 1 + \lceil \frac{1}{\zeta}\rceil + \frac{1-d_i}{2d_i} \leq \frac{1}{\zeta} + \frac{1 + 3d_i}{2d_i} \label{line:b+near}
\end{align}
 Hence, we have $\E[\lVert \mathbf{b}_t\rVert_1] \leq \BigO\left(\frac{1}{\zeta} + 1\right)$ for all $t \leq {\Stopping}$. Finally, the average \backlog{} is bounded by

\begin{align*}
\frac{1}{T}\E\left[\sum_{t=1}^T \lVert \mathbf{b}_t\rVert_1\right] 
&= \frac{1}{T}\E\left[\sum_{t=1}^{{\Stopping}
} \lVert \mathbf{b}_t\rVert_1\right]  + \frac{1}{T}\E\left[\sum_{t={\Stopping}+1}^T \lVert \mathbf{b}_t\rVert_1\right] \\
&\leq \BigO\left(\frac{1}{\zeta} + 1\right)
\label{line:build+up}
\end{align*}
The last inequality is because we have $\E[\lVert \mathbf{b}_t\rVert_1] \leq \BigO\left(\frac{1}{\zeta} + 1\right)$ for all $t \leq {\Stopping}$ from inequality \eqref{line:b+near}, {along with Claim \ref{claim:Backlog+after+stopping} (See Appendix \ref{apx+finish+ALG+1+build}). This completes the proof of \Cref{lemma+near+critical+a}.}
\hfill\halmos

{\bf Proof of \Cref{near+part+a}}. We now complete the proof of  \Cref{near+part+a}. For ease of reference, we recall the inequality \eqref{ineq:important} (see \Cref{subsec:ALG+D+pfsketch}) in the following: {for \emph{any} \allocation{} profile $\{ \hat{\DecisionVec}_{\Timeidx}\}_{\Timeidx=1}^\TotalTime$that satisfies $\hat{\DecisionVec}_{\Timeidx} \in \FeasibleSet{\Target_\Timeidx}$ for all $\Timeidx \in [\TotalTime]$ and \eqref{line:hard+constraint} (see \Cref{def:offline}),}
\begin{equation*}
\underbrace{
\vphantom{\sum_{\Timeidx=1}^\TotalTime}
\E[\NetReward(\{ \hat{\DecisionVec}_{\Timeidx}\}_{\Timeidx=1}^\TotalTime; \OverCost)]
-
\E[\NetReward(\{{\DecisionVec}_{\Timeidx}\}_{\Timeidx=1}^\TotalTime; \OverCost)]
}_{\mathclap{\textsf{(A)}}}+ 
\zeta\ServiceSlack\E\left[\sum_{\Timeidx=1}^{\Stopping - 1} \Norm{\BuildUpVec_\Timeidx}{1} \right] 
+\zeta\E[\psi(\BuildUpVec_{\Stopping})] 
\leq \BigO(\sqrt{\TotalTime} + \zeta\TotalTime) \
\end{equation*}
Note that the above inequality holds for \emph{any} $\epsilon\geq 0$, including the near-critical regime. This implies that the term $\textsf{(A)}$ is still $\BigO(\sqrt{\TotalTime} + \zeta \TotalTime)$. Let $\{{\DecisionVec}^*_\Timeidx\}_{\Timeidx=1}^\TotalTime$ denote the the optimal offline solution (\Cref{def:offline}). By utilizing the same decomposition as line \eqref{line:regret+decomp}, we bound the regret by
\begin{align*}
\E[\textsf{OPT}(\alpha, \gamma) ] - \E[\textsf{ALG}(\alpha, \gamma)] 
&\leq \underbrace{
\vphantom{\sum_{\Timeidx=1}^\TotalTime}
\E[\NetReward(\{ {\DecisionVec}^*_{\Timeidx}\}_{\Timeidx=1}^\TotalTime; \OverCost)]
-
\E[\NetReward(\{{\DecisionVec}_{\Timeidx}\}_{\Timeidx=1}^\TotalTime; \OverCost)]
}_{\mathclap{\textsf{(A)}}} + 
\underbrace{
\frac{\gamma}{T}\E\left[\sum_{t=1}^T \lVert \mathbf{b}_t\rVert_1\right]}_{\text{Lemma \ref{lemma+near+critical+a}}} \\
&\leq \BigO(\zeta T + \sqrt{T}) + \BigO\left(\frac{\gamma}{\zeta} + \gamma\right)\\
&\leq \BigO(\sqrt{\gamma T})
\end{align*}
where the last line follows from $\zeta = \Theta(\sqrt{\frac{\gamma}{T}})$ under the near-critical regime.


\majorrevcolor{
\section{Matching Lower Bounds}\label{apx:regret+lower}

We begin by clarifying an important aspect of our upper bound in Theorem~1. While our formal definition of regret compares the online algorithm’s objective to that of the benchmark $\OPT(\alpha, \gamma)$, the proof in \Cref{subsec:ALG+D+pfsketch} in fact establishes a stronger result: our regret bound holds even when benchmarked against $\OPT(\alpha, 0)$. This is because we explicitly separate the analysis of net matching reward (\Cref{lemma:ALG+D+reward}) and backlog penalty (\Cref{Lemma+ALG+D+Build}). In particular, the the net matching reward of \CA{} is lower-bounded relative to \emph{any feasible} offline solution—including $\OPT(\alpha, 0)$, which incurs no backlog cost~(\Cref{lemma:pseudo+lower}). As a result, the bound $O(\sqrt{T} + \gamma/\varepsilon)$ holds against the stronger benchmark $\OPT(\alpha, 0)$.

In this section, we show that the our regret upper-bound is asymptotically optimal with respect to $\OPT(\alpha, 0)$ by establishing a tight lower bound for any online algorithm's regret. In \Cref{apx+lbd+stable}, we show that the regret bound of \Cref{ALG+Surrogate+D}  in the stable regime (\Cref{thm:ALG+Surrogate+Dual}) is asymptotically optimal  with 
respect to both benchmarks $\OPT(\alpha, 0)$ and $\OPT(\alpha, \gamma)$. In \Cref{apx+lbd+nc}, we show that the regret bound of \Cref{ALG+Surrogate+D} in the near-critical regime where $\epsilon = O(1/\sqrt{T})$ (\Cref{near+part+a}) is asymptotically optimal with respect to $\OPT(\alpha,0)$ in the spacial case of  $\epsilon = 0$. 

Throughout, we let $\mathcal{I}$ denote an instance of our problem (see the paragraph preceding~\Cref{sec:model}), and use $\ALG_{\mathcal{I}}(\alpha, \gamma)$ to denote the objective achieved by any online algorithm under instance $\mathcal{I}$, and $\OPT_{\mathcal{I}}(\alpha, \gamma)$ the corresponding offline optimum~(\Cref{def:offline}). 

\subsection{Lower Bounds for Stable Regime}\label{apx+lbd+stable}

We first provide a lower bound for regret of any online algorithm against $\OPT(\alpha,0)$, showing our upper-bound in \Cref{thm:ALG+Surrogate+Dual} is asymptotically optimal in the stable regime. Specifically, the following result demonstrates that both the $\sqrt{T}$ and $\gamma/\varepsilon$ terms are necessary to capture the worst-case regret against $\OPT(\alpha, 0)$. \begin{proposition}[Lower Bound on Regret against \texorpdfstring{$\OPT(\alpha, 0)$}{OPT(alpha, 0)}]
\label{prop:lower+bound+OPT0}
In the stable regime~(\Cref{def:eps+regimes}), 
there exist instances for which the regret of any online algorithm with respect to $\OPT(\alpha, 0)$ is $\Omega\left(\sqrt{T} + \frac{\gamma}{\varepsilon}\right)$.
\end{proposition}

{\bf Proof of \Cref{prop:lower+bound+OPT0}}. In the following, we construct two instances, $\mathcal{I}_1$ and $\mathcal{I}_2$, and show that for instance $\mathcal{I}_1$ (resp. $\mathcal{I}_2$), any online algorithm incurs regret of at least $\Omega(\gamma/\varepsilon)$ (resp. $\Omega(\sqrt{T})$). We then  establish the claimed lower bound by taking the worst case over the two instances.

\smallskip 

\emph{(ii) Instance $\mathcal{I}_1$.} We consider an instance with one actual affiliate ($m=1$). The capacity is given by $c = 0.5T$ (i.e., $\rho = 0.5$), and the service rate is $r = 0.5 + \varepsilon$ for positive constant $\epsilon \in (0, 0.5)$. 
At each time $t$, an arrival is a tied (resp. free) case with probability $0.5 - \varepsilon$ (resp. $0.5 + \varepsilon$). The reward of all arrivals is one. Note that we have $\OPT_{\mathcal{I}_1}(\alpha, 0) = 0.5T$. This is because the offline algorithm accepts all tied cases (as required) and fills the remaining capacity using free cases—without incurring any congestion penalty.\footnote{Because $\epsilon = \Omega(1)$ in the stable regime and the arrival rate of tied cases is $\rho-\epsilon$ in this instance, there is no over-allocation cost with high probability.} 

We now upper bound the objective of any online algorithm. First, observe that any online algorithm must accept all tied cases, which arrive at rate $0.5 - \varepsilon$. Since the service rate is \( 0.5 + \varepsilon \), this creates a discrete-time single-server queue with arrival probability \( \lambda = 0.5 - \varepsilon \) and service probability \( \mu = 0.5 + \varepsilon \). The resulting backlog process induced by tied  cases evolves as a reflected random walk on the non-negative integers. This Markov chain is ergodic and admits a well-defined stationary distribution. In steady state, one can show that the expected backlog is \( \Theta(1/\varepsilon) \). By the ergodicity of the chain, the time-average backlog converges to this stationary expectation, implying that the expected backlog over a finite horizon \( T \) is \( \Theta(1/\varepsilon) + o(1) \). Because accepting free cases can only increase the backlog, it follows that the expected average backlog of any online algorithm is at least  $\Theta(1/\varepsilon)$. Combining this with the upper bound on achievable reward ($0.5T$), we conclude $\E[\ALG_{\mathcal{I}_1}(\alpha, \gamma)] \leq 0.5T - \Theta\left(\frac{\gamma}{\varepsilon}\right),$ which implies 
\begin{equation}
\E[\OPT_{\mathcal{I}_1}(\alpha, 0) - \ALG_{\mathcal{I}_1}(\alpha, \gamma)] \geq \Omega(\gamma/\varepsilon). \label{eq:i1-gap-final}
\end{equation}

\smallskip 

\emph{(ii) Instance \(\mathcal{I}_2\).} This instance is adapted from \citet{freund2019good}, with a single affiliate (\(m = 1\)) and all arrivals being free cases ($\Target_t =0$ for all $t \in [T]$). The total capacity is set to \(c = 0.5 T\). At each time $t$, an arrival type is given by  $\mathbf{A}_t = (w_t, 0)$ with the reward \(w_t \in \{1/3, 2/3, 1\}\) being drawn independently with probabilities:
\begin{equation}
\Pr[w_t = 1] = \frac{1}{2} - \frac{1}{\sqrt{T}}, \quad 
\Pr[w_t = 2/3] = \frac{1}{\sqrt{T}}, \quad 
\Pr[w_t = 1/3] = \frac{1}{2}. \label{eq:i2-reward-dist}
\end{equation}

\citet{freund2019good} analyzes a setting without congestion cost and show that, in this instance, no online algorithm can achieve reward within \(\BigO(\sqrt{T})\) of \(\OPT_{\mathcal{I}_2}(\alpha, 0)\). That is, when the objective consists solely of cumulative reward (i.e., \(\gamma = 0\)), we have the following: 

\begin{lemma}[Proposition~4 of \citet{freund2019good}]
\label{lem:freund-lower-i2}
In instance \(\mathcal{I}_2\), no online algorithm can achieve \(\BigO(\sqrt{T})\) regret relative to \(\OPT_{\mathcal{I}_2}(\alpha, 0)\).\footnote{Although this instance does not involve tied arrivals—and hence \(\alpha\) plays no meaningful role—we retain the notation \(\OPT(\alpha, \gamma)\) for consistency.} In particular, for any online algorithm,
\begin{equation}
\E[\OPT_{\mathcal{I}_2}(\alpha, 0) - \ALG_{\mathcal{I}_2}(\alpha, 0)] \geq \Omega(\sqrt{T}). \label{eq:i2-gap}
\end{equation}
\end{lemma}

Since \(\ALG_{\mathcal{I}_2}(\alpha, \gamma) \leq \ALG_{\mathcal{I}_2}(\alpha, 0)\) for any \(\gamma \geq 0\) (because congestion cost only reduces the objective), it follows that the same lower bound holds when the algorithm’s objective includes a congestion penalty. That is, 
\begin{equation}
\E[\OPT_{\mathcal{I}_2}(\alpha, 0) - \ALG_{\mathcal{I}_2}(\alpha, \gamma)] \geq \Omega(\sqrt{T}). \label{eq:i2-gap-final}
\end{equation}

Thus, combining \eqref{eq:i1-gap-final} and \eqref{eq:i2-gap-final}, the worst-case regret of any online algorithm against $\OPT(\alpha, 0)$ is 
$\Omega\left(\max\left\{\sqrt{T}, \frac{\gamma}{\varepsilon}\right\}\right) = \Omega\left(\sqrt{T} + \frac{\gamma}{\varepsilon}\right)$. 
\hfill \halmos

We now turn to our original benchmark  $\OPT(\alpha, \gamma)$ (that penalizes the objective by the congestion cost). The following result provides a lower bound that matches the regret upper bound with respect to  $\OPT(\alpha, \gamma)$ in the stable regime ($\epsilon = \Omega(1)$)(\Cref{thm:ALG+Surrogate+Dual}). 

\begin{proposition}[Lower Bound on Regret against \texorpdfstring{$\OPT(\alpha, \gamma)$}{OPT(alpha, gamma)}]
\label{prop:lower+bdd+OPT+gamma}
In the stable regime~(\Cref{def:eps+regimes}),  there exist instances for which the regret (Definition~\ref{def:regret}) of any online algorithm with respect to $\OPT(\alpha, \gamma)$ is $\Omega\left(\sqrt{T} + \gamma\right)$. 
\end{proposition}

{\bf Proof of \Cref{prop:lower+bdd+OPT+gamma}}. We construct two instances, $\mathcal{I}_2$ and $\mathcal{I}_3$, and show that for instance $\mathcal{I}_2$ (resp. $\mathcal{I}_3$), any online algorithm incurs regret against $\OPT(\alpha, \gamma)$ of at least $\Omega\left(\sqrt{T} - \gamma \right)$ (resp. $\Omega(\gamma)$). Combining these two bounds yields the desired lower bound on the regret. 

\emph{(i) Instance \(\mathcal{I}_2\).} We revisit instance \(\mathcal{I}_2\) from the proof of \Cref{prop:lower+bound+OPT0}. 
We begin by upper-bounding the gap between $\OPT_{\mathcal{I}_2}(\alpha, \gamma)$ and $\OPT_{\mathcal{I}_2}(\alpha, 0)$. 
Consider the optimal solution to $\OPT_{\mathcal{I}_2}(\alpha, 0)$, which maximizes total reward without penalizing backlog. With a slight abuse of notation, we use  $\OPT_{\mathcal{I}_2}(\alpha, 0)$ denote the total reward earned by this solution. We note that the optimal solution to $\OPT_{\mathcal{I}_2}(\alpha, 0)$ is feasible for $\OPT_{\mathcal{I}_2}(\alpha, \gamma)$. We now lower bound the expected time-average backlog incurred by the solution to $\OPT_{\mathcal{I}_2}(\alpha, 0)$. Let $T_o := \sup\{ t \in [T]: \sum_{\tau=1}^t \Indicator[w_\tau = 1] \leq 0.5T \}$ be the first time at which $0.5T$ reward-1 arrivals have occurred. By construction, $T_o \geq 0.5T$ on every sample path. 
The solution to $\OPT(\alpha, 0)$ maximizes total reward by accepting reward-1 arrivals whenever it can, and thus it accepts all reward-1 arrivals that occur before time $T_o$. 
Furthermore, for large enough $T \geq \Omega(1/\epsilon^2)$, the arrival rate of reward-1 case is at least $0.5 - \epsilon$. 
Consequently, during time  $[1, T_o]$, the backlog process $\{b_t\}$ induced by this solution evolves as a reflected birth–death process with arrival rate at least $0.5 - \epsilon$ and service rate $0.5 + \epsilon$. Following the similar argument of the analysis of instance $\mathcal{I}_1$, the expected time-average backlog up to $T_o$ satisfies
\[
\E \left[ \frac{1}{T_o} \sum_{t=1}^{T_o} b_t \right] \geq \Theta(1/\epsilon).
\]
Since $T_o \in [0.5T, T]$ for all sample paths, we have
\[
\mathbb{E}\left[\frac{1}{T} \sum_{t=1}^{T} b_t \right]
\geq \mathbb{E}\left[\frac{1}{T} \sum_{t=1}^{T_o} b_t \right]
= \mathbb{E}\left[ \frac{T_o}{T} \cdot \left( \frac{1}{T_o} \sum_{t=1}^{T_o} b_t \right) \right]
\geq \frac{1}{2} \cdot \mathbb{E} \left[ \frac{1}{T_o} \sum_{t=1}^{T_o} b_t \right]
\geq \Theta(1/\epsilon),
\]
which gives a valid lower bound on the expected time-average backlog incurred by $\OPT_{\mathcal{I}_2}(\alpha, 0)$ over the entire horizon.
It follows that, when evaluated under the objective with congestion penalty $\gamma$, we have: 
\begin{equation}
\E[\OPT_{\mathcal{I}_2}(\alpha, \gamma)] \geq \E[\OPT_{\mathcal{I}_2}(\alpha, 0)] - \Theta\left(\frac{\gamma}{\varepsilon}\right). \label{eq:i2-OPT-gamma-gap}
\end{equation}

Combining the lower bound in~\eqref{eq:i2-gap-final} with the inequality~\eqref{eq:i2-OPT-gamma-gap}, we obtain:
\begin{align}
\E[\OPT_{\mathcal{I}_2}(\alpha, \gamma) - \ALG_{\mathcal{I}_2}(\alpha, \gamma)]
&= \E[\OPT_{\mathcal{I}_2}(\alpha, 0) - \ALG_{\mathcal{I}_2}(\alpha, \gamma)] - \E[\OPT_{\mathcal{I}_2}(\alpha, 0) - \OPT_{\mathcal{I}_2}(\alpha, \gamma)] \nonumber \\
&\geq \Omega\left(\sqrt{T} - \frac{\gamma}{\varepsilon}\right). \label{eq:i2-OPTgamma-regret}
\end{align}

\smallskip 

\emph{(iii) Instance \(\mathcal{I}_3\).} We revisit the instance considered in \Cref{prop:impossibility}, where there is a single affiliate (\(m = 1\)) with capacity \(c = 0.5 T\), and all arrivals are free cases with deterministic reward of one. Following an identical argument to that in \Cref{prop:impossibility}, one can show that, under the stable regime, any online algorithm must satisfy:
\begin{equation}
\E[\OPT_{\mathcal{I}_3}(\alpha, \gamma) - \ALG_{\mathcal{I}_3}(\alpha, \gamma)] \geq \Omega(\gamma). \label{eq:i3-regret}
\end{equation}

\smallskip 

Taking the maximum of the lower bounds in~\eqref{eq:i2-OPTgamma-regret} and~\eqref{eq:i3-regret} and assuming $\epsilon = \Omega(1)$, the worst-case regret of any online algorithm against $\OPT(\alpha, \gamma)$ is at least $\Omega\left( \max\left\{ \sqrt{T} - \gamma, \gamma \right\} \right)$. Consider two cases. If $\gamma = o(\sqrt{T})$, then the bound reduces to $\Omega(\sqrt{T})$. Otherwise, if $\gamma = \Omega(\sqrt{T})$, it becomes $\Omega(\gamma)$. Thus, in combining cases, the regret against $\OPT(\alpha,\gamma)$ is $\Omega\left( \max\{\sqrt{T}, \gamma\} \right)=\Omega(\sqrt{T} + \gamma)$. This completes the proof. \hfill \halmos

\subsection{Lower Bounds for Near-Critical Regime}\label{apx+lbd+nc}

In the near-critical regime where \( \epsilon = \BigO(1/\sqrt{T}) \), the regret upper bound of \Cref{ALG+Surrogate+D} is given by \( \BigO(\sqrt{\gamma T}) \) (see \Cref{near+part+a}), even when compared against the stronger benchmark \( \OPT(\alpha, 0) \). In the following proposition, we show that this regret bound is tight with respect to \( \OPT(\alpha, 0) \) in the special case where \( \epsilon = 0 \).

\begin{proposition}\label{prop:nc+lbd}
For $\epsilon = 0$, there exists an instance for which the regret of any online algorithm with respect to $\OPT(\alpha,0)$ is $\Omega(\sqrt{\gamma T})$.
\end{proposition}

The rest of this section is devoted to proving \Cref{prop:nc+lbd}. 
For the near-critical regime, establishing the lower bound even in this special case of $\epsilon = 0$ turns out to be technically challenging due to the difficulty of characterizing a (non-stationary) finite-horizon optimal online algorithm. Consequently, our proof introduces several novel techniques: (i) characterizing the optimal policy of an auxiliary infinite-horizon Markov decision process tailored to our instance, and (ii) establishing the convergence rate of the finite-horizon optimal policy to its infinite-horizon counterpart.

\smallskip 
{\bf Proof of \Cref{prop:nc+lbd}}. 
We analyze the following instance $\mathcal{I}_4$. 
\paragraph{Instance $\mathcal{I}_4$.} We consider a single-affiliate setting  where all arrivals are free cases, and the reward $w_t \in \{0,1\}$ of each arrival is independently drawn from a Bernoulli distribution with $\Pr(w_t = 1) = 0.5$. We often write \( \textsf{Bernoulli}(0.5) \) to denote this distribution. 
We set the total capacity to $c = 0.5T$. The service rate is exactly $r = 0.5$ (thus $\epsilon = 0$). 
Because there are no tied cases, the parameter $\alpha$ plays no role in the current instance. Hence, we simplify notation by writing $\OPT(0)$ to denote $\OPT(\alpha, 0)$—i.e., the offline benchmark that maximizes total reward subject to capacity, with no congestion penalty. 

In this instance, note that $\OPT(0)$ accepts all of reward-1 arrivals (and reject the others) to fill the capacity. Thus, the value of $\OPT(0)$ on instance $\mathcal{I}_4$ is:
\begin{equation}
\mathbb{E}[\OPT_{\mathcal{I}_4}(0)] = \mathbb{E}\left[\min\left(0.5T, \sum_{t=1}^T \mathbbm
{1}[w_t = 1]\right)\right] = 0.5T - \Theta(\sqrt{T}).  \label{eq:OPT} 
\end{equation}

To analyze the regret of any online algorithm, we examine the optimal online policy \( \DP_T(\gamma) \), which maximizes \eqref{ALG+obj}—the cumulative reward minus the congestion cost—over a horizon of \( T \) periods, given a congestion penalty \( \gamma \). This policy can be formulated as a finite-horizon Markov Decision Process (MDP), which we define formally in \Cref{apx+proof+nc-lbd}. The following lemma provides an upper bound on the expected value of \( \DP_T(\gamma) \).

\begin{lemma}\label{lem:main-regret-bound}
    $\E[\DP_T(\gamma)] \leq 0.5T - \Theta(\sqrt{\gamma T})$,
\end{lemma}

Combined with \Cref{eq:OPT}, \Cref{lem:main-regret-bound} directly implies \Cref{prop:nc+lbd}. The proof of \Cref{lem:main-regret-bound} is fairly intricate; here, we provide a heuristic argument and defer the full proof to the next subsection. Suppose, for the sake of developing intuition, that $\DP_T(\gamma)$ can be well approximated by the following stationary threshold policy: accept a reward-1 arrival if and only if the current backlog $b_{t-1} \leq K-1$, for some fixed threshold $K > 0$, which we will optimize later. Under this policy, the backlog evolves similar to an $M/M/1/K$ queue with both arrival and service rates equal to $0.5$. The stationary distribution of backlog of such $M/M/1/K$ queue is uniform over $\{0, \dots, K\}$, leading to a rejection probability of $\Theta(1/K)$ and an average backlog of $\Theta(K)$. The expected objective of this threshold policy can then be roughly given by:
\[
0.5T \cdot \left(1 - \Theta\left(\frac{1}{K}\right)\right) - \gamma \cdot \Theta(K).
\]
where the first (resp. second) term is the matching rewards (resp. congestion cost). 
Optimizing over $K$ yields the choice $K^* = \Theta(\sqrt{T/\gamma})$.  With this choice of $K^*$, the resulting value of the optimum online is $\E[\DP_T(\gamma)] \approx  0.5 T - \Theta(\sqrt{\gamma T})$, as in \Cref{lem:main-regret-bound}. 

In \Cref{apx+proof+nc-lbd}, we formalize the above intuition by showing that (i) the horizon-optimal online policy \( \DP_T(\gamma) \) can indeed be well approximated by a threshold-form stationary policy with an additive gap of \( O(\sqrt{T}) \), and (ii) the value of such a stationary policy is \(0.5T - \Theta(\sqrt{\gamma T}) \).
\smallskip

\subsubsection{Proof of \Cref{lem:main-regret-bound}}\label{apx+proof+nc-lbd} \hfill 

For the proof, we first formally define the optimal online policy \( \DP_T(\gamma) \), which provides a valid upper bound on the objective value \eqref{ALG+obj} for any online algorithm. Analyzing the value of \( \DP_T(\gamma) \) directly is challenging, as the finite-horizon optimal policy is generally non-stationary. To overcome this, we define an auxiliary infinite-horizon problem closely related to \( \DP_T(\gamma) \). We then characterize the optimal stationary policy and its value for the infinite-horizon problem, and connect this value back to that of \( \DP_T(\gamma) \). We elaborate each step in the following.

\smallskip 

\textbf{Finite-Horizon Optimal Online Algorithm.} 
We define $\DP_T(\gamma)$ as a dynamic program that maximizes the total objective over $T$ periods (reward minus congestion cost), \emph{ignoring the capacity constraint}. Note that the objective value of $\DP_T(\gamma)$ is a valid upper bound on the objective value of any feasible online algorithm (which must respect the capacity constraint). At each time $t$, the state of $\DP_T(\gamma)$ is the current backlog $b_{t-1}$. We denote the horizon-optimal policy by
\[
\hat{\pi}_T = (\pi_1, \pi_2, \dots, \pi_T), \quad 
\]
where $\pi_t(b) \in \{0,1\}$ represents the decision to match an arrival at time $t$ when the backlog is $b$.

At each time $t$, let $w_t \sim 
\textsf{Bernoulli}(0.5)$ denote the reward of the arriving case.  Technically, the matching decision depends on both the backlog and the realized reward $w_t$. However, it is without loss of optimality to reject any arrival with $w_t = 0$, since matching it incurs congestion cost without increasing the matching reward. To simplify notation, we thus adopt the convention that
\[
\pi_t(b_{t-1}) := 0 \quad \text{whenever } w_t = 0.
\]
That is, we implicitly restrict attention to reward-1 arrivals, and interpret $\pi_t(b_{t-1})$ as the decision to match such an arrival when the backlog is $b_{t-1}$. Under this convention, we fine the \emph{per-period payoff} function as follows:
\begin{equation}
\label{eq:per+reward}
r(\pi_t(b_{t-1}), b_{t-1}) := \pi_t(b_{t-1}) - \frac{\gamma}{T} b_{t-1}.
\end{equation}

The total expected value of the finite-horizon dynamic program when starting from backlog $b_0 = 0$ is:
\begin{equation}
\label{eq:dp-finite}
\E[\DP_T(\gamma)] := \mathbb{E}\left[\sum_{t=1}^T r(\pi_t(b_{t-1}), b_{t-1}) \,\middle|\, b_0 = 0 \right],
\end{equation}
where the expectation is taken over the sequence of arrivals  $\{w_t\}_{t=1}^T$ and backlog states \( (b_t)_{t=1}^T \) induced by the policy \( \hat{\pi}_T \).

 \smallskip 

\textbf{Auxiliary Infinite-Horizon MDP.} We now define a parallel infinite-horizon average-reward MDP as an auxiliary problem. This MDP ignores capacity constraints and uses the same per-period congestion penalty \(\gamma/T\) as in the finite-horizon problem. Importantly, both \(\gamma\) and \(T\) are fixed in this formulation; we consider running the MDP for a growing horizon \(H \to \infty\) while keeping \(\gamma/T\) constant. This will allow us to analyze the long-run behavior of stationary policies under the same trade-off between matching reward and congestion penalty as in the original finite-horizon setting.

Let \(\pi^* \colon \mathbb{N} \to \{0,1\}\) denote a stationary policy that maps the current backlog \(b\) to a matching decision. As before, we assume without loss of generality that the policy only acts on arrivals with \(w_t = 1\), and we suppress the dependence on \(w_t\) in notation. Using 
the payoff function from \eqref{eq:per+reward},
we define an infinite-horizon optimal stationary policy as
\[
\pi^* \in \argmax_{\pi} \limsup_{H \to \infty} \frac{1}{H} \mathbb{E} \left[\sum_{t=1}^H r(\pi(b_{t-1}), b_{t-1}) \right],
\]
and its long-run average payoff is denoted by:
\[
\rho^* := \lim_{H \to \infty} \frac{1}{H} \mathbb{E} \left[\sum_{t=1}^H r(\pi^*(b_{t-1}), b_{t-1}) \right],
\]
where the expectation is taken over the stationary distribution of the backlog $b_{t-1}$ induced by \(\pi^*\) and the i.i.d.\ arrival rewards \( w_t \sim \textsf{Bernoulli}(0.5) \).

\smallskip

We now state two auxiliary results from infinite-horizon average-reward MDP theory~\citep{Puterman1994} that are instrumental in our analysis. 

First, we invoke an optimality condition for infinite-horizon  problems, which characterizes the optimal stationary policy via its long-run average payoff and associated bias function. This result will help bridge the finite-horizon optimal value \( \DP_T(\gamma) \) with the performance of a stationary policy in the associated infinite-horizon problem.
\begin{definition}[Equation (8.2.3) in \cite{Puterman1994}, Bias function]\label{def:bias}
Let \( \pi^* \) be the infinite-horizon optimal stationary policy with long-run average payoff \( \rho^* \). The bias function \( h \colon \mathbb{N} \to \mathbb{R} \) is defined by:
\[
h(b) := \sum_{t=1}^\infty  \mathbb{E} \left[ r(\pi^*(b_{t-1}), b_{t-1}) - \rho^* \Big| \, b_0 = b \right],
\]
where \( r(\cdot, \cdot) \) is the per-period payoff function defined in \eqref{eq:per+reward}. The expectation is taken over the backlog trajectory induced by \( \pi^* \), starting from backlog \( b \), and an i.i.d.\ arrival sequence \( \{w_t\}_{t=1}^\infty \).
\end{definition}

\begin{lemma}[Equation (8.4.2) in \citet{Puterman1994}, Bias-gain optimality condition]\label{lem:bias-gain}
Let \( \pi^* \) be the optimal stationary policy for the infinite-horizon problem with long-run payoff \( \rho^* \) and bias function \( h(\cdot) \). Then \( \pi^* \) satisfies the following optimality condition:
\[
\rho^* + h(b) = \max_{z \in \{0,1\}} \left\{  r(z, b) + \mathbb{E}\left[h(b') \right] \right\},
\]
for all non-negative integers $b$. Here, \( b' = (b + \pi^*(b) - s)_+ \) denotes the next backlog starting from $b$, where \( s \sim \textsf{Bernoulli}(0.5) \) represents the random service. The expectation \( \mathbb{E}[h(b')] \) is taken with respect to the random arrival \( w \sim \textsf{Bernoulli}(0.5) \) and service \( s \sim \textsf{Bernoulli}(0.5) \), and the resulting next state \( b' = (b + \pi^*(b) - s)_+ \) (induced by applying policy \( \pi^* \) at state \( b \) when \( w = 1 \)).
\end{lemma}

Furthermore, the following lemma shows that  optimal infinite-horizon policy admits a simple threshold structure.  

\begin{lemma}[\cite{yechiali1971optimal}]\label{lem:threshold}
The infinite-horizon optimal policy \( \pi^* \) is of threshold type: there exists threshold \( K^* \) such that \( \pi^*(b) = 1 \) if and only if \( b < K^* \).
\end{lemma}

\smallskip 

\textbf{Proof of \Cref{lem:main-regret-bound}.}
We are now ready to prove \Cref{lem:main-regret-bound} in four steps. The proof of all auxiliary claims is deferred to the end of this section.

 \smallskip 
\textbf{Step 1: Upper bound via bias decomposition.}

We begin by upper bounding the finite-horizon value \( \DP_T(\gamma) \) using the bias function \( h(\cdot) \) and long-run average payoff \( \rho^* \) from the infinite-horizon MDP. Recall \( \hat{\pi}_T = (\pi_1, \dots, \pi_T) \) be the finite-horizon optimal policy. Let \( \hat{b}_0 = 0 \) be the initial backlog. Let \( \hat{b}_1, \dots, \hat{b}_T \) denote the resulting backlog process under \( \hat{\pi}_T \).

\begin{claim}
\label{clm:step1}
For all \( T \) and \( \gamma \), we have:
\begin{equation}
\label{eq:step1}
\E[\DP_T(\gamma)] \le T \rho^* + h(0) - \mathbb{E}[h(\hat{b}_T)].
\end{equation}
\end{claim}

The proof follows by applying the bias–gain optimality condition (Lemma~\ref{lem:bias-gain}) recursively along the backlog trajectory induced by the finite-horizon optimal policy \( \hat{\pi}_T \).

\smallskip 

\textbf{Step 2: Asymptotic expression for \( \rho^* \).}
We now analyze the long-run average payoff of the infinite-horizon MDP under the threshold policy \( \pi^* \). 

\begin{claim}
\label{clm:rho-star}The long-run average payoff  under the infinite-horizon optimal policy satisfies:
\[
\rho^* = 0.5 - \Theta\left( \sqrt{\frac{\gamma}{T}} \right),
\]
and the optimal stationary policy uses threshold \( K^* = \Theta\left( \sqrt{T/\gamma} \right) \).
\end{claim}

To prove Claim \ref{clm:rho-star}, we crucially rely on the threshold structure of the infinite-horizon optimal policy (\Cref{lem:threshold}) and optimize the backlog threshold \( K \) by analyzing the stationary distribution induced by the corresponding threshold policy.

\smallskip

\textbf{Step 3: Bounding the Bias Magnitude.} We now upper bound the bias term \( h(0) - \mathbb{E}[h(\hat{b}_T)] \) appearing in \eqref{eq:step1}.

{
\begin{claim}
\label{clm:bias-bound}
For any fixed \( \gamma \) and \( T \), the bias gap satisfies:
\[
|\E[h(0) - h(\hat{b}_T)]| \le \BigO\left( \sqrt{T} \right).
\]
\end{claim}}

The proof of Claim~\ref{clm:bias-bound} involves several technical steps. In particular, we leverage the fact that the backlog process induced by the infinite-horizon optimal policy (characterized in Claim \ref{clm:rho-star}) mixes ``fast'' to bound the bias term \( \mathbb{E}[h(\hat{b}_T)] \) in terms of \( \mathbb{E}[\hat{b}_T] \), and analyze the worst-case expected backlog using a result on the running maximum of a random walk~\citep{comtet2005precise}.

\smallskip 

\textbf{Step 4: Putting everything together.} 
From Claim~\ref{clm:step1}, \( \E[\DP_T(\gamma)] \le T \cdot \rho^* + h(0) - \mathbb{E}[h(\hat{b}_{T+1})] \). Claim~\ref{clm:rho-star} gives \( \rho^* = 0.5 - \Theta\big( \sqrt{\gamma/T} \big) \), and Claim~\ref{clm:bias-bound} shows \( |h(0) - \mathbb{E}[h(\hat{b}_{T+1})]| \le \BigO\big( \sqrt{T} \big) \). Substituting,
\[
\E[DP_T(\gamma)] \le T \left(0.5 - \Theta\big( \sqrt{\gamma/T} \big) \right) + \BigO\big( \sqrt{T} \big) = 0.5T - \Theta(\sqrt{\gamma T}).
\]
Thus \( \E[DP_T(\gamma)] \le 0.5T - \Theta(\sqrt{\gamma T}) \), as desired.
\hfill \halmos 

\medskip 

{\bf Proof of Claim \ref{clm:step1}}.
By the bias-gain optimality condition (Lemma~\ref{lem:bias-gain}), we have, for all \( t = 1, \dots, T \),
\[
r(\pi_t(\hat{b}_{t-1}), \hat{b}_{t-1}) \le \rho^* + h(\hat{b}_{t-1}) - \mathbb{E}[h(\hat{b}_t)].
\]

Summing both sides over \( t = 1 \) to \( T \), we obtain:
\[
\sum_{t=1}^T r(\pi_t(\hat{b}_{t-1}), \hat{b}_{t-1}) \le T \rho^* + h(\hat{b}_0) - \mathbb{E}[h(\hat{b}_T)].
\]

Taking expectations and using \( \hat{b}_0 = 0 \), we conclude:
\[
\mathbb{E}\left[\sum_{t=1}^T r(\pi_t(\hat{b}_{t-1}), \hat{b}_{t-1})\right] \le T \rho^* +  \mathbb{E}[h(0)  - h(\hat{b}_T)],
\]
Because the left-hand side is equal to \( \E[\DP_T(\gamma)] \) by equation~\eqref{eq:dp-finite}, this completes the proof.\hfill\halmos

\smallskip

{\bf Proof of Claim~\ref{clm:rho-star}}.
We identify the threshold \( K^* \) used by the optimal stationary policy \( \pi^* \), and compute the corresponding average payoff \( \rho^* \).
From Lemma~\ref{lem:threshold}, policy \( \pi^* \) accepts an arrival if and only if the backlog \( b < K \), for some threshold \( K \in \mathbb{N} \). Under such a policy, the backlog process forms a lazy random walk on \( \{0, 1, \dots, K\} \).\footnote{
Under the threshold policy \( \pi^* \), the backlog process evolves as \( b_t = (b_{t-1} + \pi^*(b_{t-1}) \cdot w_t - s_t)_+ \), where \( w_t, s_t \sim \textsf{Bernoulli}(0.5) \) are independent reward-1 arrival and service indicators. For internal states \( 1 \le b < K \), the transition probabilities satisfy:
\[
P(b \to b+1) = 0.25, \quad P(b \to b-1) = 0.25, \quad P(b \to b) = 0.5.
\]
At the boundaries: \( P(0 \to 0) = 0.75, P(0 \to 1) = 0.25 \), and \( P(K \to K) = 0.5, P(K \to K-1) = 0.5 \). This defines a lazy, reflected random walk on \( \{0,1,\dots,K\} \).
} Let \( b^* \) denote the stationary backlog under this policy. The stationary distribution (and resulting expected value) of this random walk is given by~\citep{LevinPeresWilmer}:
\[
\mathbb{P}(b^* = K) = \Theta\left( \frac{1}{K} \right), \qquad \mathbb{E}[b^*] = \Theta(K).
\]

The resulting long-run average payoff is:
\[
\rho^* = \underbrace{0.5 \cdot \mathbb{P}(b^* < K)}_{\text{match reward}}-\underbrace{\frac{\gamma}{T} \cdot \mathbb{E}[b^*]}_{\text{ congestion penalty}}.
\]

Since \( \mathbb{P}(b^* < K) = 1 - \mathbb{P}(b^* = K) = 1 - \Theta(1/K) \), we have:
\[
\rho^* = 0.5 - \Theta\left( \frac{1}{K} \right) - \Theta\left( \frac{\gamma}{T} \cdot K \right).
\]

Optimizing over $K$, we conclude that the optimal the optimal threshold \( K^* \) must be given by $K^* = \Theta\left( \sqrt{ \frac{T}{\gamma} } \right).$ Substituting this into the expression for \( \rho^* \), we get:
\[
\rho^*  = 0.5 - \Theta\left( \sqrt{ \frac{\gamma}{T} } \right),
\]
as claimed. \hfill\halmos

{\bf Proof of Claim~\ref{clm:bias-bound}.} By triangular inequality and Jensen's inequality, we have 
\begin{equation}
    |\E[h(0) - h(\hat{b}_T)]| \leq \E[|h(0)|] + \E[|h(\hat{b}_T)|].
\end{equation}
In the following, we focus on showing that
\begin{equation}
    \E[|h(\hat{b}_T)|]\leq \BigO(\sqrt{T}). \label{ineq:b_T}
\end{equation}

The argument for $\E[|h(0)|]$ follows the identical steps and is omitted for brevity.

To prove inequality~\eqref{ineq:TV}, we first note that, as shown in Claim~\ref{clm:rho-star}, the Markov chain \( \{b_t^*\} \) induced by the stationary policy \( \pi^* \) is a lazy reflected random walk on the path \( \{0, 1, \dots, K^*\} \). This chain is irreducible and aperiodic, and hence admits a unique stationary distribution \( \mu \). By Theorem~4.9 of \cite{LevinPeresWilmer}, the chain satisfies geometric convergence: there exists a rate parameter \( \alpha \in (0,1) \) such that for any initial state \( b \) and any \( t \geq 0 \), the total variation distance satisfies
\begin{equation}
\left\| P^t(b, \cdot) - \mu(\cdot) \right\|_{\mathrm{TV}} \le \BigO(\alpha^t),
\label{ineq:TV}
\end{equation}
where \( P^t(b, \cdot) \) denotes the distribution of \( b_t^* \) conditional on \( b_0^* = b \), and \( \|\cdot\|_{\mathrm{TV}} \) denotes total variation distance.\footnote{For probability measures \( \nu, \mu \) on a state space \( \Omega \), the total variation distance is defined as:
\[
\|\nu - \mu\|_{\mathrm{TV}} := \frac{1}{2} \sum_{x \in \Omega} |\nu(x) - \mu(x)|.
\]
} Furthermore, it is known that the convergence rate \( \alpha \) is determined by the spectral gap \( \kappa \) of the chain, with \( \alpha = 1 - \kappa \) (see, for example, chapter 12 of \cite{LevinPeresWilmer}). Since \( \{b_t^*\} \) is a lazy random walk on a path of length \( K^* \), it follows from Exercise~13.2 in \cite{LevinPeresWilmer} that \( \kappa = \Theta((K^*)^{-2}) \), and hence \( \alpha = 1 - \Theta((K^*)^{-2}) \).


Building on the above result, we now establish an upper bound on $|h(\hat{b}_T)|$. Recall from Definition~\ref{def:bias} that the bias function is given by
\[
h(\hat{b}_T) := \sum_{t=1}^\infty \left( \mathbb{E}[r(\pi^*(b_{t-1}^*), b_{t-1}^*) \mid b_0^* = \hat{b}_T] - \rho^* \right).
\]

Define \( \tilde{r}(k) := \pi^*(k) - \frac{\gamma}{T} \cdot k \), so that
\begin{align}
 \mathbb{E}[{r}(\pi^*(b_{t-1}^*), b_{t-1}^*) \mid b_0^* = \hat{b}_T] &= \sum_{k=0}^{\max(K^*, \hat{b}_T)} \tilde{r}(k) \cdot P^{t-1}(\hat{b}_T, k), \label{eq:r+support}\\
 \rho^* &= \sum_{k=0}^{K^*} \tilde{r}(k) \cdot \mu(k) \leq \sum_{k=1}^{\max(K^*, \hat{b}_T)} \tilde{r}(k)\cdot \mu(k).\label{eq:rho+star}
\end{align}

To see why equation \eqref{eq:r+support} holds, we recall that the infinite-horizon optimal  policy \( \pi^* \) accepts a reward-1 arrival if and only if the backlog is strictly less than \( K^* \). Thus,  if the initial state is \( \hat{b}_T > K^* \), the backlog under \( \pi^* \) can only decrease and will eventually enter and remain in \( \{0, \dots, K^*\} \). Hence, the payoff at time \( t \) only depend on the values \( k \le \max(K^*, \hat{b}_T) \).

From equations \eqref{eq:r+support} and \eqref{eq:rho+star}, we have:
\begin{align}
 \left| \mathbb{E}[r(\pi^*(b_{t-1}^*), b_{t-1}^*) \mid b_0^* = \hat{b}_T] - \rho^* \right|
&\le \sum_{k=0}^{\max(K^*, \hat{b}_T)} |\tilde{r}(k)| \cdot |P^{t-1}(b, k) - \mu(k)| \\
&\le \left(\max_{0\leq k \leq \max(K^*, \hat{b}_T)}| \tilde{r} (k) |\right) \cdot 2 \cdot \| P^{t-1}(b, \cdot) - \mu \|_{\mathrm{TV}}. \label{ineq:r_k}   
\end{align}
{Note that \( |\tilde{r}(k)| \le 1 + \frac{\gamma}{T} k = \BigO\left( \frac{\gamma}{T} k \right) \)}. Combining this with \eqref{ineq:TV} and \eqref{ineq:r_k}, we obtain:
\[
\left| \mathbb{E}[r(\pi^*(b_{t-1}^*) \mid b_0^* = \hat{b}_T] - \rho^* \right| \le \BigO\left( \frac{\gamma}{T} \cdot \max(\hat{b}_T, K^*) \cdot \alpha^{t-1} \right),
\]
and thus 
\begin{equation}
    |h(\hat{b}_T)| \leq \BigO\left(\frac{\gamma}{T}\cdot \max(\hat{b}_T, K^*)\cdot \sum_{t=1}^\infty \alpha^{t-1} \right) =   \BigO\left(\frac{\gamma}{T}\cdot \max(\hat{b}_T, K^*)\cdot \frac{1}{1-\alpha} \right)
\end{equation}

Recall that $\alpha = 1 - O((K^*)^{-2})$. Furthermore, from Claim \ref{clm:rho-star}, we have $K^* = \sqrt{T/\gamma}$. Plugging this into the above bound yields:
\begin{equation}
\label{eq:bias-bound}
|h(\hat{b}_T)| \le \BigO(\max(\hat{b}_T, K^*)) \le \BigO(\hat{b}_T + K^*).
\end{equation}
Taking expectation over $\hat{b}_T$, we conclude:
\begin{equation}
\mathbb{E}[|h(\hat{b}_T)|] \le \BigO\left(\mathbb{E}[\hat{b}_T] + K^*\right). \label{eq:bias+bound+2}
\end{equation}

 We now show that $\mathbb{E}[\hat{b}_T] \le \BigO(\sqrt{T})$. Plugging this into \eqref{eq:bias+bound+2} and using the fact that $K^* = \BigO(\sqrt{T/\gamma}) \le \BigO(\sqrt{T})$ (from Claim~\ref{clm:rho-star}), we conclude:
\[
\mathbb{E}[|h(\hat{b}_T)|] \le \BigO(\sqrt{T}),
\] as desired.

To prove $\mathbb{E}[\hat{b}_T] \le \BigO(\sqrt{T})$, Consider a (trivial) stationary  policy that accepts every arrival with reward 1 (and reject the others). Let $\tilde{b}_t$ denote the backlog under this policy. Then, for any sample path,
\[
\hat{b}_T \le \tilde{b}_T.
\]
Thus, it suffices to bound $\E[\tilde{b}_T]$. 
Note that $\tilde{b}_t$ evolves as a lazy reflected random walk:
\[
\tilde{b}_t = \max(\tilde{b}_{t-1} + \delta_t, 0),
\]
where $\delta_t \in \{-1, 0, 1\}$ is an i.i.d. random variable with:
\[
\mathbb{P}[\delta_t = -1] = \frac{1}{4}, \quad
\mathbb{P}[\delta_t = 0] = \frac{1}{2}, \quad
\mathbb{P}[\delta_t = 1] = \frac{1}{4}.
\]
Let $S_t = \sum_{\tau=1}^t \delta_\tau$ be the associated (unreflected) lazy random walk. By induction, one can show that:
\[
\tilde{b}_t = \max_{1 \leq \tau \leq t} S_\tau - S_t,
\]
for any sample path. 
Since $\mathbb{E}[\delta_t] = 0$, it follows that:
\[
\mathbb{E}[\tilde{b}_t] = \mathbb{E}\left[\max_{1 \leq \tau \leq t} S_\tau\right],
\]
i.e., the expected running maximum of a symmetric lazy random walk. This is known to be $\Theta(\sqrt{t})$; see, for example, \citet{comtet2005precise}. Therefore:
\[
\mathbb{E}[\hat{b}_T] \le \mathbb{E}[\tilde{b}_T] = \BigO(\sqrt{T}),
\]
as claimed. This completes the proof.\hfill\halmos

}
\subsection{Proof of \texorpdfstring{Claim \ref{fact:average+conv}}{}}\label{apx+OMD+pf}

For any given $\boldsymbol{\nu} \in \mathcal{D}$ and $\mathbf{g}_t =\nabla f_t(\boldsymbol{\nu}_t)$, 
\begin{align}
f_t(\boldsymbol{\nu}_t) - f_t(\boldsymbol{\nu}) &\leq \mathbf{g}_t\cdot(\boldsymbol{\nu}_t - \boldsymbol{\nu}) \nonumber \\
&= \hat{\mathbf{g}}_t\cdot(\boldsymbol{\nu}_t - \boldsymbol{\nu}) + (\mathbf{g}_t - \hat{\mathbf{g}}_t)\cdot (\boldsymbol{\nu}_t - \boldsymbol{\nu}) \nonumber 
\\
&= \frac{1}{\eta_t}(\nabla h(\boldsymbol{\nu}_t) - \nabla h(\tilde{\boldsymbol{\nu}}_{t+1}))\cdot (\boldsymbol{\nu}_t - \boldsymbol{\nu}) + (\mathbf{g}_t - \hat{\mathbf{g}}_t)\cdot (\boldsymbol{\nu}_t - \boldsymbol{\nu})\label{line:trivial}\\
&= \frac{1}{\eta_t}(V_h(\boldsymbol{\nu}, \boldsymbol{\nu}_t) + V_h(\boldsymbol{\nu}_t, \tilde{\boldsymbol{\nu}}_{t+1}) - V_h(\boldsymbol{\nu},\tilde{\boldsymbol{\nu}}_{t+1})) + (\mathbf{g}_t - \hat{\mathbf{g}}_t)\cdot (\boldsymbol{\nu}_t - \boldsymbol{\nu})\label{line:three+point}\\
&\leq \frac{1}{\eta_t}(
V_h(\boldsymbol{\nu}, \boldsymbol{\nu}_t) + V_h(\boldsymbol{\nu}_t, \tilde{\boldsymbol{\nu}}_{t+1}) - 
V_h(\boldsymbol{\nu}, \boldsymbol{\nu}_{t+1})  - V_h(\boldsymbol{\nu}_{t+1}, \tilde{\boldsymbol{\nu}}_{t+1}) 
) + (\mathbf{g}_t - \hat{\mathbf{g}}_t)\cdot (\boldsymbol{\nu}_t - \boldsymbol{\nu}) \label{line:projection}\\
&= \frac{1}{\eta_t}(
V_h(\boldsymbol{\nu}, \boldsymbol{\nu}_t) - V_h(\boldsymbol{\nu}, \boldsymbol{\nu}_{t+1})
) + 
\frac{1}{\eta_t}
\underbrace{(V_h(\boldsymbol{\nu}_t, \tilde{\boldsymbol{\nu}}_{t+1}) - V_h(\boldsymbol{\nu}_{t+1}, \tilde{\boldsymbol{\nu}}_{t+1}) 
)
}_{(\star)} + (\mathbf{g}_t - \hat{\mathbf{g}}_t)\cdot (\boldsymbol{\nu}_t - \boldsymbol{\nu}) \label{line:continue}
\end{align}
The first line is because of the definition of the subgradient. Line \eqref{line:trivial} is from the update rule \eqref{line:update+temp}. Line \eqref{line:three+point} is from the three-point equality property of $V_h$ (Lemma 5.2 of \cite{bubeck2011introduction}). Line \eqref{line:projection} is due to Generalized Pythagorean Theorem (Lemma 5.3 of \cite{bubeck2011introduction}). We now bound the term $(\star)$ as 
\begin{align}
(\star) &= h(\boldsymbol{\nu}_t) - h(\boldsymbol{\nu}_{t+1}) - \nabla h(\tilde{\boldsymbol{\nu}}_{t+1})\cdot (\boldsymbol{\nu}_t - \boldsymbol{\nu}_{t+1}) \label{line:V+def} \\
&\leq (\nabla h(\boldsymbol{\nu}_{t}) - \nabla h(\tilde{\boldsymbol{\nu}}_{t+1}) )\cdot(\boldsymbol{\nu}_t - \boldsymbol{\nu}_{t+1}) - \frac{\sigma}{2}\Norm{\boldsymbol{\nu}_t - \boldsymbol{\nu}_{t+1}}{}^2 \label{line:h+str} \\
&= \eta_t \mathbf{\hat{g}}_t \cdot(\boldsymbol{\nu}_t - \boldsymbol{\nu}_{t+1}) - \frac{\sigma}{2}\Norm{\boldsymbol{\nu}_t - \boldsymbol{\nu}_{t+1}}{}^2 \label{line:trivial+again} \\
&\leq \eta_t\Norm{\mathbf{\hat{g}}_t}{*}\Norm{\boldsymbol{\nu}_t - \boldsymbol{\nu}_{t+1}}{}  - \frac{\sigma}{2}\Norm{\boldsymbol{\nu}_t - \boldsymbol{\nu}_{t+1}}{}^2\\
&\leq \frac{\eta_t^2 \Norm{
\mathbf{\hat{g}}_t
}{*}^2}{2\sigma} \label{last+line}
\end{align}
Line \eqref{line:V+def} is by the definition of $V_h$. Line \eqref{line:h+str} is because $h(\cdot)$ is $\sigma$-strongly convex with respect to $\Norm{\cdot}{}$. Line \eqref{line:trivial+again} is again due to the update rule \eqref{line:update+temp}. In the final two lines, we used the generalized Cauchy-Schartz inequality and the fact that $az - bz^2 \leq \frac{a^2}{4b}$ for any $a>0, b>0$ and $z \in \R$. \par 
Plugging the bound of $(\star)$ in \eqref{last+line} into \eqref{line:continue}, we have
\begin{equation}
f_t(\boldsymbol{\nu}_t) - f_t(\boldsymbol{\nu}) \leq \frac{1}{\eta_t}(
V_h(\boldsymbol{\nu}, \boldsymbol{\nu}_t) - V_h(\boldsymbol{\nu}, \boldsymbol{\nu}_{t+1})
) + \frac{\eta_t \Norm{
\mathbf{\hat{g}}_t
}{*}^2}{2\sigma} + (\mathbf{g}_t - \hat{\mathbf{g}}_t)\cdot (\boldsymbol{\nu}_t - \boldsymbol{\nu})
\end{equation}

We sum up the preceding inequality for $k \leq t \leq s$ to obtain
\begin{align*}
\sum_{t=k}^{s}(f_t(\boldsymbol{\nu}_t) - f_t(\boldsymbol{\nu})) &\leq \sum_{t=k}^s \frac{1}{\eta_t}(
V_h(\boldsymbol{\nu}, \boldsymbol{\nu}_t) - V_h(\boldsymbol{\nu}, \boldsymbol{\nu}_{t+1})
)  + \sum_{t=k}^s \frac{\eta_t \Norm{
\mathbf{\hat{g}}_t
}{*}^2}{2\sigma} + \sum_{t=k}^s (\mathbf{g}_t - \hat{\mathbf{g}}_t)\cdot (\boldsymbol{\nu}_t - \boldsymbol{\nu})\\
\begin{split}
&= \sum_{t=k+1}^s \left(\frac{1}{\eta_t} - \frac{1}{\eta_{t-1}}\right)V_h(\boldsymbol{\nu}, \boldsymbol{\nu}_t) + \frac{1}{\eta_k}V_h(\boldsymbol{\nu}, \boldsymbol{\nu}_k) - \frac{1}{\eta_{s}}V_h(\boldsymbol{\nu}, \boldsymbol{\nu}_{s+1}) + \sum_{t=k}^s \frac{\eta_t \Norm{
\mathbf{\hat{g}}_t
}{*}^2}{2\sigma} + \\
&\quad 
\sum_{t=k}^s (\mathbf{g}_t - \hat{\mathbf{g}}_t)\cdot (\boldsymbol{\nu}_t - \boldsymbol{\nu}) \\
\end{split} \\
&\leq 
\sum_{t=k+1}^s \left(\frac{1}{\eta_t} - \frac{1}{\eta_{t-1}}\right)V_h(\boldsymbol{\nu}, \boldsymbol{\nu}_t) + \frac{1}{\eta_k}V_h(\boldsymbol{\nu}, \boldsymbol{\nu}_k) + \sum_{t=k}^s \frac{\eta_t \Norm{
\mathbf{\hat{g}}_t
}{*}^2}{2\sigma} + \sum_{t=k}^s (\mathbf{g}_t - \hat{\mathbf{g}}_t)\cdot (\boldsymbol{\nu}_t - \boldsymbol{\nu}) 
\end{align*}
where the last inequality follows from the non-negativity of $V_h$. This completes the proof. 
\par 


\section{Missing Details and Proofs of 
\texorpdfstring{\Cref{subsec:alg+surrogate+primal}}{}}\label{apx+thm+surrogate+P}
\subsection{Formal description of \CO{}}\label{apx+P+pseudo+code}
We provide the formal description of the congestion-oblivious algorithm in \Cref{ALG+Surrogate+P}. 
\begin{algorithm}[ht]
\caption{Congestion-Oblivious Dual-Learning (\CO{}) Algorithm}\label{ALG+Surrogate+P}
\begin{algorithmic}[1]
    \STATE \textbf{Input:} $\TotalTime$, $\CapRatioVec$, constant $k>0$, and $\eta_\Timeidx \leftarrow \frac{k}{\sqrt{\Timeidx}}$ for all $\Timeidx$.
    \STATE Initialize $\DualOver_{1,\Locidx} \leftarrow \exp(-1)$, $\DualHard_{1,\Locidx} \leftarrow \exp(-1)$, and {$\Capacity_{0,\Locidx} \leftarrow \CapRatio_\Locidx \TotalTime$} for all $\Locidx \in [\Locnum]$.
      \FOR{each arrival $\{\Arrival_\Timeidx\}_{\Timeidx=1}^\TotalTime$}
    \IF{$\Target_\Timeidx \in [\Locnum]$} 
    \STATE Set $\DecisionVec_\Timeidx = \BasisVec{\Target_\Timeidx}$ 
    \ELSE 
    \IF{{$\min_{\Locidx\in[\Locnum]} \Capacity_{\Timeidx-1,\Locidx} > 0$}}
    \STATE {Set $\DecisionVec_\Timeidx  \in  \argmax_{\mathbf{z} \in \Simplex \cap \{0,1\}^m} (\Weightvec_\Timeidx - \DualOverVec_\Timeidx - \DualHardVec_\Timeidx)\cdot \DecisionVec$} \quad 
    \ELSE 
    \STATE {Set $\DecisionVec_\Timeidx = \mathbf{0}$} 
    \ENDIF
    \ENDIF
    \STATE {Update the remaining capacity: $\Capacity_{\Timeidx, \Locidx} \leftarrow \Capacity_{\Timeidx-1, \Locidx} - \DecisionEach \quad \forall \Locidx \in [\Locnum]$} 
    \STATE Update the dual variables for all $\Locidx \in [\Locnum]$:
    \begin{align}
    \DualOver_{\Timeidx+1, \Locidx} &\leftarrow \min\{\DualOver_{\Timeidx,\Locidx}\exp(\eta_\Timeidx(\DecisionEach-\CapRatio_i)), \OverCost\} \\
 \DualHard_{\Timeidx+1,\Locidx} &\leftarrow \min\left\{\DualHard_{\Timeidx,\Locidx}\exp(\eta_\Timeidx({\DecisionEach-\CapRatio_\Locidx})), \frac{1+2\OverCost}{\underline{\CapRatio}}\right\}
\end{align} 
\ENDFOR
\end{algorithmic}
\vspace{-0.1cm}
\end{algorithm}

\subsection{Preliminaries: Properties of the Static Dual Problem}\label{apx+fluid+dual}
The purpose of this section is to study the static dual problem defined in \Cref{subsec:ALG+P+static+dual}. We prove some properties of the static dual problem which will be useful throughout the proof of \Cref{thm:ALG+Surrogate+Primal} (\Cref{apx+thm+2+pf}).

Recall that we use $\boldsymbol{\nu} = (\DualOverVec, \DualHardVec) \in \R_{+}^{2\Locnum}$ to collectively denote the two dual variables. For ease of reference, we recall that the static dual problem (\Cref{def:staic+dual} in \Cref{subsec:ALG+P+static+dual}) was defined as
\begin{equation}
\begin{split}
    D(\boldsymbol{\nu})&:=\E\left[
    \max_{\DecisionVec \in \FeasibleSet{\Target}}(\Weightvec - \boldsymbol{\DualOver} - \boldsymbol{\DualHard})\cdot \DecisionVec + \CapRatioVec\cdot \boldsymbol{\DualOver} + \CapRatioVec\cdot \boldsymbol{\DualHard}
    \right] \\ 
    D(\boldsymbol{\nu}^*)&:= \min_{ \boldsymbol{\nu} \in \mathcal{V}} \ D(\boldsymbol{\nu}) 
\end{split}\label{def:nu+star}
\end{equation}
where we recall that $\mathcal{V} := [0,\OverCost]^\Locnum \times [0, \frac{1+2\OverCost}{\underline{\rho}}]$. We further defined the dual-based primal decision $\tilde{\DecisionVec}(\boldsymbol{\nu}, \mathbf{A})$ and \emph{matching rate} $\MRateVec(\boldsymbol{\nu})$ as
\begin{equation*}
\begin{split}
\tilde{\DecisionVec}(\boldsymbol{\nu}, \mathbf{A})&:= \argmax_{\DecisionVec \in \FeasibleSet{\Target}} (\Weightvec - \DualOverVec -\DualHardVec)\cdot \DecisionVec  \\ 
\MRateVec(\boldsymbol{\nu}) &:= \E_{\Arrival}[\tilde{\DecisionVec}(\boldsymbol{\nu}, \Arrival)].
\end{split}
\end{equation*}

Note that, under Assumption \ref{assump:PDF}, the dual-based primal decision is unique almost surely for each pair of arrival type and dual variable, and hence the dual-based matching rate is well-defined. 

In the following, we show some useful properties of the optimal solution $\boldsymbol{\nu}^*$ for the static dual problem.
{
\begin{restatable}[Properties of Optimal Solution for Static Dual Problem]{claim}{fluiddualclaim}\label{claim:arrival+rate}
The optimal static dual variable $\boldsymbol{\nu}^*$ defined in \eqref{def:nu+star} satisfies the following properties:
\begin{enumerate}[label=(\alph*)]
    \item $\lVert \DualOverVec^* + \DualHardVec^* \rVert_{\infty} \leq 1$.\footnote{
    {
It is worth pointing out that we can re-parameterize this dual problem by defining $\boldsymbol{\phi} = \boldsymbol{\theta} + \boldsymbol{\lambda}$. However, for our theoretical analysis, it is helpful to define $\boldsymbol{\theta}$ and $\boldsymbol{\lambda}$ separately as they play different roles. In simple terms, $\boldsymbol{\theta}$ and $\boldsymbol{\lambda}$ of our algorithms control the over-allocation cost and the stopping time $\Stopping$ (see eq. \eqref{def:stopping} for the definition), respectively — see, for example, Claim \ref{claim:theta} and \ref{claim:lambda} in Section \ref{apx:pseudo+upper}.
}
    } \label{fluid+dual+bound}
    \item Under Assumption \ref{assump:PDF}, $\MRate_i(\boldsymbol{\nu}^*)  \leq \rho_i$ with for all $i\in[m]$ with equality if $\DualOver_i^* >0$. \label{fluid+dual+rate}
    \item Under Assumption \ref{assump:PDF}, $D(\boldsymbol{\nu}^*) = \E_{\Arrival}[\Weightvec\cdot\tilde{\DecisionVec}(\boldsymbol{\nu}^*, \Arrival)]$\label{fluid+dual+value}
\end{enumerate}
\end{restatable}
}

{\bf Proof of Claim \ref{claim:arrival+rate}.} 
We first prove part \ref{fluid+dual+bound}. Suppose for a contradiction that $\DualOver_i^* + \DualHard_i^* > 1$ for some $i\in [\Locnum]$. To derive contradiction, we will construct an alternative solution $\tilde{\boldsymbol{\nu}}$ that can strictly decrease the dual objective function. Note that we can write $D(\boldsymbol{\nu})$ as 
\begin{align*}
    D(\boldsymbol{\nu}) = \sum_{j=1}^\Locnum \E[(\Weight_j - \DualOver_j - \DualHard_j)\cdot \Indicator[\Target = j]]  + \E\left[\max_{j \in [\Locnum]}(\Weight_j - \DualOver_j - \DualHard_j)_{+}\cdot \Indicator[\Target=0]\right] + \sum_{j=1}^\Locnum \rho_j(\DualOver_j + \DualHard_j).
\end{align*}

We now define $\tilde{\boldsymbol{\DualOver}} := \boldsymbol{\DualOver}^* - \delta \BasisVec{i}$ and $\tilde{\DualHardVec} := \DualHardVec^* - \delta \BasisVec{i}$ with infinitesimally small $\delta>0$. We first observe that 
\begin{equation}
\max_{j \in [m]}(w_j - \tilde{\DualOver}_j - \tilde{\DualHard}_j)_{+} = \max_{j \in [m]}(w_j - \DualOver^*_j - \DualHard^*_j)_{+} \label{line:same+adjusted}
\end{equation}
for every sample path. To see this, recall that $w_i \leq 1$ (by our assumption in \Cref{sec:model}). Because $\DualOver_i^* +\DualHard_i^* > 1$, we have $w_i - \DualOver^*_i -\DualHard^*_i <0$ and therefore $w_i - \tilde{\DualOver}_i -\tilde{\DualHard}_i <0$ by choosing sufficiently small $\delta>0$. Furthermore, $(w_j- \DualOver^*_j - \DualHard^*_j)_{+} = (w_j -\tilde{\DualOver}_j - \tilde{\DualHard}_j)_{+}$ for all $j \neq i$ by definition of $\tilde{\boldsymbol{\DualOver}}$ and $\tilde{\boldsymbol{\DualHard}}$. Combining the two facts leads to line \eqref{line:same+adjusted}. Consequently, we have:
\begin{equation*}
\begin{split}
D(\tilde{\boldsymbol{\nu}}) - D(\boldsymbol{\nu}^*)&= 2\delta\E[\mathbbm{1}[\Target = i]] + 
    \E\left[
    \Big\{\max_{j \in [m]}(w_j - \tilde{\DualOver}_j - \tilde{\DualHard}_j)_{+} - \max_{j \in [m]}(w_j - \DualOver_j^* - \DualHard^*_j)_{+}\Big\}\Indicator[\Target=0]
    \right] - 2\delta \rho_i \\ 
    &= 2\delta (\P[\Target = i ] - \rho_\Locidx) \\
    & <0.
\end{split}
\end{equation*}
The second line follows from \eqref{line:same+adjusted}. The last line is due to our assumption $\P[\Target=\Locidx] < \CapRatio_\Locidx$ for all $\Locidx \in [\Locnum]$ (see \Cref{sec:model}). Hence, we have proved Claim \ref{claim:arrival+rate}-\ref{fluid+dual+bound}. 

We now prove parts \ref{fluid+dual+rate} and \ref{fluid+dual+value}. Let $\bar{\lambda} = \frac{1+2\OverCost}{\underline{\rho}}$. By the Karush-Kuhn-Tucker condition for $\boldsymbol{\nu}^*$, there exists $\underline{u}_i \geq 0$ and $\overline{u}_i \geq 0$ such that
\begin{align}
    \frac{\partial D(\boldsymbol{\nu})}{\partial \DualHard_i}\bigg \rvert_{\nu = \boldsymbol{\nu^*}} - \underline{u}_i + \overline{u}_i &= 0 \label{eq:KKT-(1)}\\
    \underline{u}_i \DualHard_i^* &= 0 \label{eq:KKT-(2)}\\
    \overline{u}_i(\DualHard^*_i - \bar{\lambda}) &= 0.
\end{align}

Because $\bar{\lambda}>1$ (due to $\rho_i \leq 1$ for all $ i\in[m]$) and $\DualHard_i^* \leq 1$ by part \ref{fluid+dual+bound}, we must have $\overline{u}_i =0$. Furthermore, Assumption \ref{assump:PDF} implies that each $\tilde{\DecisionVec}(\boldsymbol{\nu}, \mathbf{A})$ is unique almost surely. Therefore, by Theorem 7.44 of \citep{shapiro2021lectures}, the function $D(\cdot)$ is differentiable and the partial derivative with respect to $\DualHard_\Locidx$ is 
\begin{equation}
    \frac{\partial D(\boldsymbol{\nu})}{\partial \DualHard_\Locidx}\Big|_{\boldsymbol{\nu}=\boldsymbol{\nu}^*} = \CapRatio_\Locidx - \E[\tilde{\Decision}_i(\boldsymbol{\nu}^*, \Arrival)].
\end{equation}
Plugging this into the preceding Karush-Khun-Tucker conditions (along with $\overline{u}_i =0$), we have:
\begin{align}
    \CapRatio_\Locidx - \E[\tilde{\Decision}_i(\boldsymbol{\nu}^*, \Arrival)] &\geq 0 \label{line:KKT+1}\\
    (\CapRatio_\Locidx - \E[\tilde{\Decision}_i(\boldsymbol{\nu}^*, \Arrival)])\DualHard_i^* &= 0. \label{line:KKT+2}    
\end{align}

Part \ref{fluid+dual+rate} is a direct consequence of line \eqref{line:KKT+1}. For part \ref{fluid+dual+value}, one can follow the identical steps as before to show the complementary slackness $(\CapRatio_\Locidx - \tilde{\Decision}_i(\boldsymbol{\nu}^*, \Arrival)])\DualOver_i^* = 0$ for all $i \in [\Locnum]$. Combining this with line \eqref{line:KKT+2}, the dual objective at optimal dual $\boldsymbol{\nu}^*$ must be $D(\boldsymbol{\nu}^*) = \E[\Weightvec\cdot\tilde{\DecisionVec}(\boldsymbol{\nu}^*, \Arrival)]$. This completes the proof of part \ref{fluid+dual+value}. 
\hfill\halmos
\subsection{Proof of 
\texorpdfstring{\Cref{prop:harvey}}{}}\label{apx+harvey}

\noindent{\bf General Setup.} We begin by describing a general setup of online stochastic mirror descent (OSMD). Let $D: \mathcal{V}\to \R$ be a convex function and (with a slight abuse of notation) $\boldsymbol{\nu}_t$ be a sequence of iterates such that
\begin{equation*}
\begin{split}
   \nabla h (\mathbf{y}_b) &= \nabla h(\boldsymbol{\nu}_{b-1}) - \eta_{b-1} \hat{\mathbf{g}}_{b-1}\\
    \boldsymbol{\nu}_b &= \argmin_{\boldsymbol{\nu} \in \mathcal{V}} V_h(\boldsymbol{\nu}, \mathbf{y}_{b}) \nonumber
\end{split}
\end{equation*}
where
\begin{enumerate}[label=(\roman*)]
    \item (Bounded domain): $\mathcal{V}$ is a bounded convex set \label{item:begin}
    \item (Strongly convex mirror map): $h(\cdot):\mathcal{V}\to \R$ is a $\sigma$-strongly convex on domain $\mathcal{V}$ with respect to $\Norm{\cdot}{1}$.
    \item (Unbiased and bounded Gradient): $\E[\mathbf{\hat{g}}_t|\History_{t-1}] = \nabla D(\boldsymbol{\nu}_t)$ and 
    $\Norm{\mathbf{\hat{g}}_t}{\infty} \leq G$ almost surely for a constant $G$.
    \item (Bounded Noise): The noise $\hat{\mathbf{u}}_t:= \E[\mathbf{\hat{g}}_t|\History_{t-1}] - \mathbf{\hat{g}}_t$ satisfies $\Norm{\hat{\mathbf{u}}}{1}\leq U$ almost surely for some constant $U$.
    \item $\eta_t = \frac{k}{\sqrt{t}}$ for some constant $k>0$.\label{item:end}
\end{enumerate}

The following is the main result we will establish.
\begin{theorem}[Extending Theorem 3.2 of \citet{harvey2019tight} to OSMD]\label{thm:last+iterate} Define $\boldsymbol{\nu}^*$ such that
\[\boldsymbol{\nu}^*:= \argmin_{\boldsymbol{\nu} \in \mathcal{V}} D(\boldsymbol{\nu}).\]

Under the conditions \eqref{item:begin}-\eqref{item:end} and for any given $s$, we have
\begin{equation}
    \P\left[D(\boldsymbol{\nu}_s ) - D(\boldsymbol{\nu}^*) \leq \BigO\left(\frac{\log(s)\log(1/\delta)}{\sqrt{s}}\right)\right] \geq 1-\delta.
\end{equation}
\end{theorem}

\noindent{\bf Connection to \Cref{prop:harvey}}.\label{example:last+iterate}
 To see the connection of the theorem to \Cref{prop:harvey}, recall that we aimed to study the last-iterate convergence of the dual variables $\{\boldsymbol{\nu}_{\Timeidx} \}_{\Timeidx=1}^{\Stopping}$ of \Cref{ALG+Surrogate+P}. For ease of reference, we recall that the domain of the dual variables was given by $\mathcal{V} := [0, \OverCost]^\Locnum\times  [0, \bar{\DualHard}]^m$ where $\bar{\DualHard}:= \frac{1+2\OverCost}{\underline{\CapRatio}}$. The static dual function (\Cref{def:staic+dual} in \Cref{subsec:ALG+P+static+dual}) was defined as:
\begin{equation*}
    \begin{split}
    D(\boldsymbol{\nu}) &:=  \E\left[
    {\max_{\DecisionVec \in \FeasibleSet{\Target}}(\Weightvec - \boldsymbol{\DualOver} - \boldsymbol{\DualHard}})\cdot \DecisionVec + \CapRatioVec\cdot \boldsymbol{\DualOver} + \CapRatioVec\cdot \boldsymbol{\DualHard}
    \right]. \\
    D(\boldsymbol{\nu}^*) &= \min_{\boldsymbol{\nu} \in \mathcal{V}} D(\boldsymbol{\nu}).
    \end{split}
\end{equation*}
The dual variables $\{\boldsymbol{\nu}_\Timeidx\}_{\Timeidx=1}^{\Stopping}$ is the sequence of OMSD (with the negative entropy mirror map), as we discussed in \Cref{subsec:alg+surrogate+primal}. Furthermore, the primitives $(U, G, \sigma)$ for our variant of OSMD is constant. In particular, it is straightforward to see $G = 1$, $U=2$, and $\sigma = \frac{1}{2m\max(\bar{\lambda},\OverCost)}$
Hence, applying \Cref{thm:ALG+Surrogate+Primal}, we directly obtain \Cref{prop:harvey}. 

\subsubsection{Proof of \texorpdfstring{\Cref{thm:last+iterate}}{}}\hfill\\
The proof closely follows the seminal work of \cite{harvey2019tight}, which establishes the same result for stochastic gradient descent, a special case of stochastic mirror descent with a mirror map denoted as $h(\boldsymbol{\nu}) = \Norm{\boldsymbol{\nu}}{2}^2$. Given our primary use of the multiplicative update rule, an online mirror descent employing the negative entropy mirror map, we need to extend the result to the general mirror descent setting. Although extending the proof of \cite{harvey2019tight} to non-Euclidean geometry is straightforward, the exact result we require is, to the best of our knowledge, not available in the literature. Therefore, for the sake of completeness, we will revisit the main steps of \cite{harvey2019tight} and highlight necessary modifications to prove the result for our setting. To avoid unnecessary repetitions, we clearly label any result that can be directly derived from \cite{harvey2019tight} and kindly refer readers to the original paper.\par 
Without loss of generality, we assume that the primitives $(U,G,\sigma)$ are the values corresponding to our variants of OSMD, as given in the last paragraph of \Cref{example:last+iterate}. We begin with the standard result on the average convergence guarantee, which follows from Claim \ref{fact:average+conv}.
\begin{lemma}[Average convergence]\label{lemma:average+convergence}
For any $\boldsymbol{\nu} \in \mathcal{V}$, $k$, and $s$, we have
\begin{equation*}
\sum_{t=k}^s (D(\boldsymbol{\nu}_t) - D(\boldsymbol{\nu})) \leq 
\sum_{t=k}^s \frac{\eta_t G^2}{2\sigma} + \frac{1}{\eta_k}V_h(\boldsymbol{\nu}, \boldsymbol{\nu_k}) + \sum_{t=k}^s \hat{\mathbf{u}}_t\cdot(\boldsymbol{\nu}_t - \boldsymbol{\nu}) + 
\sum_{t=k+1}^{s} \left(\frac{1}{\eta_t} - \frac{1}{\eta_{t-1}}\right)V_h(\boldsymbol{\nu}, \boldsymbol{\nu}_t)
\end{equation*}
\end{lemma}
The following lemma associates the last-iterate convergence with the average convergence as follows. 
\begin{lemma}[Lemma 8.1 of \citet{harvey2019tight}]\par \label{lemma:last<=average+noise} For any $s\geq 1$,
    \begin{align*}
    D(\boldsymbol{\nu}_s) - D(\boldsymbol{\nu}^*)\leq 
    \underbrace{ \frac{1}{s/2 + 1}\sum_{t=s/2}^s[D(\boldsymbol{\nu}_t) - D(\boldsymbol{\nu}^*)]}_{\spadesuit}
    + \BigO\left(\frac{\log(s)}{\sqrt{s}}\right) + 
    \underbrace{
       \sum_{t=s/2}^s \boldsymbol{\omega}_t \cdot \hat{\mathbf{u}}_t 
    }_{\clubsuit} 
    \end{align*}
where
\begin{align}
    \boldsymbol{\omega}_t = \sum_{j=s/2}^{t} \frac{1}{(s-j)(s-j+1)}(\boldsymbol{\nu}_t - \boldsymbol{\nu}_j). \label{eq:omega}
\end{align}
\end{lemma}

The proof mirrors that of Lemma 8.1 in \citet{harvey2019tight} and we omit the proofs for brevity. \Cref{lemma:last<=average+noise} implies that the last iterate optimality gap is related to that of the suffix average ($\spadesuit$) and the weighted sum of the noise of the gradients $(\clubsuit)$. Hence, it suffices to obtain the high-probability bound for each term. We first begin with the term $\spadesuit$.\par 
\noindent{\bf Bounding $\spadesuit$.}
\begin{lemma}[Lemma 8.2 of \citet{harvey2019tight}]\label{lemma:first+term} With probability at least $1-\delta$,
 \begin{align*}
    \frac{1}{s/2 + 1}\sum_{t=s/2}^s[D(\boldsymbol{\nu}_t) - D(\boldsymbol{\nu}^*)] \leq \BigO\left(\frac{\sqrt{\log(1/\delta)}}{\sqrt{s}}\right)
   \end{align*}
\end{lemma}

The proof again mirrors Lemma 8.2 of \citet{harvey2019tight} and we kindly refer the readers to the original paper. The idea is to observe that $\hat{\mathbf{u}}_t\cdot(\boldsymbol{\nu}_t -\boldsymbol{\nu}^*)$ is a Martingale-difference sequence with respect to the filtration $\{ \History_t\}$ because $\E[\hat{\mathbf{u}}_t|\History_{t-1}]=0$ and $\boldsymbol{\nu}_t$ is $\History_{t-1}$-measurable. Due to the assumption of boundedness of all the primitives (domain, noise, and the norm of gradients), the magnitude of the difference sequence is bounded by a constant. Hence, a direct application of Azuma's inequality leads to the lemma. \par 
\noindent{\bf Bounding $\clubsuit$.}

The main technical lemma will be:
\begin{lemma}\label{lemma:last+term}
    \begin{align*}
\sum_{t=s/2}^s \boldsymbol{\omega}_t \cdot \hat{\mathbf{u}}_t \leq \BigO\left(\frac{\log(s)\log(1/\delta)}{\sqrt{s}}\right)
    \end{align*}
with probability at least $1-\delta$.
\end{lemma}
Combining the previous two lemmas directly implies \Cref{thm:last+iterate}. The proof of \Cref{lemma:last+term} is fairly intricate. We outline the proof in the following. Similar to the previous lemma, we note that $\boldsymbol{\omega}_t\cdot \hat{\mathbf{u}}_t$ is a Marintagle difference sequence with respect to $\{\History_t\}$. Hence, the standard Azuma-type inequality would require a bound of $\sum_{t=s/2}^s \lVert \boldsymbol{\omega}_t\rVert^2$ to obtain a high-probability bound of $\clubsuit$. However, we will see that the bound of $\sum_{t=s/2}^s \lVert \boldsymbol{\omega}_t\rVert^2$ itself is related to a linear combination of the Martingale difference sequence $\boldsymbol{\omega}_t\cdot \hat{\mathbf{u}}_t$. This non-standard nature requires a more sophisticated concentration inequality than the plain Azuma's inequality. For this purpose, we extract the following concentration result from \cite{harvey2019tight}.
\begin{theorem}[Corollary C.5 of \citet{harvey2019tight}]\label{thm:concentration}
Let $d_t := \mathbf{a}_t \cdot \mathbf{b}_t$ where $\mathbf{a}_t$ is $\History_{t}$ measurable and $\mathbf{b}_t$ is $\History_{t-1}$ measurable. Suppose $\E[\mathbf{a}_t | \History_{t-1}] = 0$, $\lVert \mathbf{a}_t\rVert_2 \leq 1$ almost surely, and there exists positive values $\{\tilde{\alpha}_t\}_{t=1}^s$ and $R >0$ such that
\begin{enumerate}[label=(\alph*)]
    \item $\max_{t\in[s]} \tilde{\alpha}_t \leq \BigO(\sqrt{R})$
    \label{item:alpha}
    \item $\sum_{t=1}^s \lVert \mathbf{b}_t \rVert_2^2 \leq \sum_{t=1}^{s-1} \tilde{\alpha}_t d_t + R \sqrt{\log(1/\delta)}$ with probability at least $1-\delta$.\label{item:bound+on+b}
\end{enumerate}
Then $\sum_{t=1}^T d_t \leq \BigO(\sqrt{R}\log(1/\delta))$ with probability at least $1-\delta$.
\end{theorem}
To see the connection of Theorem \ref{thm:concentration} to our task of bounding $\clubsuit$, let $\mathbf{a}_t = \frac{1}{2}\mathbf{\hat{u}}_t$ and $\mathbf{b}_t = \boldsymbol{\omega}_t$, which are $\History_{t}$ and $\History_{t-1}$ measurable, respectively. Note that $\E[\hat{\mathbf{u}}_t|\History_{t-1}]=0$ and  $\lVert \mathbf{a}_t\rVert_2 \leq \lVert \mathbf{a}_t\rVert_1 \leq 1 $ because of $U=2$ in our variant of OSMD. Hence, if we can obtain a bound in the form of Theorem \ref{thm:concentration}-\ref{item:alpha} and \ref{item:bound+on+b}, the proof is complete. Naturally, by the definition of $\boldsymbol{\omega}_t$ in 
line \eqref{eq:omega}, this requires studying $\lVert \boldsymbol{\nu}_t - \boldsymbol{\nu}_j \rVert_2^2$ for arbitrary pair $(j,t)$. The following claim makes this point explicit. 
\begin{claim}[Claim E.3 of \citet{harvey2019tight}]\label{claim:omega}
For any $t \leq s$,
\begin{align*}
    \lVert \boldsymbol{\omega}_t \rVert_2^2 \leq  \frac{1}{s-t+1} \sum_{j=s/2}^{t-1} \alpha_j\lVert \boldsymbol{\nu}_t - \boldsymbol{\nu}_j \rVert_2^2
\end{align*}
where $\alpha_j := \frac{1}{(s-j)(s-j+1)}$.
\end{claim}
The proof is again mainly algebraic and can be found in Claim E.3 of \cite{harvey2019tight}. In light of Theorem \ref{thm:concentration} and Claim \ref{claim:omega}, we obtain the bound on $\lVert \boldsymbol{\nu}_t - \boldsymbol{\nu}_j \rVert_2^2$ for arbitrary pair $(j,t)$ in the following. 
\begin{lemma}[Extension of Lemma 7.3 of \citet{harvey2019tight} to Mirror Descent] For any $a<b$, we have
\label{lemma:bound+distance}
\begin{align*}
\frac{\sigma}{2}\lVert \boldsymbol{\nu}_a - \boldsymbol{\nu}_b \rVert_2^2 \leq \sum_{i=a}^{b-1} \frac{G^2 \eta_i^2}{\sigma} + \sum_{i=a}^{b-1}\eta_i(D(\boldsymbol{\nu}_a)-D(\boldsymbol{\nu}_i)) + \sum_{i=a}^{b-1} \eta_i\mathbf{\hat{u}}_i\cdot(\boldsymbol{\nu}_i - \boldsymbol{\nu}_a)
\end{align*}
\end{lemma}
This is the key lemma we extend from \cite{harvey2019tight} where the same result was established only for the stochastic gradient descent. The original result of \citet{harvey2019tight}  relies on the fact that the update rule is performed in the Euclidean space.\footnote{That is, the form of $\boldsymbol{\nu}_{t+1} = \Pi_{\mathcal{V}}(\boldsymbol{\nu}-\eta_t \hat{\mathbf{g}}_t)$ where $\Pi_{\mathcal{V}}(\cdot)$ the Euclidean projection onto set $\mathcal{V}$.} However, the original the proof cannot be directly applied to a general online mirror descent setting because, for general mirror descent, the dual update is performed through the Bregman distance rather than the Euclidean distance. Fortunately, the Bregman distance shares similar structural properties with the Euclidean distance. Hence, we can establish a recursive relationship between $V_h(\boldsymbol{\nu}_a, \boldsymbol{\nu}_b)$ and $V_h(\boldsymbol{\nu}_a, \boldsymbol{\nu}_{b-1})$ and use the strong convexity of $h(\cdot)$ to translate the result to that of the Euclidean distance. We prove the lemma in the following. \par 
{\bf Proof of \Cref{lemma:bound+distance}.}
Throughout the proof, we will use $\Norm{\cdot}{}$ to denote $\mathcal{L}_1$-norm as a primal norm. We also use the following standard result.
\begin{fact}[Three point equality: Lemma 5.2 of \citet{bubeck2011introduction}]\label{fact:three+point}
For any $\mathbf{a}$, $\mathbf{b}$, and $\mathbf{c}$, 
    \begin{equation*} 
    V_h(\mathbf{a},\mathbf{c}) = V_h(\mathbf{a},\mathbf{b}) + V_h(\mathbf{b}, \mathbf{c}) + (\nabla h(\mathbf{b}) - \nabla h(\mathbf{c}))\cdot(\mathbf{a} - \mathbf{b})
    \end{equation*}
\end{fact}

\begin{fact}[Bregman Projection: Lemma 5.3 of \citet{bubeck2011introduction}]\label{fact:projection}
For a convex set $C$, let \[\mathbf{x}^* = \argmin_{\mathbf{x} \in C}  V_h(\mathbf{x}, \mathbf{x}_0).\]
For any $\mathbf{y} \in C$, we have
    \begin{equation*}
V_h(\mathbf{y}, \mathbf{x}_0) \geq V_h(\mathbf{y}, \mathbf{x}^*) + V_h(\mathbf{x}^*, \mathbf{x}_0) 
    \end{equation*}
\end{fact}

\begin{fact}[Lemma 5.1 of \citet{bubeck2011introduction}]\label{fact:conjugate}
    Let $h^*(\mathbf{y}) = \sup_{\mathbf{x}\in \mathcal{D}}\{\mathbf{x}\cdot\mathbf{y} - h(\mathbf{x})\}$ be convex conjugate of a  convex function $h$. Then
    $\nabla (h^*) = (\nabla (h))^{-1}$. That is, $\nabla (h^*)$ is inverse of $\nabla (h)$.
\end{fact}\par 

\begin{fact}[Theorem 5.26 of \citet{beck2017first}]\label{fact:strong+smooth}
The following is equivalent: 
    \begin{enumerate}
        \item $h$ is strongly-convex in $\lVert \cdot \rVert$ with modulus $\sigma$. That is, $h(\mathbf{x}) \geq h(\mathbf{y}) + \nabla h(\mathbf{y})\cdot(\mathbf{y}-\mathbf{x}) + \frac{\sigma}{2}\lVert \mathbf{x} - \mathbf{y}\rVert^2$
        \item $h^*$ (convex conjugate of $h$) is smooth in $\lVert \cdot \rVert_{*}$ with modulus $1/\sigma$. That is, $\lVert \nabla h^*(\mathbf{x}) - \nabla h^*(\mathbf{y})\rVert \leq \frac{1}{\sigma} \lVert 
        \mathbf{x} - \mathbf{y}\rVert_{*}$ where $\Norm{\cdot}{*}$ is the dual norm of $\Norm{\cdot}{}$.
    \end{enumerate}
\end{fact}

Recall that we can write the update rule of the dual variable as 
\begin{align}
    \nabla h (\mathbf{y}_b) &= \nabla h(\boldsymbol{\nu}_{b-1}) - \eta_{b-1} \hat{\mathbf{g}}_{b-1} \label{eq:update+rule}\\
    \boldsymbol{\nu}_b &= \argmin_{\boldsymbol{\nu} \in \mathcal{V}} V_h(\boldsymbol{\nu}, \mathbf{y}_{b}) \nonumber
\end{align}
We first establish the following recursion:
\begin{equation}
\begin{split}
    V_h(\boldsymbol{\nu}_a, \boldsymbol{\nu}_b) &\leq 
    V_h(\boldsymbol{\nu}_a, \mathbf{y}_b) \\
    &\leq V_h(\boldsymbol{\nu}_a, \boldsymbol{\nu}_{b-1}) +
    V_h(\boldsymbol{\nu}_{b-1}, \mathbf{y}_b) + 
    (\nabla h(\boldsymbol{\nu}_{b-1}) - \nabla h(\mathbf{y}_b))\cdot
    (\boldsymbol{\nu}_a - \boldsymbol{\nu}_{b-1}) 
    \\
    &\leq V_h(\boldsymbol{\nu}_a, \boldsymbol{\nu}_{b-1}) +
    V_h(\boldsymbol{\nu}_{b-1}, \mathbf{y}_b) + 
    \eta_{b-1}\hat{\mathbf{g}}_{b-1}\cdot (\boldsymbol{\nu}_a - \boldsymbol{\nu}_{b-1}) \quad 
\end{split}\label{ineq:V}
\end{equation}
The first line follows from Fact \ref{fact:projection}. In the second line, we used Fact \ref{fact:three+point}. The last line follows from Line \eqref{eq:update+rule}.
We now bound the second term in the following. 
\begin{align}
 V_h(\boldsymbol{\nu}_{b-1}, \mathbf{y}_b) &\leq 
 V_h(\boldsymbol{\nu}_{b-1}, \mathbf{y}_b) + 
 V_h(\mathbf{y}_b, \boldsymbol{\nu}_{b-1}) \nonumber \\
&= (\nabla h(\boldsymbol{\nu}_{b-1}) - \nabla h(\mathbf{y}_b))\cdot(\boldsymbol{\nu}_{b-1}- \mathbf{y}_b) \label{line:2nd}\\
&=  \eta_{b-1}\hat{\mathbf{g}}_{b-1}\cdot (\boldsymbol{\nu}_{b-1} - \mathbf{y}_b) \label{line:3rd} \\
&\leq \eta_{b-1} G \lVert \boldsymbol{\nu}_{b-1} - \mathbf{y}_b \rVert \label{line:4th} \\
&= \eta_{b-1} G \lVert \nabla h^* (\nabla h (\boldsymbol{\nu}_{b-1})) - \nabla h^* (\nabla h(\mathbf{y}_b)) \rVert \label{line:5th}\\
&\leq \frac{\eta_{b-1}G}{\sigma} \lVert \nabla h (\boldsymbol{\nu}_{b-1}) - \nabla h(\mathbf{y}_b) \rVert_{*} \quad \label{line:6th} \\
&\leq \frac{\eta_{b-1}^2 G^2}{\sigma} 
\end{align}

The first line is due to the non-negativity of Bregman distance. Line \eqref{line:2nd} follows from Fact \ref{fact:three+point}.\footnote{
That is, $0 = V_h(\boldsymbol{\nu}_{b-1}, \boldsymbol{\nu}_{b-1}) = V_h(\boldsymbol{\nu}_{b-1}, \mathbf{y}_{b}) + V_h(\mathbf{y}_{b}, \boldsymbol{\nu}_{b-1}) 
+ (\nabla h(\mathbf{y}_{b}) - \nabla h(\boldsymbol{\nu}_{b-1}))\cdot (\boldsymbol{\nu}_{b-1} - \mathbf{y}_{b})$ 
} Line \eqref{line:3rd} comes from the update rule \eqref{eq:update+rule}. In line \eqref{line:4th}, we used Cauchy Schwartz and the bounded gradient assumption ($\lVert \hat{\mathbf{g}}_{b-1} \rVert_{*} \leq G$). For lines \eqref{line:5th} and \eqref{line:6th}, we applied Facts \ref{fact:conjugate} and \ref{fact:strong+smooth}, respectively. The final line is again due to the update rule \eqref{eq:update+rule} and the bounded gradient assumption.
Hence, from the last line of \eqref{ineq:V}, We have
\begin{align}
V_h(\boldsymbol{\nu}_a, \boldsymbol{\nu}_{b}) \leq V_h(\boldsymbol{\nu}_a, \boldsymbol{\nu}_{b-1}) +
 \frac{\eta_{b-1}^2 G^2}{\sigma} + 
    \eta_{b-1}\hat{\mathbf{g}}_{b-1}\cdot (\boldsymbol{\nu}_a - \boldsymbol{\nu}_{b-1}) 
\end{align}
By induction, we obtain 
\begin{align*}
V_h(\boldsymbol{\nu}_a, \boldsymbol{\nu}_{b}) &\leq 
\sum_{i=a}^{b-1} \frac{\eta_{i}^2 G^2}{\sigma} + 
\sum_{i=a}^{b-1}\eta_{i}\hat{\mathbf{g}}_{i}\cdot (\boldsymbol{\nu}_a - \boldsymbol{\nu}_{i})\\ 
&=\sum_{i=a}^{b-1} \frac{\eta_{i}^2 G^2}{\sigma}+ 
\sum_{i=a}^{b-1}\eta_{i}{\mathbf{g}}_{i}\cdot (\boldsymbol{\nu}_a - \boldsymbol{\nu}_{i}) + 
\sum_{i=a}^{b-1}\eta_{i}\hat{\mathbf{u}}_{i}\cdot (\boldsymbol{\nu}_i - \boldsymbol{\nu}_{a}) \\
&\leq 
\sum_{i=a}^{b-1} \frac{\eta_{i}^2 G^2}{\sigma}+ 
\sum_{i=a}^{b-1}\eta_{i}(D(\boldsymbol{\nu}_a) - D(\boldsymbol{\nu}_i)) + 
\sum_{i=a}^{b-1}\eta_{i}\hat{\mathbf{u}}_{i}\cdot (\boldsymbol{\nu}_i - \boldsymbol{\nu}_{a})
\end{align*}

The last line is because $\mathbf{g}_i$ is subgradient of $D$ at $\boldsymbol{\nu} = \boldsymbol{\nu}_i$. The proof is complete by noting that (i) $V_h(\boldsymbol{\nu}_a, \boldsymbol{\nu}_{b}) \geq \frac{\sigma}{2}\lVert \boldsymbol{\nu}_a - \boldsymbol{\nu}_b\rVert_{1}^2 $ by strong convexity of $h$ with modulus $\sigma$ (recall that the primal norm is $\mathcal{L}_1$ norm) with respect to $\lVert \cdot \rVert_1$ and (ii) $\lVert\cdot \rVert_1^2 \geq \lVert \cdot \rVert_2^2$. \hfill\halmos

Now that we have successfully extended the above lemma to the general mirror descent setting, we can apply Theorem \ref{thm:concentration} in the following form.  

\begin{lemma}[Claims E.4-E.6 of \citet{harvey2019tight}]\label{lemma:final+bound}
Let $d_t = \frac{1}{2}\hat{\mathbf{u}}_t \cdot \boldsymbol{\omega}_t$. There exists $R = \BigO(\log^2(s)/{s})$ and positive values $\tilde{\alpha}_t > 0$ such that
\begin{enumerate}
    \item $\max_{s/2 \leq t \leq s} \tilde{\alpha}_t \leq \BigO(\sqrt{R})$
    \item $\sum_{t=s/2}^s \lVert \boldsymbol{\omega}_t \rVert_2^2 \leq  \sum_{t=s/2}^{s-1} \tilde{\alpha}_t d_t + R \sqrt{\log(1/\delta)}$ with probability at least $1-\delta$. 
\end{enumerate}
\end{lemma}
The proof is mainly algebraic and can be found in Lemmas E.4-E.6 of \cite{harvey2019tight}.
Here we only give a sketch of the proof. From Claim \ref{claim:omega} and Lemma \ref{lemma:bound+distance}, we have $\sum_{t=s/2}^s \lVert 
\omega_t\rVert_2^2 \leq \Lambda_1 + \Lambda_2 + \Lambda_3$ where
\begin{align*}
    \Lambda_1 &:= \frac{2G^2}{\sigma^2}\sum_{t=s/2}^s \frac{1}{s-t+1}\sum_{j=s/2}^{t-1}\alpha_j \sum_{i=j}^{t-1} \eta_i^2 \\
    \Lambda_2 &:= \frac{2}{\sigma} \sum_{t=s/2}^{s}\frac{1}{s-t+1}\sum_{j=s/2}^{t-1} \alpha_j \sum_{i=j}^{t-1} \eta_i (D(\boldsymbol{\nu}_j) - D(\boldsymbol{\nu}_i))\\
    \Lambda_3 &:= \frac{2}{\sigma} \sum_{t=s/2}^{s}\frac{1}{s-t+1}\sum_{j=s/2}^{t-1}\alpha_j \sum_{i=j}^{t-1} \eta_i \mathbf{\hat{u}}_i(\boldsymbol{\nu}_i - \boldsymbol{\nu}_j).
\end{align*} 
With some delicate algebras, we can show that $\Lambda_1 \leq O\Big(\frac{\log^2(s)}{s}\Big)$ (see Claim E.4 of \cite{harvey2019tight}) and $\Lambda_2 \leq O\Big(\frac{\log^2(s)\sqrt{\log(1/\delta)}}{s}\Big)$ with proabability $1-\delta$ (see Claim E.5 of \cite{harvey2019tight}). Interestingly, to prove the bound for $\Lambda_2$, we use the result in Lemma \ref{lemma:first+term} again. Finally, note that $\Lambda_3$ is essentially the scaled version of term $\clubsuit$ in Lemma \ref{lemma:last<=average+noise}, which we were aiming to bound. In fact, one can follow the same algebras taken in Claim E.6 of \cite{harvey2019tight} to show $\Lambda_3 \leq \frac{2}{\sigma}\sum_{i=s/2}^{s-1} {\alpha}_i (\mathbf{\hat{u}}_i \cdot \boldsymbol{\omega})$ where ${\alpha}_i =\frac{4}{\sigma} \eta_i \sum_{j=i+1}^s \frac{1}{s-i+1}$ is the extra scaling factor. Note that ${\alpha}_i\leq \BigO(\frac{\log(s)}{\sqrt{s}})$ for all $\frac{s}{2}\leq i\leq s$. Combining the bounds for $\Lambda_1$, $\Lambda_2$, and $\Lambda_3$ will give the desired result. 

Finally, the direct application of Theorem \ref{thm:concentration} and Lemma \ref{lemma:final+bound} leads to Lemma \ref{lemma:last+term}. This completes the proof of \Cref{thm:last+iterate}. 

\subsection{Proof of 
\texorpdfstring{\Cref{thm:ALG+Surrogate+Primal}}{}}\label{apx+thm+2+pf}

We first begin by formally defining the endogenous matching rate of \CO{} given the dual variable:
\begin{definition}[{Dual-based Matching Rate}]\label{def:M}
For a given $\boldsymbol{\nu} = (\DualOverVec, \DualHardVec)$, the dual-based matching rate is defined as \begin{equation*}
\MRateVec(\boldsymbol{\nu}):= \E_{\Arrival}[\tilde{\mathbf{z}}(\boldsymbol{\nu}, \Arrival)]
\end{equation*}
where we recall from \Cref{def:staic+dual} that $ \tilde{\mathbf{z}}(\boldsymbol{\nu}, \Arrival) := \argmax_{\mathbf{z} \in \FeasibleSet{\Target}} \{(\Weightvec - \DualOverVec - \DualHardVec)\cdot \DecisionVec \} $.
\end{definition}

Note that, under Assumption \ref{assump:PDF}, the primal-based decision $\tilde{\mathbf{z}}(\boldsymbol{\nu}, \Arrival)$ is uniquely determined for each $\boldsymbol{\nu}$ and $\Arrival$ almost surely, and hence the dual-based matching rate is well defined.

Given all the mentioned ingredients, we now outline the main steps to prove \Cref{thm:ALG+Surrogate+Primal}. Similar to \Cref{subsec:ALG+D+pfsketch}, we prove \Cref{thm:ALG+Surrogate+Primal} based on the the same regret decomposition as in line \eqref{line:regret+decomp} and obtain bounds on $\textsf{(A)}$, the loss regarding the expected net \allocation{} reward, and $\textsf{(B)}$, the expected average \backlog{}. The following lemma first establishes the bound on $\textsf{(A)}$.
\begin{lemma}[Bounding Loss of Net \Allocation{} Reward of \CO{}]\label{lemma:ALG+P+reward} For any arrival distribution $\ArrivalDist$ and service slack $\ServiceSlack \geq 0$, 
\CO{} satisfies
\begin{equation*}
\E[\NetReward(\{\DecisionVec^*_\Timeidx\}_{\Timeidx=1}^\TotalTime; \OverCost)]
-
\E[\NetReward(\{\DecisionVec_\Timeidx\}_{\Timeidx=1}^\TotalTime; \OverCost)]
\leq \BigO(\sqrt{\TotalTime})
\end{equation*}  
where we recall that $\NetReward(\cdot;\OverCost)$ is the total net \allocation{} reward and $\{\DecisionVec_\Timeidx^*\}_{\Timeidx=1}^\TotalTime$ is the optimal offline solution (\Cref{def:offline}).
\end{lemma}
The proof mirrors that of \Cref{lemma:ALG+D+reward} and we present the proof in Appendix \ref{apx+P+reward}. 

In \Cref{subsubsec:ALG+P+pf+sketch}, we stated \Cref{Lemma+ALG+P+Build} that bounds $\textsf{(B)}$, the expected average \backlog{}. For ease of reference, we restate the lemma in the following.

\COBacklog*



The proof consists of three steps. In the first step, we establish that the good event $\mathcal{B}_\TotalTime$ (see \eqref{line:good+event+def} in \Cref{subsubsec:ALG+P+pf+sketch} for its definition) occurs with sufficiently high probability using \Cref{prop:harvey} and Assumption \ref{assump:PDF}. In the second step, we use the definition of the good event to show an $\BigO(1/\ServiceSlack)$ bound on the average \backlog{} in expectation up to the stopping time. In the last step, we prove that the average build-up after the stopping time is $\BigO(1)$. We elaborate each step in the following.

\noindent{\bf Step 1: Establishing a High-probability Bound of the Good Event.}\par 
The following lemma shows that the good event occurs with a high probability.\footnote{More precisely, there exists a constant $\underline{K}$, which only depends on (i) the service slack parameter $\ServiceSlack>0$, (ii) the Lipschitz constant $L>0$ in Assumption \ref{assump:PDF}, and (iii) the input of the algorithm other than $\TotalTime$ (i.e., $\boldsymbol{\rho}$ and $k >0$), such that $\P[\mathcal{B}_{\TotalTime}] \geq 1 - \frac{1}{\TotalTime^2}$ for all $\TotalTime \geq \underline{K}$. }
\begin{lemma}[High-probability Bound of Good Event] \label{lemma:good+event}\label{lemma:B}$\Pr[\mathcal{B}_{\TotalTime}] \geq 1 - \BigO(1/\TotalTime^2)$.
\end{lemma}
The proof of this lemma is presented in \Cref{apx+B+finish}. For the proof, we translate the high-probability convergence in terms of the dual objective (established in \Cref{prop:harvey}) to that of the matching rate for each $\Timeidx \in \{\sqrt{\TotalTime}, \Stopping\}$ using Assumption \ref{assump:PDF}. We then take the union bound of the preceding high probability bound across $ \Timeidx \in \{\sqrt{\TotalTime}, \Stopping\}$ to prove the desired result.  

\noindent{\bf Step 2: Using the Good Event to Bound the Average Backlog up to the Stopping time.}\par 
To utilize the good event for our task of bounding the average \backlog{}, we start with the following decomposition of the expected cumulative \backlog{}.
\begin{equation}
\E\Big[\sum_{\Timeidx=1}^\TotalTime\Norm{\BuildUpVec_\Timeidx}{1}\Big] = 
\E\Big[\sum_{\Timeidx=1}^{\sqrt{\TotalTime}}\Norm{\BuildUpVec_\Timeidx}{1}\Indicator[\mathcal{B}_\TotalTime]\Big]  + 
\underbrace{
\E\Big[\sum_{\Timeidx=\sqrt{\TotalTime}+1}^{\Stopping}\Norm{\BuildUpVec_\Timeidx}{1}\Indicator[\mathcal{B}_\TotalTime]\Big]}_{\mathclap{\textsf{(B-1)}}} + 
\underbrace{
\E\Big[\sum_{\Timeidx=\Stopping + 1}^{\TotalTime}\Norm{\BuildUpVec_\Timeidx}{1}\Indicator[\mathcal{B}_\TotalTime]\Big]}_{\mathclap{\textsf{(B-2)}}} + 
\E\Big[\sum_{\Timeidx=1}^{\TotalTime}\Norm{\BuildUpVec_\Timeidx}{1}\Indicator[\mathcal{B}^c_\TotalTime]\Big]
\label{line:P+build+decompose}
\tag{\textsf{Decomposition}}
\end{equation}
The first three terms are the cumulative \backlog{} under the good event. The last term is the one under the complement of the good event. We first note that the first and the last terms can bounded by $\BigO(\TotalTime)$ and $\BigO(1)$, respectively, because (i) $\Norm{\BuildUpVec_\Timeidx}{1} \leq \BigO(\Timeidx)$ for every sample path and (ii) $\P[\mathcal{B}^c_\TotalTime]\leq \BigO(1/\TotalTime^2)$ from \Cref{lemma:B}. Hence, the remaining task is to obtain an upper bound of the terms $\textsf{(B-1)}$ and $\textsf{(B-2)}$. In the following, we bound $\textsf{(B-1)}$ by $\BigO(\TotalTime/\ServiceSlack)$.
\begin{lemma}[Bounding \textsf{(B-1)}]\label{lemma:B-1} The term $\textsf{(B-1)}$ in line \eqref{line:P+build+decompose} is $\BigO(\TotalTime/\ServiceSlack)$.
\end{lemma}
The proof of \Cref{lemma:B-1} can be found in Appendix \ref{apx+B-1}. For the proof, we combine our drift lemma (\Cref{lemma:drift}) with the definition of the good event.

\noindent{\bf Step 3: Bounding the Average \backlog{} after the Stopping Time.}\par 

The last step is to establish that the term $\textsf{(B-2)}$ in \eqref{line:P+build+decompose}, the cumulative \backlog{} after the stopping time, is $\BigO(\TotalTime)$. 
We prove this result in the following lemma:
\begin{lemma}[Bounding \textsf{(B-2)}]\label{lemma:B-2}The term $\textsf{(B-2)}$ in line \eqref{line:P+build+decompose} is $\BigO(\TotalTime)$.\end{lemma}
The proof of this lemma follows the identical steps taken in \Cref{apx+finish+ALG+1+build}. The only modification in the proof is establishing an upper bound on $\E[\Norm{\BuildUpVec_{\Stopping}}{1}^2]$, which we prove through the following claim.

\begin{claim}\label{claim:b+P} For \CO{}(\Cref{ALG+Surrogate+P}),
$\E[\Norm{\BuildUpVec_{\Stopping}}{1}^2] \leq O(\TotalTime)$
\end{claim}

 We prove Claim \ref{claim:b+P} in Appendix \ref{apx+pf+B2+part1}. The remainder of the proofs proceed identically to those in \Cref{apx+finish+ALG+1+build}, and we omit them for conciseness.

\subsubsection{Proof of \texorpdfstring{\Cref{lemma:ALG+P+reward}}{}}\label{apx+P+reward}\hfill\\
The proof mirrors the same step of \Cref{lemma:ALG+D+reward}. Here, we only give an outline of the steps and highlight a main difference compared to the proof of \Cref{lemma:ALG+D+reward}. Define the pseudo-reward $\{K_\Timeidx\}$ as:
\begin{equation}
    K_t := \Weightvec_\Timeidx\cdot\DecisionVec_\Timeidx + \DualOverVec_\Timeidx\cdot(\CapRatioVec-\DecisionVec_\Timeidx) + \DualHardVec_\Timeidx\cdot(\CapRatioVec-\DecisionVec_\Timeidx) \label{line:def+K+P}
\end{equation}

We also define the stopping time $\Stopping$ as line in \eqref{def:stopping} for \CA{} (see \Cref{subsec:ALG+D+pfsketch}). The following lemma lower bounds the expected pseudo-rewards up to the stopping time.
\begin{claim}[Lower bound on Pseudo-rewards for \CO{}] 
{For any \allocation{} profile $\{\hat{\DecisionVec}_\Timeidx\}_{\Timeidx=1}^\TotalTime$ that satisfies $\hat{\DecisionVec}_\Timeidx \in \FeasibleSet{\Target_\Timeidx}$ for all $\Timeidx \in [\TotalTime]$ and \eqref{line:hard+constraint} (see \Cref{def:offline})}, we have
    \begin{equation*}
    \E\left[\sum_{\Timeidx=1}^{\Stopping} K_t\right] \geq \E[\NetReward(\{\hat{\DecisionVec}_\Timeidx\}_{\Timeidx=1}^\TotalTime; \OverCost)]  - (\TotalTime-\Stopping) .
    \end{equation*}
\end{claim}
The proof of this claim follows the identical steps taken in the proof of \Cref{lemma:pseudo+lower}. We omit the proof for brevity. \par 
The next lemma upper bounds the cumulative pseudo-rewards via the total net matching reward of \CO{}. 
\begin{claim}[Upper bound on Pseudo-rewards for \CO{}]\label{claim:pseudo+upper+P} For every sample path $\{\Arrival_\Timeidx, \ServiceVec_\Timeidx\}_{\Timeidx=1}^\TotalTime$, we have
    \begin{equation*}
   \sum_{\Timeidx=1}^{\Stopping} K_t \leq \NetReward(
   \{\DecisionVec_\Timeidx\}_{\Timeidx=1}^\TotalTime
   ;\OverCost) - (T- \Stopping) + \BigO(\sqrt{\TotalTime}) 
    \end{equation*}
\end{claim}
{\bf Proof of Claim \ref{claim:pseudo+upper+P}.}
The proof follows the similar steps as that of \Cref{lemma:pseudo+upper} except for a minor detail arising from the time-varying learning rates. To avoid repetitions, we only highlight the difference. By applying the similar line of algebras in lines \eqref{line:b0=0}-\eqref{line:cauchy} to the new pseudo-reward \eqref{line:def+K+P}, we can deduce that
\begin{equation}
\begin{split}
\sum_{\Timeidx=1}^{\Stopping} K_t &\leq 
\underbrace{
\sum_{\Timeidx=1}^{\Stopping}
\Weightvec_\Timeidx\cdot\DecisionVec_\Timeidx  - \OverCost \sum_{\Locidx=1}^\Locnum \PositivePart{\sum_{\Timeidx=1}^\TotalTime \DecisionEach - \Capacity_\Locidx}}_{
\NetReward(
   \{\DecisionVec_\Timeidx\}_{\Timeidx=1}^\TotalTime
   ;\OverCost)
} + 
\underbrace{
\sum_{\Timeidx=1}^{\TotalTime}\DualOverVec_\Timeidx\cdot(\CapRatioVec-\DecisionVec_\Timeidx) - \sum_{\Timeidx=1}^{\TotalTime}\DualOverVec^\star\cdot(\CapRatioVec-\DecisionVec_\Timeidx)
}_{R_{\DualOver}} +  \\
&\underbrace{
\sum_{\Timeidx=1}^{\Stopping}\DualHardVec_\Timeidx\cdot(\CapRatioVec-\DecisionVec_\Timeidx)}_{R_{\DualHard}} + 2\OverCost(\TotalTime - \Stopping). \label{line:upper+P}
\end{split}    
\end{equation}
for some $\DualOverVec^\star \in [0,\OverCost]^\Locnum$.
Similar to Claim \ref{claim:theta}, one can invoke the adversarial online learning guarantee of the online mirror descent (Claim \ref{fact:average+conv}) to show that
\begin{equation}
R_{\DualOver} \leq 
2\OverCost \sum_{\Timeidx=1}^{\TotalTime} \eta_\Timeidx+ \frac{1}{\eta_1}V_h(\DualOverVec^\star,\DualOverVec_1) + \sum_{\Timeidx=2}^{\TotalTime} \left(\frac{1}{\eta}_t - \frac{1}{\eta_{t-1}}\right)V_h(\DualOverVec^\star, \DualOverVec_1) \label{eq:temp1} \\
\end{equation}
where we recall that (i) $\eta_\Timeidx = \frac{k}{\sqrt{\Timeidx}}$ with the input constant $k>0$ (ii) $h(\cdot)$ is the negative entropy function, and (iii) $V_h(\cdot, \cdot)$ is the Bregman distance with respect to $h$. Because $\DualOverVec_\Timeidx \in [0,\OverCost]^\Locnum$ for all $\Timeidx \in [\TotalTime]$, we can bound $\max_{\Timeidx\in [\TotalTime]}V_h(\DualOverVec^\star, \DualOverVec_\Timeidx) \leq \bar{V}$ for some constant $\bar{V}$. Utilizing the non-negativity of the Bregman divergence and the fact that $\eta_\Timeidx$ is non-increasing in $\Timeidx$, we can further bound the right-hand side of the inequality \eqref{eq:temp1} as
\begin{equation*}
R_{\DualOver} \leq 2\OverCost\sum_{\Timeidx=1}^{\TotalTime} \eta_\Timeidx + \frac{\bar{V}}{\eta_\TotalTime}.
\end{equation*}
The right hand side of the above inequality is $\BigO(\sqrt{\TotalTime})$ because $\sum_{k=1}^T \frac{1}{\sqrt{\Timeidx}} \leq 2\sqrt{\TotalTime}$ and $1/\eta_\TotalTime = \Theta(\sqrt{\TotalTime})$. Similarly, One can follow the same algebras and the steps taken in Claim \ref{claim:lambda} (\Cref{apx:pseudo+upper}) to show that $R_{\DualHardVec} \leq O(\sqrt{\TotalTime}) + (1+2\OverCost)(\Stopping-\TotalTime)$. Plugging these bound on $R_{\DualOver}$ and $R_{\DualHard}$ into line \eqref{line:upper+P} completes the proof.
\hfill\halmos

Finally, we note that \Cref{lemma:ALG+P+reward} is a direct consequence of the above two claims, by setting $\hat{\DecisionVec}_\Timeidx = \DecisionVec^*_\Timeidx$ where $\DecisionVec^*_\Timeidx$ is the optimal offline solution (\Cref{def:offline}) for arrival $\Timeidx$ for each sample path. This completes the proof of \Cref{lemma:ALG+P+reward}. 
\subsubsection{Proof of 
\texorpdfstring{\Cref{lemma:B}}{}
}\label{apx+B+finish}\hfill\\
Our first step is to establish that the dual-based matching rate is $L$-Lipschitz continuous using Assmumption \ref{assump:PDF}, which we prove in the following claim.
\begin{claim}[Lipschitz Continuity of Dual-based Matching Rate]\label{lemma:M}
Under Assumption \ref{assump:PDF}, we have $\Norm{\MRateVec(\boldsymbol{\nu}) -\MRateVec(\boldsymbol{\nu}')}{\infty} \leq  L\Norm{\boldsymbol{\nu} - \boldsymbol{\nu}'}{1}$ for all $\boldsymbol{\nu}, \boldsymbol{\nu}' \in \mathcal{V}$.
\end{claim}
{\bf Proof of Claim \ref{lemma:M}}.
Let $\boldsymbol{\xi} := \boldsymbol{\nu}' - \boldsymbol{\nu}$. We will show that, for all $\Locidx \in [m]$, 
\begin{equation*}
    |\MRate_\Locidx(\boldsymbol{\nu}) - \MRate_\Locidx(\boldsymbol{\nu'})| \leq L \Norm{\boldsymbol{\nu} - \boldsymbol{\nu}'}{1}.
\end{equation*}

First, observe that 
\begin{equation}
\MRate_\Locidx(\boldsymbol{\nu}) = \P[\Target=0]\times \P\Big[(\Weight_\Locidx - \nu_\Locidx)_{+} \geq (\Weight_j - \nu_j)_{+},\ \forall j \neq i\Big|\Locidx=0\Big] + \P[\Target = \Locidx]. \label{line:rate+nu}
\end{equation}

Let $f(\cdot)$ denote the PDF of the reward distribution conditional that $\Target= 0$. we then have: 
\begin{equation*}
\begin{split}
&\P\Big[(\Weight_\Locidx - \nu_\Locidx)_{+} \geq (\Weight_j - \nu_j)_{+},\ \forall j \neq i\Big|\Locidx=0\Big]\\
&= \mathrlap{\int_{(\Weight_\Locidx - \nu_\Locidx)_{+} \geq (\Weight_j - \nu_j)_{+},\ \forall j \neq i}}\quad f(\Weightvec)d\Weightvec\\
&=  \mathrlap{\int_{(\Weight_\Locidx - \nu'_\Locidx+\xi_\Locidx)_{+} \geq (\Weight_j - \nu'_j + \xi_j)_{+},\ \forall j \neq i}}\quad f(\Weightvec)d\Weightvec\\
&= \mathrlap{\int_{(\Weight'_\Locidx - \nu'_\Locidx)_{+} \geq (\Weight'_j - \nu'_j)_{+},\ \forall j \neq i}}\quad f(\Weightvec' - \boldsymbol{\xi})d\Weightvec'\\
&\leq   \mathrlap{\int_{(\Weight'_\Locidx - \nu'_\Locidx)_{+} \geq (\Weight'_j - \nu'_j)_{+},\ \forall j \neq i}}\quad f(\Weightvec')d\Weightvec' \hspace{2cm}
+\quad L\Norm{\boldsymbol{\xi}}{1} \mathrlap{\int_{(\Weight'_\Locidx - \nu'_\Locidx)_{+} \geq (\Weight'_j - \nu'_j)_{+},\ \forall j \neq i}} \quad  d\Weightvec'
\\
&\leq \mathrlap{\int_{(\Weight'_\Locidx - \nu'_\Locidx)_{+} \geq (\Weight'_j - \nu'_j)_{+},\ \forall j \neq i}}\quad f(\Weightvec')d\Weightvec' \hspace{2cm} +  L\Norm{\boldsymbol{\xi}}{1} \\
&= \P\Big[(\Weight_\Locidx - \nu'_\Locidx)_{+} \geq (\Weight_j - \nu'_j)_{+},\ \forall j \neq i\Big|\Locidx=0\Big] + L\Norm{\boldsymbol{\xi}}{1}
\end{split}
\end{equation*}
The fourth line follows from the variable transformation $\Weightvec' = \Weightvec + \boldsymbol{\xi}$. In the fifth line, we used Assumption \ref{assump:PDF} that the PDF $f$ is $L$-Lipschitz with respect to $\ell_1$ norm. The sixth line follows because the support of the reward vector is $[0,1]^m$ by our assumption. The rest of the lines are mainly algebraic. Plugging the preceding bound into line \eqref{line:rate+nu}, we obtain
\begin{equation*}
\begin{split}
\MRate_\Locidx(\boldsymbol{\nu}) &\leq \P[\Target=0]\times \Big(\P\Big[(\Weight_\Locidx - \nu'_\Locidx)_{+} \geq (\Weight_j - \nu'_j)_{+},\ \forall j \neq i\Big|\Locidx=0\Big] + L\Norm{\boldsymbol{\xi}}{1} \Big)+ \P[\Target = \Locidx].\\
&= \MRate_\Locidx(\boldsymbol{\nu'}) + \P[\Target=0]L\Norm{\boldsymbol{\xi}}{1} \\
&\leq \MRate_\Locidx(\boldsymbol{\nu'}) + L\Norm{\boldsymbol{\xi}}{1} 
\end{split}
\end{equation*}
By switching the role of $\boldsymbol{\nu}$ and $\boldsymbol{\nu}'$, one can also show that $\MRate_\Locidx(\boldsymbol{\nu'}) - \MRate_\Locidx(\boldsymbol{\nu}) \leq  L\Norm{\boldsymbol{\xi}}{1}$. Hence, we have $|\MRate_\Locidx(\boldsymbol{\nu}) - \MRate_\Locidx(\boldsymbol{\nu'})| \leq L \Norm{\boldsymbol{\nu} - \boldsymbol{\nu}'}{1} $ for all $\Locidx \in [\Locnum]$. This completes the proof.
\hfill\halmos 

Building on the preceding claim, we now translate the last iterate convergence in \Cref{prop:harvey} to that of the matching rate. Precisely, we prove the following. 

\begin{claim}\label{claim:last+M}
For any $\boldsymbol{\nu} \in \mathcal{V}$, we have
\begin{equation*}
    \max_{\Locidx \in [m]} \{\MRate_\Locidx(\boldsymbol{\nu}) - \CapRatio_\Locidx \}\leq  \sqrt{2L(D(\boldsymbol{\nu}) - D(\boldsymbol{\nu}^*))}.
\end{equation*}

\end{claim}
{\bf Proof of Claim \ref{claim:last+M}}.
We first observe that Claim \ref{lemma:M} implies $L$-smoothness of $D(\cdot)$. That is, for any $\boldsymbol{\nu}_1, \boldsymbol{\nu}_2 \in \mathcal{V}$, we have
\begin{equation}
\Norm{\nabla D(\boldsymbol{\nu}_1) - \nabla D(\boldsymbol{\nu}_2)}{\infty} \leq L\Norm{\boldsymbol{\nu}_1 - \boldsymbol{\nu}_2}{1}.\label{line:smoothness}
\end{equation}
To see this, we first recall that Assumption \ref{assump:PDF} implies that each $\tilde{\DecisionVec}(\boldsymbol{\nu}, \mathbf{A})$ is unique almost surely. Therefore, by Theorem 7.44 of \citep{shapiro2021lectures}, the function $D(\cdot)$ is differentiable and the partial derivative of $D(\cdot)$ is 
\begin{equation}
\begin{split}
    \frac{\partial D(\boldsymbol{\nu})}{\partial \DualOver_\Locidx} =  \frac{\partial D(\boldsymbol{\nu})}{\partial \DualHard_\Locidx} = \CapRatio_\Locidx - \MRate_\Locidx(\boldsymbol{\nu})
    \label{line:gradient+theta}
\end{split}
\end{equation}
which directly implies line \eqref{line:smoothness}.

From Theorem 5.8-(iii) of \cite{beck2017first}, the $L$-smoothness in line \eqref{line:smoothness} is equivalent to
\begin{equation*}
    D(\boldsymbol{\nu}_1) - D(\boldsymbol{\nu}_2) \geq \nabla D(\boldsymbol{\nu}_2)\cdot(\boldsymbol{\nu}_1 - \boldsymbol{\nu}_2) + \frac{1}{2L}\Norm{\nabla D(\boldsymbol{\nu}_1) - \nabla D(\boldsymbol{\nu}_2)}{\infty}^2
\end{equation*}
for all $\boldsymbol{\nu}_1, \boldsymbol{\nu}_2 \in \mathcal{V}$. In particular, letting $\boldsymbol{\nu}^* \in \argmin_{\boldsymbol{\nu} \in \mathcal{V}} D(\boldsymbol{\nu})$,
\begin{equation*}
D(\boldsymbol{\nu}) - D(\boldsymbol{\nu}^*) \geq \nabla D(\boldsymbol{\nu}^*)\cdot(\boldsymbol{\nu} - \boldsymbol{\nu}^*) + \frac{1}{2L}\Norm{\nabla D(\boldsymbol{\nu}) - \nabla D(\boldsymbol{\nu}^*)}{\infty}^2 \geq  \frac{1}{2L}\Norm{\nabla D(\boldsymbol{\nu}) - \nabla D(\boldsymbol{\nu}^*)}{\infty}^2 
\end{equation*}
for all $\boldsymbol{\nu} \in \mathcal{V}$. Here the last line follows from the first-order condition of $\boldsymbol{\nu}^*$ that minimizes $D(\cdot)$ over domain $\mathcal{V}$ (see, for example, Lemma 5.4 of \cite{bubeck2011introduction}). 
Hence, from equation \eqref{line:gradient+theta}, we deduce that, for all $\Locidx \in [m]$,
\begin{equation*}
\MRate_\Locidx(\boldsymbol{\nu}) \leq \MRate_\Locidx(\boldsymbol{\nu}^*) + \sqrt{2L(D(\boldsymbol{\nu}) - D(\boldsymbol{\nu}^*))}.
\end{equation*}
The proof is complete by noting that $\MRate_\Locidx(\boldsymbol{\nu}^*) \leq \CapRatio_\Locidx$ for all $\Locidx \in [\Locnum]$ from Claim \ref{claim:arrival+rate}-\ref{fluid+dual+rate}. 
\hfill\halmos

We now use the above claim to obtain the high-probability bound of good event $\mathcal{B}_\TotalTime$. For ease of reference, we recall the (equivalent) definition of good event $\mathcal{B}_\TotalTime$ from line \eqref{line:good+event+def}:
\begin{equation*}
   \mathcal{B}_{\TotalTime} := \Big\{ \max_{i \in [\Locnum]} \{\MRate_\Locidx(\boldsymbol{\nu}_\Timeidx) - \rho_i\} \leq \frac{\epsilon}{2} \ \text{for all $\sqrt{\TotalTime} \leq \Timeidx \leq \Stopping $}\Big\}.
\end{equation*}

\begin{claim}\label{claim:good+B+M}
There exists a constant $\underline{K}$, which only depends on (i) the input of the algorithm (other than $\TotalTime$), (ii) $L>0$ in Assumption \ref{assump:PDF}, and (iii) $\ServiceSlack = \Omega(1)$, such that 
for all $\TotalTime \geq \underline{K}$, we have
$\P[\mathcal{B}_{\TotalTime}] \geq 1 - \tfrac{1}{\TotalTime^2}$.
\end{claim}
{\bf Proof of Claim \ref{claim:good+B+M}}.
By \Cref{prop:harvey}, there exists constant $\kappa>0$ such that, for any given $\Timeidx \leq \Stopping$ and $\delta_\Timeidx\in (0,1)$,
\begin{equation*}
\P\left[
D(\boldsymbol{\nu}_\Timeidx) - D(\boldsymbol{\nu}^*) \leq \kappa\frac{\log(\Timeidx)\log(1/\delta_\Timeidx)}{\sqrt{\Timeidx}}\right] \geq 1-\delta_\Timeidx.
\end{equation*}
where the constant $\kappa$ only depends on the input of \Cref{ALG+Surrogate+P} (other than $\TotalTime$).
We now set $\delta_t = \frac{1}{\TotalTime^3}$ for all $\Timeidx \geq \sqrt{\TotalTime}$ and use Claim \ref{claim:last+M} to obtain
\begin{equation*}
\P\left[\max_{\Locidx \in [\Locnum]} \{\MRate_\Locidx(\boldsymbol{\nu}_\Timeidx) - \CapRatio_\Locidx\} \leq \frac{\sqrt{6L \kappa\log(\Timeidx)\log(\TotalTime)}}{\Timeidx^{1/4}}\right] \geq 1- \frac{1}{\TotalTime^3}
\end{equation*}
for all $\Timeidx \geq \sqrt{\TotalTime}$. Note that  $f(t) = \frac{\log(t)}{t^{1/2}}$ is decreasing in $t \geq e^2$. Hence, for any $\Timeidx \geq \sqrt{\TotalTime} \geq e$, we have
\begin{equation*}
\frac{\sqrt{6L\kappa\log(\Timeidx)\log(\TotalTime)}}{\Timeidx^{1/4}} \leq \frac{\sqrt{3L\kappa}\log(\TotalTime)}{\TotalTime^{1/8}} 
\end{equation*}
The right-hand side converges to zero as $\TotalTime \to \infty$, meaning that there exists $f(\kappa, L, \ServiceSlack)$ for which the right-hand side is at most $\epsilon/2$ for all  $T\geq f(\kappa, L, \ServiceSlack)$ (under the stable regime, we have $f(\kappa, L, \ServiceSlack) = \Omega(1)$). Hence, defining $\underline{K}:= \max\{e^4, f(\kappa, L, \ServiceSlack)\}$, we conclude that for all $\TotalTime \geq \underline{K}$ and $\Timeidx \geq \sqrt{\TotalTime}$,
\begin{equation}
\P\left[\max_{\Locidx \in [\Locnum]} \{\MRate_\Locidx(\boldsymbol{\nu}_\Timeidx) - \CapRatio_\Locidx\} \leq \frac{\ServiceSlack}{2}\right] \geq  1 - \frac{1}{\TotalTime^3}. \label{line:high+prob+final}
\end{equation}
 Finally, we use the above bound to lower-bound the good event $\mathcal{B}_\TotalTime$. For any given $\Stopping$, we have:
\begin{equation*}
\begin{split}
\P[\mathcal{B}_{\TotalTime}] &= 1 - \P\left[\max_{\Locidx \in [\Locnum]} \{\MRate_\Locidx(\boldsymbol{\nu}_\Timeidx) - \CapRatio_\Locidx\} > \frac{\ServiceSlack}{2} \quad \text{for some $\Timeidx \in [\sqrt{\TotalTime}, \Stopping]$}\right] \\
&\geq 1 - \sum_{\Timeidx = \sqrt{\TotalTime}}^{\Stopping} \P\left[
\max_{\Locidx \in [\Locnum]} \{\MRate_\Locidx(\boldsymbol{\nu}_\Timeidx) - \CapRatio_\Locidx\} > \frac{\ServiceSlack}{2}
\right] \\
&\geq 1- \frac{1}{\TotalTime^2}.
\end{split}
\end{equation*}
In the second line, we used the union bound. The third line follows from inequality \eqref{line:high+prob+final} and the fact that $\Stopping \leq \TotalTime$ for every sample path.\footnote{
Note that our high-probability results hold for any value of $\Stopping$. 
} This completes the proof.

\hfill\halmos

\subsubsection{Proof of 
\texorpdfstring{\Cref{lemma:B-1}}{}}\label{apx+B-1}\hfill\\
Recall from \Cref{lemma:drift} that
    $\psi(\BuildUpVec_\Timeidx) - \psi(\BuildUpVec_{\Timeidx-1})\leq \BuildUpVec_{\Timeidx-1} \cdot (\DecisionVec_\Timeidx - \ServiceVec_\Timeidx) + \BigO(1)$. Hence, for any history $\History_{\Timeidx-1}$, 
\begin{equation}
\E[\psi(\BuildUpVec_\Timeidx) - \psi(\BuildUpVec_{\Timeidx-1})|\History_{\Timeidx-1}] 
 \leq \BuildUpVec_{\Timeidx-1}\cdot \left(\E[\DecisionVec_\Timeidx|\History_{\Timeidx-1}] - \CapRatioVec - \ServiceSlackVec\right) + \BigO(1) \label{line:drift+cond}
\end{equation}
where we used the fact that (i) $\BuildUpVec_{\Timeidx-1}$ is $\History_{\Timeidx-1}$-measurable and (ii) the $\ServiceVec_\Timeidx$ is i.i.d random variables with mean $\CapRatioVec - \ServiceSlackVec$. Recall that, up to $\Timeidx \leq \Stopping$, the primal decision is $\DecisionVec_\Timeidx = \tilde{\DecisionVec}(\Arrival
_\Timeidx, \boldsymbol{\nu}_\Timeidx)$. Because the dual variables $\boldsymbol{\nu}_\Timeidx$ is $\History_{\Timeidx-1}$-measurable, we further have 
\begin{equation*}
    \E[\DecisionVec_\Timeidx|\History_{\Timeidx-1}] = \MRateVec(\boldsymbol{\nu}_\Timeidx).
\end{equation*}
Hence, one can re-write line \eqref{line:drift+cond} as 
\begin{equation*}
\E[\psi(\BuildUpVec_\Timeidx) - \psi(\BuildUpVec_{\Timeidx-1})|\History_{\Timeidx-1}] 
 \leq \BuildUpVec_{\Timeidx-1}\cdot \left(\MRateVec(\boldsymbol{\nu}_\Timeidx) - \CapRatioVec - \ServiceSlackVec\right) + \BigO(1) 
\end{equation*}
for any history $\History_{\Timeidx-1}$. We now consider two cases. Under the good event $\mathcal{B}_\TotalTime$, we have
\begin{equation*}
  \BuildUpVec_{\Timeidx-1}\cdot \left(\MRateVec(\boldsymbol{\nu}_\Timeidx) - \CapRatioVec - \ServiceSlackVec\right) \leq -\frac{\ServiceSlack}{2}\Norm{\BuildUpVec_{\Timeidx-1}}{1}  
\end{equation*}
by the non-negativity of the \backlog{} and by definition of the good event $\mathcal{B}_\TotalTime$. Otherwise, we have $\BuildUpVec_{\Timeidx-1}\cdot \left(\MRateVec(\boldsymbol{\nu}_\Timeidx) - \CapRatioVec - \ServiceSlackVec\right) \leq  \Norm{\BuildUpVec_{\Timeidx-1}}{1}$ by Cauchy-Schwartz inequality because $\mu_i(\boldsymbol{\nu}_t) \in [0,1]$ and $r_i \in (0,1]$ for all $t \in [T]$ and $i$. Combining these two cases, we have
\begin{equation*}
\E[\psi(\BuildUpVec_\Timeidx) - \psi(\BuildUpVec_{\Timeidx-1})|\History_{\Timeidx-1}] \leq \Indicator[\mathcal{B}_\TotalTime]\left(
-\frac{\ServiceSlack}{2}\Norm{\BuildUpVec_{\Timeidx-1}}{1}
\right) + \Indicator[\mathcal{B}_\TotalTime^c]\Norm{\BuildUpVec_{\Timeidx-1}}{1} + \BigO(1).
\label{psi+drift+apx}
\end{equation*}
Summing the preceding inequality over $\sqrt{\TotalTime}\leq \Timeidx \leq \Stopping$ and taking the outer expectation, we obtain 
\begin{equation}
\begin{split}
    \E\left[
    \sum_{\Timeidx= 
    \sqrt{\TotalTime}}^{\Stopping} \E[\psi(\BuildUpVec_{\Timeidx})  - \psi(\BuildUpVec_{\Timeidx-1})| \History_{\Timeidx-1}] 
    \right] &\leq 
    -  \frac{\ServiceSlack}{2} \E\left[
    \sum_{\Timeidx=\sqrt{\TotalTime}-1}^{\Stopping-1} 
    \lVert \BuildUpVec_\Timeidx\rVert_1 \cdot \Indicator[\mathcal{B}_\TotalTime] \right]+ \\
    \quad&\E\left[
    \sum_{\Timeidx=\sqrt{\TotalTime}-1}^{\Stopping-1} 
    \lVert \BuildUpVec_\Timeidx\rVert_1 \cdot \Indicator[\mathcal{B}_\TotalTime^c]
    \right] + \BigO(\E[\Stopping - \sqrt{\TotalTime}]).\label{line:optional+2}
\end{split}
\end{equation}

Now define $X_t := \psi(\BuildUpVec_{\Timeidx})  - \psi(\BuildUpVec_{\Timeidx-1})$ and $Y_t = X_t - \E[X_t | \History_{\Timeidx-1}]$ with $Y_0 = 0$. It is straightforward to see that $\{Y_t\}_{t=1}^\TotalTime$ is a martingale difference sequence with respect to $\{\History_{\Timeidx}\}_{t=1}^\TotalTime$. Since ${\Stopping}$ is a bounded stopping time with respect to $\{\History_{\Timeidx}\}$, the optional stopping theorem implies 
\begin{equation*}
    \E\left[
    \sum_{t = 
    \sqrt{\TotalTime}}^{\Stopping} \E[\psi(\BuildUpVec_{\Timeidx})  - \psi(\BuildUpVec_{\Timeidx-1})| \History_{\Timeidx-1}] 
    \right] = \E\left[\sum_{\Timeidx = \sqrt{\TotalTime}}^{\Stopping} 
    [\psi(\BuildUpVec_{\Timeidx})  - \psi(\BuildUpVec_{\Timeidx-1})] \right] = \E[\psi(\mathbf{b}_{\Stopping}) - \psi(\mathbf{b}_{{\sqrt{\TotalTime} -1}})]. 
\end{equation*}
Combining this with \eqref{line:optional+2}, we have 
\begin{equation}
    \frac{\ServiceSlack}{2}
    \E\left[
    \sum_{\Timeidx=\sqrt{\TotalTime}-1}^{\Stopping-1} 
    \lVert \BuildUpVec_\Timeidx\rVert_1 \cdot \Indicator[\mathcal{B}_\TotalTime]
    \right] \leq
\E[\psi(\mathbf{b}_{{\sqrt{\TotalTime} -1}}) - \psi(\mathbf{b}_{\Stopping}) ] + 
    \E\left[
    \sum_{\Timeidx=\sqrt{\TotalTime}-1}^{\Stopping-1} 
    \lVert \BuildUpVec_\Timeidx \rVert_1 \cdot \Indicator[\mathcal{B}_\TotalTime^c]
    \right] + \BigO(\E[\Stopping - \sqrt{\TotalTime}]). \label{ineq:drift+sum}
\end{equation}

Each term on the right-hand side is $\BigO(\TotalTime)$. To see this, by taking the worst case growth of the \backlog{}, we have $\BuildUpVec_\Timeidx \leq \BigO(\Timeidx)$ and therefore $\psi(\mathbf{b}_{{\sqrt{\TotalTime} -1}}) \leq \BigO(\TotalTime)$. The second term is again $\BigO(\TotalTime)$ because $\Pr[\mathcal{B}_\TotalTime^c] \leq \BigO(1/\TotalTime)$ from \Cref{lemma:good+event} (along with the worst case growth of the build-up). Finally, the last term in line \eqref{ineq:drift+sum} is $\BigO(\TotalTime)$ since $\Stopping \leq \TotalTime$ for every sample path. The proof is complete by diving both sides of line \eqref{ineq:drift+sum} by $(\ServiceSlack/2)$.
\subsubsection{Proof of \texorpdfstring{Claim \ref{claim:b+P}}{}}\label{apx+pf+B2+part1}\hfill \\
We will use the inequality \eqref{ineq:drift+sum} in Appendix \ref{apx+B-1}. For ease of reference, we restate it in the following:
\begin{equation*}
    \frac{\ServiceSlack}{2}
    \E\left[
    \sum_{\Timeidx=\sqrt{\TotalTime}-1}^{\Stopping-1} 
    \lVert \BuildUpVec_\Timeidx\rVert_1 \cdot \Indicator[\mathcal{B}_\TotalTime]
    \right] \leq
\E[\psi(\mathbf{b}_{{\sqrt{\TotalTime} -1}}) - \psi(\mathbf{b}_{\Stopping}) ] + 
    \E\left[
    \sum_{\Timeidx=\sqrt{\TotalTime}-1}^{\Stopping-1} 
    \lVert \BuildUpVec_\Timeidx \rVert_1 \cdot \Indicator[\mathcal{B}_\TotalTime^c]
    \right] + \BigO(\E[\Stopping - \sqrt{\TotalTime}]).
\end{equation*}

Note that the left-hand side is non-negative. Hence, moving $\E[\psi(\BuildUpVec_{\Stopping})]$ to the left-hand side, we have 
\begin{equation*}
\E[\psi(\BuildUpVec_{\Stopping})]\leq 
\E[\psi(\mathbf{b}_{{\sqrt{\TotalTime} -1}})] + 
\E\left[
\sum_{\Timeidx=\sqrt{\TotalTime}-1}^{\Stopping-1} 
    \lVert \BuildUpVec_\Timeidx \rVert_1 \cdot \Indicator[\mathcal{B}_\TotalTime^c]
    \right] + \BigO(\E[\Stopping - \sqrt{\TotalTime}]).
\end{equation*}

In the proof of \Cref{lemma:B-1}, we have already shown that each term in the right-hand side is $\BigO(\TotalTime)$ (see the paragraph following inequality \eqref{ineq:drift+sum}). The proof is complete by recalling the definition of $\psi(\BuildUpVec_{\Stopping}) = \frac{1}{2}\Norm{\BuildUpVec_{\Stopping}}{2}^2$ and $\Norm{\BuildUpVec_{\Stopping}}{1}^2 \leq m \Norm{\BuildUpVec_{\Stopping}}{2}^2 $ by Cauchy-Schwartz inequality. 

\section{Missing Details and Proof of  
\texorpdfstring{\Cref{thm:near+critical}}{}}\label{apx+near+thm+proof}
{\subsection{Numerical Illustration of \Cref{thm:near+critical}}
\begin{figure}[t]
  \centering
  \begin{subfigure}{0.45\textwidth}
    \includegraphics[width=\linewidth]{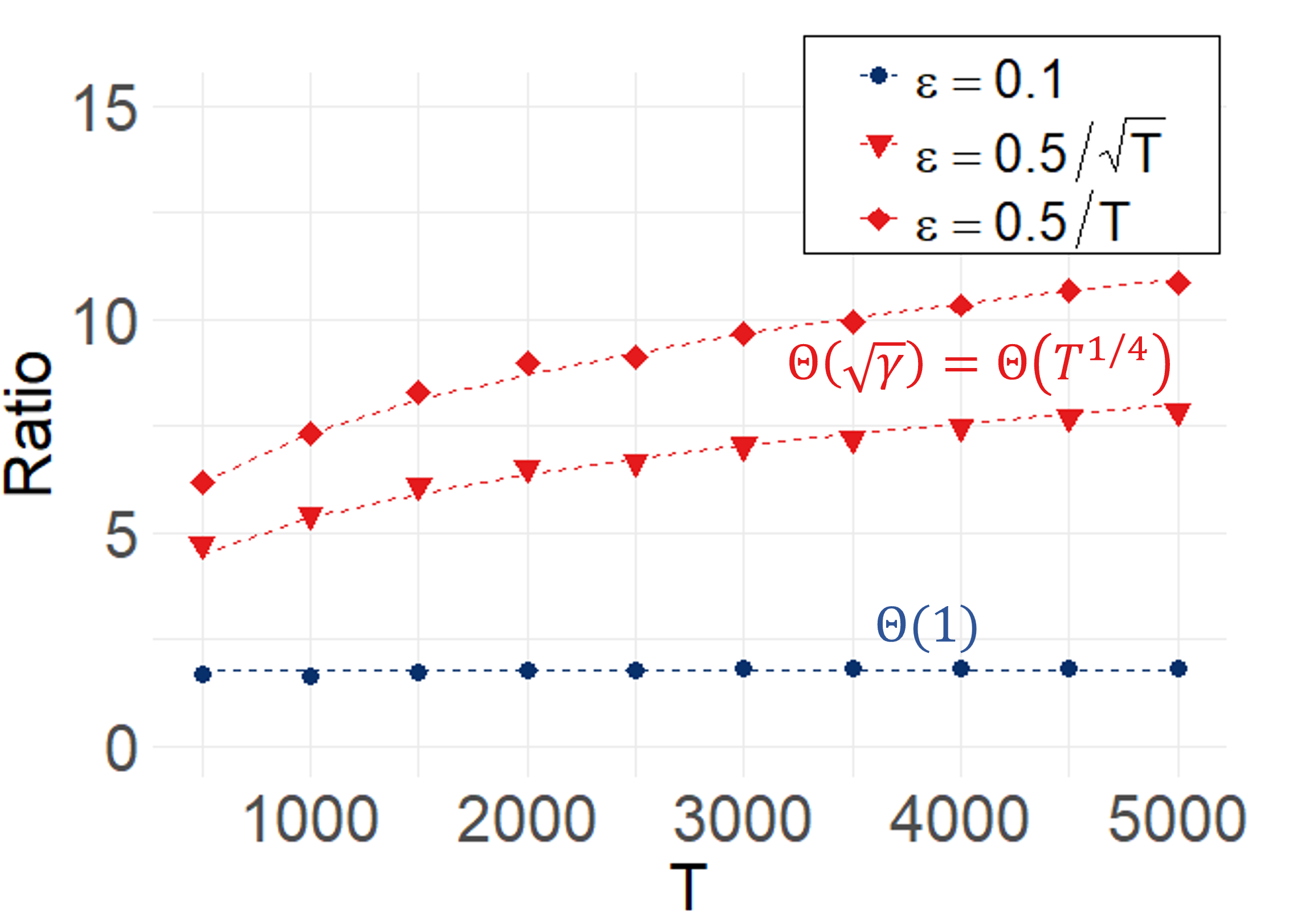}
    \caption{$\textsf{Diff}^{\CO{}}/\textsf{Diff}^{\CA{}}$}
    \label{fig:critical+subfig1}
  \end{subfigure}
  \hfill
  \begin{subfigure}{0.45\textwidth}
    \includegraphics[width=\linewidth]{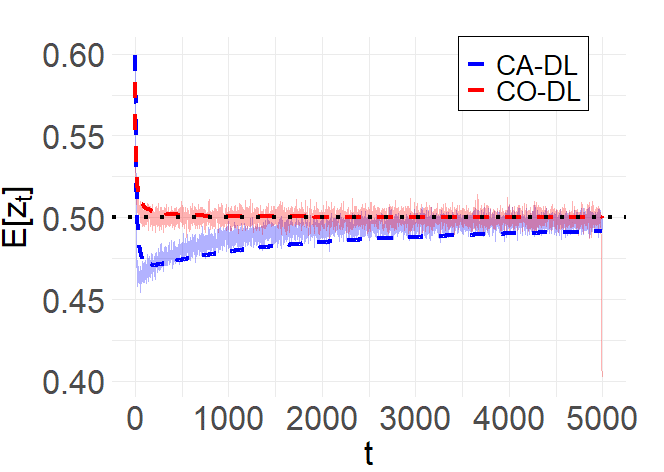}
    \caption{$\E[\Decision_\Timeidx]$ for $\TotalTime=5000$ and $\ServiceSlack=0.5/\sqrt{\TotalTime}$}
    \label{fig:critical+subfig2}
  \end{subfigure}

  \caption{Numerical Illustrations for \Cref{thm:near+critical} with 1000 sample paths and $\boldsymbol{\BuildUpCost = \sqrt{\TotalTime}}$. We use notation $\boldsymbol{\textsf{Diff}^{\pi}}:= \boldsymbol{\E[\OPT(\BuildUpCost)] - \E[\textsf{ALG}^{\pi}(\BuildUpCost)]}$. The learning rate of \CA{} is $\boldsymbol{\eta = \frac{1}{\sqrt{\TotalTime}}}$, with $\boldsymbol{\zeta = \frac{10}{\sqrt{\TotalTime}}}$ ($\boldsymbol{\zeta = \sqrt{\frac{\BuildUpCost}{\TotalTime}}}$, resp.) under the stable regime (the near-critical regime, resp.). The learning rate of \CO{} is $\boldsymbol{\eta_\Timeidx = \frac{1}{\sqrt{\Timeidx}}}$ under both regimes. }
\end{figure}


In this section, we provide numerical illustration for the lower bound of regret in \Cref{thm:near+critical} by simulating our two algorithms. We consider the same one-affiliate instance as in \Cref{example:stable}. For each $\TotalTime \in \{500,1000,...,5000\}$, we use $\BuildUpCost = \sqrt{\TotalTime}$ and $\ServiceSlack \in \{0.1, \frac{0.5}{\sqrt{\TotalTime}}, \frac{0.5}{{\TotalTime}}\}$, where the first value of $\ServiceSlack$ belongs to the stable regime and the other two to the near-critical regime. By using the notation of $\textsf{Diff}^{\pi} := \E[\OPT(\BuildUpCost)] - \E[\textsf{ALG}^{\pi{}}(\BuildUpCost)]$ for an algorithm $\pi$ {(note that $\OverCost$ does not play any role in this instance because there are no tied cases)}, 
we display $\textsf{Diff}^{\CO{}}/\textsf{Diff}^{\CA{}}$ in Figure \ref{fig:critical+subfig1} (as a function of $\TotalTime$).  
Consistent with \Cref{thm:near+critical} and \Cref{near+part+a}, we note that the ratios (two red curves) diverge in the order of $\sqrt{\BuildUpCost} = \TotalTime^{1/4}$ under the near-critical regimes. In contrast, the ratio under the stable regime (blue curve) remains constant, which is also consistent with Theorems~\ref{thm:ALG+Surrogate+Dual} and \ref{thm:ALG+Surrogate+Primal}. However, we observe that \CA{} still achieves a lower regret than \CO{} by a constant factor (roughly 70\%). \par 

The main reason behind the performance difference lies in the endogenous arrival rates that each algorithm's decision induces. More specifically, \Cref{fig:critical+subfig2} illustrates the empirical average of $\Decision_\Timeidx$ (solid line) for $\TotalTime=5000$ and $\ServiceSlack = 0.5/\sqrt{\TotalTime}$, which serves as a proxy for the (expected) endogenous arrival rates. The arrival rate of \CO{} quickly converges to $\CapRatio=0.5$, as evident from the cumulative time-average (dashed red line). In contrast, \CA{} induces an arrival rate strictly less than $\CapRatio=0.5$ (dashed blue line). Under the stable regime, maintaining an induced arrival rate of $\CapRatio$ ensures a constant average backlog. However, in the near-critical regime, we can show that the approach of \CO{} results in a time-average backlog of $\Omega(\TotalTime^{1/2})$ for this specific instance (see \Cref{lemma:all+time+build}). To gain some intuition, consider momentarily that the algorithm ``exactly'' induces the matching rate of $\CapRatio$ at every period. Because $\ServiceSlack=\BigO(1/\sqrt{\TotalTime})$, a simple anti-concentration result implies that, with constant probability, there remains $\Omega(\sqrt{\Timeidx})$ \backlog{} at each period $\Timeidx$. By contrast, by penalizing the current backlog level, \CA{} achieves a better order of the backlog (and thus, a better order of regret).} 

\subsection{Proof of \Cref{thm:near+critical}}\label{apx+near+proof+new}
{\bf Proof Sketch of \Cref{thm:near+critical}}: For the proof, we revisit the instance considered in Example \ref{example:stable} (we again omit the subscript of $\Locidx=1$ for  brevity). That is, we have $m=1$ and ${\CapRatio} = 0.5$. The reward $\Weight_{t}$ is an i.i.d sample from the uniform distribution on interval $(0,1)$ and there is no tied case.

The proof consists of two steps. First, we establish that the value of \Cref{ALG+Surrogate+P} (i.e. \CO{}) on this instance is at most $\frac{3}{8}\TotalTime - \Theta(\BuildUpCost \sqrt{\TotalTime})$. To prove this, we show that (i) the expected reward of \CO{} is at most $\frac{3}{8}\TotalTime$ and (ii) the average \backlog{} of \CO{} is $\Omega(\sqrt{\TotalTime})$ in expectation. In the second step, we show that the expected value of the optimal offline (\Cref{def:offline}) is at least $\frac{3}{8}\TotalTime - \Theta(\sqrt{\BuildUpCost\TotalTime} +\BuildUpCost)$. Combining (and omitting the lower-order terms), we complete the proof of \Cref{thm:near+critical}. In the following, we outline each step of the proof. (Throughout the proof, we use $z_\Timeidx$ and $b_\Timeidx$ to denote the matching decision and \backlog{} of \CO{} at time $\Timeidx$, respectively.) 

\noindent{\bf Step 1: Upper Bound the Value of \CO{}}.\\
The following lemma establishes an upper bound on the expected total reward of \CO{}.
\begin{lemma}\label{lemma:near+alg+reward*}
$\E\Big[\sum_{\Timeidx=1}^\TotalTime w_\Timeidx z_\Timeidx \Big] \leq \frac{3}{8}\TotalTime$
\end{lemma}
{\bf Proof of \Cref{lemma:near+alg+reward*}}. The proof is a direct consequence of Claim \ref{claim:arrival+rate}-\ref{fluid+dual+value} (Appendix \ref{apx+fluid+dual}). The optimal solution for the static dual problem (\Cref{def:staic+dual}) is such that $\theta^* + \lambda^*=0.5$ at which a per-period expected reward is $\E[w_\Timeidx \Indicator[w_\Timeidx \geq 0.5]] = \frac{3}{8}$. Hence, the result follows. \hfill\halmos

Next, we lower-bound the average \backlog{} of \CO{} by $\Omega(\sqrt{\TotalTime})$ in expectation. The following lemma establishes a high probability bound on the total number of matched cases, which we prove in Appendix \ref{apx+all+time+reward+new}. 
\begin{lemma}\label{lemma:Z+lower}
There exists a constant $\kappa >0$ such that, for any fixed $\Timeidx \leq 0.5\TotalTime$ and $\delta \in (0,1)$, 
\begin{equation*}
    \P\left[
     \sum_{\Tauidx=1}^\Timeidx \Decision_{\Tauidx} \geq  0.5\Timeidx - \kappa\sqrt{t\log(1/\delta)}
    \right] \geq 1-\delta - \BigO(1/\Timeidx).
\end{equation*}
\end{lemma}

We now use \Cref{lemma:Z+lower} to obtain a lower bound on the expected \backlog{}. 
\begin{lemma}\label{lemma:all+time+build}
    There exists constants $a>0$ and $t(a)$, such that, whenever $\TotalTime \geq 2t(a)$ and $\ServiceSlack \leq \frac{a}{4\sqrt{\TotalTime}}$, we have:
    \begin{equation*}
    \E[\BuildUp_\Timeidx] \geq \frac{a\sqrt{\Timeidx}\Phi(-3\sqrt{2}a)}{16}
    \end{equation*}
for all $t(a) \leq t \leq \frac{\TotalTime}{2}$. Here $\Phi(\cdot)$ is the cumulative distribution function of the standard normal random variable. The exact form of the constant $t(a)$, as a function of the constant $a>0$, is given in the proof.
\end{lemma}

The proof of this lemma is presented in Appendix \ref{apx+all+time+build}. For the proof, we use the following simple anti-concentration result. Let $S_\Timeidx$ be the total number of periods that the server is available up to time $\Timeidx$. Under the near-critical regime, there is a constant probability that $S_\Timeidx \leq 0.5\Timeidx - \Theta(\sqrt{\Timeidx})$. Combining this with the lower bound on the total number of matched cases established in \Cref{lemma:Z+lower}, we can show that at least $\Theta(\sqrt{\Timeidx})$ cases remain in the backlog in expectation for each period $t(a)\leq \Timeidx \leq \frac{\TotalTime}{2}$. Finally, we note that summing the inequality \Cref{lemma:all+time+build} for all $t(a) \leq \Timeidx \leq \frac{\TotalTime}{2}$ implies the $\Omega(\sqrt{\TotalTime})$ average \backlog{} of \Cref{ALG+Surrogate+P} in expectation. 
\smallskip

\noindent{\bf Step 2: Lower-bounding the Value of the Optimal Offline.}
The following lemma lower-bound the value of the offline. 
\begin{lemma}\label{lemma:near+opt+value}
    $\E[\textsf{OPT}(\BuildUpCost)] \geq \frac{3}{8}\TotalTime - \Theta(\sqrt{\BuildUpCost\TotalTime} + \BuildUpCost)$
\end{lemma}
The proof of this lemma can be found in Appendix \ref{apx+near+opt}. For the proof, we study the expected objective value of the following feasible solution for each sample path: let $\delta := \sqrt{\BuildUpCost/\TotalTime}$ and define $\Decision^*_\Timeidx = \Indicator[\Weight_\Timeidx \geq 0.5 + \delta]$ up to $\Timeidx \leq \Stopping$ and $\Decision^*_\Timeidx =0$ for all $\Timeidx > \Stopping$,\footnote{Note that this is $\delta = o(1)$ because we focus on $\BuildUpCost = o(\TotalTime)$ in light of \Cref{prop:impossibility}} where the stopping time $\Stopping$ is defined as $\Stopping = \min \{\Timeidx \leq \TotalTime: \sum_{\Tauidx=1}^\Timeidx \Decision^*_\Timeidx \geq 0.5\TotalTime\}$. We then show that (i) the total expected reward of this feasible solution is at least $\frac{3}{8}\TotalTime - \Theta(\sqrt{\BuildUpCost\TotalTime} + \BuildUpCost)$ and (ii) the expected time-average \backlog{} is $O(\sqrt{\TotalTime/\BuildUpCost} + 1)$.

\subsubsection{Proof of 
\texorpdfstring{\Cref{lemma:Z+lower}}{}}\label{apx+all+time+reward+new} \hfill  

For the proof, we first establish the following high probability bound. 
\begin{proposition}\label{prop:total+arrival}
Suppose the arrival distribution is such that there are no tied cases. Then there exists a constant $M>0$ such that, for any fixed $\delta \in (0,1)$ and $\Timeidx \leq \Stopping$, the matching profile $\{\DecisionVec_{\Tauidx}\}_{\Tauidx=1}^\Timeidx$ of \CO{} satisfies that, with probability at least $1-\delta$,
\begin{equation}
0 
\leq 
\sum_{\Locidx=1}^\Locnum \nu_{\Timeidx, \Locidx}^*\left(
 \sum_{\Tauidx=1}^\Timeidx \Decision_{\Tauidx, \Locidx} - \Timeidx\CapRatio_\Locidx
\right)  + M\sqrt{t\log(1/\delta)}
\end{equation}
where $ \boldsymbol{\nu}^*_\Timeidx = (\nu_{\Timeidx,\Locidx})_{\Locidx=1}^\Locnum$ is a random variable defined by
\begin{equation}
    \boldsymbol{\nu}^*_\Timeidx \in \argmin_{\boldsymbol{\nu} \in \R_{+}^\Locnum} \sum_{\Timeidx=1}^\TotalTime \{ 
    \max_{\DecisionVec \in \FeasibleSet{\Target}} (\Weightvec_\Timeidx - \boldsymbol{\nu})\cdot \DecisionVec + \CapRatioVec\cdot\boldsymbol{\nu}
    \} \label{def:sample+dual}.
\end{equation}
\end{proposition}
The proof is fairly intricate, and we defer it to the end of this section. But first, we complete the proof of \Cref{lemma:Z+lower} using \Cref{prop:total+arrival}. Recall that the instance considered in \Cref{example:stable} consists of one actual affiliate.  Hence, we apply \Cref{prop:total+arrival} to this instance and deduce that, for any $\Timeidx \leq 0.5\TotalTime$,
\begin{equation}
    0 \leq \nu_{\Timeidx}^*\left(
 \sum_{\Tauidx=1}^\Timeidx \Decision_{\Tauidx} - 0.5\Timeidx
\right)  + M\sqrt{t\log(1/\delta)} \label{line:deduce}
\end{equation}
with probability at least $1-\delta$ and some constant $M>0$. Furthermore, because the reward follows the uniform distribution on $(0,1)$, the random variable $\nu_{\Timeidx}^*$ is the sample median of $t$ i.i.d. samples from the uniform distribution on $(0,1)$ (see Footnote \ref{footnote:sample+median}). From \cite{casella2021statistical} Example 5.4.5., the expectation and variance of $\nu_{\Timeidx}^*$ is given by $\E[\nu_{\Timeidx}^*] = 1/2$ and $\mathbbm{V}[\nu_{\Timeidx}^*] = \Theta(1/\Timeidx)$, respectively. Hence, a straightforward application of ChebyShev's inequality implies that $\P[\nu_{\Timeidx}^* \geq \frac{1}{4}] \geq 1-\BigO(1/\Timeidx)$. Combining this with line \eqref{line:deduce} with the union bound, we conclude that, with probability at least $1-\delta - \BigO(1/\Timeidx)$, we have 
\begin{equation}
    0 \leq \left(
 \sum_{\Tauidx=1}^\Timeidx \Decision_{\Tauidx} - 0.5\Timeidx
\right)  + 4M\sqrt{t\log(1/\delta)},
\end{equation}
which leads to the desired result of \Cref{lemma:Z+lower} by taking $\kappa = 4M$.

Finally, we conclude this section with the proof of \Cref{prop:total+arrival}.

{\bf Proof of \Cref{prop:total+arrival}}.
First, we provide some preliminaries that will be useful throughout the proof. 
With a slight abuse of notations, we define the following static problem. 
\begin{equation}
D(\boldsymbol{\nu}):= \E\left[
\max_{\DecisionVec \in \FeasibleSet{\Target}}(\Weightvec - \boldsymbol{\nu})\cdot \DecisionVec + \CapRatioVec\cdot \boldsymbol{\nu}
    \right]  \nonumber
\end{equation}
\begin{equation}
D(\boldsymbol{\nu}^*) = 
\min_{\boldsymbol{\nu}\in \R_{+}^m } D(\boldsymbol{\nu}). \label{fluid:dual+def}
\end{equation}
Given $\boldsymbol{\nu} \in \R_{+}^\Locnum$, we similarly define the dual-based primal decision with respect to arrival type $\Arrival$ as follows.
\begin{equation*}
\begin{split}
\tilde{\DecisionVec}(\boldsymbol{\nu}, \mathbf{A})&:= \argmax_{\DecisionVec \in \FeasibleSet{\Target}} (\Weightvec - \boldsymbol{\nu})\cdot \DecisionVec  \\ 
\MRateVec(\boldsymbol{\nu}) &:= \E[\tilde{\DecisionVec}(\boldsymbol{\nu}, \Arrival)].
\end{split}
\end{equation*}    
By following a similar step to the proof of claim \ref{claim:arrival+rate} (Appendix \ref{apx+fluid+dual}), we can show the following properties of the above static problem. The proof mirrors that of Claim \ref{claim:arrival+rate} and we omit it for brevity.

\begin{claim}\label{claim:arrival+rate+2}
For $D(\cdot)$ and $\boldsymbol{\nu^*}$ defined in line \eqref{fluid:dual+def}, the following property holds:
\begin{enumerate}[label=(\alph*)]
\item $\lVert \boldsymbol{\nu}^*\rVert_{\infty} \leq 1$.\label{fluid+dual+bound+2}
\item $\MRate_\Locidx(\boldsymbol{\nu}^*)  \leq \rho_i$ for all $i\in[m]$ with equality if $\nu_i^* >0$. \label{fluid+dual+rate+2}
\item $D(\boldsymbol{\nu}^*) = \E[\Weightvec\cdot\tilde{\DecisionVec}(\boldsymbol{\nu}^*, \Arrival)]$ \label{fluid+dual+value+2}
\end{enumerate}
\end{claim}
We will also use the following result on the sample average approximation of the stochastic optimization problem. 
\begin{lemma}[Proposition 1 of \cite{guigues2017non}]\label{lemma:SAA}
Let $\mathcal{X}$ be a nonempty bounded convex set. Consider the following stochastic optimization problem:
\begin{align*}
    f^* = \min_{\mathbf{x}\in \mathcal{X}} \E_{\mathbf{A} \sim \mathcal{F}}[f(\mathbf{x}, \mathbf{A})]
\end{align*}
where $\mathbf{A}$ is a random vector from probability distribution $\mathcal{F}$ on support $\ArrivalSupp$. Given i.i.d samples $\mathbf{A}_1,...,\mathbf{A}_\Timeidx \sim \ArrivalDist$, consider the following sample-average approximation of $f^*$:
\begin{align*}
    \hat{f}_\Timeidx := \min_{\mathbf{x} \in \mathcal{X}} \frac{1}{\Timeidx}\sum_{\Tauidx=1}^\Timeidx f(\mathbf{x}, \mathbf{A}_\Tauidx). 
\end{align*}
Assume that there exists positive constant $M_1$ such that $|f(\mathbf{x}, \mathbf{A}) - \E_{\mathbf{A} \sim \ArrivalDist}[f(\mathbf{x}, \mathbf{A})]|\leq M_1$ for all $\mathbf{x} \in \mathcal{X}$ and $\mathbf{A} \in \ArrivalSupp$. Then, for any given $\delta \in (0,1)$,
\begin{align*}
    \P\left[\hat{f}_\Timeidx \leq f^* + 2M_1 \sqrt{\frac{\log(1/\delta)}{\Timeidx}}\right] \geq 1-\delta.
\end{align*}
\end{lemma}

\smallskip
\noindent{\bf Main Proof.}
Fix an arbitrary time $\Timeidx \leq \Stopping$ and a sample path of arrivals $(\Arrival_1, ..., \Arrival
_\Timeidx)$.\footnote{Recall that we consider an arrival distribution with no tied cases. Further, the service sequence never affects the decision of \CO{} (as it is oblivious to the service sequence).} 
Recall that we use $\DecisionVec_\Tauidx$ to denote the decision of \CO{} for arrival $\Arrival_\Tauidx$ for $\Tauidx \in [\TotalTime]$. 
Because \CO{} is unconstrained until $\Timeidx \leq \Stopping$, for any arbitrary time $\Tauidx \leq \Timeidx$, we can write it as:
\begin{equation}
\DecisionVec_\Tauidx = \argmax_{\DecisionVec \in \FeasibleSet{\Target_{\Tauidx}}} (\Weightvec_\Tauidx - \DualOverVec_\Tauidx - \DualHardVec_\Tauidx)\cdot \DecisionVec. \label{line:mod+opt}
\end{equation}
We will compare  $\DecisionVec_\Tauidx$ against an alternative decision vector $\DecisionVec^*_\Tauidx$ defined as follows:\begin{equation}
    \DecisionVec^*_\Tauidx := \argmax_{\DecisionVec \in \FeasibleSet{\Target_{\Tauidx}}} (\Weightvec_\Tauidx - \boldsymbol{\nu}^*)\cdot \DecisionVec = \DecisionVec(\boldsymbol{\nu}^*, \Arrival_\Tauidx)\label{line:z+tau+star}
\end{equation}
where $\boldsymbol{\nu}^*$ is the minimizer of the static problem \eqref{fluid:dual+def}.
Note that $\DecisionVec^*_\Tauidx$ is i.i.d. random variables because $\boldsymbol{\nu}^*$ only depends on the arrival distribution. By the optimality criterion of the modified algorithm \eqref{line:mod+opt}, we have
\begin{equation}
\sum_{\Tauidx=1}^\Timeidx (\Weightvec_\Tauidx - \DualOverVec_\Tauidx - \DualHardVec_\Tauidx)\cdot\DecisionVec_\Tauidx + \CapRatioVec\cdot(\DualOverVec_\Tauidx + \DualHardVec_\Tauidx) \geq 
\sum_{\Tauidx=1}^\Timeidx (\Weightvec_\Tauidx - \DualOverVec_\Tauidx - \DualHardVec_\Tauidx)\cdot\DecisionVec_\Tauidx ^* + \CapRatioVec\cdot(\DualOverVec_\Tauidx + \DualHardVec_\Tauidx). \label{ineq:sum+to+t}
\end{equation}

From Claim \ref{fact:over} (Appendix \ref{apx+thm+surrogate+D}), there exists $\DualOverVec^\star \in [0,\OverCost]^\Locnum$ for which
\begin{equation*}
\sum_{\Tauidx=1}^\Timeidx \DualOverVec^\star \cdot (\DecisionVec_\Tauidx- \CapRatioVec) = \OverCost \sum_{\Locidx=1}^\Locnum \left(\sum_{\Tauidx=1}^\Timeidx \Decision_{\Tauidx, \Locidx} - \Timeidx\CapRatio_\Locidx\right)_{+}.
\end{equation*}

Hence, a straightforward manipulation of line \eqref{ineq:sum+to+t} leads to
\begin{align}
\begin{split}
\OverCost \sum_{\Locidx=1}^\Locnum \left(\sum_{\Tauidx=1}^\Timeidx \Decision_{\Tauidx, \Locidx} - \Timeidx\CapRatio_\Locidx\right)_{+}   &\leq 
\underbrace{
\sum_{\Tauidx=1}^\Timeidx \Weightvec_\Tauidx\cdot\DecisionVec_\Tauidx - \sum_{\Tauidx=1}^\Timeidx \Weightvec_\Tauidx\cdot \DecisionVec_\Tauidx^*
}_{\mathcal{\textsf{(A)}}} + 
\underbrace{
\sum_{\Tauidx=1}^\Timeidx \DualOverVec_\Tauidx\cdot(\DecisionVec_\Tauidx^*- \CapRatioVec) + 
\sum_{\Tauidx=1}^\Timeidx \DualHardVec_\Tauidx\cdot(\DecisionVec_\Tauidx^*- \CapRatioVec)
}_{\mathcal{\textsf{(B)}}} + \\
\quad & 
\underbrace{
\sum_{\Tauidx=1}^\Timeidx \DualOverVec_\Tauidx\cdot (\CapRatioVec-\DecisionVec_\Tauidx) - \sum_{\Tauidx=1}^\Timeidx \DualOverVec^\star\cdot (\CapRatioVec-\DecisionVec_\Tauidx)
 + 
\sum_{\Tauidx=1}\DualHardVec_\Tauidx\cdot (\CapRatioVec- \DecisionVec_\Tauidx)
}_{\mathcal{\textsf{(C)}}}
\end{split}
\label{line:lemma+over}
\end{align}

We prove the main result through four claims. In the first three claims (Claim \ref{claim:A}-\ref{claim:bound+C}), we obtain the high-probability bound of each term $\textsf{(A)}-\textsf{(C)}$, respectively. 

\begin{claim}[Bound $\textsf{(A)}$]\label{claim:A}
There exists constant $M_{\textsf{A}}$ such that, for any given $\delta \in (0,1)$,
\begin{equation*}
\textsf{(A)}\leq 
\sum_{\Locidx=1}^\Locnum \nu^*_{\Timeidx, i}\left(
 \sum_{\Tauidx=1}^\Timeidx \Decision_{\Tauidx, \Locidx} - \Timeidx\CapRatio_\Locidx 
\right)
+ M_{\textsf{A}}\sqrt{t\log(1/\delta)}
\end{equation*}
with probability at least $1-\delta$ where $\boldsymbol{\nu}_\Timeidx^*$ is defined in line \eqref{def:sample+dual}.
\end{claim}
{\bf Proof of Claim \ref{claim:A}}.
Consider the following linear program defined on the arrival sequence $(\Arrival_1, ...,\Arrival_\Timeidx)$:
\begin{equation}
\overline{\textsf{OPT}}_\Timeidx(t\CapRatioVec):=  
\max_{
     \DecisionVec_\Tauidx \in \FeasibleSet{\Target_\Tauidx}
     }   
 \sum_{\Tauidx=1}^\Timeidx 
     \Weightvec_\Tauidx\cdot \DecisionVec_\Tauidx  
\quad {\text{s.t.}} \quad  
\sum_{\Tauidx=1}^{\Timeidx}\DecisionVec_\Tauidx\leq t\CapRatioVec  \quad \quad [\boldsymbol{\nu}_\Timeidx^*] 
\label{line:const}
\end{equation}
We use subscript $\Timeidx$ in the optimal objective value to emphasize that the optimization problem is defined on the arrival up to time $\Timeidx$. In the similar spirit, the optimal value of the above offline depends on the right-hand side of the capacity constraint, which we also emphasized as the argument of $\overline{\textsf{OPT}}_\Timeidx(\cdot)$. The dual variable corresponding to the constraint \eqref{line:const} (at the right hand side value $t\CapRatioVec$) is denoted by $\boldsymbol{\nu}_\Timeidx^* \geq 0$. 
The above program has a feasible solution for every sample path because we consider the arrival distribution without tied cases.

Note that, by definition, $\{\DecisionVec_\Tauidx\}_{\Tauidx=1}^\Timeidx$ is a feasible solution for $\overline{\textsf{OPT}}_t\Big(\sum_{\Tauidx=1}^\Timeidx \DecisionVec_\Tauidx\Big)$. 
Hence, we have 
\begin{equation}
\begin{split}
\sum_{\Tauidx=1}^\Timeidx 
     \Weightvec_\Tauidx\cdot \DecisionVec_\Tauidx &\leq 
\overline{\textsf{OPT}}_\Timeidx\left(
\sum_{\Tauidx=1}^\Timeidx \DecisionVec_{\Tauidx}
\right)  
     \\     
&= \overline{\textsf{OPT}}_\Timeidx\left(
\Timeidx \CapRatioVec + 
\sum_{\Tauidx=1}^\Timeidx \DecisionVec_{\Tauidx} - \Timeidx\CapRatioVec
\right)  \\
&
\leq 
\overline{\textsf{OPT}}_\Timeidx(\Timeidx\CapRatioVec) + \sum_{\Locidx=1}^\Locnum \nu_{\Timeidx, \Locidx}^*\left(
 \sum_{\Tauidx=1}^\Timeidx \Decision_{\Tauidx, \Locidx} - \Timeidx\CapRatio_\Locidx 
\right)
\label{line:lower}
\end{split}
\end{equation}
where in the last line we used the fact that (i) $\overline{\textsf{OPT}}_\Timeidx(\cdot)$ is concave and increasing in each coordinate of its argument (see, for example, Section 5.6.2 of \cite{boyd2004convex}) and (ii) 
$\boldsymbol{\nu}_\Timeidx^*$ is a gradient of $\overline{\textsf{OPT}}_\Timeidx(\cdot)$ evaluated at the right-hand side value of $t\CapRatioVec$ in line \eqref{line:const}.\par 
We now turn our attention to lower-bounding $\sum_{\Tauidx=1}^{\Timeidx}\Weightvec_\Tauidx\cdot \DecisionVec_\Tauidx^*$. Here, we will crucially rely on the properties of the static problem in Claim \ref{claim:arrival+rate+2}. By definition of $\DecisionVec_\Tauidx^*$ and i.i.d nature of the arrival sequence, each $\Weightvec_\Tauidx\cdot \DecisionVec_\Tauidx^*$ is i.i.d random variables with $|\Weightvec_\Tauidx\cdot \DecisionVec_\Tauidx^*|\leq 1$. Furthermore, from Claim \ref{claim:arrival+rate+2}-\ref{fluid+dual+value+2}, the identical mean is $\E[\Weightvec_\Tauidx\cdot \DecisionVec_\Tauidx^*]= D(\boldsymbol{\nu}^*)$ for all $\Tauidx\in[\Timeidx]$. Hence, the Azuma-Hoeffding inequality implies that, with probability at least $1-\delta$, 
\begin{equation*}
\sum_{\Tauidx=1}^{\Timeidx}\Weightvec_\Tauidx\cdot \DecisionVec_\Tauidx^*  \geq \Timeidx D(\boldsymbol{\nu}^*) - \sqrt{\frac{\Timeidx}{2}\log(1/\delta)}.
\end{equation*}
Combining the above bound with line \eqref{line:lower}, we have
\begin{equation}
\textsf{(A)}\leq 
\underbrace{
\overline{\textsf{OPT}}_\Timeidx(\Timeidx\CapRatioVec) - \Timeidx 
D(\boldsymbol{\nu}^*)}_{\clubsuit} +
\sum_{\Locidx=1}^\Locnum \nu_{\Timeidx, \Locidx}^*\left(
 \sum_{\Tauidx=1}^\Timeidx \Decision_{\Tauidx, \Locidx} - \Timeidx\CapRatio_\Locidx 
\right)
+ \sqrt{\frac{\Timeidx}{2}\log(1/\delta)}.\label{line:clubsuit}
\end{equation}

To complete the proof of the claim, we obtain the high-probability bound on $\clubsuit$ by invoking \Cref{lemma:SAA}. First, let us first define
\begin{equation*}
    D(\boldsymbol{\nu};\Arrival) := \max_{\DecisionVec \in \FeasibleSet{\Target}}(\Weightvec - \boldsymbol{\nu})\cdot \DecisionVec + \CapRatioVec\cdot \boldsymbol{\nu}.
\end{equation*}
Given this notation, we first note that $D(\boldsymbol{\nu}^*)$ defined in line \eqref{fluid:dual+def} can be written as 
\begin{equation*}
    D(\boldsymbol{\nu}^*) = \min_{
    \boldsymbol{\nu} \in [0,1]^m
    }
    \E[D(\boldsymbol{\nu};\Arrival)]
\end{equation*}
where we used Claim \ref{claim:arrival+rate+2}-\ref{fluid+dual+bound+2} to restrict the domain to $[0,1]^m$ without loss of optimality. On the other hand, we have: 

\begin{equation}
\frac{\overline{\textsf{OPT}}_\Timeidx(\Timeidx\CapRatioVec)}{\Timeidx} =
\min_{
\boldsymbol{\nu}_\Timeidx \in \R_{+}^m
}
    \frac{1}{\Timeidx}\sum_{\Tauidx=1}^\Timeidx D(\boldsymbol{\nu}_\Timeidx;\Arrival_\Tauidx)
= \min_{
\boldsymbol{\nu}_\Timeidx \in [0,1]^m
}
    \frac{1}{\Timeidx}\sum_{\Tauidx=1}^\Timeidx D(\boldsymbol{\nu}_\Timeidx;\Arrival_\Tauidx).\label{line:SAA}
\end{equation}

The first equality follows from writing the dual of $\overline{\textsf{OPT}}_\Timeidx(\Timeidx\CapRatioVec)$. For the second equality, we again used the fact that the optimal value $\boldsymbol{\nu}_\Timeidx^*$ is always in $[0,1]^\Locnum$ without loss of optimality.
\footnote{Consider the minimization problem $\min_{
\boldsymbol{\nu}_\Timeidx \in \R_{+}^m
}\sum_{\Tauidx=1}^\Timeidx D(\boldsymbol{\nu}_\Timeidx;\Arrival_\Tauidx)$. Let $\boldsymbol{\nu_\Timeidx}^*$ be the optimal solution of this problem. Suppose for a contradiction that there exists $i$ such that $\nu_{\Timeidx,i}^* > 1$. Such location $i$ is not assigned any free case since the reward is at most one. Hence, one can always consider an alternative solution that reduces $\nu_{\Timeidx,i}^*$ infinitesimally and strictly decreases the objective value. }

Hence, because the arrivals are i.i.d samples, we observe that $\frac{\overline{\textsf{OPT}}_\Timeidx(\Timeidx\CapRatioVec)}{\Timeidx}$ is the sample average approximation for $D(\boldsymbol{\nu}^*)$ with $\Timeidx$ i.i.d samples $(\Arrival_1, ..., \Arrival_\Timeidx)$. Finally, it is straightforward to show that $|D(\boldsymbol{\nu};\Arrival)| \leq 3$ for all $\boldsymbol{\nu}\in [0,1]^\Locnum$, and $\Arrival\in \ArrivalSupp$.
Hence, $|D(\boldsymbol{\nu};\Arrival) - \E[D(\boldsymbol{\nu};\Arrival)]| \leq 6$ for all $\boldsymbol{\nu} \in [0,1]^m$ and $\Arrival\in\ArrivalSupp$. Invoking \Cref{lemma:SAA} with $M_1 = 6$, we obtain
\begin{equation*}
\P\left[\frac{\overline{\textsf{OPT}}_\Timeidx(\Timeidx\CapRatioVec)}{\Timeidx} \leq D(\boldsymbol{\nu}^*) + 12\sqrt{\frac{\log(1/\delta)}{\Timeidx}}\right] \geq 1-\delta.
\end{equation*}
This equivalently implies that, with probability at least $1-\delta$, the term $\clubsuit$ in line \eqref{line:clubsuit} is at most $12\sqrt{t\log(1/\delta)}$. Plugging this bound in line \eqref{line:clubsuit} and taking the union bound, with probability at least $1-2\delta$, we have
\begin{equation*}
    \textsf{(A)} \leq 
    \sum_{\Locidx=1}^\Locnum \nu_{\Timeidx, \Locidx}^*\left(
 \sum_{\Tauidx=1}^\Timeidx \Decision_{\Tauidx, \Locidx} - \Timeidx\CapRatio_\Locidx 
\right)
    + M_{\textsf{A}}\sqrt{t\log(1/\delta)}
\end{equation*}
where $M_{\textsf{A}}= 12 + \sqrt{\frac{1}{2}}$. The proof is complete by replacing $\delta$ with $\delta/2$.
\hfill\halmos

The following lemma shows the high-probability bound of term \textsf{(B)} in line \eqref{line:lemma+over}.
\begin{claim}[Bound \textsf{(B)}]\label{claim:bound+B}
There exists constant $M_{\text{B}}$ such that, for any given $\delta \in (0,1)$, we have
\begin{equation*}
    \P[\textsf{(B)}\leq M_B\sqrt{t\log(1/\delta)}] \geq 1-\delta.
\end{equation*}
\end{claim}
{\bf Proof of Claim \ref{claim:bound+B}.}
Let $X_{s} := \DualOverVec_s\cdot(\DecisionVec_s^*- \CapRatioVec) +  \DualHardVec_s\cdot(\DecisionVec_s^*- \CapRatioVec)$ and $Y_\Tauidx := \sum_{s=1}^{\Tauidx} X_s$, with $X_0 := 0$ and $Y_0 := 0$. Note that $Y_{\Tauidx}$ is $\History_\Tauidx$-measurable. We first note that $Y_{\Tauidx}$ is supermartingale with respect to $\History_\Tauidx$ because:
\begin{equation*}
\begin{split}
\E[X_{\tau}|\History_{\Tauidx-1}] &= 
\DualOverVec_\Tauidx\cdot(\E[\DecisionVec_\Tauidx^*|\History_{\Tauidx-1}]- \CapRatioVec) + 
\DualHardVec_\Tauidx\cdot(\E[\DecisionVec_\Tauidx^*|\History_{\Tauidx-1}]- \CapRatioVec) \\
&= \DualOverVec_\Tauidx\cdot(\E[\DecisionVec_\Tauidx^*]- \CapRatioVec) + 
\DualHardVec_\Tauidx\cdot(\E[\DecisionVec_\Tauidx^*]- \CapRatioVec) \\
&\leq 0.
\end{split}
\end{equation*}
The first line is because $\DualOverVec_\Tauidx$ and $\DualHardVec_\Tauidx$ is $\History_{\Tauidx-1}$-measurable. The second line is because each $\DecisionVec_\Tauidx^*$ is i.i.d random variable. The final line is because the identical mean is $\E[\DecisionVec_\Tauidx^*] \leq \CapRatioVec$ from Claim \ref{claim:arrival+rate+2}-\ref{fluid+dual+rate+2}. Furthermore, the Cauchy-Schwartz inequality gives  $|X_{\Tauidx}| \leq 2(\alpha + \bar{\lambda})$ almost surely where $\bar{\lambda} := \frac{1+2\OverCost}{\underline{\rho}}$ because the dual variables $(\boldsymbol{\theta}, \boldsymbol{\lambda})$ are restricted to $\mathcal{V} = [0,\alpha]^m \times [0, \bar{\lambda}]^m$. Hence, we use Azuma-Hoeffding inequality for super martingale to obtain
\begin{equation*}
    \P\Big[Y_\Timeidx \leq \sqrt{2t(\alpha +\bar{\lambda})\log(1/\delta)}\Big] \geq 1-\delta
\end{equation*}
for any given $\delta \in (0,1)$. This completes the proof. 
\hfill\halmos\par
The following lemma obtains a deterministic bound on term \textsf{(C)} of line \eqref{line:lemma+over}.
\begin{claim}[Bound $\textsf{(C)}$]\label{claim:bound+C}
There exists a constant $M_{\textsf{C}}$ for which $\textsf{(C)} \leq M_{\textsf{C}}\sqrt{t}$ for every sample path. 
\end{claim}
{\bf Proof of Claim \ref{claim:bound+C}.}
The proof directly follows from the adversarial regret guarantee of the online mirror descent (Claim \ref{fact:average+conv}). 
In particular, let 
\begin{equation*}
\begin{split}
    R_{\DualOverVec}(\Timeidx)&:= \sum_{\Tauidx=1}^\Timeidx \DualOverVec_\Tauidx\cdot (\CapRatioVec-\DecisionVec_\Tauidx) - \sum_{\Tauidx=1}^\Timeidx \DualOverVec^\star\cdot (\CapRatioVec-\DecisionVec_\Tauidx) \\
    R_{\DualHard}(\Timeidx)&:= \sum_{\Tauidx=1}^\Timeidx\DualHardVec_\Tauidx\cdot (\CapRatioVec- \DecisionVec_\Tauidx)
\end{split}
\end{equation*}
One can follow the same line of proofs as Claim \ref{claim:pseudo+upper+P} (Appendix \ref{apx+P+reward}) to show that
\begin{equation*}
\begin{split}
R_{\DualOverVec}(\Timeidx) &\leq 2\OverCost\sum_{\Tauidx=1}^{\Timeidx} \eta_\Tauidx + \frac{\bar{V}_{\DualOverVec}}{\eta_\Timeidx} \\
R_{\DualHardVec}(\Timeidx) &\leq 2\bar{\DualHard}\sum_{\Tauidx=1}^{\Timeidx} \eta_\Tauidx + \frac{\bar{V}_{\DualHard}}{\eta_\Timeidx} 
\end{split}
\end{equation*}
where $\bar{\lambda}:= \frac{1+2\alpha}{\underline{\CapRatio}}$, $h(\cdot)$ is the negative entroypy function, $\bar{V}_{\DualOver}= \max_{\Tauidx\in [\Timeidx]}V_h(\DualOverVec^\star, \DualOverVec_\Tauidx)$, and $\bar{V}_{\DualHard}= \max_{\Tauidx\in [\Timeidx]}V_h(\mathbf{0}, \DualHardVec_\Tauidx)$. Because the dual variables are all restricted to the bounded domain $\mathcal{V} = [0,\alpha]^m \times [0, \bar{\lambda}]^m$, both $\bar{V}_{\DualOver}$ and $\bar{V}_{\DualHard}$ are bounded by some constant $\bar{V}$. Furthermore, recall that the step size of \Cref{ALG+Surrogate+P} is given by $\eta_\Tauidx = \frac{k}{\sqrt{\Tauidx}}$ for input constant $k>0$. Hence, both terms above are upper bounded by $M_{\textsf{C}}\sqrt{\Timeidx}$ where a positive constant $M_{\textsf{C}}$ only depends on the input of \CO{} (other than $\TotalTime$). This completes the proof. 
\hfill \halmos\par 

To complete the proof of \Cref{prop:total+arrival}, we plug the high-probability bounds in the previous claims into line \eqref{line:lemma+over} to obtain that, with probability at least $1-\delta$, 
\begin{equation}
\begin{split}
\OverCost \sum_{\Locidx=1}^\Locnum \left(\sum_{\Tauidx=1}^\Timeidx \Decision_{\Tauidx, \Locidx} - \Timeidx\CapRatio_\Locidx\right)_{+} 
&\leq 
\sum_{\Locidx=1}^\Locnum \nu_{\Timeidx, \Locidx}^*\left(
 \sum_{\Tauidx=1}^\Timeidx \Decision_{\Tauidx, \Locidx} - \Timeidx\CapRatio_\Locidx 
\right)
+ M\sqrt{t\log(1/\delta)} \label{line:upto} 
\end{split}
\end{equation}
where the constant $M>0$ only depends on $(M_{\textsf{A}},M_{\textsf{B}}, M_{\textsf{C}})$ specified in Claims \ref{claim:A}-\ref{claim:bound+C}. The proof is complete by non-negativity of the left-hand side.
\hfill\halmos

\subsubsection{Proof of 
\texorpdfstring{\Cref{lemma:all+time+build}}{}}\label{apx+all+time+build} \hfill \\
Recall that we are analyzing an instance with $m=1$. Hence, for brevity, we omit the subscript for $\Locidx=1$.  Let $S_t = \sum_{\Tauidx=1}^t s_\Tauidx$ and $Z_\Timeidx = \sum_{\Tauidx=1}^\Timeidx z_\Tauidx$. We first obtain a simple bound on the \backlog{}. 
\begin{claim}\label{claim:build+up+lower+bound} $b_t \geq (Z_t - S_t)_+$ for all $t \geq 1$ for every sample path.
\end{claim}
{\bf Proof of Claim \ref{claim:build+up+lower+bound}}.
The proof is by induction on $t$. The base case for $t=1$ is trivial because $b_0 = 0$. Now suppose that the claim is true for $t \geq 1$. Then we have
\[b_{t+1} = (b_t + z_{t+1} - s_{t+1})_{+} \geq (Z_t - S_t + z_{t+1} - s_{t+1})_{+} = (Z_{t+1} - S_{t+1})_{+}\]
where the first inequality follows from the induction hypothesis. This completes the proof.
\hfill\halmos\par 
Fix a positive constant $a>0$, which we will specify later. For given constant $a>0$ and $\Timeidx$, define an event
\begin{equation*}
\begin{split}
    \mathcal{Z}_{a,\Timeidx} &= \{Z_\Timeidx \geq 0.5\Timeidx - a\sqrt{\Timeidx} \} \\
    \mathcal{S}_{a,\Timeidx} &= \{0.5 \Timeidx - a\sqrt{\Timeidx} -S_\Timeidx\geq \frac{a}{4}\sqrt{\Timeidx} \}\\
    \mathcal{E}_{a,\Timeidx} &:= \mathcal{Z}_{a,\Timeidx}\cap \mathcal{S}_{a,\Timeidx}
\end{split}
\end{equation*}
First, from Claim \ref{claim:build+up+lower+bound}, we note that
\begin{equation*}
\E[b_t] \geq \E[(Z_\Timeidx - S_\Timeidx)_{+}] \geq \P[\mathcal{E}_{a,\Timeidx}]\E[(Z_\Timeidx - S_\Timeidx)_{+}|\mathcal{E}_{a,\Timeidx}] \geq \P[\mathcal{E}_{a,\Timeidx}]\times \frac{a\sqrt{\Timeidx}}{4}.
\end{equation*}
where the last inequality follows from the definition of $\mathcal{E}_{a,\Timeidx}$.
Hence, it suffices to lower-bound $\P[\mathcal{E}_{a,\Timeidx}]$ by 
$\frac{\Phi(-3\sqrt{2}a)}{4}$ for some constant $a>0$. First, note that the decision of \CO{} \emph{does not} depend on the service (and therefore backlog) process. Hence, the event $\mathcal{Z}_{a,\Timeidx}$ and $\mathcal{S}_{a,\Timeidx}$ must be independent for \Cref{ALG+Surrogate+P}, implying that $\P[\mathcal{E}_{a,\Timeidx}] = \P[\mathcal{Z}_{a,\Timeidx}] \P[\mathcal{S}_{a,\Timeidx}]$.


Thus, it suffices to separately lower-bound the probability of $\mathcal{Z}_{a,\Timeidx}$ and $\mathcal{S}_{a,\Timeidx}$.
We first obtain a lower bound on the former. 
From \Cref{lemma:Z+lower}, there exists constant $\kappa>0$ such that, for any given $a>0$ and $\Timeidx \leq 0.5\TotalTime$, we have
\begin{equation}
    \P[\mathcal{Z}_{a,\Timeidx}] \geq 1 - \exp\left(-\frac{a^2}{\kappa^2}\right) - \BigO(1/\Timeidx).\label{ineq:prob+Z}
\end{equation}

For a sufficiently large constant $a>0$ and another constant $g(a)$, the right-hand side is at least $\frac{1}{2}$ for all $t\geq g(a)$. For such constant $a>0$, we now turn our attention to lower-bounding $\P[\mathcal{S}_{a,\Timeidx}]$ as a function of the constant $a>0$.  By Berry-Esseen Theorem (see, for example, Theorem 3.4.17 of \citet{durrett2019probability}), there exists a constant $d >0$ such that for all $\Timeidx \geq 1$, 
\begin{equation*}
    \P\Big[S_\Timeidx \leq \E[S_\Timeidx] - 3\sqrt{2}a\sqrt{\Var[S_\Timeidx]}\Big] \geq \Phi(-3\sqrt{2}a) - \frac{d}{\sqrt{t}}.
\end{equation*}
where $\E[S_\Timeidx] = (0.5+\ServiceSlack)\Timeidx$, $\Var[S_\Timeidx] = (0.25 - \ServiceSlack^2)\Timeidx$, and $\Phi(\cdot)$ is the cumulative distribution function of the standard normal random variable. Moving the terms, the preceding inequality is equivalent to 
\begin{equation*}
    \P\left[0.5 t - a\sqrt{t} - S_t \geq    3\sqrt{2}a\times \sqrt{\Var[S_t]} - a\sqrt{t} - \epsilon t\right] \geq \Phi(-3\sqrt{2}a) - \frac{d}{\sqrt{t}}. \label{ineq:b}
\end{equation*}
Let us define a constant $t(a)$ as 
\begin{equation}
 t(a) := \max\Big\{\frac{4d^2}{\Phi^2(-3\sqrt{2}a)}, \frac{a^2}{2}, g(a)\Big\}    \label{line:t(a)}
\end{equation}
We now use the assumption $\epsilon \leq \frac{a}{4\sqrt{\TotalTime}}$ and $T \geq 2t(a)$ to prove the following:
\begin{equation*}
\begin{split}
3\sqrt{2}a\times \sqrt{\Var[S_t]} - a\sqrt{t} - \epsilon t &= 3\sqrt{2}a\sqrt{(0.25 - \epsilon^2)t} - a \sqrt{t} - \epsilon t\\
&\geq 
3\sqrt{2}a\sqrt{\left(0.25 - \frac{a^2}{16T}\right)t} - a \sqrt{t} - \frac{at}{4\sqrt{T}} \\
&\geq  \frac{a}{2}\sqrt{t} - \frac{at}{4\sqrt{T}} \\
&\geq \frac{a}{4}\sqrt{t}
\end{split}
\end{equation*}
In the second line, we used the assumption $\epsilon \leq \frac{a}{4\sqrt{T}}$. In the third line, we used the assumption $\TotalTime \geq t(a) \geq \frac{a^2}{2}$. The last line follows from $\TotalTime \geq \Timeidx$. Furthermore, for all $t \geq t(a) \geq \frac{4d^2}{(\Phi(-3\sqrt{2}a))^2}$, it is straightforward to check that $\Phi(-3\sqrt{2}a) - \frac{d}{\sqrt{t}} \geq \frac{\Phi(-3\sqrt{2}a)}{2}$. Hence, for any $t(a) \leq t \leq \frac{1}{2}T$, we conclude that
\begin{equation*}
\P[\mathcal{S}_{a,\Timeidx}]=\P\left[0.5 t - a\sqrt{t} - S_t \geq \frac{a}{4}\sqrt{t}\right] \geq 
     \P\left[0.5 t - a\sqrt{t} - S_t \geq    3\sqrt{2}a\times \sqrt{\Var[S_t]} - a\sqrt{t} - \epsilon t\right] \geq 
    \frac{\Phi(-3\sqrt{2}a)}{2} 
\end{equation*}\par 
Combining, we conclude that, there exists a constant $a>0$ and $t(a)$ such that, for all $\Timeidx \in [t(a), \TotalTime/2]$,
\begin{equation*}
    \P[\mathcal{E}_{a,\Timeidx}] 
    = \P[\mathcal{Z}_{a,\Timeidx}]\P[
    \mathcal{S}_{a,\Timeidx}
    ]
    \geq \frac{\Phi(-3\sqrt{2}a)}{4}.
\end{equation*}
whenever $\TotalTime \geq 2t(a)$ and $\ServiceSlack \leq \frac{a}{4\sqrt{\TotalTime}}$.
 This completes the proof.
\par

\subsubsection{Proof of \Cref{lemma:near+opt+value}}\label{apx+near+opt}\hfill\\
Let $\delta := \sqrt{\BuildUpCost/\TotalTime}$. Consider the following feasible solution: $\Decision^*_\Timeidx = \Indicator[\Weight_\Timeidx \geq 0.5 + \delta]$ up to $\Timeidx \leq \Stopping$ and $\Decision^*_\Timeidx =0$ for all $\Timeidx > \Stopping$, where the stopping time $\Stopping$ is defined as $\Stopping = \min \{\Timeidx \leq \TotalTime: \sum_{\Tauidx=1}^\Timeidx \Decision^*_\Timeidx \geq 0.5\TotalTime\}$. We analyze the the expected objective value of this feasible solution, which will be the lower bound on the expected value of the optimal offline solution.

We first analyze the reward of this policy. Because (i) $\Stopping$ is a bounded stopping time with respect to $\{\History_\Timeidx\}$ and (ii) each $\Weight_\Timeidx \Decision_\Timeidx$ is independent with identical mean for $\Timeidx \leq \Stopping$, Wald's equation implies that
\begin{equation}
    \E\Big[\sum_{\Timeidx=1}^{\Stopping} \Weight_\Timeidx \Decision_\Timeidx^*\Big] = \E[\Stopping]\E[\Weight_1 \Decision_1^*] = \E[\Stopping]\int_{0.5+\delta}^1 w dw =  \E[\Stopping]\Big(\frac{3}{8} -\frac{1}{2}\delta -\frac{1}{2}\delta^2\Big).\label{line:opt+reward+1}
\end{equation}
where in the second equality we used the fact that reward follows the uniform distribution on interval $(0,1)$. To further lower-bound $\E[\Stopping]$, we establish a concentration inequality of $\TotalTime -\Stopping$. By Hoeffding's inequality, we have, for any $\Timeidx \leq \Stopping$,
\begin{equation*}
    \P\Big[\sum_{\Tauidx=1}^\Timeidx \Decision_\Tauidx^* \leq (0.5-\delta)\Timeidx + \sqrt{\frac{t}{2}\log(1/x)}\Big] \geq 1 - x
\end{equation*}
for all $x \in (0,1)$. By plugging $t(x) = \TotalTime - \frac{1}{0.5}\sqrt{\frac{\TotalTime}{2}\log(1/x)}$, we deduce that $
\P[\sum_{\Timeidx=1}^{t(x)} z_{t(x)}^* \leq 0.5\TotalTime] \geq 1 - x $
for any $x \in (0,1)$. Equivalently, for any $x \in (0,1)$, we have
\begin{equation}
   \P\Big[\Stopping \geq \TotalTime - \frac{1}{0.5}\sqrt{\frac{\TotalTime}{2}\log(1/x)}\Big] \geq 1-x \label{line:stop+concentration}.
\end{equation}
From this inequality, we now observe that
\begin{equation}
\E[(\TotalTime-\Stopping)^2] = \int_{0}^{T^2} \P[(\TotalTime - \Stopping)^2 \geq y] d y  = \int_{0}^{T^2} e^{-\frac{8y}{T}} dy \leq  \Theta(\TotalTime).  \label{ineq:temp2}
\end{equation}
Here, the third equality follows from inequality \eqref{line:stop+concentration}. By Jensen's inequality, we conclude from inequality \eqref{ineq:temp2} that $\E[(\TotalTime -\Stopping)] \leq \Theta(\sqrt{\TotalTime})$, or equivalently, $\E[\Stopping] \geq \TotalTime - \Theta(\sqrt{\TotalTime})$.

Plugging this bound into line \eqref{line:opt+reward+1} (with $\delta = \sqrt{\BuildUpCost/\TotalTime}$), we obtain\footnote{
Because $\BuildUpCost = o(\TotalTime)$, for sufficiently large $\TotalTime$, we have $\frac{3}{8}-\frac{1}{2}\delta - \frac{1}{2}\delta^2 \geq 0$. 
}
\begin{equation*}
    \E\Big[\sum_{\Timeidx=1}^{\Stopping} \Weight_\Timeidx \Decision_\Timeidx^*\Big] \geq (\TotalTime - \Theta(\sqrt{\TotalTime}))\Big(\frac{3}{8} - \frac{1}{2}\delta - \frac{1}{2}\delta^2\Big) = \frac{3}{8}\TotalTime - \Theta(\sqrt{\BuildUpCost\TotalTime} + \BuildUpCost).
\end{equation*}

We now turn our attention to bounding the average backlog of the proposed feasible solution. Let $\{\BuildUp_\Timeidx^*\}_{\Timeidx=1}^\TotalTime$ be the induced backlog. From our drift lemma (\Cref{lemma:drift}), for all $\Timeidx \leq \Stopping$, we have:
\begin{equation}
    \E\Big[\frac{(\BuildUp_\Timeidx^*)^2 -(\BuildUp_{\Timeidx-1}^*)^2  }{2}\Big| \History_{\Timeidx-1}\Big] \leq \BuildUp^*_{\Timeidx-1}(\E[\Decision_\Timeidx^*|\History_{\Timeidx-1}] - 0.5) + O(1) \leq -\delta  \BuildUp^*_{\Timeidx-1} + O(1) \label{line:opt+build+1}
\end{equation}
where in the last inequality we used $\E[\Decision_\Timeidx^*|\History_{\Timeidx-1}] = \E[\Decision_\Timeidx^*] = \P[\Weight_\Timeidx \geq 0.5+\delta] = 0.5-\delta$ because the reward follows the uniform distribution on interval $(0,1)$. Summing up line \eqref{line:opt+build+1} for $\Timeidx \leq \Stopping$ and taking the outer expectation, we obtain that 
\begin{equation}
    \delta \E\Big[\sum_{\Timeidx=1}^{\Stopping-1} \BuildUp_\Timeidx^* \Big] \leq -\frac{1}{2}\E[(\BuildUp_{\Stopping}^*)^2] + O(\TotalTime). \label{line:opt+build+2}
\end{equation}
Hence, we have
\begin{equation*}
\begin{split}
    \frac{\E[\sum_{\Timeidx=1}^\TotalTime \BuildUp_\Timeidx^*]}{\TotalTime} &= 
    \frac{\E[\sum_{\Timeidx=1}^{\Stopping - 1} \BuildUp_\Timeidx^*]
         + \E[\sum_{\Timeidx=\Stopping}^{\TotalTime} \BuildUp_\Timeidx^*]
    }{\TotalTime} \\
    &=     \frac{\E[\sum_{\Timeidx=1}^{\Stopping - 1} \BuildUp_\Timeidx^*]
         + \E[(\TotalTime - \Stopping + 1)\BuildUp_{\Stopping-1}^*]
    }{\TotalTime} \\
    &= \frac{\E[\sum_{\Timeidx=1}^{\Stopping - 1} \BuildUp_\Timeidx^*]
         + \sqrt{\E[(\TotalTime - \Stopping + 1)^2]}\sqrt{\E[(\BuildUp_{\Stopping-1}^*)^2]}
    }{\TotalTime} \\
    &\leq O(\frac{1}{\delta}+1).
\end{split}
\end{equation*}
In the second line, we used the fact that $b_\Timeidx \leq b_{\Stopping}$ for all $\Timeidx \geq \Stopping$ (the backlog cannot increase after $\Timeidx \geq \Stopping$ by definition of the feasible solution $z^*_t$). In the third line, we used the Cauchy-Schwartz inequality. In the last line, we used (i) $\E[\sum_{\Timeidx=1}^{\Stopping - 1} \BuildUp_\Timeidx^*] \leq O(\TotalTime/\delta)$ from line \eqref{line:opt+build+2}, (ii) $\E[(\BuildUp_{\Stopping}^*)^2] \leq O(\TotalTime)$ from line \eqref{line:opt+build+2}, and (iii) $\E[(\TotalTime - \Stopping)^2] \leq O(\TotalTime)$ from the high-probability bound \eqref{line:stop+concentration}. Combining with $\delta=\sqrt{\BuildUpCost/\TotalTime}$, we have $\frac{\BuildUpCost}{\TotalTime}\E[\sum_{\Timeidx=1}^\TotalTime \BuildUp_\Timeidx^*] = O(\sqrt{\BuildUpCost\TotalTime} + \BuildUpCost)$.

Combining, we conclude that the objective value of the feasible solution $\{z_t\}_{t=1}^T$ is 
\begin{equation}
       \E\Big[\sum_{\Timeidx=1}^{\Stopping} \Weight_\Timeidx \Decision_\Timeidx^*\Big]  - \frac{\BuildUpCost}{\TotalTime}\E[\sum_{\Timeidx=1}^\TotalTime \BuildUp_\Timeidx^*] \geq \frac{3}{8}\TotalTime - \Theta(\sqrt{\BuildUpCost\TotalTime} + \BuildUpCost).
\end{equation}
This completes the proof. 
\section{Supplementary Results for Case Study (\Cref{sec:numerics})}\label{apx+case+study}
\subsection{Further Discussion on Learning Within-year Arrival Patterns}\label{apx+within+year+variations}

In \Cref{sec:intro}, we motivated the design of distribution-free algorithms through \Cref{fig:tied+proportion}, which shows the significant year-to-year variations of the refugee pool composition. In light of the annual quota, we reiterate that the decision-making horizon of refugee matching is one year. We take 2015 as a running example. At the beginning of 2015, a new process begins for dynamic matching decisions. One approach (taken in prior work — see Sections \ref{subsec:intro+background} and \ref{subsec:lit+review}) is to assume that 2015 resembles earlier years, such as 2014, and use them as “distributional knowledge.” However, \Cref{fig:tied+proportion} highlights the significant differences across the years, motivating design of algorithms without distributional knowledge, i.e., past years’ data. To this end, our learning-based algorithms are designed to learn the arrival pattern \emph{within the year} through the dual variables $(\boldsymbol{\theta}^*, \boldsymbol{\lambda}^*)$ in Proposition 3.1. Note that $(\boldsymbol{\theta}^*, \boldsymbol{\lambda}^*)$ are defined for the one-year problem. For example, for year 2015, we have  $(\boldsymbol{\theta}_{2015}^*, \boldsymbol{\lambda}_{2015}^*)$. Since we do not rely on the data of 2014 to learn $(\boldsymbol{\theta}_{2015}^*, \boldsymbol{\lambda}_{2015}^*)$, the difference between 2014's and 2015's pool composition is irrelevant to learning $(\boldsymbol{\theta}_{2015}^*, \boldsymbol{\lambda}_{2015}^*)$.


That said, learning the dual variables for the one-year problem requires stationarity  in arrivals within that year. For example, if the arrival patterns in the first half of 2015 significantly differ from those in the second half, we cannot learn $(\theta_{2015}^*, \lambda_{2015}^*)$ effectively. In this section, we provide  numerical evidence against such drastic changes. In particular, we show that the variation in arrival patterns \emph{within the year} is smaller compared to the variation \emph{across years}. 

In the following, similar to \Cref{fig:tied+proportion}, we focus on the normalized number of tied cases. Specifically, we fix a year $y \in \{2014, 2015, 2016\}$ and use $N^y_{t,i}$ to denote the number of tied cases up to period $t$ (i.e., the $t$-th arrival in the actual arrival sequence) at affiliate $i$. Letting $T^y$ denote the total number of arrivals in year $y$, the normalized number of tied cases in year $y$ at affiliate $i$ is then given by $N^{y}_{T^y,i}/T^y$. These normalized values were presented in \Cref{fig:tied+proportion}. For each of comparison, we present them again in \Cref{fig:across+years}. On the same figure, we also present the coefficient of variation of $N^{y}_{T,i}/T^y$ across years $y \in  \{2014, 2015, 2016\}$ (referred to as the \emph{across-years CV} hereafter), as a measure of across-year variation of the number of tied cases. 

To further investigate the within-year variations, we repeat the same process but this time within a year. Specifically, we divide a year into 4 quarters and compute the ``quarterly'' cumulative normalized number of tied cases. Formally, we define $m^y_{q,i}$ ($q \in [4]$) as
\begin{equation}
    m^y_{q,i} :=\dfrac{N^y_{\lceil q\TotalTime^y/4 \rceil, i}}{
    \lceil q\TotalTime^y/4 \rceil
    }, \quad q \in [4]
\end{equation}
and report $\{m^y_{q,i}\}_{q \in [4]}$ for each affiliate $i$. If the arrival within year $y$ is stationary, the variation of $\{m^y_{q,i}\}_{q \in [4]}$ (across $q$) should be small. Motivated by this observation, we calculate and present the coefficient of variation for $\{m^y_{q,i}\}_{q \in [4]}$ for each year $y$, referring to this as the \emph{within-year CV} (of year $y$). \Cref{fig:within+years} presents $\{m^y_{q,i}\}_{q \in [4]}$ for each affiliate $i$ and year $y$ considered in \Cref{fig:across+years} along with the corresponding within-year CV.

\begin{figure}[htp]
  \begin{subfigure}{\linewidth}
    \centering
    \includegraphics[width=0.8\linewidth]{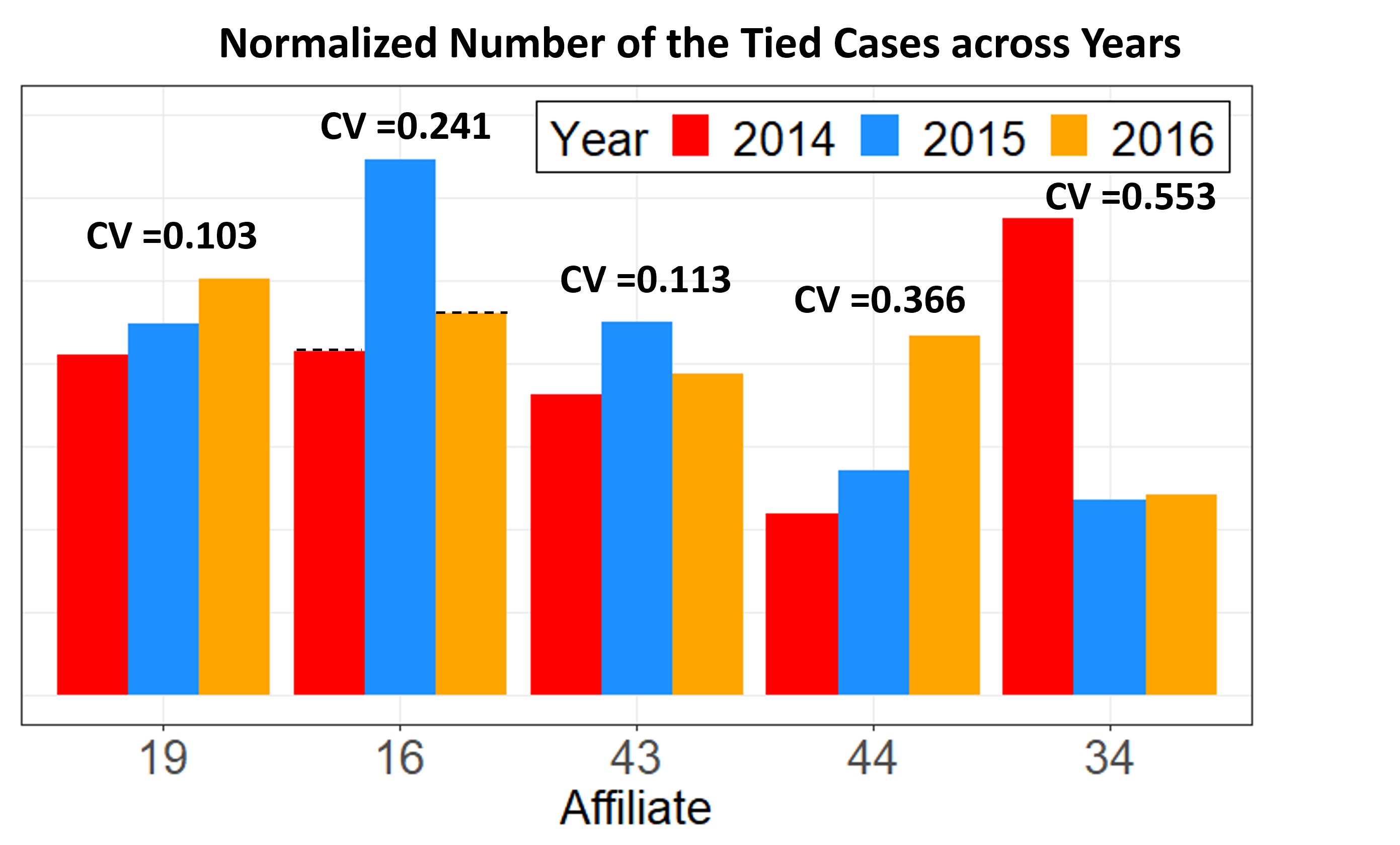}
    \caption{Across-years variation of the normalized number of tied cases}
    \label{fig:across+years}
  \end{subfigure}

    \begin{subfigure}{\linewidth}
    \centering
    \includegraphics[width=\linewidth]{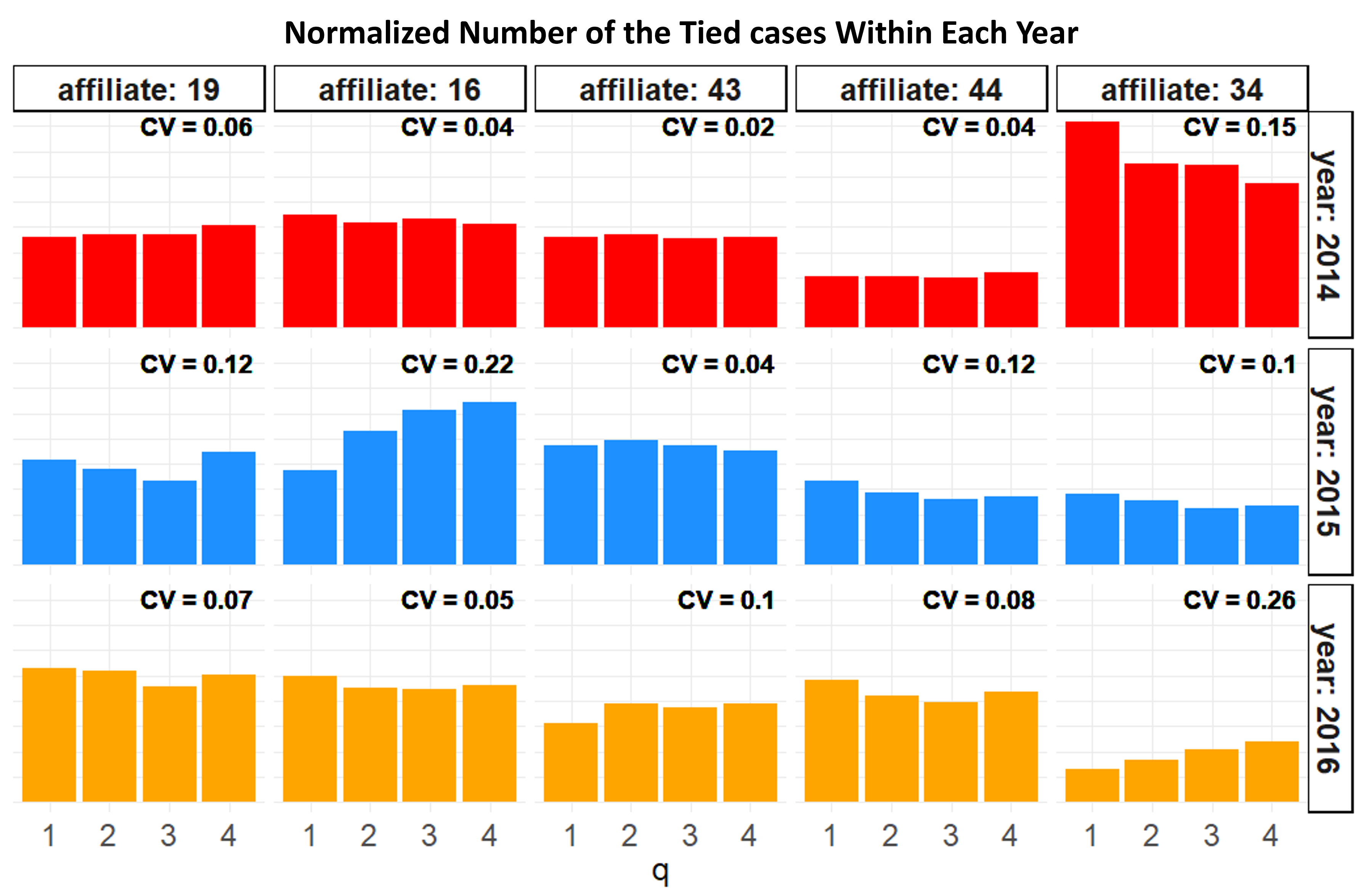}
    \caption{Within-year variation of the normalized number of tied cases}
    \label{fig:within+years}
  \end{subfigure}

    \caption{Across-years versus within-year variation of the normalized number of tied cases. 
    }
  \label{fig:variations}
\end{figure}

By comparing Figures \ref{fig:across+years} and \ref{fig:within+years}, we observe that within-year variation is generally smaller than across-years variation. For example, when visually inspecting affiliate 44, we notice that the normalized number of tied cases varies significantly across the years 2014-2016, yet the tied cases arrived in a relatively stationary manner within each year. This observation is further supported by the significantly larger across-years CV compared to the within-year CV. We also note that some affiliates do not exhibit ideal stationarity in certain years (e.g., affiliate 34 in 2016). However, even for these cases, the within-year CV is generally smaller than the across-years CV. This suggests that our partner agency could benefit from algorithms that learn the arrival pattern \emph{within the year}, rather than relying on past years' data. As explained in detail in \Cref{sec:numerics}, indeed our distribution-free algorithm (that only relies on the current year's data to make decisions) outperforms the algorithms that rely on past years' data.

\subsection{Implementation Details of \texttt{RO-Learning} and \texttt{CO-DL}}\label{apx+case+ALG+D}

In the following, we present more details on the implementation of \texttt{RO-Learning} for our case study. Specifically, Recall that \Cref{ALG+Surrogate+D} (which \texttt{RO-Learning} is based on) requires the parameters $\eta$ and $\zeta$ (i.e., the step-sizes for $(\DualOverVec, \DualHardVec)$ and $\DualBuildVec$, respectively). We set these step sizes as a function of the penalty parameters as follows:
 \begin{align}
    \eta &= \frac{s_{\eta} \log(\alpha+1)}{\sqrt{T}} \nonumber \\
    \zeta &= \frac{s_{\zeta} \gamma}{\sqrt{T}} \nonumber
\end{align}   
Intuitively, this makes the algorithm more conservative in terms of the cost metrics as the penalty parameters $\OverCost$ and $\BuildUpCost$ increase.

We adopt a data-driven approach of tuning the parameters $(s_{\eta}, s_{\zeta})$, to select values that lead to acceptable outcomes for both over-allocation and average \backlog{}. Specifically, we run the algorithm on the dataset of year 2014, obtaining the over-allocation and average \backlog{} for each {$1 \leq s_\eta \leq 5$} and $0.1 \leq s_{\zeta} \leq 1$. The resulting \emph{acceptable region} for $(s_{\eta}, s_{\zeta})$ is defined by values that satisfy two criteria: (i) achieving an over-allocation at most $1.2\times$ (the minimum value across all parameter settings) and (ii) maintaining an average \backlog{} at most $1.1\times$ (the value achieved by the actual placement in the year 2014). 
\begin{figure}[t]
    \centering
    \includegraphics[scale=0.6]{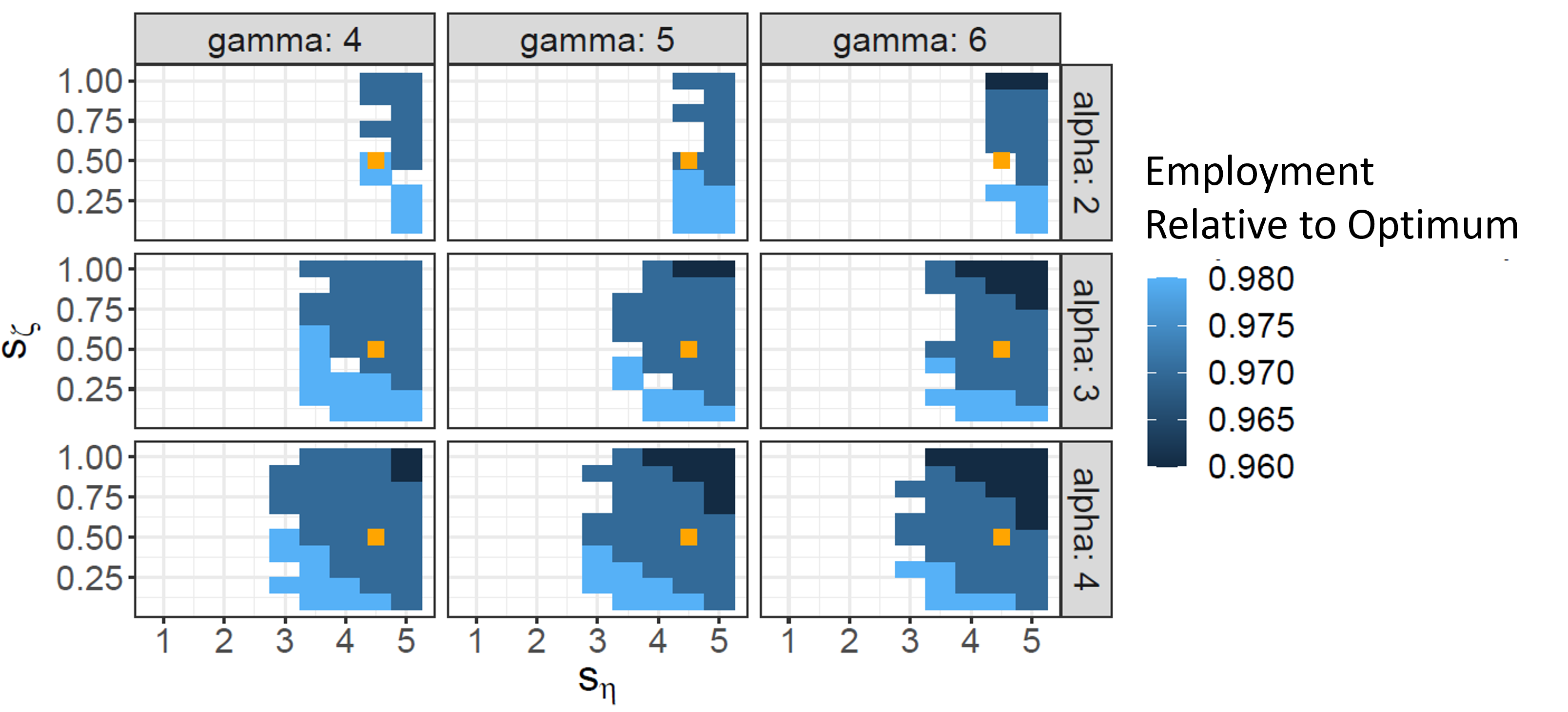}
    \caption{Acceptable region of $\boldsymbol{(s_{\eta}, s_{\zeta})}$ based on data of year 2014}
    \label{fig:acceptable+region}
\end{figure}
Figure \ref{fig:acceptable+region} illustrates the acceptable region for some penalty parameters, with the ratio of employment outcome compared to the optimal level of surrogate primal (\Cref{def:primal+surrogate}). Upon visual examination, we identify that values within the ranges {$4 \leq s_\eta \leq 5$} and $0.1 \leq s_{\zeta} \leq 1$ generally fall within the acceptable region across a broad range of the penalty parameters. Consequently, we opt for {$s_\eta=4.5$} and $s_{\zeta} = 0.5$, roughly representing a central point within the established acceptable region. This selection aims to manage cost metrics without compromising employment outcomes too severely. The choice is further motivated to ensure the robustness of our case study results with respect to variations in step sizes.

\majorrevcolor{
We applied the same data-driven procedure to tune the step size $\eta_t$ for implementing \CO{}. Specifically, we set the time-varying step size as $\eta_t = \frac{k \log(\alpha+1)}{\sqrt{t}}$, and selected $k \in [1,10]$ based on the 2014 data, using the same acceptability criteria described above. The resulting choice was $k = 4$.
}
\subsection{Implementation Details of \texttt{Sampling}}\label{apx+case+Sampling}
In the following section, we present the details of \texttt{Sampling} for the sake of completeness. The original version of \texttt{Sampling} \citep{bansak2024outcome} was designed for a setting without tied cases. Here, we formally describe the extension of the original version to accommodate scenarios with tied cases.\par 
As we discussed in \Cref{sec:numerics}, \texttt{Sampling} \emph{requires} a sample trajectory of future arrivals. The future arrivals are drawn from a \emph{sampling pool}, denoted by $\mathcal{P}$, with replacement. Following \citet{bansak2024outcome}, we use the previous year's arrival data as the sampling pool. At each time $t$, let $\Capacity_{\Timeidx,\Locidx}$ denote the remaining  capacity for each affiliate $i$. Upon arrival of case $t$ and observing $(\Weightvec_\Timeidx, \Target_\Timeidx)$, \texttt{Sampling} first samples a future arrival $(\Weightvec_\Tauidx, \Target_\Tauidx)_{\Tauidx=\Timeidx+1}^\TotalTime$ from its sampling pool $\mathcal{P}$. In our simulation, this means that we sample future arrivals of time $\Timeidx+1$ and onward from the previous year's arrival data. Given this sampled trajectory of future arrivals, \texttt{Sampling} solves the following optimization problem to obtain the matching decision:
\footnote{We also note that, when $\OverCost \geq 1$ (i.e., when the over-allocation penalty parameter is greater than the maximum employment outcome of each case), \texttt{Sampling} does not over-allocate free cases with respect to the current capacity due to the nature of the offline program it solves. Hence, as observed in \Cref{fig:more+case} in \Cref{sec:numerics}, the outcome of \texttt{Sampling} does not change with respect to the values of $\OverCost$.  }
\begin{align}
\max_{\{\DecisionVec_\Tauidx\}_{\Tauidx=t}^T} &\quad \tilde{\Weightvec}_t \cdot \DecisionVec_t + \sum_{\Tauidx = t+1}^T \Weightvec_\Timeidx\cdot\DecisionVec_t - \OverCost\sum_{i=1}^m \left(\sum_{\Tauidx=t}^T \Decision_{\Timeidx, \Locidx}- \Capacity_{\Timeidx,\Locidx} \right)_{+} \nonumber \\
    \text{s.t}\quad  & \sum_{\Tauidx=t}^T \Indicator[\Target_\Timeidx = 0 ]\Decision_{\Timeidx, \Locidx}\leq 
    {\left(\Capacity_{\Timeidx, \Locidx} - \sum_{\Tauidx = \Timeidx}^\TotalTime \Indicator[\Target_\Tauidx = \Locidx]\right)_{+}}
    \nonumber
\end{align}
where 
\begin{align}
   \tilde{\Weight}_{\Timeidx, \Locidx} = {\Weight}_{\Timeidx, \Locidx} - \frac{\BuildUpCost}{T}\left \lceil{\frac{\BuildUp_{\Timeidx-1,\Locidx}- \CapRatio_\Locidx}{\CapRatio_\Locidx}}\right \rceil \Indicator[\BuildUp_{\Timeidx-1,\Locidx}>0]. \label{eq:waiting+time}
\end{align}
That is, \texttt{Sampling} solves an offline problem that maximizes the net matching reward given the current remaining capacity with a tweak: to further control the congestion, as in equation \eqref{eq:waiting+time}, it penalizes the (observed) employment probability at affiliate $i$ for the \emph{current} case $t$ by the waiting time the case will experience upon placement computed based on the first-come-first-served manner (see Lemma 1 of \citet{bansak2024outcome} for discussion on such penalty function), with the penalty increasing with $\BuildUpCost$ (the penalty parameter for congestion). The \texttt{Sampling} algorithm repeats this process for $K$ simulated trajectories of future arrivals and assigns case $t$ to the location where it was matched most often. In our case study, we set $K=5$ for these simulations. 
\subsection{Comparison of \CO{} and \texttt{RO-Learning} Across Penalty Parameters}
\label{apx+CO}
\SLdelete{In \Cref{table:performance+CO}, we report the numerical result of \CO{}. For the sake of brevity, we focus on the results based on the same penalty parameter value considered in \Cref{sec:numerics}. We tune the step size of the algorithms in the same way as described in \Cref{apx+case+ALG+D}. {In \Cref{table:performance+CO}, we observe that \CO{}, despite not using backlog information, achieves substantial gains in employment and backlog relative to \texttt{Actual}. However, we observe that \CO{} performs worse than \texttt{RO-Learning}, which is based on \CA{}, with higher over-allocation and average backlog.}}

\majorrevcolor{
To further investigate the impact of penalty parameters on the performance of \CO{}, we report  the relative difference in objective value (as defined in Equation~\eqref{ALG+obj}) between \texttt{RO-Learning} and \CO{} across a broad range of penalty parameters. The heatmaps in Figure~\ref{fig:CO+CA} report the relative improvement of \texttt{RO-Learning} in terms of the objective compared to \CO{}. We observe that \texttt{RO-Learning} consistently achieves a higher objective value than \CO{} for all parameter values considered. This improvement naturally stems from \texttt{RO-Learning} using backlog information and thus being able to better mitigate congestion. However, the performance gap narrows when the congestion penalty parameter $\gamma$ is small; for instance, in year 2016, the relative improvement is less than 5\% across all values of $\alpha$ when $\gamma = 1$.

\begin{figure}
    \centering
    \includegraphics[width=0.95\linewidth]{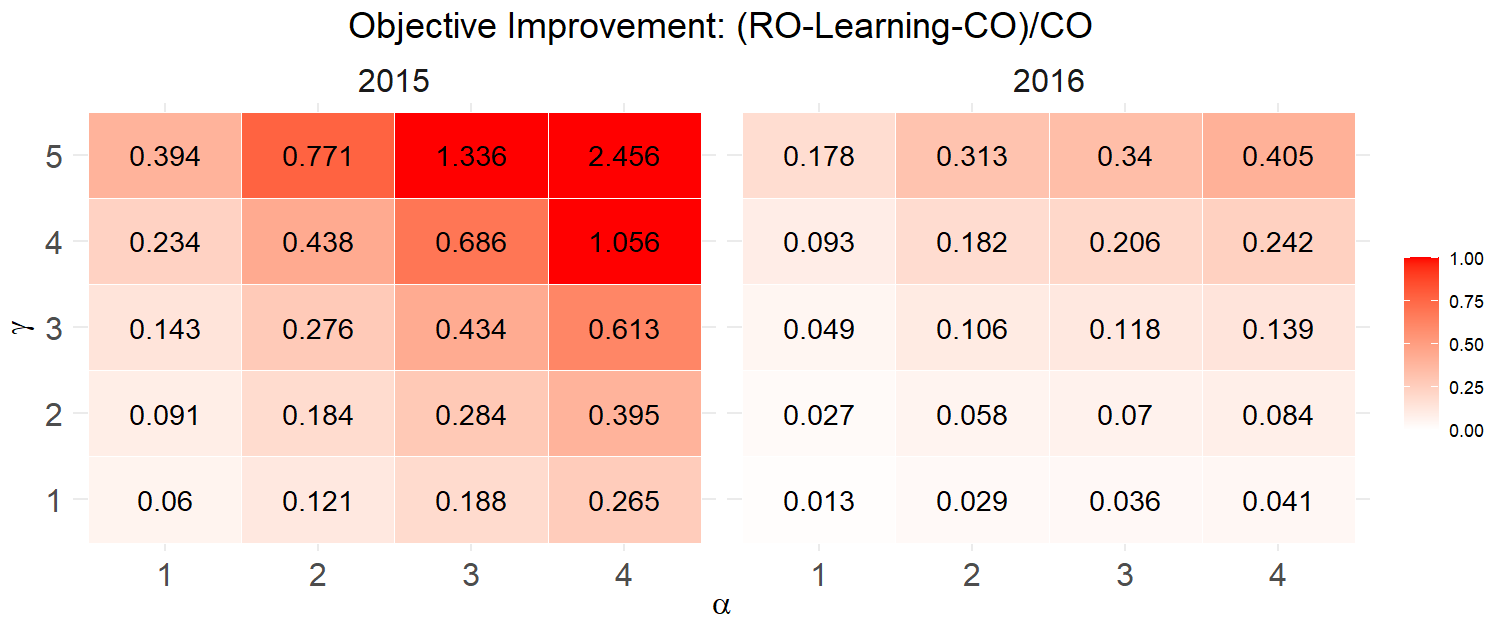}
    \caption{Relative objective improvement of \texttt{RO-Learning} over \CO{} across a grid of penalty parameters $(\alpha, \gamma)$. Each cell reports the relative improvement in the  objective (Equation~\eqref{ALG+obj}) for the corresponding pair of penalty parameters. Results are shown separately for years 2015 and 2016.}
    \label{fig:CO+CA}
\end{figure}
}

\subsection{Tradeoff Curves Across a Range of Penalty Parameters}\label{appx:full-param-sweep}
\Cref{fig:more+case} presents additional numerical results for years 2015 and 2016, covering all combinations of penalty parameters $\OverCost \in \{1,2,\ldots,5\}$ and $\BuildUpCost \in \{0,\ldots,10\}$. Each figure includes two panels per value of $\OverCost$: employment rate versus over-allocation (left) and employment rate versus average backlog (right). Results are shown for \texttt{RO-Learning}, \texttt{Sampling}, and \texttt{Actual}, with \texttt{Opt Emp} included as an additional  benchmark. Overall, we observe similar qualitative patterns to those shown in \Cref{fig:frontier-2016-summary}, including the dominance of \texttt{RO-Learning} over \texttt{Sampling} across a wide range of penalty parameters.

\begin{figure}[htp]
  \begin{subfigure}{\linewidth}
    \centering
    \includegraphics[width=\linewidth]{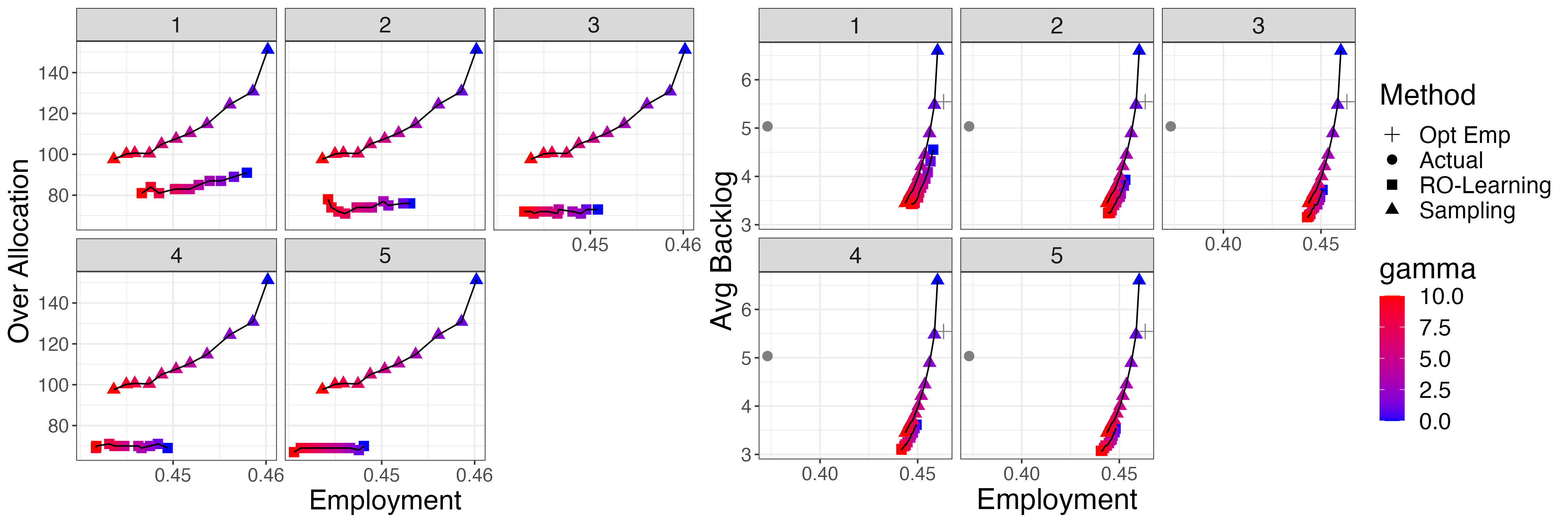}
    \caption{Year 2015}
    \label{fig:more+case:subfig1}
  \end{subfigure}

    \vspace{0.5cm}
  \begin{subfigure}{\linewidth}
    \centering
    \includegraphics[width=\linewidth]{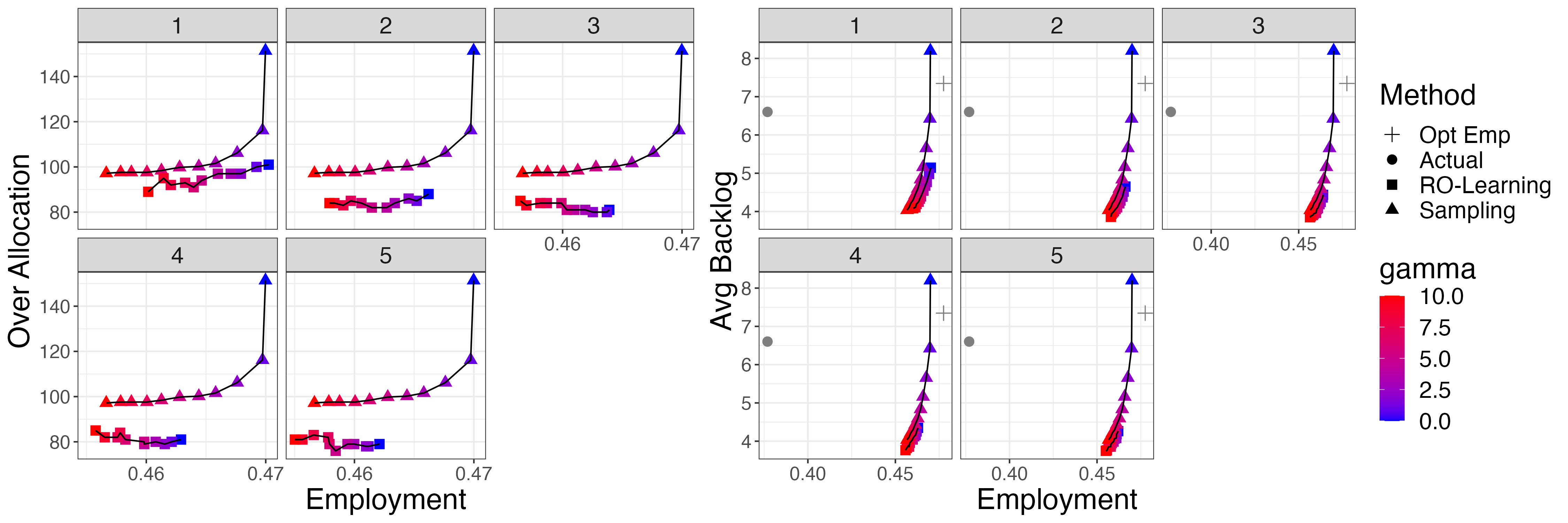}
    \caption{Year 2016}
    \label{fig:more+case:subfig2}
  \end{subfigure}

  \caption{Numerical results with  $\boldsymbol{\OverCost \in \{1,2,..,5\}}$ and $\boldsymbol{\BuildUpCost \in \{0,1,...,10\}}$. In each panel, we display the trajectory of the outcome metrics of fixed $\boldsymbol{\OverCost}$ (label of the penal) and varying $\boldsymbol{\BuildUpCost}$. The left (the right, resp.) set of panels shows the over-allocation (the average backlog, resp.) versus the employment rate. \texttt{Opt Emp} on the right panels is the outcome of the surrogate primal problem (\Cref{def:primal+surrogate}).  }
  \label{fig:more+case}
\end{figure}



\section{Generalized Model with Multiple Knapsack Constraints}\label{apx+non+unit+size}
\SLcomment{Changed the title slightly from ``Generalized Model with A Multiple Types of Resources'' to ``Generalized Model with A Multiple Knapsack Constraints'' (wanted to highlight knapsack (non-unit capacity consumption..)}

In our main model, for simplicity of exposition, we assumed that (i) each affiliate has one type of static resource (i.e., annual quota) and (ii) each case consumes one unit of the static resource at the matched affiliate. While these assumptions are largely aligned with the practice of our partner agency, in other contexts, a case may consume more than one type of resource. For example, for a family with children, there may be a constraint in terms of the number of children that can be enrolled at the local school (a static resource). Additionally, a case may have more than one child, thus consuming multiple units. Furthermore, enrolling children may require different forms of assistance, implying that it also requires a separate post-allocation service. In this section, we describe how our model, algorithms, and their performance guarantee seamlessly generalize in these settings.


\subsection{Generalized Model}
\label{subsec:generalized+model}

\smallskip 

\noindent{\emph{ (i) Static Resource (Capacity):}} 
In this generalized model,\footnote{We note that a similar model with multiple knapsack constraints was studied in \cite{ahani2021placement} and \cite{delacretaz2023matching}. However, these papers studies the static (rather than dynamic) matching problem and does not model the dynamic resource.} each affiliate $\Locidx \in [\Locnum]$ can have $l$ types of  \emph{static} resources, referred to as the (static) resource-type $(i,j)$ for $j \in [l]$. Each case $\Timeidx$ is characterized by its type $\Arrival_\Timeidx = (\Weightvec_\Timeidx, \Target_\Timeidx, \boldsymbol{n}_\Timeidx)$, where $\boldsymbol{n}_\Timeidx = (n_{\Timeidx, \Locidx, j})_{\Locidx \in [\Locnum], j \in [l]} \in \R_{+}^{ml}$ is the number of units consumed by case $\Timeidx$ for resource-type $(\Locidx, j)$ when matched to affiliate $\Locidx$. We assume that $(n_{\Timeidx, \Locidx, j})_{\Locidx \in [\Locnum], j \in [l]}$ is bounded by a constant denoted $\bar{n}$. 
Upon arrival of case $t$, the agency chooses a matching decision $\mathbf{z}_t \in \mathcal{Z}(\Target_t)$ where $\FeasibleSet{\Target} = \Simplex$ if $\Target = 0$ and $\FeasibleSet{\Target} =\{\BasisVec{\Target}\}$ otherwise.\footnote{While the type-feasibility set allows for fractional allocation, our algorithms will always make an integral decision. For example, when a refugee case is a family of multiple members, the case will be matched to the same affiliate.} 
Similar to our base model, we assume that $\Arrival_\Timeidx$ is generated from an unknown i.i.d. distribution $\mathcal{F}$. 

Each affiliate $\Locidx$ is endowed with a fixed capacity $\Capacity_{\Locidx,j}$ for the static resource-type $(i,j)$. We use $\rho_{i,j} = c_{i,j}/\TotalTime$ to denote the capacity ratio of the static resource-type $(i,j)$. Similar to our main model in \Cref{sec:model}, we use $\underline{\rho} := \min_{i \in [m], j \in [l]} \rho_{i,j}$ to denote the minimum capacity ratio and impose mild assumptions on these ratios: (i) $\sum_{i=1}^m \rho_{i,j} \leq 1$ for all $j \in [l]$ and (ii) $\rho_{i,j} - \E[n_{\Timeidx,\Locidx, j}\Indicator[\Target_\Timeidx = \Locidx]] = \Theta(1)$ for all $i \in[m]$ and $j \in [l]$, which means that, in expectation, we have $\Theta(\TotalTime)$ capacity for free cases for each static resource-type $(\Locidx, j)$. Similar to our main base model, we impose the hard constraint that, for each static resource-type $(i,j)$, no over-allocation can be made from free cases. Formally, 
\begin{equation*}
    \sum_{\Tauidx=1}^\Timeidx \Indicator[\Target_\Tauidx=0]n_{\Tauidx, \Locidx, j}\Decision_{\Tauidx, \Locidx} \leq \left(
    \Capacity_{\Locidx, j} - 
    \sum_{\Tauidx=1}^\Timeidx \Indicator[\Target_\Tauidx=\Locidx]n_{\Tauidx, \Locidx, j}
    \right)_{+}, \quad \forall \Locidx\in[\Locnum], j \in [l], \Timeidx \in [\TotalTime].
\end{equation*}


\smallskip 

\noindent{\emph{ (ii) Dynamic Resource (Server):}} 
To incorporate the post-allocation service, we endow each static resource-type $(i,j)$ with a dedicated server, referred to as the dynamic resource-type  $(i,j)$. Each server's availability $\Service_{\Timeidx, \Locidx, j}$ for period $\Timeidx$ follows i.i.d Bernoulli process with service rate $r_{\Locidx, j}$. 
Similar to our base model, we impose a stability condition that $r_{\Locidx, j} = \CapRatio_{\Locidx, j} + \ServiceSlack$ where $\ServiceSlack\geq0$ is a service slack. After case $\Timeidx$ is matched to affiliate $\Locidx$, it will add $n_{\Timeidx,\Locidx,j}$ unit of workload to the corresponding server. Hence, the backlog for the dynamic resource-type $(\Locidx,j)$ at time $\Timeidx$ will be given by
\begin{equation}
\BuildUp_{\Timeidx,\Locidx,j} = (\BuildUp_{\Timeidx - 1,\Locidx,j} + n_{\Timeidx,\Locidx,j}\Decision_{\Timeidx, \Locidx} - \Service_{\Timeidx, \Locidx, j})_{+}.  \label{eq:backlog+generalized}
\end{equation}

\smallskip

\noindent{\emph{ (iii) Benchmark and Regret:}} 
Analogous to \eqref{ALG+obj}, for given penalty parameters $\OverCost$ and $\BuildUpCost$, the objective of online algorithm $\pi$ is given by
\begin{equation}
  \textsf{ALG}^\pi(\OverCost, \BuildUpCost) := \sum_{\Timeidx=1}^\TotalTime\sum_{\Locidx=1}^\Locnum \WeightEach \DecisionEach^\pi
    - \OverCost\sum_{\Locidx=1}^{\Locnum} \sum_{j=1}^l\PositivePart{\sum_{\Timeidx=1}^\TotalTime \size_{\Timeidx, \Locidx, j }\DecisionEach^\pi - \Capacity_{\Locidx, j } }
    - \frac{\BuildUpCost}{\TotalTime} \sum_{\Timeidx=1}^\TotalTime\sum_{\Locidx=1}^{\Locnum}\sum_{j=1}^l b_{t,i,j}^\pi
\end{equation}
where $\{\DecisionVec_t^\pi\}_{t=1}^T$ is the matching profile of online algorithm $\pi$ and $\{{b}_{t,i,j}^\pi\}_{t=1}^T$ is the corresponding backlog induced by the backlog dynamics \eqref{eq:backlog+generalized}.

The optimal offline benchmark (generalization of  \Cref{def:offline}) is given by
\begin{align*}
\OPT(\OverCost, \BuildUpCost) &:= \underset{
    \substack{
\DecisionVec_\Timeidx \in 
\FeasibleSet{\Target_\Timeidx}\\
\BuildUpVec_\Timeidx \geq \mathbf{0}
}}{\max}    
\sum_{\Timeidx=1}^\TotalTime\sum_{\Locidx=1}^\Locnum \WeightEach \DecisionEach
    - \OverCost\sum_{\Locidx=1}^{\Locnum} \sum_{j=1}^l\PositivePart{\sum_{\Timeidx=1}^\TotalTime \size_{\Timeidx, \Locidx, j }\DecisionEach - \Capacity_{\Locidx, j } }
    - \hspace{-4em}
    & 
    \frac{\BuildUpCost}{\TotalTime} \sum_{\Timeidx=1}^\TotalTime\sum_{\Locidx=1}^{\Locnum}
    \sum_{j=1}^l b_{t,i,j}\\
    \textrm{s.t.} \ &
\sum_{\Timeidx=1}^\TotalTime\Indicator[\Target_\Timeidx=0]n_{\Timeidx, \Locidx, j}\Decision_{\Timeidx, \Locidx}
    \leq
    \left(
    \Capacity_{\Locidx, j} - 
    \sum_{\Timeidx=1}^\TotalTime \Indicator[\Target_\Timeidx=\Locidx]n_{\Timeidx, \Locidx, j}
    \right)_{+}, 
    &
    \forall \Locidx\in[\Locnum], j \in [l] \\ 
& \BuildUp_{\Timeidx, \Locidx, j} \geq \BuildUpti{\Timeidx-1}{\Locidx, j} + \size_{\Timeidx, \Locidx, j}\DecisionEach - {\Service_{\Timeidx, \Locidx, j }}, & \forall t\in[T], i\in [m], j \in [l]
\end{align*}
where we define $b_{0,\Locidx, j}=0$ for all $\Locidx \in [\Locnum]$ and $j \in [l]$ by convention. 

Given this benchmark, we define the same notion of the regret as \Cref{def:regret}. That is, we evaluate an algorithm by its worst-case performance over all instances, where each instance $\mathcal{I}$ consists of (i) the set of recourse types $(i,j) \in [\Locnum]\times [l]$, (ii) capacity ratios $({\CapRatio}_{i,j})_{(i,j) \in [\Locnum]\times [l]}$, and (iii) the (unknown) arrival type distribution $\ArrivalDist$. The regret of an online algorithm $\OnlineALG$ is then given by
\begin{equation}
   \Regret^\OnlineALG := \sup_{\mathcal{I}}~ \E\left[\OPT(\OverCost,\BuildUpCost) - \ALGObj{\OnlineALG}{\OverCost} {\BuildUpCost}\right].
\end{equation}
where the expectation is over the arrival distribution $\ArrivalDist$, the Bernoulli service process with a service rate vector $\mathbf{\ServiceRate}$, and (potential) randomness of the algorithm itself.

\subsection{Algorithms and Analysis}\label{subsec+generalized+algos}
Having described the set up of the generalized model, in this section, we explain how we modify our algorithms and their analysis for this generalized model. To avoid repetition, we mainly focus on our first algorithm (\Cref{ALG+Surrogate+D}) and provide the details.

\noindent\textbf{Congestion-Aware Dual Learning Algorithm (\CAM{}):} To understand how to modify \CA{} (\Cref{ALG+Surrogate+D}) for the generalized model, we write the dual of the optimal offline benchmark. Define $G_\TotalTime$ be the event where sample path $(\Arrival_t, \ServiceVec_t)_{t=1}^T$ satisfies $\sum_{\Timeidx=1}^\TotalTime \Indicator[\Target_\Timeidx =\Locidx]n_{t,i,j}\leq c_{i,j}$ for all $(i,j) \in [m]\times[l]$. Because we assumed that $\bar{n} = \Theta(1)$ and $\min_{\Locidx \in [\Locnum], j \in [l]} \CapRatio_{\Locidx,j} - \P[n_{\Timeidx,\Locidx, j}\Indicator[\Target_\Timeidx = \Locidx]] = \Theta(1)$, Azuma-Hoeffding inequality (and the union bound) implies that $G_\TotalTime$ occurs with probability $1-\BigO(e^{-\TotalTime})$ (we again suppress the dependence on $ml$ because they are constant by our assumption). Hence, in the same vein of \Cref{prop:offline}, we can characterize the dual program of the optimal offline benchmark as follows:

\begin{proposition}\label{prop:offline+2}
Define a per-period dual function $D_\Timeidx$ for case $\Timeidx$ as:
\begin{equation*}
    D_\Timeidx(\DualOverVec, \DualHardVec,\DualBuildVec) := 
\max_{\DecisionVec_\Timeidx \in \FeasibleSet{\Target_\Timeidx}}
\sum_{i=1}^\Locnum
\left\{
\left(
\WeightEach - 
\sum_{j=1}^l
{\size}_{\Timeidx, \Locidx, j}(\DualOver_{\Locidx, j} + \DualHard_{\Locidx, j} + \DualBuild_{\Timeidx, \Locidx, j})
\right)\DecisionEach
+\sum_{j=1}^l (\CapRatio_{\Locidx, j}\DualOver_{\Locidx,j} + 
\CapRatio_{\Locidx, j}\DualHard_{\Locidx, j} + 
\Service_{\Timeidx, i, j}\DualBuild_{\Timeidx, i, j})
\right\}
\end{equation*}

Further, for any given sample path $(\Arrival_t, \ServiceVec_t)_{t=1}^\TotalTime$, Consider the following dual program:
\begin{align}
\textsf{Dual}(\OverCost,\BuildUpCost) &:= \underset{\DualOverVec, \DualHardVec,\DualBuildVec_\Timeidx\geq \mathbf{0}}{\min}  \sum_{\Timeidx=1}^\TotalTime D_\Timeidx(\DualOverVec,
 \DualHardVec, \DualBuildVec_\Timeidx) \\
 \text {s.t.} \ 
& \DualOver_{\Locidx,j} \leq \alpha, & \forall i \in [m], j \in [l]  \\ 
 &\DualBuild_{\Timeidx,\Locidx, j} - \DualBuild_{\Timeidx+1,\Locidx, j} \leq \frac{\BuildUpCost}{\TotalTime} & \forall \Timeidx \in [\TotalTime-1], \Locidx\in[\Locnum], j \in [l] \\
& \DualBuild_{\TotalTime,\Locidx, j} \leq \frac{\BuildUpCost}{\TotalTime} & \forall\Locidx\in[\Locnum], j \in [l]
\end{align}

Then, we have the following strong duality: $\textsf{OPT}(\OverCost, \BuildUpCost)\Indicator[G_\TotalTime] = \textsf{Dual}(\OverCost, \BuildUpCost)\Indicator[G_\TotalTime]$ where $G_T$ is the event that sample path $(\Arrival_t, \ServiceVec_t)_{t=1}^T$ satisfies $\sum_{t=1}^\TotalTime \Indicator[\Target_t = i]n_{t,i,j} \leq c_{i,j}, \forall i\in[m], j\in[l]$.
\end{proposition}

The proof of \Cref{prop:offline+2} follows the same steps as \Cref{prop:offline} and is therefore omitted. As in \Cref{prop:offline}, the matrix $\DualOverVec \in \R^{m \times l}$ in \Cref{prop:offline+2} represents the dual variables for over-allocation costs, where $\DualOver_{i,j}$ corresponds to the resource-type $(i, j)$. The interpretation of other dual variables can be similarly understood in parallel with \Cref{prop:offline}. From \Cref{prop:offline+2}, we again observe that the optimal offline benchmark matches case $t$ to the affiliate that maximizes the adjusted score. However, the main difference from \Cref{prop:offline} is that adjustment for each affiliate $i$ is $\sum_{j=1}^l n_{t,i,j}(\theta_{i,j} + \lambda_{i,j} + \beta_{t,i,j})$. In words, when matched with affiliate $i$, the case consumes multiple units of multiple resource types, and therefore the opportunity cost of matching case $t$ to that affiliate should reflect that. 

Given \Cref{prop:offline+2}, we can now follow the same steps outlined in \Cref{sec:alg+design} to adapt \Cref{ALG+Surrogate+D} for the generalized model. Specifically, we consider the surrogate dual problem (analogous to \Cref{def:dual+surrogate}) and then apply the same online learning update rules to each dual variable in the surrogate dual problem.  We call this modified algorithm \CAM{}. For brevity, we focus on describing the key differences between \CAM{} and \Cref{ALG+Surrogate+D} below:

\smallskip
\begin{description}
    \item[\bf Primal Phase:] \CAM{} matches case $\Timeidx$ to affiliate $\Locidx$ with the maximum dual-adjusted score $\WeightEach - 
\sum_{j=1}^l
{\size}_{\Timeidx, \Locidx, j}(\DualOver_{\Timeidx, \Locidx, j} + \DualHard_{\Timeidx, \Locidx, j} + \eta \BuildUp_{\Timeidx-1,\Locidx,j})$ (breaking ties arbitrarily), as long as (i) the adjusted score is non-negative and (ii) every affiliate $i$ has remaining capacity for the static resource-type $(i,j)$. Otherwise, the case is matched to the dummy affiliate. Here the scaled build-up $ \zeta\BuildUp_{\Timeidx-1,\Locidx, j}$ again implicitly plays a role of the dual variable for the dynamic resource-type $(i,j)$. 

    \item[\bf Dual Phase:] The dual update rule (equation \eqref{eq:update+theta} in \Cref{ALG+Surrogate+D}) is replaced with 
    \begin{equation}
    \begin{split}
    \DualOver_{\Timeidx+1, \Locidx, j} &= \min\{\DualOver_{\Timeidx,\Locidx, j}\exp(\eta(\size_{\Timeidx, \Locidx, j}\DecisionEach-\CapRatio_{i, j})), \OverCost\} \\
    \DualHard_{\Timeidx+1,\Locidx, j} &= \min\left\{\DualHard_{\Timeidx,\Locidx, j}\exp(\eta(\size_{\Timeidx, \Locidx, j}\DecisionEach-\CapRatio_{\Locidx,})), \frac{1+2\OverCost}{\underline{\CapRatio}}\right\}.
    \end{split}\label{CA+new+dual+update}
\end{equation}
Analogous to equation \eqref{eq:beta+dynamics}, the dual variables for the dynamic resource are implicitly updated as
$\DualBuild_{\Timeidx+1, \Locidx, j} = \PositivePart{\DualBuild_{\Timeidx,\Locidx, j} + \zeta(\size_{\Timeidx, \Locidx, j}\DecisionEach - \Service_{\Timeidx, \Locidx, j})}$ with $\DualBuild_{0, \Locidx, j}=0$. Similar to our base model, we can show that $\DualBuild_{\Timeidx+1, \Locidx, j} = \zeta \BuildUp_{\Timeidx, \Locidx, j}$ using an induction argument.
\end{description}

With these modifications, the regret guarantees of \CA{} (\Cref{thm:ALG+Surrogate+Dual} and \Cref{near+part+a}) extend to \CAM{}. We summarize these analogous results in the following corollary.
\begin{corollary}[Regret of \CAM{}]\label{cor:CA+new}
Let $\eta = \Theta(1/\sqrt{\TotalTime})$ and $\zeta = \Theta(1/\sqrt{\TotalTime})$. Under the stable regime (\Cref{def:eps+regimes}), the regret of \CAM{} is $\BigO(\sqrt{\TotalTime} + \frac{\BuildUpCost}{\ServiceSlack})$. Furthermore, under the near-critical regime (\Cref{def:eps+regimes}), the regret of \CAM{} is $\BigO(\sqrt{\BuildUpCost\TotalTime})$ by setting $\eta = \Theta(1/\sqrt{\TotalTime})$ and $\zeta = \Theta(\sqrt{\BuildUpCost/\TotalTime})$. 
\end{corollary}

We remark that even though the regret bounds given in \Cref{cor:CA+new} have the same order as the ones given in \Cref{thm:ALG+Surrogate+Dual} and \Cref{near+part+a}, they are not the same because the new parameters $\bar{n}$ (the maximum number of units consumed for each static resource type by a case) and $l$ (the number of resource types) impact the bounds. However, under the natural assumption that these parameters are constant (i.e., do not increase with $\TotalTime$), they do not change the order of the regret.

The proof of \Cref{cor:CA+new} follows nearly identical steps to those outlined in \Cref{subsec:ALG+D+pfsketch}. The only change in the proof is in the definition of the pseudo-rewards due to the modification in the dual update rules. Specifically, for the backlog vector $\mathbf{b}_t = (b_{t,i,j})_{i \in [m], j \in [l]}$, we consider the quadratic potential function and its drift:
\begin{equation}
    \psi(\BuildUpVec_\Timeidx) := \frac{1}{2}\sum_{i=1}^m \sum_{j=1}^l b_{t,i,j}^2, \quad \textrm{}\quad 
    D_\Timeidx := \psi(\BuildUpVec_\Timeidx) - \psi(\BuildUpVec_{\Timeidx-1}).
\end{equation}

For the proof of \Cref{cor:CA+new}, we replace  the pseudo-rewards $K_t$ in equation \eqref{eq:K_t} with the following:
\begin{equation}
    K_t := \mathbf{w}_t \cdot \mathbf{z}_t  + \sum_{i=1}^m \sum_{j=1}^l \theta_{t,i,j}(\rho_{i,j} - n_{t,i,j}z_{t,i}) + \sum_{i=1}^m \sum_{j=1}^l \lambda_{t,i,j}(\rho_{i,j} - n_{t,i,j}z_{t,i}) - \zeta D_t.
\end{equation}

With the above definition of the pseudo-rewards, one can repeat the proofs of \Cref{lemma:pseudo+lower} and \Cref{lemma:pseudo+upper} to establish the lower and upper bound of the expected sum of the pseudo-rewards, leading to an analogous inequality to \eqref{ineq:important} (see \Cref{subsec:ALG+D+pfsketch}). The details are omitted for brevity. 



\textbf{Congestion-Oblivious Dual Learning Algorithm:}
One can apply the similar ideas in \Cref{subsec:ALG+P+design} to modify \CO{}  (\Cref{ALG+Surrogate+P}) for the generalized model. The modified algorithm, which we call \COM{}, is again identical to the \CAM{} except that (i) \COM{} only maintains the dual variables $\theta_{t,i,j}$ and $\lambda_{t,i,j}$ for each static resource-type $(i,j)$ and (ii) the fixed step size $\eta$ for the dual update rules in equation \eqref{CA+new+dual+update} is replaced with the time-varying step size $\eta_t = \Theta(1/\sqrt{t})$ for each period $t$. The regret guarantee of \Cref{thm:ALG+Surrogate+Primal} extends to \COM{}. That is, the regret  of \COM{} is $\BigO(\sqrt{\TotalTime} + \frac{\BuildUpCost}{\epsilon})$ under the stable regime. The proof again follows the identical steps as the proof of \Cref{thm:ALG+Surrogate+Primal} and hence is omitted for brevity.


\majorrevcolor{
\section{Numerical Results with Mid-year Disruption and Capacity Revisions}\label{apx+disrpution}

Our main case study in Section~\ref{sec:numerics} was based on data from 2014-2016. However, in the middle of fiscal year 2017, a change in federal administration led to significant shifts in refugee resettlement policy, including a sharp reduction and changes to the composition of arriving cases. These changes introduced substantial disruptions in the refugee arrival process. In this section, we examine the applicability of our proposed algorithms in the presence of such policy disruptions. To do so, we first provide brief background on the major changes to the U.S. refugee policy that took place in 2017. We then describe how we construct synthetic data that reflects those changes and simulate refugee arrivals under this disruption. Finally, we evaluate the performance of our algorithms on this synthetic data and show that they continue to yield meaningful improvements relative to the  benchmarks.

\smallskip 

{\bf Background}. 
In fiscal year 2017, a change in federal administration brought about significant disruptions to the U.S. refugee admissions system. The presidential ceiling on refugee admissions was significantly reduced, and this reduction was accompanied by a sharp decline in the number of free cases. For example, Executive Order 13769 restricted the admission of refugees from certain countries, with exceptions primarily granted to those with existing U.S. ties~\citep{howe2017travelban, advancingjustice2019muslimban}. As a result, both the total number and composition of refugee arrivals changed substantially during 2017. 


When such policy changes occur, resettlement agencies typically revise capacity for each affiliate based on the newly announced ceiling. While this revision process involves informal negotiation with the U.S. State Department, capacity adjustments are generally made in proportion to the original annual quota across affiliates. 

\smallskip 

{\bf Synthetic Data of Disruption}. 
Motivated by the above background, we construct a synthetic dataset that mimics the policy disruption that occurred in year 2017. To do so, we apply a synthetic “shock” to the 2016 arrival sequence. Let $T$ denote the total number of cases in the original dataset. We preserve the overall setup of our base case study (with unit-size case;  \Cref{subsec:data,subsec:results}). We use the actual number of cases resettled at each affiliate as its initial capacity $c_i$, and set the normalized capacity $\rho_i = c_i / T$ for each affiliate $i$. To simulate the disruption, we select a disruption time \( t_d \) and randomly remove 50\% of the free cases (those that were to arrive after \( t_d \)), while preserving the original arrival order among the remaining cases. This models both the drop in overall arrivals and the shift in composition similar to the 2017 policy changes. Let $\hat{T}$ denote the total number of arrivals after the disruption (including those before $t_d$). We assume that the algorithm is informed of $\hat{T}$ at the time of disruption. This reflects institutional practice, where a revised ceiling is publicly announced and used to revise capacities. Affiliate capacities are then proportionally updated to $\hat{c}_i = \rho_i \hat{T}$, in line with our partner agency’s typical approach to revising capacity proportionally to the initial ones.

\smallskip

{\bf Algorithm Evaluation and Numerical Results.} 
We focus on comparing the performance of \texttt{Sampling} and \texttt{RO-Learning}.\footnote{Because our simulation is based on synthetic data set, we do not know an actual placement that would have been made by our partner.} Both algorithms are run exactly as in the base case study~(\Cref{subsec:data}), with one exception: upon the disruption at time $t_d$, we replace the original capacity \( c_i \) with the revised capacity \( \hat{c}_i \). (For \texttt{RO-Learning}, the learning rates are left unchanged.) We set $t_d = T/3$ (resp.\ $t_d = 2T/3$) to model early (resp.\ late) disruption. 

\Cref{table:disruption-results} presents the numerical results. Comparing \texttt{RO-Learning} with \texttt{Sampling}, we note that \texttt{RO-Learning} achieves employment outcomes comparable to \texttt{Sampling}, while further reducing both over-allocation and average backlog. The main source of this performance improvement lies in the robust learning process of \texttt{RO-Learning} (as highlighted in \Cref{subsec:results}): after the disruption at \( t_d \), although the dual variables initially reflect pre-disruption arrivals, the underlying adversarial learning algorithm exhibits a powerful self-correcting property that quickly adapts to changes in the arrival pattern.\footnote{
\majorrevcolor{In fact, we implemented an alternate version of the algorithm that explicitly ``restarts'' at the time of disruption by resetting the dual variables to their default initial values (i.e., \( e^{-1} \)), and found that its performance rarely differs from the original version. From a theoretical standpoint, the initial dual variables affects regret guarantees only through a constant multiplicative factor and thus not impact the asymptotic order of regret~\citep{hazan2016introduction}.}} This fast adaptation is what enables \texttt{RO-Learning} to maintain strong performance even under sudden structural shifts. 

 
\begin{table}[htp]
    \centering
    \begingroup
   \majorrevcolor{
    \caption{
   \majorrevcolor{Numerical performance under synthetic disruption (early vs. late reduction in free cases, 2016). We use the same penalty cost parameter $\alpha =3 $ and $\gamma = 5$ as in \Cref{fig:performance}. Early (late, resp.) disruption randomly reduces the remaining number of free cases by half at $t_d = T/3$ (resp, $t_d = 2T/3$). Each cell shows the result under early disruption with the late disruption result in parentheses. The results are averaged over 10 simulation runs (each involving both the random disruption and the internal randomness of \texttt{Sampling}).} 
    }
    \label{table:disruption-results}
    \begin{tabular}{@{}lccc@{}}
        \toprule
        & \makecell{\textbf{Employment}  \textbf{Rate (\%)}} 
        & \makecell{\textbf{Total} \textbf{Over-} \textbf{allocation}} 
        & \makecell{\textbf{Average}  \textbf{Backlog}} \\
        \midrule
        \texttt{Sampling}    & 43.7 (45.5) & 191.1 (149.4)  & 149.9 (179.3) \\
        \texttt{RO-Learning} & 43.7 (45.3) & 190.2 (129.8)  & 140.4 (160.2) \\
        \bottomrule
    \end{tabular}
    }
    \endgroup
\end{table}

}

\section{Incorporating Batched Arrivals}\label{apx+batching}

In our main model, we have assumed that refugee cases arrive one by one (which is true for Switzerland or the Netherlands). However, in some countries (such as the U.S.), the resettlement agency can encounter a periodic batch of refugee arrivals (e.g., weekly), for which matching decisions can be made simultaneously. In this section, we discuss how our main algorithm can be modified to incorporate batched arrivals and explore the impact of such modifications. 

Before going forward, we highlight that our main algorithms (in particular \texttt{RO-Learning} in \Cref{sec:numerics}) can still be implemented in practice for batched arrivals by simply ignoring batching and matching the cases within the batch one-by-one. We do not expect the loss from making decisions one by one to be substantial, partly because, in practice, batch sizes are relatively small. Specifically, even though our partner manages a weekly batch, the placement officer typically splits it into 20-30 cases. This is because finalizing a matching decision often requires communication with the corresponding affiliate, making it overwhelming for the officer to handle a larger number of cases at once.

That said, batched matching with even moderate batch sizes may offer an opportunity for performance improvement. To explore this opportunity, we propose a modification of \texttt{RO-Learning} --- an adaptation of \CA{} for our partner agency (see \Cref{sec:numerics}) --- to incorporate batching. Before proceeding, we remark that developing a primal-dual algorithm with adversarial online learning for batched arrivals is not straightforward and has not been studied in the literature (to the best of our knowledge).

\subsection{Model}\label{apx+batching+model}
 We first begin by describing our model under batching. In describing the model and algorithm, we specifically restrict our attention to the same setup as in the numerical case study (\Cref{sec:numerics}). In particular, we consider a deterministic service flow with service rate $\CapRatioVec$ (which is currently used by our partner to measure congestion). 

We consider a discrete-time model with $\TotalTime$ periods. There is a sequence of $\TotalTime$ cases, which arrive in $K$ batches. For each batch $k \in [K]$, there are $B_k$ cases. We index each arrival in batch $k$ by $\mathcal{T}_k = \{t_k, t_k+1, \ldots, t_k + B_k -1\}$, where $t_1 = 1$ and $t_k = \sum_{j=1}^{k-1} B_j$ for $k \geq 2$. At the beginning of period $t_k$, batch $k$ arrives, and we observe their types $\{\Arrival_t\}_{t \in \mathcal{T}_k}$. The agency can make matching decisions $\{\mathbf{z}_t\}_{t \in \mathcal{T}_k}$ for cases within batch $k$ simultaneously. However, {for simplicity,} we assume that cases arrive at their matched affiliate sequentially and in the order of their indices; that is, case $t \in \mathcal{T}_k$ arrives at period $t$. This assumption is aligned with practice {as it is unlikely that several cases can simultaneously arrive at the affiliates.} 
%
Hence, the backlog dynamic $\Timeidx$ remains the same and is given by
\begin{equation}
    b_{t,i} = (b_{t-1,i} + z_{t,i} - \rho_i)_{+}
\end{equation}
for each $t \in [T]$ and $i \in [m]$. The objective is  again given by equation \eqref{ALG+obj}.

\subsection{Warm-up: \texttt{RO-Learning-B}}\label{subsubsec:naive}
We begin with a natural yet naive modification of \texttt{RO-Learning}. Recall that \texttt{RO-Learning} finds the matching decision for case $t$ by maximizing its adjusted score, subject to the capacity constraint. The adjusted score was defined by the reward subtracted by (i) the current estimate of dual variables for the static resource and (ii) the scaled level of the current backlog. Hence, a natural modification, which we call \texttt{RO-Learning-B} (with B standing for batching), is as follows: find the matching decisions for batch $k$ by maximizing the sum of the adjusted scores within the batch (subject to the capacity constraint), given the current estimate of dual variables. Formally, let $\mathbf{c}_k = (c_{k,i})_{i \in [m]}$ denote the current capacity at time $t_k$ (i.e., upon the arrival of batch $k$). We further use $\boldsymbol{\theta}$ and $\boldsymbol{\lambda}$ to denote the current estimate of dual variables for the static resource (for brevity, we omit the time index for the dual variables). Then \texttt{RO-Learning-B} implements the following primal and dual phases:

\smallskip

\begin{description}
    \item[{\bf Primal Phase:}] Find the matching decisions $\{\mathbf{z}_t\}_{t \in \mathcal{T}_k}$ for batch $k$ by solving:\footnote{It is straightforward to see that the original capacity constraint (i.e., \eqref{hard+constraint:t} for all $t \in \mathcal{T}_k$) is satisfied if and only if the constraint \eqref{batching-cap-constraint} is satisfied. }
\noindent\begin{align}
        \{{\DecisionVec}_t\}_{t \in \mathcal{T}_k} &= 
        \argmax_{
        \substack{
    \DecisionVec_t \in \mathcal{X}(\Target_\Timeidx), \\    \forall t \in \mathcal{T}_k}
        } 
{\sum_{t \in \mathcal{T}_k}}
(\mathbf{w}_t - \boldsymbol{\theta} - \boldsymbol{\lambda} - \Indicator[t = t_k]\mathbf{b}_{t-1})\cdot {\DecisionVec}_{t,i}  \label{batching-obj}\\ 
        \text{s.t.} \ &   
        \sum_{t \in \mathcal{T}_k}
        \Indicator[\Target_t = 0]\Decision_{t,i} \leq \Big(c_{k,i} - \sum_{t \in \mathcal{T}_k} \Indicator[\Target_t = i] \Big)_{+},  \quad \forall \Locidx \in [m]. \label{batching-cap-constraint} \end{align}
where where $\mathcal{X}(\Target) = \Simplex$ if $\Target = 0$ and $\mathcal{X}(\Target) =\{\BasisVec{\Target}\}$ otherwise (recall that \texttt{RO-Learning} matches every case to an actual affiliate --- see \Cref{sec:numerics}).

\item[{\bf Dual Phase:}] Based on the primal matching decision, we obtain the ``batched’’ gradient, which is simply the summation of the gradient information $\mathbf{z}_t - \CapRatioVec$ over all arrival $\Timeidx \in \mathcal{T}_k$. We then perform the multiplicative update (similar to \Cref{ALG+Surrogate+D}) using this batched gradient:
    \begin{equation}
    \begin{split}
    \DualOver_{i} &\leftarrow \min\Big\{\DualOver_{\Locidx}\exp\Big(\eta \sum_{\Timeidx \in \mathcal{T}_k}({z}_{t,i}-\CapRatio_i)\Big), \OverCost\Big\} \\
    \DualHard_{\Locidx} &\leftarrow \min\Big\{\DualHard_{\Locidx}\exp\Big(\eta \sum_{\Timeidx \in \mathcal{T}_k}({z}_{t,i}-\CapRatio_i)\Big), \frac{1+2\OverCost}{\underline{\CapRatio}}\Big\}   
    \end{split} \label{batching-dual-update}
\end{equation}
\end{description}

\smallskip

We evaluate \texttt{RO-Learning-B} in \Cref{table:performance+batch}. \SLedit{For the numerical evaluation, we preserve the overall setup of our base case study (with unit-size case;  Section~\ref{sec:numerics}), but  we use $B_k=30$ (similar to the current practice)} and the same penalty parameters and the step sizes in \Cref{sec:numerics} (see \Cref{fig:performance}). 
Notably, the heuristic performs significantly worse than our original algorithm (which ignores batching) in terms of the objective (equation \eqref{ALG+obj}), primarily due to a higher over-allocation and an increased average backlog.

{In fact, despite its intuitive appeal, the above heuristic has two potential issues. First, the primal and dual variables are not updated sufficiently. To understand how this can lead to underperformance, consider the extreme case of having only one batch. In this case, the current backlog $\mathbf{b}_0$ is simply zero, and the initial dual variables $(\boldsymbol{\theta}, \boldsymbol{\lambda})$ are arbitrarily set. As a result, not only can the primal decisions from the primal phase  be highly inefficient, but the updated dual variables from the dual phase are also never used to update these inefficient primal decisions. This suggests the need for multiple iterations of the primal-dual phase within each batch, as typically done in the batch gradient descent \citep{ruder2016overview}. Second, the objective function of \eqref{batching-obj} does not account for the backlog dynamics within each batch. In summary, these two issues suggest that iterating the primal-dual phases multiple times per batch, as well as incorporating the backlog dynamics within each batch, may improve performance.}




\subsubsection{Main Heuristic: \texttt{RO-Learning-B-Iterate}}\label{apx+batch+b+iterate}\hfill\\
Motivated by the above observations, we propose a more sophisticated modification of \texttt{RO-Learning}, called \texttt{RO-Learning-B-Iterate}. The formal description of the algorithm is presented in \Cref{ALG+batching}. For brevity, we only describe the main changes from \texttt{RO-learning-B}. The most significant changes are that we account for the backlog induced within the current batch and iterate the primal and dual phases $L>1$ times. We further introduce the following modifications in the primal and dual phases:

\smallskip

\begin{description}
    \item[{\bf Primal Phase (line \ref{line:batching+primal}-\ref{line:batching+primal_end}):}]  The objective function for finding the ``best'' matching $\{\mathbf{z}_t^{(l)}\}_{t \in \mathcal{T}_k}$ (for the $l$-th iteration of batch $k$) is now the sum of the adjusted scores, penalized by the cost of over-allocation and average backlog within the batch (see line \ref{line:batching+primal_end}). In other words, we leverage the knowledge of the arrivals within the current batch (along with the deterministic service flow) to incorporate the average backlog induced by their matching decisions.
\item[\bf Dual Phase (line \ref{batching+dual+phase}):] We perform the same dual updates as described in equation \eqref{batching-dual-update}. However, we divide the original step size $\eta$ by the number of iterations $L$ to account for the multiple iterations within the batch.
\end{description}

\smallskip

Finally, we set the actual matching decisions for batch $k$ to be $\{\mathbf{z}_t^{(L)}\}_{t \in \mathcal{T}_k}$, the final iterate of the primal phase after $L$ iterations (line \ref{batching:round}). Specifically, we select the maximum coordinate of $\mathbf{z}_t^{(L)}$ as the actual affiliate to which case $t$ will be matched.


\begin{algorithm}[htp]
\caption{\texttt{RO-Learning-B-Iterate}}\label{ALG+batching}
\begin{algorithmic}[1]
    \STATE {\textbf{Input:} $\TotalTime$, $\CapRatioVec$, $\eta$, $\zeta$, and $L$.} 
    \textcolor{blue}{\texttt{// $L$ is the number of iterations for primal-dual phase}}
    \STATE Initialize $\DualOver_{\Locidx} = \exp(-1)$, $\DualHard_{\Locidx} = \exp(-1)$, 
    $\Capacity_{0,\Locidx} = \CapRatio_\Locidx \TotalTime$,   $\BuildUpti{0}{\Locidx}=0$ for all $\Locidx \in [\Locnum]$, and $t_1 = 1$.
    
    \textcolor{blue}{\texttt{// For brevity, we omit time index for the dual variables }}

    \FOR{Each batch $k \in [K]$}
    \STATE Set the arrival index for batch $k$ as $\mathcal{T}_k = \{t_k, t_k+1, ..., t_k+B_k-1\}$
    \FOR{Each iteration $l \in [L]$} 
    \STATE \label{line:batching+primal} \textbf{Primal Phase}: Set the dual-adjusted score as 
    \begin{equation}
      \overline{\mathbf{w}}_t := \mathbf{w}_t - \DualOverVec - \DualHardVec - \zeta \BuildUpVec_{\Timeidx-1}\Indicator[t = t_k], \quad \forall t \in \mathcal{T}_k. \label{EC_line:adjusted+score} 
    \end{equation}
    \vspace{-1cm}
    \STATE \label{line:batching+primal_end} Solve the following program:
\noindent\begin{align*}
        \{\tilde{\DecisionVec}_t^{(l)}, \tilde{\BuildUpVec}_t^{(l)}\}_{t \in \mathcal{T}_k} &= 
        \argmax_{
        \substack{
    \DecisionVec_t^{(l)} \in \mathcal{X}(\Target_\Timeidx),     \forall t \in \mathcal{T}_k \\{\BuildUpVec}_t^{(l)} \geq \mathbf{0},     \forall t \in \mathcal{T}_k
        }
        } 
\sum_{t \in \mathcal{T}_k}\overline{\mathbf{w}}_t\cdot {\DecisionVec}_{t,i}^{(l)} 
        -\OverCost\Big(\sum_{t \in \mathcal{T}_k}\sum_{i \in [m]} \Decision_{t,i}^{(l)} - c_{k,i}\Big)_{+} 
        \hspace{-0.6cm}
         & - \frac{\BuildUpCost}{B_k}\sum_{t \in \mathcal{T}_k}
            \sum_{i \in [m]} b_{t,i} \\ 
        \text{s.t.} \ &   
        \sum_{t \in \mathcal{T}_k}
        \Indicator[\Target_t = 0]\Decision_{t,i}^{(l)} \leq \Big(c_{k,i} - \sum_{t \in \mathcal{T}_k} \Indicator[\Target_t = i] \Big)_{+}, & \forall \Locidx \in [m]  \\
        &  b_{t,i}^{(l)} \geq b_{t-1,i}^{(l)} + z_{t,i}^{(l)} - \rho_i,  & \forall \Timeidx \in \mathcal{T}_k, \Locidx \in [m] 
        \end{align*}
where $\mathcal{X}(\Target) = \Simplex$ if $\Target = 0$ and $\mathcal{X}(\Target) =\{\BasisVec{\Target}\}$ otherwise. \textcolor{blue}{\texttt{// Every case is matched to an actual affiliate consistent with the case study (\Cref{sec:numerics})}}
    \STATE  \label{batching+dual+phase}\textbf{Dual Phase}: update the dual variables as 
    \begin{equation}
    \begin{split}
     \DualOver_{\Locidx} &\leftarrow \min\Big\{\DualOver_{\Locidx}\exp\Big(\frac{\eta}{L}\sum_{\Timeidx \in \mathcal{T}_k}(\tilde{z}_{t,i}^{(l)}-\CapRatio_i)\Big), \OverCost\Big\} \\
    \DualHard_{\Locidx} &\leftarrow \min\Big\{\DualHard_{\Locidx}\exp\Big(\frac{\eta}{L}\sum_{\Timeidx \in \mathcal{T}_k}(\tilde{z}_{t,i}^{(l)}-\CapRatio_i)\Big), \frac{1+2\OverCost}{\underline{\CapRatio}}\Big\}   
    \end{split}
\end{equation}
    \ENDFOR
    \STATE \label{batching:round} Set the primal decisions: ${z}_{t,i} = \Indicator[i = \argmax_{j \in [m]} \tilde{z}_{t,j}^{(L)}] \quad \forall t \in \mathcal{T}_k, i \in [m]$ \textcolor{blue}{\texttt{// Use the primal decision of the last iterate to make the actual matching decision.}}
    \STATE Update the remaining capacity: $\Capacity_{k, \Locidx} \leftarrow \Capacity_{k-1, \Locidx} - \sum_{t \in \mathcal{T}_k}\DecisionEach \quad \forall \Locidx \in [\Locnum]$
\ENDFOR
\end{algorithmic}
\vspace{-0.1cm}
\end{algorithm}

The numerical performance of \texttt{RO-Learning-B-Iterate} is presented in \Cref{table:performance+batch}. We use $L=10$ and the same value of $\eta$ used in \Cref{sec:numerics}. We note that \texttt{RO-Learning-B-Iterate} performs slightly better than \texttt{RO-Learning} in terms of the objective value (by 3\% for both years 2015 and 2016). This mild improvement is driven by a reduction in the average backlog (by 4-7\%) and/or the over-allocation, with a small decrease in the employment rate (less than 2\%). 
This numerical result suggests that leveraging batched arrivals in the setting of our case study may yield a small performance gain. However, we emphasize that the algorithm required to achieve such a gain is significantly more complex and less interpretable. This level of complexity appears necessary for performance improvement: as discussed in \Cref{subsubsec:naive}, our more straightforward yet naive modification of \texttt{RO-Learning} (\texttt{RO-Learning-B}) indeed hurts performance. Given this, it is unclear whether the additional complexities are justified by such a small gain.


\begin{table}[htp]
    \centering
    \caption{
    Numerical Performance for year 2015 (2016, resp.)
    of \texttt{RO-Learning}, with and without batched matching. We use penalty parameters $\boldsymbol{\OverCost=3}$ and $\boldsymbol{\BuildUpCost=5}$ (as used for \Cref{fig:performance} in \Cref{sec:numerics}) to evaluate the objective defined in equation \eqref{ALG+obj}. We use the batch size of 30 cases per batch. For \texttt{RO-Learning-B-Iterate}, we use $\boldsymbol{L=10}$ and the same value of $\boldsymbol{\eta}$ as used for \texttt{RO-Learning} in \Cref{apx+case+ALG+D}. The total number of cases is $\boldsymbol{T=3819}$ (4950, resp.) for year 2015 (2016, resp.).
    }. 
    \label{table:performance+batch}
    \begin{tabular}{@{}lcccc@{}}
        \toprule
        & \makecell{\textbf{Employment} \\ \textbf{Rate (\%)}} 
        & \makecell{\textbf{Total} \\ \textbf{Over-} \textbf{allocation}} 
        & \makecell{\textbf{Average} \\ \textbf{Backlog}} 
        & \makecell{\textbf{Objective} \\
        \textbf{(in eq. \eqref{ALG+obj}})} \\
        \midrule
        \texttt{RO-Learning} & 44.6 (46) & 71 (81) & 151 (199) & 737 (1054.3) \\
        \texttt{RO-Learning-B} & 44.4 (45.6) & 104 (120) & 192.7 (248.6) & 423.2 (667.4) 
        \\
        \texttt{RO-Learning-B-Iterate} & 43.9 (45.5) & 73 (72) & 140.5 (191.4) & 757.6 (1090.9) \\
        \bottomrule
    \end{tabular}
\end{table}


\majorrevcolor{
\section{An Alternative Service Model with Server Idleness}\label{apx+server+idleness}

In our main model, we have assumed that the service availability sequence $\{\mathbf{s}_t\}_{t=1}^T$ consists of i.i.d. Bernoulli random variables with mean $r_i = \rho + \epsilon$. In particular, this assumption implies that the server does not necessarily remain idle if it becomes available when no case is waiting. This modeling choice is relevant in settings where servers may initiate other tasks when the system is empty. For example, case workers in our partner may assist other teams when there is no case to serve.\SLcomment{\@ EP: Is the last sentence a fair example?}

In other contexts, however, a server may remain idle when the queue is empty, allowing them to serve the next arriving case without delay. In this section, we describe an alternative service model that explicitly allows for such idleness, and we extend the  impossibility result of \Cref{prop:impossibility}. Section~\ref{sec:alternative-model} formally defines the idle-server model. At first glance, one might hope that this idleness  would help an online algorithm better manage its backlogs—perhaps weakening the impossibility result stated in Proposition~\ref{prop:impossibility}. However, in Section~\ref{sec:alternative-lower-bound}, we show that even under this more favorable regime, no online algorithm \SLedit{making integral decisions} can achieve sublinear regret when the congestion penalty parameter $\gamma$ is $\Omega(T)$.
Finally, in \Cref{subsec:upper+idle}, we show that the regret upper bounds for \CA{} (\Cref{thm:ALG+Surrogate+Dual} and \Cref{apx+near+a}) and \CO{} extend to the alternative model with server idleness. Thus, \CA{} continues to achieve sublinear regret in all regimes where it is possible (i.e., when \( \gamma = o(T) \)) under the alterative service model.

\SLcomment{
The impossibility result in this section relies—albeit in a subtle way—on the algorithm making integral allocation decisions (i.e., \( z_t \in \{0,1\} \)). I attempted to extend the argument to fractional decisions but wasn’t able to complete it. For now, I’ve clarified in footnote~\ref{footnote:integral} that the result is stated for integral algorithms only.
}
\subsection{Model}
\label{sec:alternative-model}

{\bf Model}. We now formally describe an extension of our base model that incorporates server idleness. As in the base model, a server at each affiliate $i$ becomes newly available with probability $r_i = \rho_i + \epsilon$, modeled by an i.i.d.\ Bernoulli random variable $s_{t,i} \sim \texttt{Ber}(r_i)$. We refer to this as a \emph{fresh service availability}. If a server becomes newly available when there is no case to serve, it enters the idle state and remains idle until a new case arrives. 
To capture extra availability from idleness, we introduce a binary idleness state variable $\ell_{t,i} \in \{0,1\}$, where $\ell_{t,i} = 1$ if the server has remained idle and thus is immediately available for service for case $t$. 
 We emphasize that $s_{t,i}$ and $\ell_{t,i}$ refer to different modes of availability for the same server—not to separate servers.

We now formalize the service dynamics. 
Each period $t$ begins with the arrival of a case $t$ with type $\mathbf{A}_t = (\mathbf{w}_t, \Target_t)$. The algorithm immediately selects a matching decision $\DecisionVec_t \in \mathcal{Z}^\text{int}(\Target_t)$ for case $t$. We define a \emph{pre-processed backlog} $q_{t,i}$, which is updated as
\[
q_{t,i} := b_{t-1,i} + z_{t,i}.
\]
Next, for each affiliate $i$, we draw a fresh service availability $s_{t,i} \sim \text{Bernoulli}(\rho_i + \epsilon)$, indicating whether the server at affiliate $i$ becomes newly available during period~$t$. 
A server is able to serve a case at period~$t$ if it has become newly available (i.e., $s_{t,i} = 1$) or has already been idle (i.e., $\ell_{t,i} = 1$). When both conditions $s_{t,i} = 1$ and $\ell_{t,i} = 1$ hold, the case is served using the server’s idle status, and the fresh service opportunity $s_{t,i}$ does not carry over. This naturally reflects that an idle server immediately becomes busy once it serves an arriving case.\footnote{One could equivalently define the model by drawing $s_{t,i}$ only when $\ell_{t,i} = 0$. However, for simplicity of exposition—particularly to define the offline benchmark in \Cref{def:offline+idle} in a consistent way—we draw $s_{t,i}$ at every time step, regardless of the server’s idle status. Alternatively, one can interpret $s_{t,i}$ as an indicator that the server \emph{would have become newly available} at time $t$ had it not already been idle.} Formally, we define the \emph{effective availability} as:
\begin{equation}
u_{t,i} := \ell_{t,i} + (1 - \ell_{t,i}) \cdot s_{t,i},    \label{eq:u_t}
\end{equation}
and update the backlog (after processing case $t$) as:
\begin{equation}
b_{t,i} = (q_{t,i} - u_{t,i})_{+} = (b_{t-1,i} + z_{t,i} - u_{t,i})_{+}. \label{backlog:dynamics+idle}
\end{equation}

We now describe how the server’s idleness state is updated. A server enters or remains idle at period $t+1$ if and only if (i) it is not used to serve a case at time $t$ and (ii) is available in some form—either because it was already idle, or because it has just become newly available and found no case awaiting service. On the other hand, a server's idle status will reset to zero whenever it serves a case. Concretely, we have:
\begin{itemize}
    \item[$\bullet$] If the server was idle ($\ell_{t,i} = 1$) and there is no case to serve ($q_{t,i} = 0$), it remains idle ($\ell_{t+1,i} = 1$).
    \item[$\bullet$] 
    If no case was waiting ($q_{t,i} = 0$) and the server became newly available ($s_{t,i} = 1$), it enters the idle state ($\ell_{t+1,i} = 1$).
    \item[$\bullet$] If an idle server serves an arriving case ($\ell_{t,i} = 1$, $q_{t,i} \geq 1$), it becomes unavailable and the idle status resets to zero ($l_{t+1,i} = 0$).
\end{itemize}
Combining the above, we can succinctly write the idleness dynamics as:
\begin{equation}
  \ell_{t+1,i} = \mathbbm{1}[\text{$q_{t,i} = 0$ and $(\ell_{t,i}=1$ or $s_{t,i}=1)$}]. \label{eq:idle+dynamics}  
\end{equation}


\smallskip 

{\bf Decision, Benchmark, and Regret.} An online algorithm \( \pi \) makes an immediate and irrevocable matching decision \( \mathbf{z}_t^\pi \in \mathcal{Z}(\Target_t) \) for each arriving case \( t \). This decision may depend on the history \( \mathcal{H}_{t-1} = \{ \mathbf{A}_\tau, \mathbf{s}_\tau, \mathbf{z}_\tau^\pi \}_{\tau=1}^{t-1} \) and the current arrival type \( \mathbf{A}_t \), but not on the current realization of fresh service availability \( \mathbf{s}_t \). We further allow the algorithm to observe the server's idle status \( \boldsymbol{\ell}_t \) when making its decision, which is well defined since \( \boldsymbol{\ell}_t \) is determined by prior state variables $\ell_{t-1,i}$ and $\mathcal{H}_{t-1}$ through \eqref{eq:idle+dynamics}. The objective value of algorithm $\pi$ on a given sample path of arrivals and services, with penalty parameters $(\alpha, \gamma)$, is denoted by $\textsf{ALG}^\pi(\alpha, \gamma)$ as in~\eqref{ALG+obj}.

Similar to \Cref{def:offline}, we define the offline benchmark as an algorithm that observes the full arrival sequence $\{\mathbf{A}_t\}_{t=1}^T$ and the fresh service availability sequence $\{\mathbf{s}_t\}_{t=1}^T$. A key difference from the base model in \Cref{sec:model}, however, is that the backlog dynamics now depend on the idle status of servers, which is endogenously determined by the offline's own matching decisions. To capture this dependency, we augment the offline optimization program to include idle-state dynamics, as formally defined below. 

\begin{definition}[Offline Benchmark with Server Idleness]\label{def:offline+idle}
Given the full sample path $\{\mathbf{A}_t, \mathbf{s}_t\}_{t=1}^T$ of arrivals and fresh service availabilities, the offline benchmark solves the following optimization problem:
\begin{alignat}{4}
&\OPT(\alpha, \gamma) :=   
& \! \max_{\substack{
\mathbf{z}_t \in \mathcal{Z}(\Target_t) \\
\mathbf{b}_t \geq \mathbf{0} 
}}
& \sum_{t=1}^T \mathbf{w}_t \cdot \mathbf{z}_t
    - \alpha \sum_{i=1}^{m}
    \left( \sum_{t=1}^T z_{t,i} - c_i \right)_{+}
    - \frac{\gamma}{T} \sum_{t=1}^T \sum_{i=1}^m b_{t,i}
\nonumber \\
&
&\textrm{s.t.} \quad
& \sum_{t=1}^T \Indicator[\Target_t = 0] z_{t,i} \leq \left( c_i - \sum_{t=1}^T \Indicator[\Target_t = i] z_{t,i} \right)_{+}
\quad \forall i \in [m] \tag{\textsf{Capacity Feasibility}} \\
&
& 
& b_{t,i} \geq b_{t-1,i} + z_{t,i} - u_{t,i}
\quad \forall t \in [T], \; i \in [m] \tag{\textsf{Backlog Dynamics}} \\
&
&
& u_{t,i} = \ell_{t,i} + (1 - \ell_{t,i}) \cdot s_{t,i} \tag{\textsf{Effective Availability}} \\
&
&
& \ell_{t+1,i} = \mathbbm{1}[\text{$q_{t,i} = 0$ and $(\ell_{t,i}=1$ or $s_{t,i}=1)$}]
\quad \forall t < T, \; i \in [m] \tag{\textsf{Idle-State Update}}
\end{alignat}
with initial conditions $b_{0,i} = 0$ and $\ell_{1,i} = 0$ for all $i \in [m]$.
\end{definition}

With the above definition of the offline benchmark, we define a regret of an online algorithm $\pi$ the same way as in  \Cref{def:regret}.

\subsection{Impossibility Result under Alternative Service Model}\label{apx+alternative+result}

We now extend the impossibility result in \Cref{prop:impossibility} to the  model with server idleness. 

\label{sec:alternative-lower-bound}

\begin{proposition}[Lower Bound on Achievable Regret with Server Idleness]\label{prop:impossibility+idle}
For $\BuildUpCost = \Omega(\TotalTime)$ and any service slack parameter \( \epsilon \geq 0 \) such that the resulting service rates satisfy \( r_i \in [\rho_i + \epsilon,\ 1) \) for all \( i \in [m] \), there exists an instance for which the regret of any online algorithm {making integral decisions} is $\Omega(\TotalTime)$ under the model with server idleness.\footnote{While our original impossibility result (Proposition~1) does not rely on the integrality assumption, we assume integral decisions (\( z_t \in \{0,1\} \)) for technical clarity in Proposition~2. This  ensures that the backlog evolves in discrete units, which simplifies our analysis—particularly in deriving certain inequalities such as \eqref{ineq:integrality+ref+1} and \eqref{ineq:integrality+ref+2}. With more refined analysis, we conjecture that a similar result would hold even without the integrality assumption.\label{footnote:integral}} 
\end{proposition}\par

To prove \Cref{prop:impossibility+idle}, we consider the same instance used in the proof of \Cref{prop:impossibility} (which we re-iterate below for completeness). Under the alternative service model, the server remains idle when it becomes available and finds no case to serve. This appears to offer an online  algorithm additional flexibility, especially when it can also observe the idle status. However, the potential benefit of idleness is fundamentally constrained by a key tradeoff: idle periods can only arise when the algorithm chooses \emph{not} to match a case (to an actual affiliate). Yet, to avoid linear regret, the algorithm must match almost as many cases as the benchmark. These two objectives—preserving idle periods and collecting enough reward—are directly in conflict. Building on this intuition, we show that if the algorithm matches enough cases to avoid linear regret, then idle periods become too scarce to meaningfully reduce congestion, and the algorithm must once again incur a constant average backlog in the similar vein of \Cref{prop:impossibility}. 

\smallskip 

{\bf Proof of \Cref{prop:impossibility+idle}}. 
We consider an instance with $m=1$. For brevity, we omit the subscript for $\Locidx$. There are $\TotalTime$ arrivals with deterministic reward $\Weight_\Timeidx=1$ for all $\Timeidx \in [\TotalTime]$. There are no tied cases and the capacity is $\Capacity = 0.5\TotalTime$. The service rate is $\ServiceRate = 0.5 + \ServiceSlack$ where $\ServiceSlack \in [0, 0.5)$. Note that we require $\epsilon < 0.5$ to exclude a trivial regime where service rate is $1$. Thus, we assume that $0.5-\epsilon$ is a constant (independent of $T$) 
bounded away from zero.  We further recall that $\alpha$ does not play any role here. Hence, we denote the objective value of offline benchmark (\Cref{def:offline+idle}) and algorithm by $\OPT(\BuildUpCost)$ and $\textsf{ALG}(\gamma)$, respectively.

Similar to Claim-\ref{claim:hardness+opt}, we first lower-bound the value of the offline benchmark.
\begin{claim}\label{claim:hardness+opt+idle}
For any $\epsilon \in [0, 0.5)$, the value of offline benchmark under the idle-server model (\Cref{def:offline+idle}) satisfies
\[
\E[\OPT(\gamma)] \geq 0.5T - \Theta(\sqrt{T}).
\]
\end{claim}

{\bf Proof of Claim \ref{claim:hardness+opt+idle}}
The proof mirrors Claim~\ref{claim:hardness+opt}, with a minor adjustment. Consider the following solution: match a case at time $t$ to the actual affiliate ($z_t=1$) if and only if $s_t = 1$, until reaching the capacity limit $\sum_t z_t \leq 0.5T$. This is feasible by construction. Note that from equation \eqref{eq:u_t}, a  server is always available when $s_t = 1$: either via fresh availability or by having remained idle from an earlier period. Thus, this feasible solution does not incur any backlog. The rest of the proof follows the same line of argument in Claim-\ref{claim:hardness+opt} and is thus omitted. 
\hfill\halmos

Fix any online algorithm \SLedit{with integral decisions}, and let \SLedit{$z_t \in \{0,1\}$ denote its matching decision at time $t$. Since $z_t$ is binary, the backlog evolves in integer increments and remains a non-negative integer.} By Claim-\ref{claim:hardness+opt+idle}, the offline benchmark collects approximately $0.5T$ in reward with zero backlog. Hence, to achieve sublinear regret, the algorithm must accept at least $0.5T - o(T)$ cases in expectation. Thus, without loss of generality, we assume:
\begin{equation}
0.5T - o(T) \leq \E\left[\sum_{t=1}^T \mathbbm{1}[z_t = 1] \right] \leq 0.5T, \label{eq:reward}
\end{equation}
where the last inequality is simply just due to the capacity constraint. 

We now show that under \eqref{eq:reward}, any such online  algorithm must incur a constant average backlog:
\begin{equation}
\frac{1}{T} \sum_{t=1}^T \E[b_t] \geq \Omega(1). \label{eq:const-backlog}
\end{equation}

Combining \eqref{eq:reward}, \eqref{eq:const-backlog}, and  Claim-\ref{claim:hardness+opt+idle}, we obtain:
\begin{equation}
\textsf{OPT}(\gamma) - \textsf{ALG}(\gamma) \geq \Omega(\gamma - \sqrt{T}).
\end{equation}
Thus, whenever $\gamma = \Omega(T)$, the regret of any online algorithm is $\Omega(T)$.

We now prove \eqref{eq:const-backlog} by contradiction. Assume instead that the expected total backlog is sublinear:
\begin{equation}
\sum_{t=1}^T \E[b_t] \leq o(T). \label{eq:vanishing+backlog}
\end{equation}

Assuming both \eqref{eq:reward} and \eqref{eq:vanishing+backlog}, we derive a contradiction in four steps. First, we show that most accepted cases must occur when the backlog is empty (Claim~\ref{claim:idle+step1}). Second, building on Step 1, we establish a lower bound on the number of matches that must be served via idle servers (Claim~\ref{claim:idle+step2}). Third, we derive an upper bound on the total number of such matches (Claim~\ref{claim:idle+step3}). Finally, we show that these lower and upper bounds are incompatible unless \( \epsilon \geq 0.5 \), which contradicts our assumption that \( \epsilon < 0.5 \) (ensuring the service rate remains strictly below one). This contradiction implies that any online algorithm satisfying \eqref{eq:reward} must incur a total backlog of \( \Omega(T) \). We elaborate on each step below. The proofs of all auxiliary claims are deferred to the end of this section.

\paragraph{Step 1.} The following claim shows that the majority of matches must occur when the backlog is empty in order to satisfy both assumptions~\eqref{eq:reward} and~\eqref{eq:vanishing+backlog}.

\begin{claim}\label{claim:idle+step1}
Under assumptions \eqref{eq:reward} and \eqref{eq:vanishing+backlog}, we have:
\begin{equation}
\sum_{t=1}^T \E\left[\Indicator[b_{t-1} = 0,\, z_t = 1]\right] \geq 0.5T - o(T).
\end{equation}
\end{claim}
The proof builds on a simple algebraic property of the backlog and the integrality of decisions. 

\paragraph{Step 2.} 
We build on Step~1 to lower bound the number of idle periods that an online algorithm must utilize. Specifically, the following claim shows that most matched cases must be served by an idle server.

\begin{claim}\label{claim:idle+step2}
Under assumptions \eqref{eq:reward} and \eqref{eq:vanishing+backlog}, we have:
\begin{equation}
\sum_{t=1}^T \E\left[\Indicator[b_{t-1} = 0,\, z_t = 1,\, \ell_t = 1]\right] \geq 0.5T - o(T).
\end{equation}
\end{claim}

The proof uses the fact that matches made without idle servers create a backlog unless fresh availability occurs (i.e., \( s_t = 1 \)). Because an online algorithm does not observe the realization of \( s_t \), such matches lead to backlog with constant probability. We use this to argue that if a linear number of matches occur without idle servers, the total backlog would grow linearly in \( T \)—contradicting assumption~\eqref{eq:vanishing+backlog}.

\paragraph{Step 3.} We now establish an upper bound on the number of idle periods that an online algorithm can utilize. The following claim shows that idle periods are `‘too scarce’' to serve many matches.

\begin{claim}\label{claim:idle+step3}
Under assumptions~\eqref{eq:reward} and~\eqref{eq:vanishing+backlog}, for any $\epsilon \in [0, 0.5)$, we have:
\begin{equation}
\sum_{t=1}^T \E\left[\Indicator[b_{t-1} = 0,\, z_t = 1,\, \ell_t = 1]\right] \leq \left(0.5 + \epsilon\right) \cdot 0.5T + o(T).
\end{equation}
\end{claim}

The proof uses a simple charging argument: each idle period must be preceded by a time when the algorithm chose not to match a case, despite having no backlog. By Step~1, any algorithm satisfying assumptions~\eqref{eq:reward} and~\eqref{eq:vanishing+backlog} must match most cases during such zero-backlog periods—leaving few opportunities for servers to remain idle. We use this intuition to upper bound the total number of idle periods that an online algorithm can utilize.

\paragraph{Step 4. Putting everything together.}
We now combine the previous claims to prove \eqref{eq:const-backlog}. 
Combining Claims \ref{claim:idle+step2} and \ref{claim:idle+step3}, we obtain:
\[
0.5T - o(T) \leq 0.5(0.5 + \epsilon)T + o(T).
\]

By dividing by $T$ and letting $T \to \infty$, we have:
\[
0.5 \leq 0.5(0.5 + \epsilon),
\]
which implies $\epsilon \geq 0.5$. However, this contradicts our requirement that $\epsilon < 0.5$. It follows that our initial assumption \eqref{eq:vanishing+backlog} must be false. Hence, every algorithm that accepts at least $0.5T - o(T)$ cases must incur $\Omega(T)$ total backlog, which proves \eqref{eq:const-backlog}. This completes the proof.\hfill\halmos

\medskip 

{\bf Proof of Claim~\ref{claim:idle+step1}}. 
Observe that for any sample path, we have:
\begin{equation}
b_t = (b_{t-1} + z_t - u_t)_+ \geq (2 - u_t) \cdot \Indicator[b_{t-1} \geq 1,\, z_t = 1] \geq \Indicator[b_{t-1} \geq 1,\, z_t = 1], \label{ineq:integrality+ref+1}    
\end{equation}
where final inequality uses that \( u_t = l_t + (1-l_t)\cdot s_t \leq 1 \). 

Taking expectations and summing over time $t \in [T]$:
\begin{align*}
\E\left[\sum_{t=1}^T b_t\right] 
&\geq \E\left[\sum_{t=1}^T \Indicator[b_{t-1} \geq 1,\, z_t = 1]\right] \\
&= \E\left[\sum_{t=1}^T \Indicator[z_t = 1] - \sum_{t=1}^T \Indicator[b_{t-1} = 0,\, z_t = 1]\right] \\
&\geq 0.5T - o(T) - \sum_{t=1}^T \E\left[\Indicator[b_{t-1} = 0,\, z_t = 1]\right],
\end{align*}
where the last inequality uses assumption \eqref{eq:reward}. Now combining this with the assumption \eqref{eq:vanishing+backlog}, we conclude:
\[
\sum_{t=1}^T \E\left[\Indicator[b_{t-1} = 0,\, z_t = 1]\right] \geq 0.5T - o(T).
\]\hfill \halmos

{\bf Proof of Claim~\ref{claim:idle+step2}.} We begin by decomposing the indicator:
\[
\sum_{t=1}^T \E\left[\Indicator[b_{t-1} = 0,\, z_t = 1,\, \ell_t = 1]\right]
= \sum_{t=1}^T \E\left[\Indicator[b_{t-1} = 0,\, z_t = 1]\right]
- \sum_{t=1}^T \E\left[\Indicator[b_{t-1} = 0,\, z_t = 1,\, \ell_t = 0]\right].
\]
By Claim~\ref{claim:idle+step1}, the first term is at least \( 0.5T - o(T) \). Therefore, it suffices to show that:
\[
\sum_{t=1}^T \E\left[\Indicator[b_{t-1} = 0,\, z_t = 1,\, \ell_t = 0]\right] \leq o(T).
\]

To prove this, note that if \( b_{t-1} = 0 \), \( z_t = 1 \), and \( \ell_t = 0 \), then the case is served only if \( s_t = 1 \). Otherwise, it contributes to the backlog. Thus, for every sample path, we have:
\begin{equation}
b_t \geq \Indicator[b_{t-1} = 0,\, z_t = 1,\, \ell_t = 0] \cdot (1 - s_t).   \label{ineq:integrality+ref+2} 
\end{equation}

Taking expectations:
\begin{align*}
\E[b_t] 
&\geq \E\left[\Indicator[b_{t-1} = 0,\, z_t = 1,\, \ell_t = 0] \cdot (1 - s_t)\right] \\
&= \E\left[\Indicator[b_{t-1} = 0,\, z_t = 1,\, \ell_t = 0]\cdot \E[1 - s_t \mid \mathcal{H}_{t-1}]\right] \\
&= (0.5 - \epsilon) \cdot \E\left[\Indicator[b_{t-1} = 0,\, z_t = 1,\, \ell_t = 0]\right].
\end{align*}
In the first inequality, we applied the tower property of expectation: conditioned on the algorithm's history $\mathcal{H}_{t-1}$, the indicator $\Indicator[b_{t-1} = 0,\, z_t = 1,\, \ell_t = 0]$ is deterministic and the random variable $s_t$ is independent of $\mathcal{H}_{t-1}$ and follows the Bernoulli distribution with success probability $0.5 + \epsilon$.

Summing over all \( t \in [T] \) gives:
\[
\sum_{t=1}^T \E[b_t] \geq (0.5 - \epsilon) \cdot \sum_{t=1}^T \E\left[\Indicator[b_{t-1} = 0,\, z_t = 1,\, \ell_t = 0]\right].
\]

Using \eqref{eq:vanishing+backlog} and because $0.5-\epsilon$ is a strictly positive constant, it follows that:
\[
\sum_{t=1}^T \E\left[\Indicator[b_{t-1} = 0,\, z_t = 1,\, \ell_t = 0]\right] \leq  o(T).
\]
This completes the proof.
\hfill\(\halmos\)

{\bf Proof of Claim~\ref{claim:idle+step3}.} Define the indicator:
\[
\texttt{newly\_idle}_t := \Indicator[q_{t} = 0,\, \ell_{t} = 0,\, s_{t} = 1],
\]
which captures the event that the server transitions from non-idle to idle at period $t$. 

By construction, every match served by an idle server (i.e., when $b_{t-1} = 0$, $z_t = 1$, and $\ell_t = 1$) must be preceded by a time when the server entered the idle state. Thus, we can charge each such match to a previous $\texttt{newly\_idle}_t$ event, yielding:
\[
\sum_{t=1}^T \E\left[\Indicator[b_{t-1} = 0,\, z_t = 1,\, \ell_t = 1]\right] \leq \sum_{t=1}^T \E[\texttt{newly\_idle}_t].
\]

Next, observe that a server can become newly idle at time $t$ only if $q_{t}=0$ --- or equivalently $b_{t-1} = 0$ and $z_t = 0$ --- and $s_t = 1$. Therefore:
\[
\texttt{newly\_idle}_t \leq \Indicator[b_{t-1} = 0,\, z_t = 0,\, s_t = 1],
\]
and so:
\[
\E[\texttt{newly\_idle}_t] \leq \E[\Indicator[b_{t-1} = 0,\, z_t = 0]] \cdot (0.5 + \epsilon),
\]
where we used the tower rule and  independence of $s_t$ from the algorithm’s history $\mathcal{H}_{t-1}$.

Now summing over $t$, we obtain:
\begin{align*}
\sum_{t=1}^T \E[\texttt{newly\_idle}_t] 
&\leq (0.5 + \epsilon) \cdot \sum_{t=1}^T \E[\Indicator[b_{t-1} = 0,\, z_t = 0]] \\
&= (0.5 + \epsilon) \cdot \left( \sum_{t=1}^T \E[\Indicator[b_{t-1} = 0]] - \sum_{t=1}^T \E[\Indicator[b_{t-1} = 0,\, z_t = 1]] \right) \\
&\leq (0.5 + \epsilon) \cdot (T - 0.5T + o(T)) \\
& = 0.5(0.5 + \epsilon)T + o(T),
\end{align*}
where the last inequality follows from Claim~\ref{claim:idle+step1} (and by a trivial bound $\E[\Indicator[b_{t-1} = 0]] \leq 1$). This completes the proof.
\hfill\(\halmos\)

\subsection{Regret Upper bounds under Alternative Service Model}\label{subsec:upper+idle}
We now explain how our regret upper bounds for \CA{} (\Cref{thm:ALG+Surrogate+Dual}, \Cref{near+part+a}) and \CO{} (\Cref{thm:ALG+Surrogate+Primal}) seamlessly extend to the alternative service model with server idleness. 
Both of our proposed algorithms naturally adapt to the alternative model. Specifically, \CA{} is identical to \Cref{ALG+Surrogate+D} except that it now uses the backlog updated via equation~\eqref{backlog:dynamics+idle}—that is, based on the effective availability \( u_{t,i} \) defined in equation~\eqref{eq:u_t}, which accounts for server idleness. The following corollary shows that \CA{} achieves the same regret upper bounds under this alternative model.

\begin{corollary}[Regret of \CA{} under Server Idleness]\label{cor:CA+idle}
Let $\eta = \Theta(1/\sqrt{\TotalTime})$ and $\zeta = \Theta(1/\sqrt{\TotalTime})$. Under the stable regime (\Cref{def:eps+regimes}), the regret of \CA{} under the service model with server idleness is $\BigO(\sqrt{\TotalTime} + \frac{\BuildUpCost}{\ServiceSlack})$. Furthermore, under the near-critical regime (\Cref{def:eps+regimes}), the regret of \CA{} is $\BigO(\sqrt{\BuildUpCost\TotalTime})$ by setting $\eta = \Theta(1/\sqrt{\TotalTime})$ and $\zeta = \Theta(\sqrt{\BuildUpCost/\TotalTime})$. 
\end{corollary}

{\bf Proof of \Cref{cor:CA+idle}}. 
The proof mirrors that of \Cref{thm:ALG+Surrogate+Dual} (stable regime) and \Cref{near+part+a} (near-critical regime); we only highlight a minor change. The only difference lies in inequality \eqref{line:third} for the proof of \Cref{lemma:pseudo+lower} (\Cref{apx+pseudo+lower}).  Under the new backlog update rule~\eqref{backlog:dynamics+idle}, we can establish an analogous drift inequality to \Cref{lemma:drift}, that is,  \( D_t \leq \BuildUpVec_{t-1} \cdot (\DecisionVec_t - \mathbf{u}_t) + \mathcal{O}(1) \), where \( D_t \) is the drift of the quadratic Lyapunov function \( \psi \) (see equation~\eqref{psi}). This change requires replacing \( \mathbf{s}_t \) in line~\eqref{line:third} with \( \mathbf{u}_t \), and verifying \( \E[u_{t,i} \mid \mathcal{H}_{t-1}] \geq \rho_i + \epsilon \) for all $i \in [m]$. Since \( u_{t,i} = s_{t,i} + \ell_{t,i}(1 - s_{t,i}) \geq s_{t,i}\) for all sample paths (see equation~\eqref{eq:u_t}), we have \( \E[u_{t,i} \mid \mathcal{H}_{t-1}] \geq \E[s_{t,i} \mid \mathcal{H}_{t-1}] \geq \rho_i + \epsilon \). That is, allowing server idleness can only increase the service rate. 
The rest of the argument is unchanged. Specifically, inequality~\eqref{ineq:important} still holds under the alternative service model, thereby extending \Cref{thm:ALG+Surrogate+Dual} and \Cref{near+part+a}.
\hfill\halmos

For \CO{}, since it does not use backlog information, the algorithm remains unchanged. The regret guarantee of \Cref{thm:ALG+Surrogate+Primal} extends to the alternative service model. The proof again follows analogous arguments to those used in the proof of \Cref{cor:CA+idle} and is omitted for brevity.

}

\end{document}